\documentclass{article}
\pdfoutput=1

\usepackage{microtype}
\usepackage{graphicx}
\usepackage{subcaption}
\usepackage{booktabs} %
\usepackage{mauraisMacros2}
\usepackage{makecell}

\usepackage{hyperref}

\usepackage[accepted]{icml2024}
\makeatletter
  \def\my@tag@font{\normalsize}
  \def\maketag@@@#1{\hbox{\m@th\normalfont\my@tag@font#1}}
  \let\amsmath@eqref\eqref
  \renewcommand\eqref[1]{{\let\my@tag@font\relax\amsmath@eqref{#1}}}
\makeatother

\usepackage{amsmath}
\usepackage{amssymb}
\usepackage{mathtools}
\usepackage{amsthm}
\usepackage{float} 

\usepackage[capitalize,noabbrev]{cleveref}

\theoremstyle{plain}
\newtheorem{theorem}{Theorem}[section]

\theoremstyle{definition}

\theoremstyle{remark}

\usepackage[textsize=tiny]{todonotes}
\definecolor{darkgreen}{rgb}{.15,.55,0}

\newcommand{\rev}[1]{\textcolor{black}{#1}}

\icmltitlerunning{Sampling in Unit Time with Kernel Fisher--Rao Flow}
\graphicspath{{figures/}}

\begin{document}

\twocolumn[
\icmltitle{Sampling in Unit Time with Kernel Fisher--Rao Flow}

\begin{icmlauthorlist}
\icmlauthor{Aimee Maurais}{mit}
\icmlauthor{Youssef Marzouk}{mit}
\end{icmlauthorlist}

\icmlaffiliation{mit}{Center for Computational Science and Engineering, Massachusetts Institute of Technology, Cambridge, MA, USA}

\icmlcorrespondingauthor{Aimee Maurais}{maurais@mit.edu}

\icmlkeywords{sampling, optimal transport, gradient flow, Fisher-Rao geometry, Bayesian inference, interacting particle systems}

\vskip 0.3in
]

\printAffiliationsAndNotice{}  %

\renewcommand{\thefootnote}{\fnsymbol{footnote}}
\begin{abstract}
{
We introduce a new mean-field ODE and corresponding interacting particle systems (IPS) for sampling from an unnormalized target density. The IPS are gradient-free, available in closed form, and only require the ability to sample from a reference density and compute the (unnormalized) target-to-reference density ratio. The mean-field ODE is obtained by solving a Poisson equation for a velocity field that transports samples along the geometric mixture of the two densities, $\pi_0^{1-t} \pi_1^t$, which is the path of a particular Fisher--Rao gradient flow. We employ a 
RKHS ansatz for the velocity field, which makes the Poisson equation tractable and enables discretization of the resulting mean-field ODE over finite samples. The mean-field ODE can be additionally be derived from a discrete-time perspective as the limit of successive linearizations of the Monge--Amp\`ere equations within a framework known as sample-driven optimal transport. We introduce a stochastic variant of our approach and demonstrate empirically that our IPS can produce high-quality samples from varied target distributions, outperforming comparable gradient-free particle systems and competitive with gradient-based alternatives.}
    \end{abstract}

\section{Introduction}
	In this work we consider the problem of \textit{sampling via transport}: given a target distribution $\pi_1$ on $\R^d$ and a reference distribution $\pi_0$ on $\R^d$ from which we can sample, our goal is to find $T: \R^d \to \R^d$ such that $T_\sharp \pi_0 = \pi_1$, i.e., $X_0 \sim \pi_0 \Rightarrow T(X_0) \sim \pi_1$. We assume that $\pi_0$ and $\pi_1$ both admit densities and that we can evaluate the (unnormalized) \textit{density ratio}\footnote{For the remainder of this paper, the terms ``density'' and ``{density} ratio'' refer to unnormalized quantities unless otherwise stated.} $\frac{\pi_1}{\pi_0}$ but that we do not have samples of $\pi_1$ with which to train the map, or access to gradients (including scores) of $\pi_1$ or $\pi_0$. The density ratio is available when the (unnormalized) density of $\pi_1$ is known and $\pi_0$ is chosen to be some ``standard'' reference (e.g., Gaussian). It is also accessible in the Bayesian setting so long as the likelihood function $\ell$ is known:  \rev{therein $\pi_0$ is the prior distribution of some  $\R^d$-valued parameter $X$ and $\pi_1 \propto \ell \, \pi_0 $ is the posterior distribution of $X$ given $Y = y^*$, with $\ell(\cdot) = \pi_{Y|X}(y^* \vert \cdot)$. The ratio in the Bayesian setting is $\frac{\pi_1}{\pi_0} \propto \ell$, but we will use the term ``likelihood'' to refer to the ratio $\frac{\pi_1}{\pi_0}$ %
 outside of the Bayesian setting as well. In applications of Bayesian inference such as data assimilation \citep{reich2015probabilistic} %
 it is frequently the case that $\pi_0$ is only known through samples; hence, while the target density $\pi_1 \propto \pi_0 \ell$ cannot be evaluated, the likelihood $\ell$  often can. Furthermore, in many other scientific applications, $\pi_1$ or $\ell$ may contain a complicated physical model whose gradients are inaccessible; %
    hence gradient-free sampling is a necessity.}  %
	
	The canonical sampling approach employing the likelihood is importance sampling \citep{mcbook}, which transforms an unweighted ensemble of samples of $\pi_0$ into a \textit{weighted} ensemble, enabling the estimation of expectations under $\pi_1$. Importance sampling is the foundation for sequential Monte Carlo (SMC) methods \citep{del2006sequential}, but is frequently plagued by issues of weight degeneracy and ensemble collapse, necessitating large ensemble sizes \citep{snyder2008obstacles} or interventions such as resampling \citep{kunschRecursiveMonteCarlo} and MCMC rejuvenation.%

Alternatively, many sampling approaches use \textit{dynamics} to define a transport incrementally, e.g., via the flow map induced by trajectories of an ODE or the stochastic mapping induced by sample paths of an SDE. In either case, the idea is to apply dynamics which will transform some initial state $X_0 \sim \pi_0$ to a state $X_S \sim \pi_{X_S} \approx \pi_1$ for some time $S > 0$.
This approach underlies flow, diffusion, and bridge techniques for generative modeling, e.g., \citet{kuangSampleBasedOptimalTransport2019, song2020score, debortoliDiffusionSchrodingerBridge2021, liuFlowStraightFast2022, lipmanFlowMatchingGenerative2023,xuOptimalTransportFlow2023, albergoStochasticInterpolantsUnifying2023}, wherein samples from both $\pi_0$ and $\pi_1$ are almost always required for training (with \citet{vargas2023denoising, hengDiffusionSchrDinger2023} being recent exceptions). In the setting where $\pi_1$ is known only through its unnormalized density, there are a number of dynamic sampling algorithms which have their grounding as \textit{gradient flows} of functionals on spaces of probability measures. There are several geometries in which one may define gradient flows on probability measures (see \citet{chenGradientFlowsSampling2023} for a helpful review), but most well-known algorithms in this vein (e.g., \citet{liuSteinVariationalGradient2016, eks, aldi, reichFokkerPlanckParticle2021}) use some form of the Wasserstein geometry to define dynamics which must, in principle, be run for \textit{infinite time} in order to ensure correct sampling from $\pi_1$.  

 In this work we develop a dynamic sampling approach based on an ODE which transports samples from $\pi_0$ to $\pi_1$ in \textit{unit time} such that the time-dependent distribution of the samples is the geometric mixture $\pi_t \propto \pi_0^{1-t}\pi_1^t = \pi_0(\frac{\pi_1}{\pi_0})^t$, $t \in [0,1]$. Although our algorithms are gradient-free and only require the likelihood $\frac{\pi_1}{\pi_0}$, the path of distributions $\pi_t$ corresponds to the {Fisher--Rao gradient flow} of the expected negative log likelihood. 
The underlying dynamics are described by a \textbf{mean-field ODE model}, which we show is the limit of two different \textbf{interacting particle systems}. These interacting particle systems, which we generally refer to as \emph{Kernel Fisher--Rao Flow}, are obtained in two distinct but related ways. %
On one hand, in continuous time, the mean-field ODE can be obtained from the weak formulation of a Poisson equation for a \emph{velocity field} defined in a reproducing kernel Hilbert space (RKHS), from which we obtain a finite-particle ODE system by approximating expectations via Monte Carlo (\cref{sec:methodology_poisson}). On the other, in discrete time, one can approximate the optimal transport \textit{map} which pushes $\pi_t$ to $\pi_{t+\Delta t}$ via linearization of the Monge--Amp\`ere equations discretized over finitely many kernel basis functions and finitely many samples (\cref{sec:sot}). This linearization yields an interacting particle system which in discrete time is distinct from, but in continuous time identical to, that obtained via the RKHS approach to Poisson's equation. 
	\section{Background}
	\label{sec:bg}
	Sampling via measure transport is an active area of research, with many computational approaches \citep{marzouk2016sampling, kobyzev2020normalizing,papamakarios2021normalizing,trillos2023optimization} %
	appearing in recent years. Most practical transport maps are parameterized, and thus a crucial part of realizing them is selecting an appropriately rich function class within which to search for the map. Common map approximation classes include polynomials \citep{jaini2019sum},
 radial basis functions \citep{spantiniCouplingTechniquesNonlinear2022}, composed simple transformations \citep{rezende2015variational, kobyzev2020normalizing, papamakarios2021normalizing}, neural networks \citep{bunne2022supervised, taghvaeiOptimalTransportFormulation2022, baptistaConditionalSamplingMonotone2023}, and reproducing kernel Hilbert spaces \citep{liuSteinVariationalGradient2016,kuangSampleBasedOptimalTransport2019, katzfussScalableBayesianTransport2023}. Determining an appropriate basis to represent a transport map can be challenging, especially when the target and reference distributions are high-dimensional or differ from each other considerably. For this reason it may be necessary to employ, e.g., adaptive feature selection algorithms \citep{baptista2020representation} or dimension reduction techniques \citep{spantini2018inference, chen2019projected, brennan2020greedy, daiSlicedIterativeNormalizing2021}.
	
	As an alternative to searching for a single, potentially highly complex transport map which pushes the reference $\pi_0$ directly to the target $\pi_1$, one can instead prescribe a \textit{path} of distributions $(\pi_t)_{t \in [0,1]}$ having the target and reference as endpoints and seek a sequence of maps $T_1, \dots, T_N$ which push samples along a discretization of the path, as depicted in \cref{fig:incrementalMaps}.
	The composed map $T = T_N \circ T_{N-1} \circ \cdots \circ T_1$ pushes forward $\pi_0$ to $\pi_1$. In continuous time this approach becomes one of finding a \textit{velocity field} $v_t: \R^d \to \R^d$ such that the solution to the {initial value problem}
 \[
 \dot X_t = v_t(X_t), \quad X_0 \sim \pi_0
 \]
 has distribution $\pi_t$. 
	Samplers of unnormalized densities which employ this homotopy approach frequently take $\pi_t$ to be the geometric mixture 
	\begin{equation} 
	\pi_t \propto \pi_0^{1-t}\pi_1^t = \pi_0 \left( \ltfrac{\pi_1}{\pi_0} \right)^t, \quad t \in [0,1],
 \label{eq:temperedLikelihood}
	\end{equation}  
	which interpolates between $\pi_0$ and $\pi_1$ in unit time. This mixture may be referred to as the ``power posterior'' path and appears, for example, in annealed importance sampling \citep{neal2001annealed,brekelmansAnnealedImportanceSampling2020, korba2022adaptive, goshtasbpour2023adaptive} and parallel tempering \citep{geyer1991markov, earl2005parallel, syed2021parallel}. In Bayesian computation, this path is sometimes referred to as ``tempered likelihood'' and has been used as the basis for algorithms which generate (approximate) posterior samples \citep{reichDynamicalSystemsFramework2011,daum2013particle,iglesiasEnsembleKalmanMethods2013, dingEnsembleKalmanInversion2021} or posterior densities \citep{diaContinuationMethodBayesian2023}.  %

 \newcommand{\graphicsheight}{1.0cm}
\newcommand{\tallergraphicsheight}{1.1cm}
 
 	\begin{figure}[H]
		\begin{tikzpicture}[
			statenode/.style={rectangle, draw=black!0, fill=blue!0, very thick, minimum size=7mm, font=\small},
			node distance=0.1cm and 0.3cm
			]
			\node[statenode]      (state0)                              {\includegraphics[height=\graphicsheight]{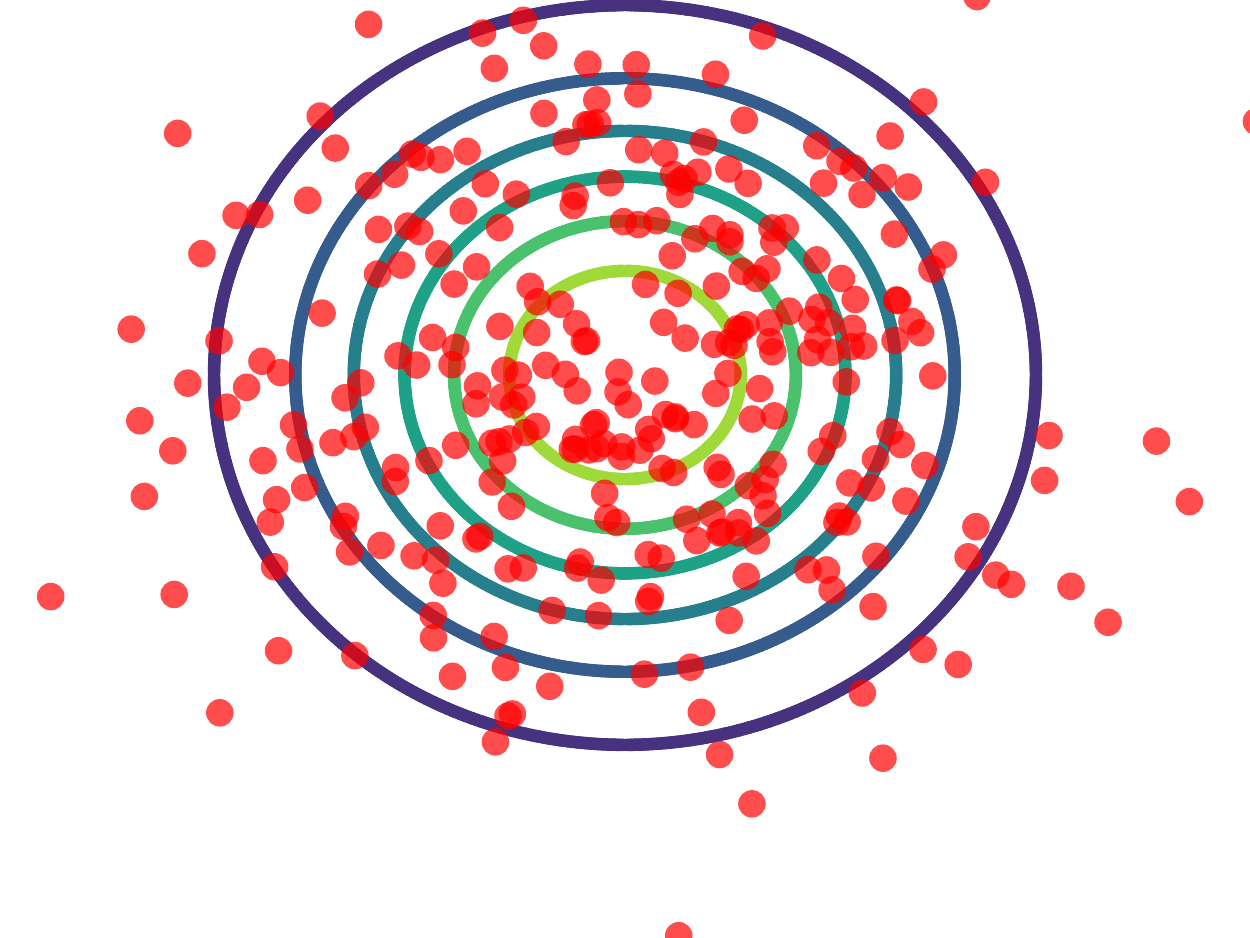}};
			\node[statenode] (desc0) [below=of state0] {\normalsize $\pi_0$};
			\node[statenode] (samps0) [above=of state0] { $\{X_0^{(j)}\}_{j=1}^J$};
			\node[statenode]      (state1)     [right=of state0]{\includegraphics[height=\graphicsheight]{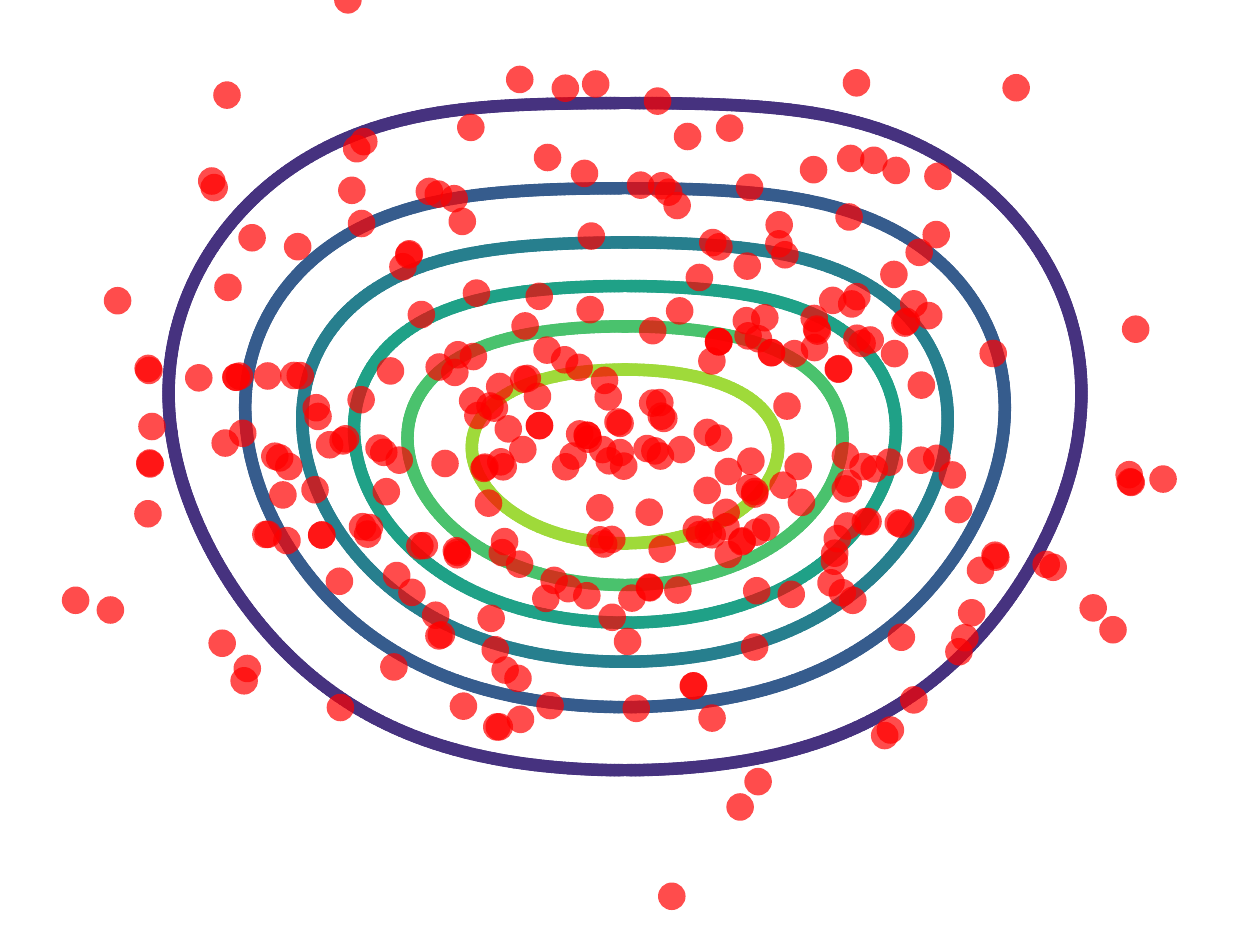}};
			\node[statenode] (desc1) [below=of state1] {\normalsize $\pi_0(\frac{\pi_1}{\pi_0})^{t_1}$};
			\node[statenode] (samps1) [above=of state1] { $\{X_{t_1}^{(j)}\}_{j=1}^J$};
			\node[statenode]      (state2)     [right=of state1, xshift=1.0cm]{\includegraphics[height=\tallergraphicsheight]{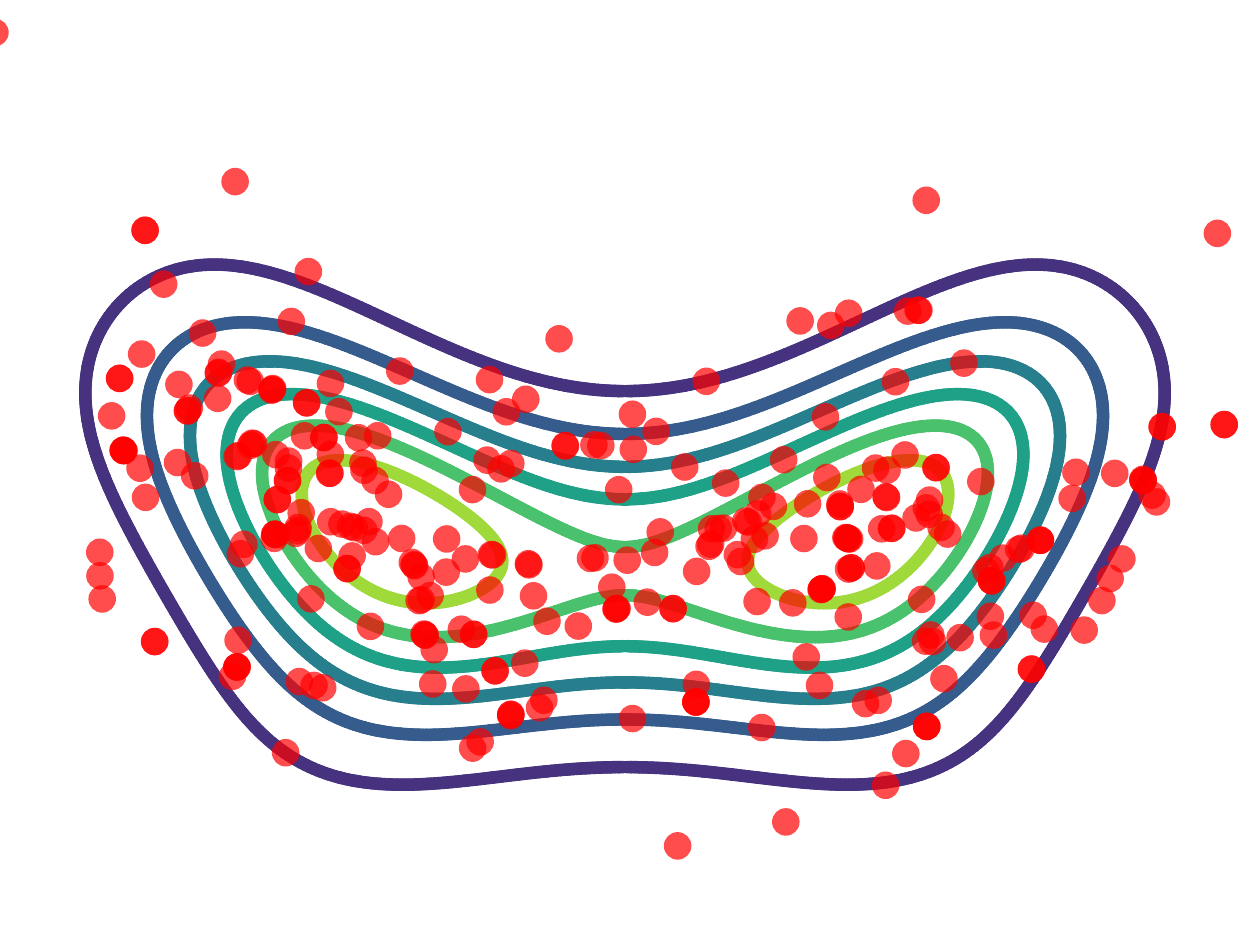}};
			\node[statenode] (desc2) [below=of state2] {\normalsize  $\pi_0(\frac{\pi_1}{\pi_0})^{t_{N-1}}$};
			\node[statenode] (samps2) [above=of state2, yshift=-0.07cm] { $\{X_{t_{N-1}}^{(j)}\}_{j=1}^J$};
			\node[statenode]      (state3)     [right=of state2]{\includegraphics[height=\tallergraphicsheight]{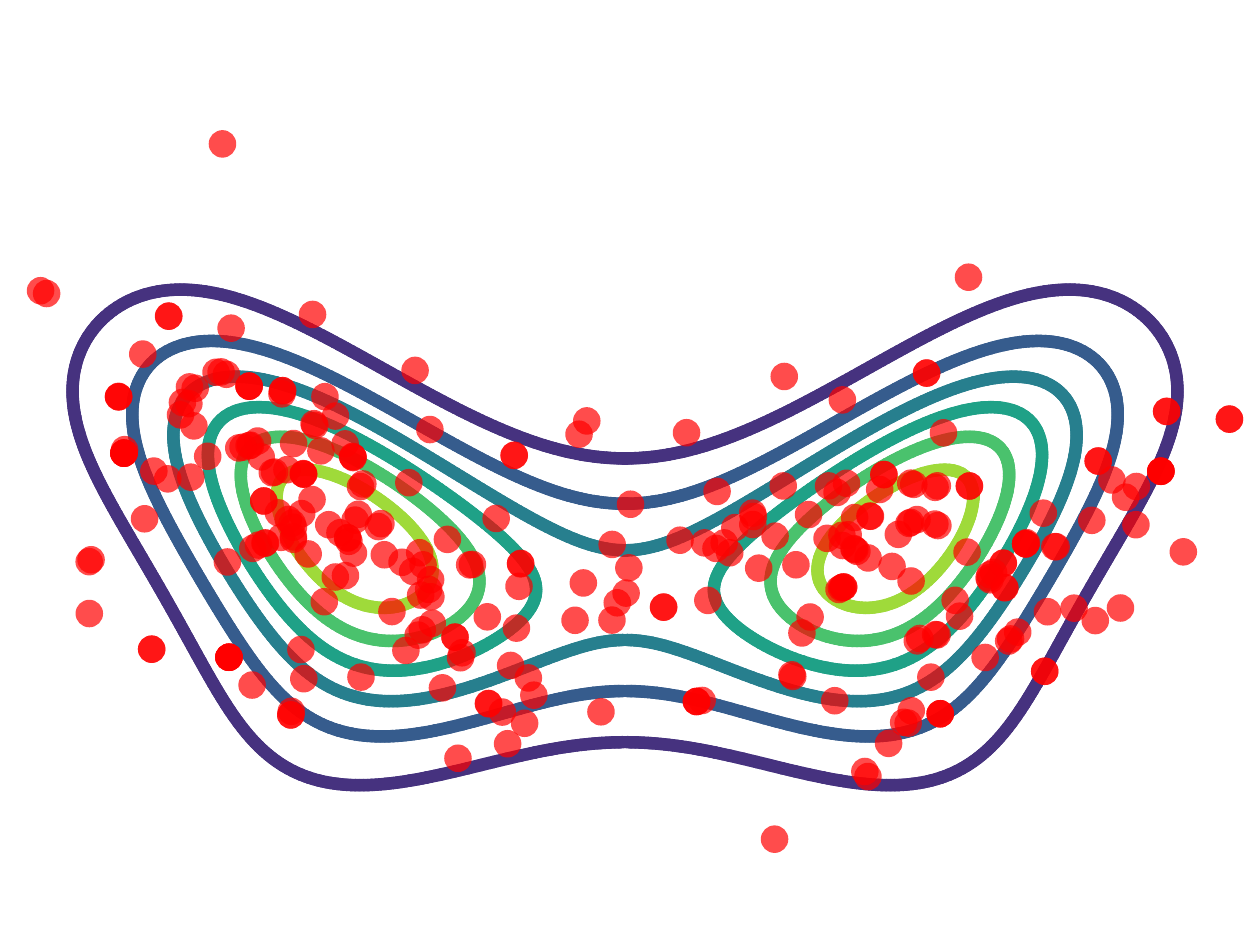}}; 
			\node[statenode] (desc3) [below=of state3] {\normalsize  $\pi_{1} $};
			\node[statenode] (samps3) [above=of state3] { $\{X_{1}^{(j)}\}_{j=1}^J$};
			
			\draw[very thick, ->] (state0.east) -- node[above, yshift=0.1cm]{\small $T_1$} (state1.west);
			\draw[very thick, dashed, ->] (state1.east) -- node[above, yshift=0.1cm]{\small $T_2, \dots, T_{N-1}$} (state2.west);
			\draw[very thick, ->] (state2.east) -- node[above, yshift=0.1cm]{\small $T_{N}$} (state3.west);
		\end{tikzpicture}
		\caption{We employ a homotopy-based sampling scheme in this work, deriving a mean-field ODE which approximately transports a reference $\pi_0$ to a target $\pi_1$ in unit time. In discrete time this approach amounts to obtaining incremental maps $T_1, \dots, T_N$.} 
		\label{fig:incrementalMaps} 
	\end{figure}
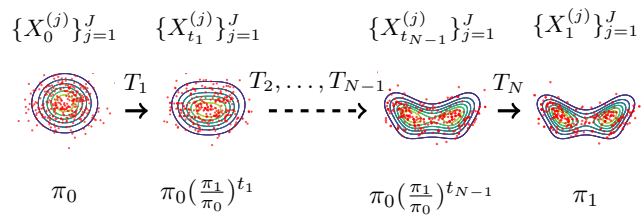 

 \subsection{Related Work}
 The idea of using dynamics with the tempered likelihood to sample the posterior distribution in Bayesian inference was, to our knowledge, first developed by \citet{daum2011particle} and \citet{reichDynamicalSystemsFramework2011} in the context of filtering. 
 In \citet{daum2011particle,daum2013particle} and related works by the same authors, the tempered likelihood is used to derive ODEs and SDEs for nonlinear {filtering} in full generality, and as such these systems require gradients and even Hessians of the likelihood and prior. The algorithms for propagating samples from prior to posterior along the tempered likelihood in  \citet{reichDynamicalSystemsFramework2011} are of ensemble-Kalman type and %
 employ a Gaussian approximation for each $\pi_t$, meaning that their expressivity is limited. A similar methodology to \citet{reichDynamicalSystemsFramework2011}, now known as Ensemble Kalman Inversion (EKI), was proposed for computing point estimates in Bayesian inverse problems in \citet{iglesiasEnsembleKalmanMethods2013}. The iteration underlying EKI is run for infinite time, but, as noted by, e.g., \citet{dingEnsembleKalmanInversion2021}, when the iteration is stopped at $t = 1$ samples from an approximation to the posterior are obtained. Even in the limit of infinite particles and continuous time, however, ensemble Kalman methods are not consistent samplers for general (i.e., non-Gaussian) posteriors.  

The tempered likelihood path is frequently employed in sequential Monte Carlo (SMC) methods, which rely on a series of interwoven importance re-sampling and mutation steps to gradually transform samples from $\pi_0$ into approximate samples from $\pi_1$. Several recent works have sought to use transport within SMC to either replace or reduce the frequency of multinomial resampling. The basic idea, which was introduced in \citet{reichNonparametricEnsembleTransform2013} and motivates our development in \cref{sec:sot}, is to use importance weights to obtain a transport between $\pi_t$ and $\pi_{t+\Delta t}$. %
In \citet{ruchiTransformbasedParticleFiltering2019,ruchiFastHybridTempered2021,myersSequentialEnsembleTransform2021}, 
discrete optimal transport couplings are used to define linear transformations which replace the importance resampling steps typically employed in SMC. 
In a similar vein, the annealed flow transport Monte Carlo of \citet{arbel2021annealed} uses transport as a preconditioner for SMC; on each SMC iteration a parametric transport map between $\pi_t$ and $\pi_{t+\Delta t}$ is learned and applied to samples from $\pi_t$ before the standard resampling and mutation steps are performed. In each of these works the transport steps remain embedded within SMC schemes, while our algorithms are based on dynamics and do not require resampling or application of mutation kernels. %

Concurrently with our work, \citet{wangMeasureTransportKernel2024} %
developed similar mean-field ODE systems and algorithms for transporting samples along the tempered likelihood. Their algorithms are motivated from a kernel mean embedding perspective, rather than one of optimal transport and Fisher--Rao gradient flows, and can be made to match ours by specifying particular choices of the parameters $v_t^0$ and $C_t$ in their setup. In our work we introduce both deterministic and stochastic interacting particle systems for sampling the tempered likelihood, but the algorithms considered in \citet{wangMeasureTransportKernel2024} are purely deterministic. 
\paragraph{Notation} We use $K(\cdot, \cdot): \R^d \times \R^d \to \R$ to represent a symmetric positive definite kernel on $\R^d$ and denote by $(\calH_K, \langle \cdot, \cdot \rangle_{\calH_K})$ the reproducing kernel Hilbert space \citep{rkhs} associated with $K$. We assume that $K(\cdot, x)$ is $C^2$ and use $\nabla_1 K(\cdot, \cdot)$ to refer to the gradient of $K$ with respect to the first argument. $\calP_{\rm ac}(\R^d)$ denotes the space of probability measures on $\R^d$ %
which admit densities. 
	\section{Methodology: Poisson Equation in Reproducing Kernel Hilbert Space}
	\label{sec:methodology_poisson}
	Our goal is to find a time-varying velocity field $v_t: \R^d \to \R^d$ such that the distribution of $X_t$ evolving according to 
	\begin{equation}
		\dot X_t = v_t(X_t), \quad X_0 \sim \pi_0
		\label{eq:dynamics}
	\end{equation} 
	is the geometric mixture \eqref{eq:temperedLikelihood}. Had we access to such a velocity field, we could obtain samples from $\pi_1$ by sampling $\pi_0$ and simulating the dynamics \eqref{eq:dynamics} for unit time. It can be shown that $\pi_t$ {in \eqref{eq:temperedLikelihood}} satisfies 
	\begin{equation*} 
		\partial_t \pi_t = \pi_t\left(\log\ltfrac{\pi_1}{\pi_0} - \E_{\pi_t}\left[\log\ltfrac{\pi_1}{\pi_0}\right]\right),
	\end{equation*}
	which is the Fisher--Rao gradient flow of the functional $\calF: \calP_{\rm ac}(\R^d) \to \R$ defined as %
	\[
	\calF(\mu) = -\E_{\mu}\left[\log\ltfrac{\pi_1}{\pi_0} \right].
	\]
	$\calF(\mu)$ is the expected negative log likelihood under $\mu$. By the continuity equation, a velocity field in \eqref{eq:dynamics} yielding $X_t \sim \pi_t$ must then satisfy 
	\begin{equation} 
		-\nabla \cdot(\pi_t v_t) = \pi_t\left(\log\ltfrac{\pi_1}{\pi_0} - \E_{\pi_t}\left[\log\ltfrac{\pi_1}{\pi_0}\right]\right).
		\label{eq:vt_must_satisfy}
	\end{equation}
	There are many possible solutions to the PDE \eqref{eq:vt_must_satisfy}, but if, as in \citet{taghvaeiSurveyFeedbackParticle2023, reichDynamicalSystemsFramework2011}, we insist that in the limit $\Delta t \to 0$ the expected transportation cost $\frac{1}{\Delta t^2}\E_{\pi_t}[\|X_t - X_{t+\Delta t} \|^2]$ is minimized for each $t$, we obtain a constrained optimization problem for each $v_t$ with a unique solution,%
	\begin{gather*}
	\min_{v_t: \R^d \to \R^d} \int_{\R^d} \| v_t \|^2\, \rmd\pi_t \\[0.1cm] 
 \text{ s.t. } -\nabla \cdot(\pi_t v_t) = \pi_t\left(\log\ltfrac{\pi_1}{\pi_0} - \E_{\pi_t}\left[\log\ltfrac{\pi_1}{\pi_0}\right]\right). 
	\end{gather*}
	It can be shown using \rev{the geometry of optimal transport (e.g., \citet[Theorem 1.3.19]{logConcaveSampling}) or} calculus of variations \citep{reichDynamicalSystemsFramework2011} that the solution to this problem is $v_t = \nabla u_t$, where $u_t$ satisfies the Poisson equation 
	\begin{equation}
			-\nabla \cdot(\pi_t \nabla u_t) = \pi_t \left(\log\ltfrac{\pi_1}{\pi_0} - \E_{\pi_t}\left[\log\ltfrac{\pi_1}{\pi_0}\right]\right).
		\label{eq:poisson_u}
	\end{equation}
	We make \eqref{eq:poisson_u} tractable by {searching for $u_t$ in the RKHS $\calH_K$}, i.e., taking $u_t(\cdot) = \int_{\R^d} K(\cdot, x)f_t(x) \,  \rmd \pi_t(x)$ for some $f_t: \R^d \to \R$, and enforcing the weak form of \eqref{eq:poisson_u}, as in \citet{laugesen2015poisson}, for kernel test functions $K(\cdot, x)$, 
	\begin{multline} 
		\int_{\R^d} \langle \nabla_1 K(y, x), \; \nabla u_t(y) \rangle \, \rmd \pi_t(y) = \\ \int_{\R^d} K(y, x)\left (\log \ltfrac{\pi_1}{\pi_0}(y) - \E_{\pi_t}\left[\log \ltfrac{\pi_1}{\pi_0} \right] \right)\, \rmd\pi_t(y).
		\label{eq:test_kernel}
	\end{multline}
	We require \eqref{eq:test_kernel} to hold for all $x \in \R^d$. Substituting the form of $u_t$ into \eqref{eq:test_kernel}, we have
	\begin{multline} 
  \iint_{\R^d\times \R^d}  f_t(z) \left\langle \nabla_1 K(y, \cdot), \;  \nabla_1 K(y, z) \right\rangle \, \rmd \pi_t(y)\, \rmd \pi_t(z)  \\ 
		= \int_{\R^d} K(\cdot, y)\left (\log \ltfrac{\pi_1}{\pi_0}(y) - \E_{\pi_t}\left[\log \ltfrac{\pi_1}{\pi_0} \right] \right)\, \rmd\pi_t(y). %
  \label{eq:Mpit_Kpit}
	\end{multline}
    We write the relationship \eqref{eq:Mpit_Kpit} succinctly as $M_{\pi_t}f_t(x) =  K_{\pi_t} (\log \frac{\pi_1}{\pi_0} - \E_{\pi_t}[\log \frac{\pi_1}{\pi_0} ] )(x)$,
	where the integral operator $M_{\pi_t}$ maps functions $g: \R^d \to \R$ to $M_{\pi_t}g(\cdot) = $ 
	\begin{multline*}
			\int_{\R^d} g(z) \E_{\pi_t} [\langle\nabla_1 K(X_{t}, \cdot), \, \nabla_1 K(X_{t}, z)\rangle ] \, \rmd \pi_t(z) \\
			= \iint_{\R^d \times \R^d} g(z)  \langle\nabla_1 K(y, \cdot), \, \nabla_1 K(y, z) \rangle \, \rmd\pi_t(y) \rmd\pi_t(z)
	\end{multline*} 
	and the kernel integral operator $K_{\pi_t}$ maps $g$ to $K_{\pi_t} g(\cdot) = \int_{\R^d} g(z) K(\cdot, z) \, \rmd\pi_t(z)$.
	Under the condition that $M_{\pi_t}$ is invertible, $f_t$ is given by 
 \[
 f_t = M_{\pi_t}\inv K_{\pi_t}\left (\log \tfrac{\pi_1}{\pi_0} - \E_{\pi_t}\left[\log \tfrac{\pi_1}{\pi_0} \right] \right)
 \]
	and we have $v_t(\cdot) = \nabla u_t(\cdot) = $  
	\begin{equation} 
        \int_{\R^d}  \!\! \nabla_1 K(\cdot, x) M_{\pi_t}\inv K_{\pi_t} \!\left(\log \tfrac{\pi_1}{\pi_0} - \E_{\pi_t}\!\left[\log \tfrac{\pi_1}{\pi_0} \right] \right)\!(x)\, \rmd\pi_t(x). 
		\label{eq:vt}
	\end{equation}
	Therefore, starting from $X_0 \sim \pi_0$, the \textbf{mean-field ODE} %
\begin{small} 
 \begin{multline}
		\dot X_t  =  v_t(X_t) = \\[0.1cm] 
  \E_{\rho_t}\! \left [ \nabla_1 K(X_{t}, X') M_{\rho_t}\inv K_{\rho_t}\left( \log\tfrac{\pi_1}{\pi_0} - \E_{\rho_t}\!\left[\log \tfrac{\pi_1}{\pi_0}\right]\right)(X') \right ],  %
		\label{eq:meanfield} 
	\end{multline}
 \end{small} 
 $\!\!$can be used to evolve samples from $\pi_0$ such that $\rho_t = \mathrm{Law}(X_t)$ is approximately $\pi_t \propto \pi_0^{1-t}\pi_1^t$, and hence at $t = 1$ they are approximately distributed as $\pi_1$. 
 
 We note that the potential $u_t$ obtained by solving a weak-form Poisson equation over the RKHS,
 {\small
 \[
 u_t = \int_{\R^d} K(\cdot, x) M_{\pi_t}\inv K_{\pi_t} \left (\log \tfrac{\pi_1}{\pi_0} - \E_{\pi_t} \left[\log \tfrac{\pi_1}{\pi_0} \right] \right)(x)\, \rmd\pi_t(x)
 \]}
 \rev{$\!\!$is only an approximation of the solution of \eqref{eq:poisson_u}, since we are enforcing the weak form \eqref{eq:test_kernel} for a specific (limited) class of test functions.}
 For this reason we distinguish between the target path of distributions $\pi_t \propto \pi_0^{1-t}\pi_1^t$ and $\rho_t = \mathrm{Law}(X_t)$. Understanding the difference between these paths and its dependence on the choice of RKHS $\calH_K$ is an important area for future work. Empirically we find the quality of samples generated by running a discretization of \eqref{eq:meanfield} to be good; see \cref{sec:numerics}.

 Because the quantities appearing in $v_t$ can be written as expectations with respect to $\rho_t$, the mean-field model \eqref{eq:meanfield} can be approximated for finite samples as an \textbf{interacting particle system} (IPS),
		\begin{small} 
			\begin{multline} 
				\dot X_t^{(j)} = \left( \nabla_1 K(X_{t}^{(j)}, X_{t}^{(1)}) \; \cdots \; \nabla_1 K(X_{t}^{(j)}, X_{t}^{(J)}) \right) M_t\inv \cdot \\ 
				\ltfrac{1}{J} \! \sum_{k=1}^J\! \left(\log\tfrac{\pi_1}{\pi_0}(X_t^{(k)}) - \ltfrac{1}{J} \!\sum_{i=1}^J \log\tfrac{\pi_1}{\pi_0} (X_t^{(i)}) \right)\!\!\! \begin{pmatrix} K(X_{t}^{(k)}\!\!,\, X_{t}^{(1)})\\ \vdots \\ K(X_{t}^{(k)}\!\!,\, X_{t}^{(J)}) \end{pmatrix}\!,\\[0.2cm]
				\label{eq:IPS_ode}
			\end{multline} 
		\end{small} 
		with $j \in \{1,\dots,J\}$, $t \in [0,1]$, $\{X_0^{(j)}\}_{j=1}^J \overset{\rm i.i.d.}{\sim} \pi_0$, and $M_t \in \R^{J \times J}$ given by 
		\begin{multline*}
		(M_t)_{\ell, m} = \fdfrac{1}{J}\sum_{i=1}^J  \langle\nabla_1 K(X_{t}^{(i)}\!\!,\, X_{t}^{(\ell)}), \, \nabla_1 K(X_{t}^{(i)}\!\!,\, X_{t}^{(m)}) \rangle, \\ \ell, m=1,\dots,J.
		\end{multline*} 
\rev{This IPS \eqref{eq:IPS_ode} is obtained by approximating the expectations in \eqref{eq:meanfield} with Monte Carlo; see \cref{app:proofMeanField} for a detailed derivation.} 
We refer to \eqref{eq:IPS_ode} as \textbf{Kernel Fisher--Rao Flow (KFRFlow)} and offer a few observations:
	\begin{itemize}
		\item We only require the ability to sample $\pi_0$ and compute the log-density ratio $\log \frac{\pi_1}{\pi_0}$ in order to simulate the ODE \eqref{eq:IPS_ode}. In particular contrast to \rev{Stein variational gradient descent (SVGD), unadjusted or Metropolis-adjusted Langevin samplers (ULA and MALA), and some recent Langevin-based %
  Bayesian inference approaches \citep{aldi, reichFokkerPlanckParticle2021}, we do {not} require gradients or scores of $\pi_0$ or $\pi_1$. Furthermore, quantities of the form $\log\frac{\pi_1}{\pi_0} - \E_{\rho}[\log \frac{\pi_1}{\pi_0}]$ appearing in \cref{eq:meanfield,eq:IPS_ode} are invariant under scaling of $\frac{\pi_1}{\pi_0}$ by constant factors. Thus, we do {not} require knowledge of the normalizing constants of $\pi_1$ or $\pi_0$ to apply KFRFlow. } %
		
		\item KFRFlow is a simple, closed-form ODE and does not require use of numerical optimization to estimate a score or velocity field. This feature stands in contrast to many comparable finite-time dynamic approaches for sampling from unnormalized densities,
  \rev{including normalizing flows \citep{rezende2015variational}, diffusion  or score-based models \citep{hengDiffusionSchrDinger2023,vargas2023denoising}, and recent approaches which also employ the tempered likelihood %
\citep{vargasTransportMeetsVariational2023,tianLiouvilleFlowImportance2024}.}  
		
		\item Similarly to SVGD \citep{liuSteinVariationalGradient2016}, which can be viewed as following a Wasserstein gradient flow with a kernelized ODE, KFRFlow is deterministic and can be viewed as following a Fisher--Rao gradient flow with a kernelized ODE. %

  \item $M_t$ is the finite-particle analogue of $M_{\pi_t}$. A necessary condition for invertibility of $M_t$, $J \leq dJ$, is satisfied by construction in the IPS \eqref{eq:IPS_ode}. This fact can be seen by noting that $M_t = \frac{1}{J}\sum_{i=1}^J \nabla \bfK_t(X_{t}^{(i)} ) \nabla \bfK_t(X_{t}^{(i)})\t$, where $\bfK_t: \R^d \to \R^J$ is the concatenation $\mathbf{K}_t(\cdot) = (K(\cdot, X_t^{(1)}), \dots, K(\cdot, X_t^{(J)}))\t$ and $\nabla \bfK_t\in \R^{J \times d}$ is the Jacobian of $\bfK_t$. 
	\end{itemize}
	
	Although  it is intriguing to view the mean-field ODE model \eqref{eq:meanfield} as resulting from kernelization of a Fisher--Rao gradient flow, we can recover it separately as the limit of a \textit{discrete-time} interacting particle system obtained using sample-driven optimal transport \citep{kuangSampleBasedOptimalTransport2019}. %
	We discuss this perspective in the following section.
	
	\section{Discrete-Time Interpretation: Sample-Driven Optimal Transport} 
	\label{sec:sot}
	In discrete time, the problem of finding a velocity field $v_t$ such that the flow $\dot X_t = v_t(X_t)$ has distribution $\pi_t \propto \pi_0^{1-t}\pi_1^t$ becomes one of finding transport maps $T_1, \dots, T_N$ which push samples from $\pi_0$ along a discretization of $\pi_t$. While we we can obtain such maps by discretizing the IPS \eqref{eq:IPS_ode}, for example taking $X_{t+\Delta t} = X_t + \Delta t\cdot v_t(X_t)$, we can alternately search for the maps \textit{directly} via a framework introduced as sample-driven optimal transport in \citet{trigilaDataDrivenOptimalTransport2016, kuangSampleBasedOptimalTransport2019}, modified for our setting in which target samples are unavailable. 
	
	Suppose that at time $t \in [0,1)$ we have samples $\{X_t^{(j)}\}_{j=1}^J \sim \pi_t$ which we would like to push forward to $\pi_{t + \Delta t} \propto \pi_t (\frac{\pi_1}{\pi_0})^{\Delta t}$. Given that $\pi_t$ and $\pi_{t+\Delta t}$ both admit densities, there are many maps $T: \R^d \to \R^d$ satisfying $T_\sharp \pi_t = \pi_{t+\Delta t}$. The optimal transport approach \citep{villaniTopicsOptimalTransportation2021}, which we will approximate, is to seek the map which {minimizes expected transport cost}, 
	\begin{equation}
		\min_{T_\sharp \pi_t = \pi_{t+\Delta t}} \E_{\pi_t}[\|T(X_t)  - X_t\|^2].
		\label{eq:ot_w2}
	\end{equation}
	Owing to the choice of quadratic cost, it can be shown that the optimal map in \eqref{eq:ot_w2} is the unique convex gradient which pushes forward $\pi_t$ to $\pi_{t+\Delta t}$ \citep{brenier1991polar}. That is, if we find $T = \nabla \phi$ satisfying $T_\sharp \pi_t = \pi_{t+\Delta t}$ with $\phi: \R^d \to \R$ convex, we have found the optimal transport map. Thus, we can search for the optimal transport map by seeking $\phi: \R^d \to \R$ convex such that $\nabla\phi_\sharp \pi_t = \pi_{t+\Delta t}$. The push-forward condition $\nabla\phi_\sharp \pi_t = \pi_{t+\Delta t}$ can be written as a Monge--Amp\`ere PDE \citep{evans1997partial}
	\[
	\pi_{t+\Delta t}(\nabla\phi(x))\det\left(H_\phi (x)\right)  = \pi_t(x),  
	\]
	where $H_\phi$ is the Hessian of $\phi$, and interpreted in weak form as for all $f: \R^d \to \R$ continuous
	\begin{equation}
		\int_{\R^d} f(\nabla \phi(x)) \, \rmd\pi_t(x) = \int_{\R^d} f(y) \, \rmd\pi_{t + \Delta t}(y).
		\label{eq:weakform}
	\end{equation}  
	Given that we only have finitely many samples of $\pi_t$ and access to the ratio $\frac{\pi_1}{\pi_0}$, we arguably do not have enough information to find a map $T = \nabla \phi$ which exactly satisfies $T_\sharp \pi_t = \pi_{t+\Delta t}$. Thus we apply a Galerkin approximation to \eqref{eq:weakform} over a basis of kernel test functions located at each particle, \{$K(\cdot, X_t^{(j)}):   j=1,\dots,J$\}, seeking a map
	\begin{equation}
		\nabla \phi_\bfs(x) = x + \sum_{j=1}^J s_j \nabla_1 K(x, X_t^{(j)}),
		\label{eq:gradPhi}
	\end{equation}  
	and discretizing the weak form \eqref{eq:weakform} with 
	\begin{equation}
		\int_{\R^d} K(\nabla_\bfs \phi(x), X_t^{(j)}) \, \rmd\pi_t(x) = \int_{\R^d} K(y, X_t^{(j)}) \, \rmd\pi_{t + \Delta t}(y),
		\label{eq:weakform_galerkin}
	\end{equation} 
	$j=1,\dots,J$.
	Approximating \eqref{eq:weakform_galerkin} via Monte Carlo, we seek map coefficients $\bfs = (s_1, \dots, s_m)$ such that
	\begin{equation} 
		\sum_{j=1}^J \ltfrac{1}{J} \mathbf{K}_t(\nabla\phi_\bfs(X_t^{(j)})) = \sum_{j=1}^J w_t^{(j)} \mathbf{K}_t(X_t^{(j)}),  
		\label{eq:sot}
	\end{equation} 
	where the $w_t^{(j)}$ are self-normalized importance weights,
 \[
 w_t^{(j)} =  \frac{(\frac{\pi_1}{\pi_0}(X_{t}^{(j)}))^{\Delta t}}{\sum_{i=1}^J (\frac{\pi_1}{\pi_0}(X_{t}^{(i)}))^{\Delta t}}, \quad j = 1,\dots, J
 \]
 and $\mathbf{K}_t(\cdot) = (K(\cdot, X_t^{(1)}), \dots, K(\cdot, X_t^{(J)}))\t$ is as before. \citet{kuangSampleBasedOptimalTransport2019} refer to the relationship \eqref{eq:sot} as \textit{sample equivalence} and denote it by $\{\nabla \phi_\bfs(X_t^{(j)})\}_{j=1}^J \sim \{ w_t^{(j)} X_t^{(j)}\}_{j=1}^J$. Because we have discretized the Monge--Amp\`ere equations \eqref{eq:weakform} over finite samples and feature functions, a solution $\bfs$ to \eqref{eq:sot} is not guaranteed to yield a unique or optimal map. The \textit{sample-driven} OT problem then, as formulated in \citet{kuangSampleBasedOptimalTransport2019}, is to find a minimum cost map $\nabla \phi_\bfs$ which satisfies sample-equivalence,
	\begin{equation}
		\min_{\{\nabla \phi_\bfs(X_t^{(j)})\}_{j=1}^J \sim \{ w_t^{(j)} X_t^{(j)}\}_{j=1}^J} \sum_{j=1}^J \left\|X_t^{(j)} - \nabla \phi_\bfs(X_t^{(j)}) \right\|^2.
		\label{eq:sampleOT}
	\end{equation}
	
	Returning to \eqref{eq:sot}, admissible choices of $\bfs$ in \eqref{eq:sampleOT} can be identified via root-finding: denote by $\bfa$ and $\bfb$ the means of $\bfK_t$ over the unweighted and weighted reference ensembles,
	\[
	\bfa = \frac{1}{J} \sum_{j=1}^J \bfK_t(X_t^{(j)}), \quad \bfb = \sum_{j=1}^J w_t^{(j)}\bfK_t(X_t^{(j)})\; \in \R^J.
	\]
	For $\bfs \in \R^J$, define $G: \R^J \to \R^J$ to be the sample mean of $\bfK_t$ over $\{\nabla \phi_{\bfs}(X_t^{(j)})\}_{j=1}^J$, 
	\begin{align*}
	G(\bfs)  %
 = \frac{1}{J} \sum_{j=1}^J \bfK_t(X_t^{(j)} +  \nabla \bfK_t(X_t^{(j)})\t \bfs). 
	\end{align*}
	In order for sample-equivalence to be satisfied, we need to find $\bfs^*$ such that $G(\bfs^*) = \bfb$. 
	
	\citet{kuangSampleBasedOptimalTransport2019} demonstrate that if the Jacobian of $G$ at $\bfs = 0$ 
	\[
	\left. \nabla G(\bfs) \right |_{\bfs = 0} = \sdfrac{1}{J}\sum_{i=1}^J \nabla \bfK_t(X_{t}^{(i)} ) \nabla \bfK_t(X_{t}^{(i)})\t \equiv M_t
	\]
	is nonsingular (for which the necessary condition $J \leq dJ$ is automatically satisfied), $G$ is a bijection from a neighborhood $U$ about $\bfs = \bf 0$ to a neighborhood $V$ about $G(\mathbf{0}) = \bfa$. If $\bfb \in V$, then the potential $\phi_\bfs$ parameterized with $\bfs^* = G\inv(\bfb)$ gives the \textit{global minimum} of the sample-based OT problem \eqref{eq:sampleOT} restricted to maps of the form \eqref{eq:gradPhi}. Furthermore, \citet{kuangSampleBasedOptimalTransport2019} show that if the kernels are $C^2$, then $\phi_{\bfs^*}$ is locally convex. %
	
	For sufficiently small $\Delta t$, the system $G(\bfs^*) = \bfb$ \eqref{eq:sot} will be close to linear. Thus, for the sake of efficiency we may approximate $\bfs^*$ with a single Newton step, setting 
	\begin{multline}
		\bfs^* \approx -\left (\sdfrac{1}{J}\sum_{i=1}^J \nabla \bfK_t(X_{t}^{(i)} ) \nabla \bfK_t(X_{t}^{(i)})\t \right)\inv \cdot \\  \sum_{k=1}^J (\ltfrac{1}{J} - w_t^{(k)})\bfK_t(X_{t}^{(k)}), 
  \label{eq:sstar}
	\end{multline}
	{to} arrive at the update 
		\begin{multline} 
			X_{t+\Delta t}^{(j)} = X_{t}^{(j)} - \\ \nabla\bfK_t(X_t^{(j)})\t  
   M_t\inv  \sum_{k=1}^J \left(\ltfrac{1}{J} - w_t^{(k)}\right) \bfK_t(X_t^{(k)}), 
			\label{eq:IPS_importance}
		\end{multline} 
with $j  \in \{1,\dots,J\}$, $t \in [0,1]$, and $\{X_0^{(j)}\}_{j=1}^J \overset{\rm i.i.d.}{\sim} \pi_0$.
Although \eqref{eq:IPS_importance} is distinct from \eqref{eq:IPS_ode} in discrete time, in \textit{continuous} time the two interacting particle systems are equivalent: 
	\begin{theorem}
		In the limit $\Delta t \to 0$, \cref{eq:IPS_importance} approaches \cref{eq:IPS_ode}. %
  \rev{Thus the IPS obtained via sample-driven optimal transport \eqref{eq:IPS_importance} can be viewed as arising from mean-field model \eqref{eq:meanfield}.}   
		\label{thm:ctsTime}
	\end{theorem}
    The proof of this result follows from simple calculus and is contained in \cref{app:proofCtsTime}. Owing to this equivalence in continuous time, we refer to the interacting particle system \eqref{eq:IPS_importance} as \textbf{KFRFlow-Importance (KFRFlow-I)}. %
    
	\Cref{thm:ctsTime} highlights the fact that linearizing a Monge--Amp\`ere equation for \textit{static} optimal transport between $\pi_t$ and $\pi_{t+\Delta t}$ results in a Poisson equation, and demonstrates that as $\Delta t \to 0$ this linearization yields the correct velocity field for the controlled \textit{dynamic} minimum-energy transport problem of \cref{sec:methodology_poisson}. Furthermore, it elucidates connections between SMC approaches based on tempered self-normalized importance sampling and Fisher--Rao gradient flows.

    \section{Implementation}
    \label{sec:implementation}
         KFRFlow \eqref{eq:IPS_ode} %
  can be discretized in time, for example, via {the explicit Euler method}
	\begin{small} 
		\begin{multline} 
			X_{t + \Delta t}^{(j)} = X_t^{(j)} +  \nabla \bfK_t(X_t^{(j)})\t M_t\inv \cdot \\ 
			\ltfrac{\Delta t}{J} \sum_{k=1}^J \left(\log\tfrac{\pi_1}{\pi_0}(X_t^{(k)}) - \ltfrac{1}{J}\sum_{i=1}^J \log\tfrac{\pi_1}{\pi_0} (X_t^{(i)}) \right) \bfK_t(X_t^{(k)}), %
			\label{eq:ode_rehash}	
		\end{multline} 
    \end{small}
$\!\!$or any other standard ODE {integration scheme}, while KFRFlow-I \eqref{eq:IPS_importance} already has the form of a discrete-time iteration. 	In either case we start from $\{X_0^{(j)}\}_{j=1}^J \overset{\rm i.i.d.}{\sim} \pi_0$ and simulate for unit time to obtain $\{X_1^{(j)}\}_{j=1}^J \sim \pi_{X_1} \approx \pi_1$. KFRFlow-I \eqref{eq:IPS_importance} and the Euler discretization of KFRFlow \eqref{eq:ode_rehash} are almost identical, %
 and in fact \eqref{eq:IPS_importance} can be recovered, up to multiplication by a constant close to one, from \eqref{eq:ode_rehash} \rev{by applying the approximation $\Delta t \log y = \log y^{\Delta t} \approx y^{\Delta t} - 1$ to all instances of $\Delta t \log\frac{\pi_1}{\pi_0}$ in \eqref{eq:ode_rehash}}. 

\rev{In practice, the performances of KFRFlow-I \eqref{eq:IPS_importance} and discretizations of KFRFlow \eqref{eq:IPS_ode} are often similar, but we have noticed that for large $\Delta t$ or particularly challenging sampling tasks, such as those in high dimensions, KFRFlow-I tends to be more stable. %
One possible explanation for this advantage is that in KFRFlow-I we evaluate $(\frac{\pi_1}{\pi_0})^{\Delta t}$ rather than $\Delta t \log \frac{\pi_1}{\pi_0}$, which is a more stable computation when the ratio $\frac{\pi_1}{\pi_0}$ is small. 
Another consideration is that the update in KFRFlow-I \eqref{eq:IPS_importance} is derived as an approximate transport \textit{map} between $\pi_{t}$ and $\pi_{t+\Delta t}$, whereas KFRFlow updates as in \eqref{eq:ode_rehash} are discretizations of a {continuous} flow %
\eqref{eq:IPS_ode}. 
These discretizations may not be good transport maps, especially %
for large $\Delta t$, in which case KFRFlow-I, which is explicitly designed for discrete-time transport, may perform better.}

\subsection{Numerical Stability}
Choice of KFRFlow versus KFRFlow-I aside, in our experiments we have noticed that the matrix $M_t$ in KFRFlow \eqref{eq:IPS_ode} and KFRFlow-I \eqref{eq:IPS_importance} may at times be poorly conditioned, leading to numerical instability or poor quality samples. We have found two helpful tactics for mitigating this issue: inflating the diagonal of $M_t$ and introducing noise. 

\subsubsection{Regularization of $M_t$}
Issues of ill-conditioning of $M_t$ can be ameliorated by replacing $M_t$ in \cref{eq:IPS_importance,eq:ode_rehash} with $M_{t, \lambda} = M_t + \lambda I$ for some $\lambda > 0$; this is essentially a Tikhonov regularization of \eqref{eq:Mpit_Kpit} or \eqref{eq:sstar}. %
Inflating the diagonal of $M_t$ 
does not require additional information about $\pi_1$ and $\pi_0$ but does require finding an appropriate $\lambda$ and potentially alters the time-dependent distribution $\rho_t = \mathrm{Law}(X_t)$. 

\subsubsection{Stochastic Modification}
As noted similarly in, e.g., \citet{song2020score, albergoStochasticInterpolantsUnifying2023}, the continuity equation \eqref{eq:vt_must_satisfy} can be written equivalently for any $\epsilon > 0$ as a {Fokker--Planck equation} 
\begin{equation} 
 \partial_t\pi_t = -\nabla \cdot (\pi_t (v_t + \epsilon \nabla \log \pi_t)) + \epsilon \nabla^2 \pi_t,
 \label{eq:FPE}
 \end{equation} 
 by making use of the identity $\nabla \log \pi_t = \frac{\nabla \pi_t}{\pi_t}$ \rev{(we use $\nabla^2$ to denote the Laplacian)}. The Fokker--Planck equation \eqref{eq:FPE} corresponds to an {SDE} for $X_t$, 
\begin{equation}
	\rmd X_t = (v_t(X_t) + \epsilon \nabla \log \pi_t(X_t)) \,\rmd t + \sqrt{2\epsilon} \,\rmd W_t, \quad t \in [0,1], 
	\label{eq:stochasticKFRFlow}
\end{equation}
which possesses the same marginal distributions as the ODE \eqref{eq:dynamics}. The SDE \eqref{eq:stochasticKFRFlow} can hence be used with the velocity field $v_t$ \eqref{eq:vt} as the basis for a \textit{stochastic} interacting particle system, which we refer to as Kernel Fisher--Rao Diffusion (KFRD), for traversing the geometric mixture $\pi_t$; see \cref{app:KFRD} for further details. Simulation of \eqref{eq:stochasticKFRFlow} does require access to the score of $\pi_t$ and hence KFRD is not gradient-free, but in the case that gradients of $\pi_0$ and $\pi_1$ are available, the score of $\pi_t$ is simply  
\[
\nabla \log \pi_t = (1-t) \nabla \log \pi_0 + t \nabla \log \pi_1.
\]
We find that the introduction of noise through \eqref{eq:stochasticKFRFlow} often increases numerical stability and \rev{enhances sample quality, owing to the incorporation of gradient information.} %

\subsection{Computational Cost}
\label{sec:computationalCost}
\rev{The na{i}ve complexity of computing the right-hand side of the KFRFlow ODE \eqref{eq:IPS_ode} or one step of KFRFlow-I \eqref{eq:IPS_importance} is $\mathcal{O}(J^3)$, %
as we require a solve with a $J \times J$ symmetric matrix. %
While symmetric linear solves are well-optimized computations, the cost of KFRFlow could be lowered in practice, for instance, by use of random features \citep{rahimiRandomFeaturesLargeScale2007} or other kernel dimension-reduction techniques. %
We only demonstrate the ``vanilla'' $\mathcal{O}(J^3)$ implementation of KFRFlow in this work for clarity of presentation; more sophisticated implementation strategies, including use of random features, are part of ongoing work.}
\section{Numerical Examples}
	\label{sec:numerics}
Now we present proof-of-concept examples demonstrating
the efficacy of KFRFlow, KFRFlow-I, and KFRD in generating samples from various target distributions. %
\rev{We also compare the performances of our algorithms to those of ensemble Kalman inversion stopped at $t=1$ (EKI, \citet{iglesiasEnsembleKalmanMethods2013}), the ensemble Kalman sampler (EKS, \citet{eks}), consensus-based sampling (CBS, \citet{carrilloConsensusbasedSampling2022}), %
Stein variational gradient descent (SVGD, \citet{liuSteinVariationalGradient2016}), and the unadjusted Langevin algorithm (ULA, e.g., \citet{roberts1996exponential}). Of the competing algorithms, EKI, EKS, and CBS are gradient-free, and hence we employ them as bases for comparison for KFRFlow and KFRFlow-I, while SVGD and ULA require $\nabla \log \pi_1$, so we compare their performances to those of KFRD. Like KFRFlow(-I) and KFRD, EKI, EKS, CBS, and SVGD are interacting particle systems which evolve ensembles of $J$ particles together such that their collective distribution approaches $\pi_1$. By contrast, ULA is a Markov chain algorithm which does not make use of interaction, but to obtain a similarly structured sampler to the IPS algorithms we use ULA in ``parallel mode,'' simulating $J$ independent chains initialized at points randomly drawn from $\calN(0,I_d)$
and retaining the final state of each chain to form a set of $J$ samples from the target. } %

\rev{In our experiments we take the kernel in KFRFlow(-I), KFRD, and SVGD to be inverse multiquadric (IMQ) 
	\begin{equation} 
	K(x, x') = \left(1 + \ltfrac{\| x - x'\|^2}{h^2}\right)^{-1/2} %
 \label{eq:IMQ}
	\end{equation} 
with bandwidth $h > 0$ selected at each step of the iterations according to the median heuristic \citep{liuSteinVariationalGradient2016}. %
We assess sample quality using kernel Stein discrepancy (KSD) \citep{gorhamMeasuringSampleQuality2020} with the IMQ kernel \eqref{eq:IMQ} with bandwidth $h = 1$. The reference distribution $\pi_0$ is always standard Gaussian. %
We perform all experiments in Julia using the package \texttt{DifferentialEquations.jl} \citep{rackauckas2017differentialequations} to integrate the ODEs and SDEs associated with KFRFlow, KFRD, EKI, EKS, CBS, SVGD, and ULA. }
Code for the experiments is available at at \url{https://github.com/amaurais/KFRFlow.jl}.%
    \newcommand{\samplewidth}{0.17\linewidth}
    \subsection{Two-Dimensional Bayesian Posteriors}
    \label{sec:2Ddists}
         \begin{figure}[h]
		\centering
			\begin{subfigure}{\samplewidth}
				\includegraphics[width=\linewidth]{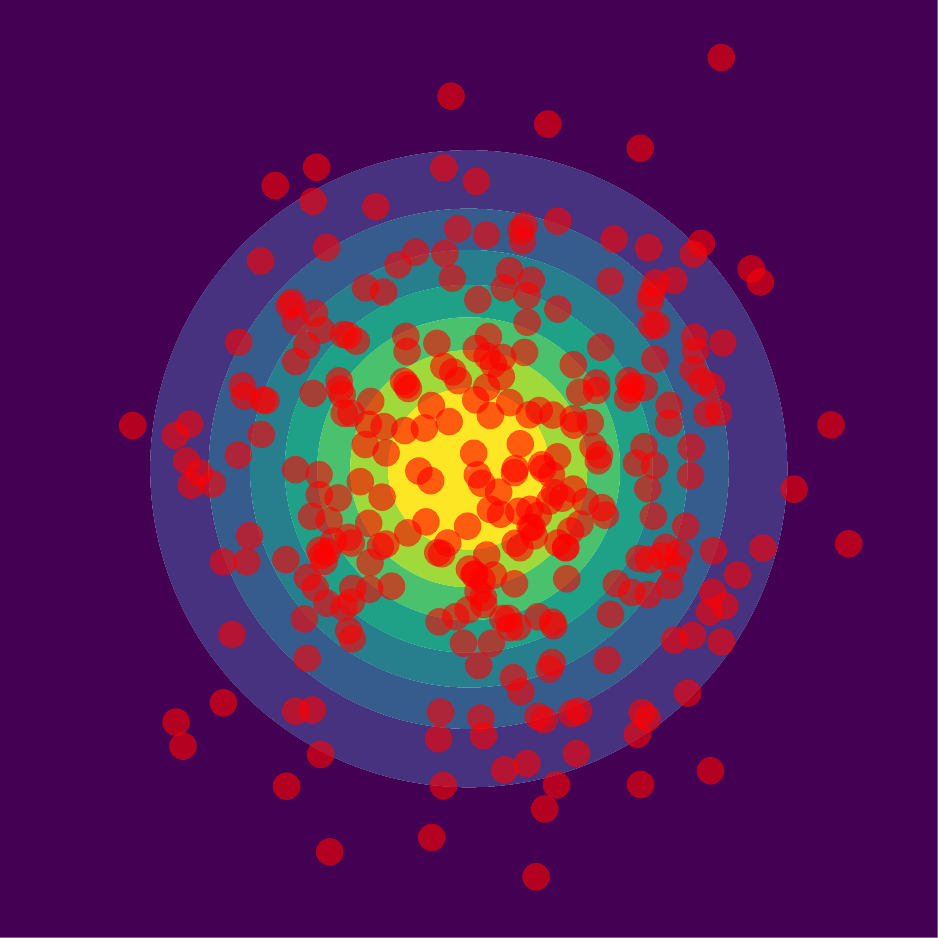}
			\end{subfigure}
			\begin{subfigure}{\samplewidth}
				\includegraphics[width=\linewidth]{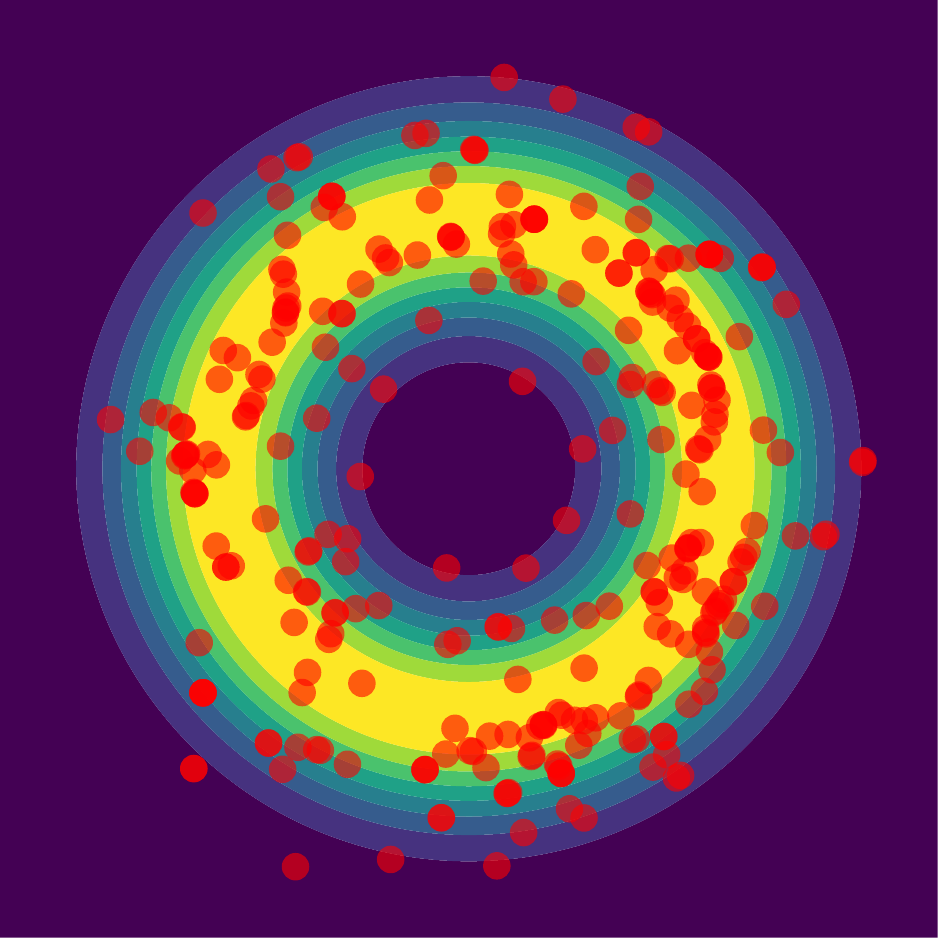}
			\end{subfigure}
			\begin{subfigure}{\samplewidth}
				\includegraphics[width=\linewidth]{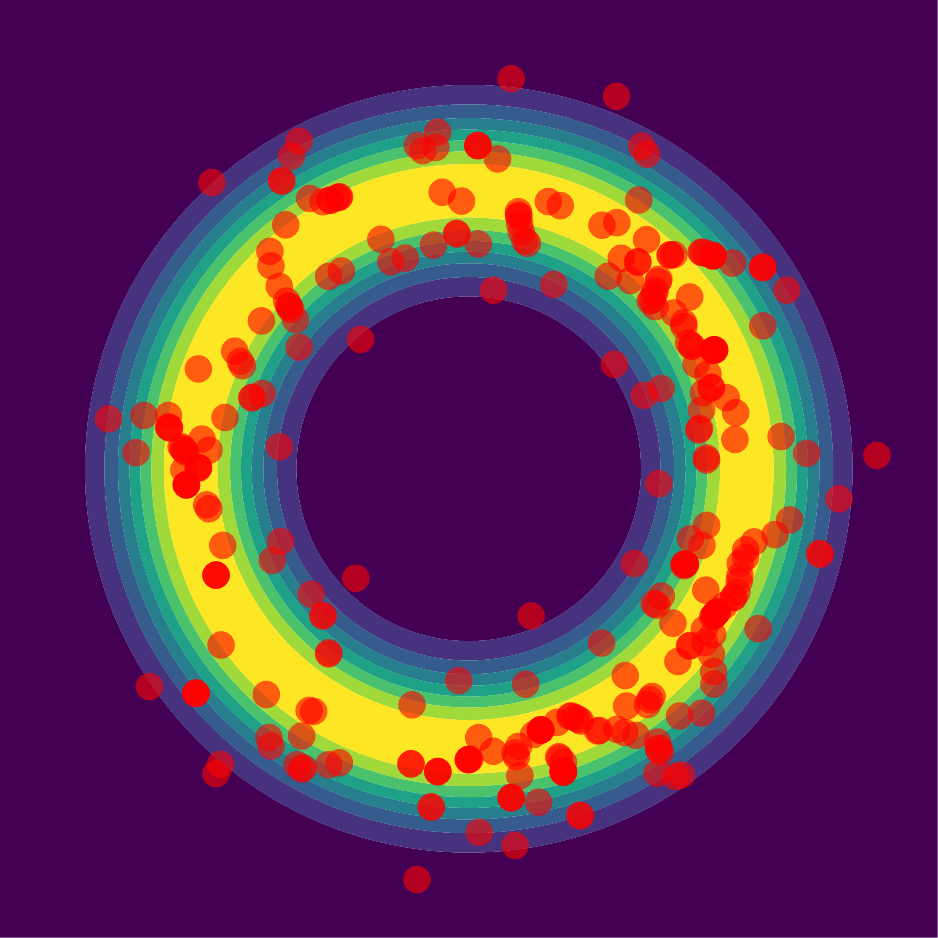}
			\end{subfigure}
			\begin{subfigure}{\samplewidth}
				\includegraphics[width=\linewidth]{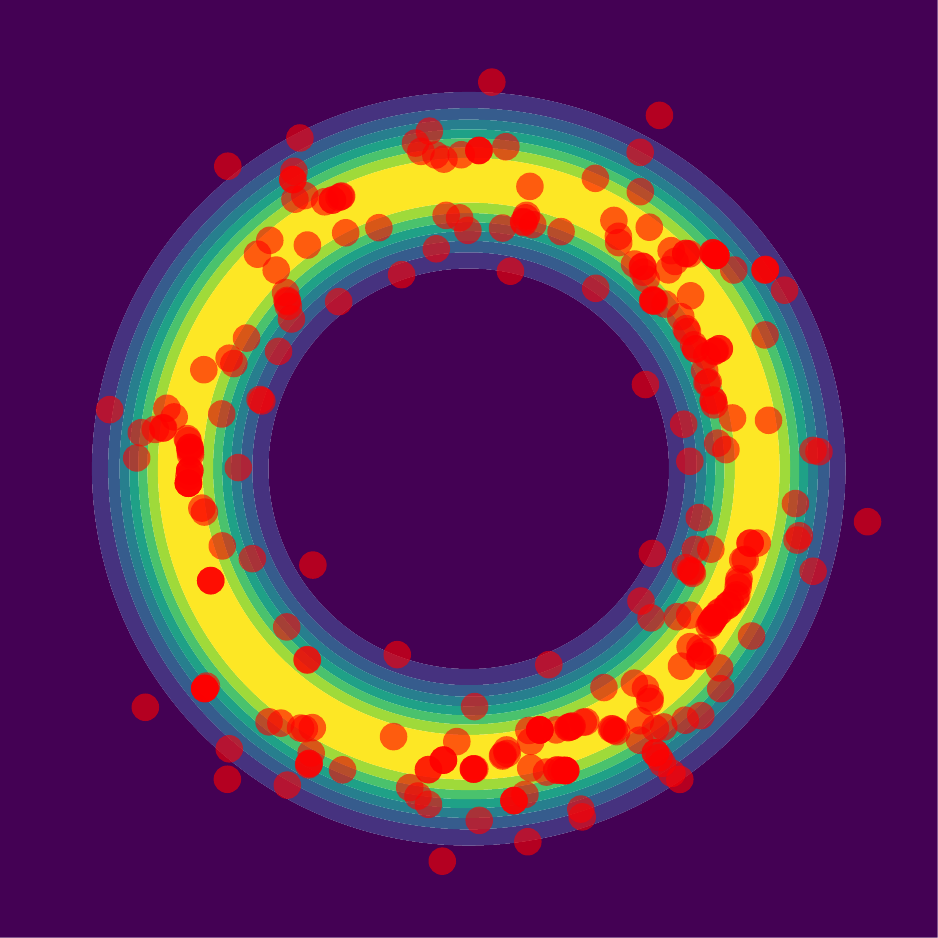}
			\end{subfigure}
			\begin{subfigure}{\samplewidth}
				\includegraphics[width=\linewidth]{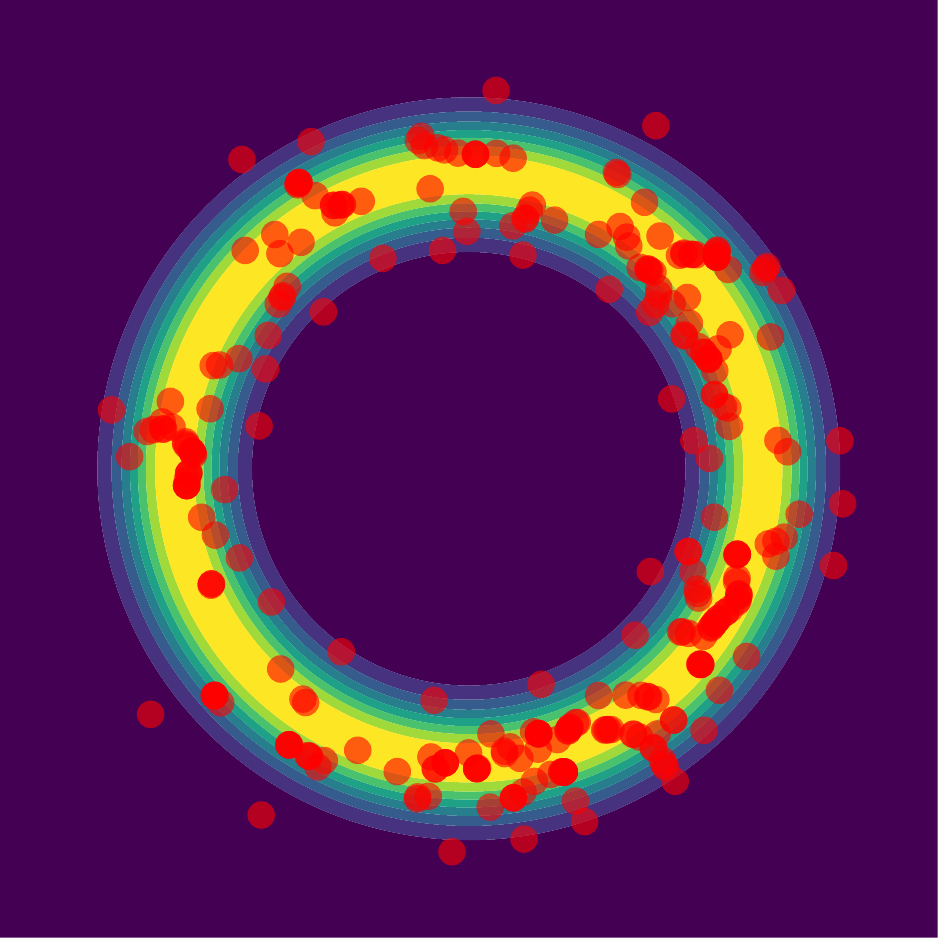}
			\end{subfigure}
			\\
			\begin{subfigure}{\samplewidth}
				\includegraphics[width=\linewidth]{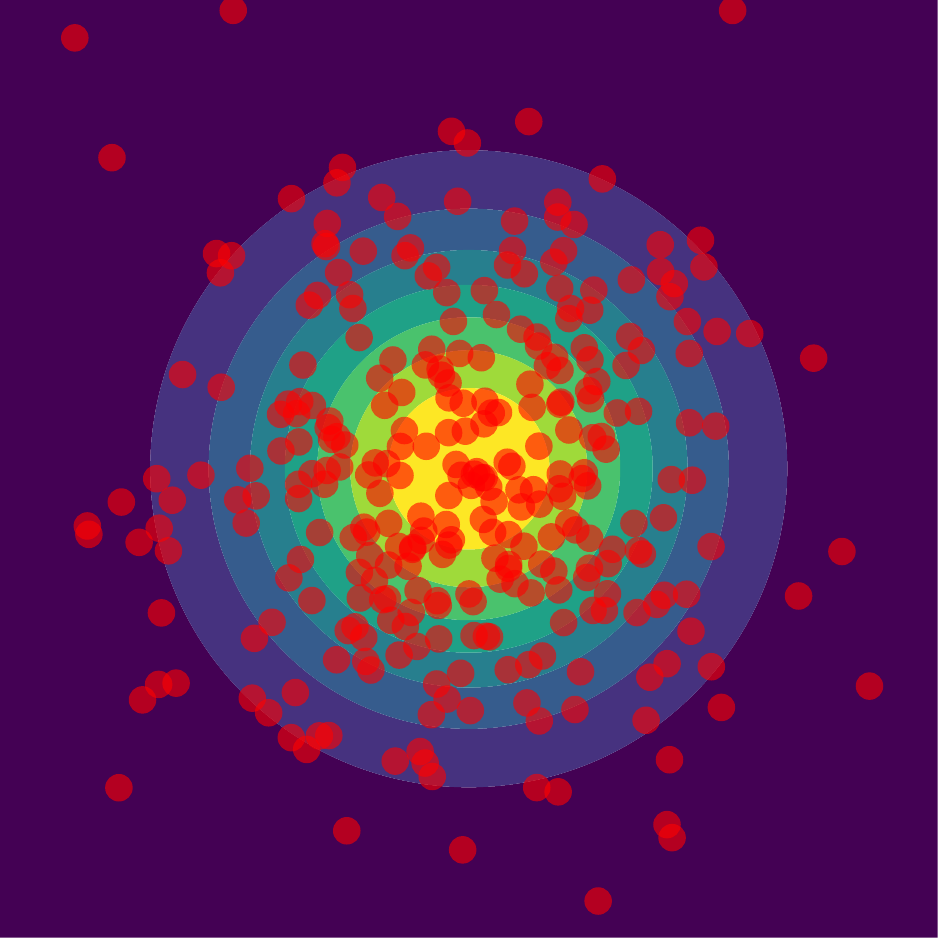}
			\end{subfigure}
			\begin{subfigure}{\samplewidth}
				\includegraphics[width=\linewidth]{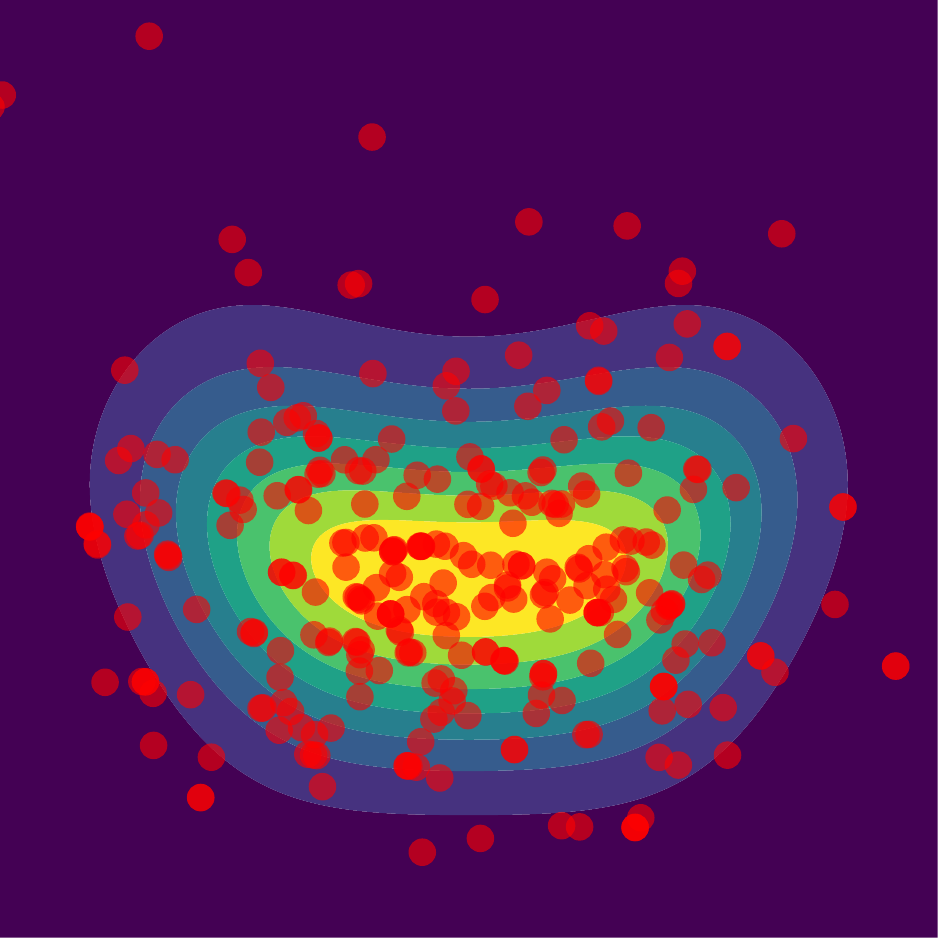}
			\end{subfigure}
			\begin{subfigure}{\samplewidth}
				\includegraphics[width=\linewidth]{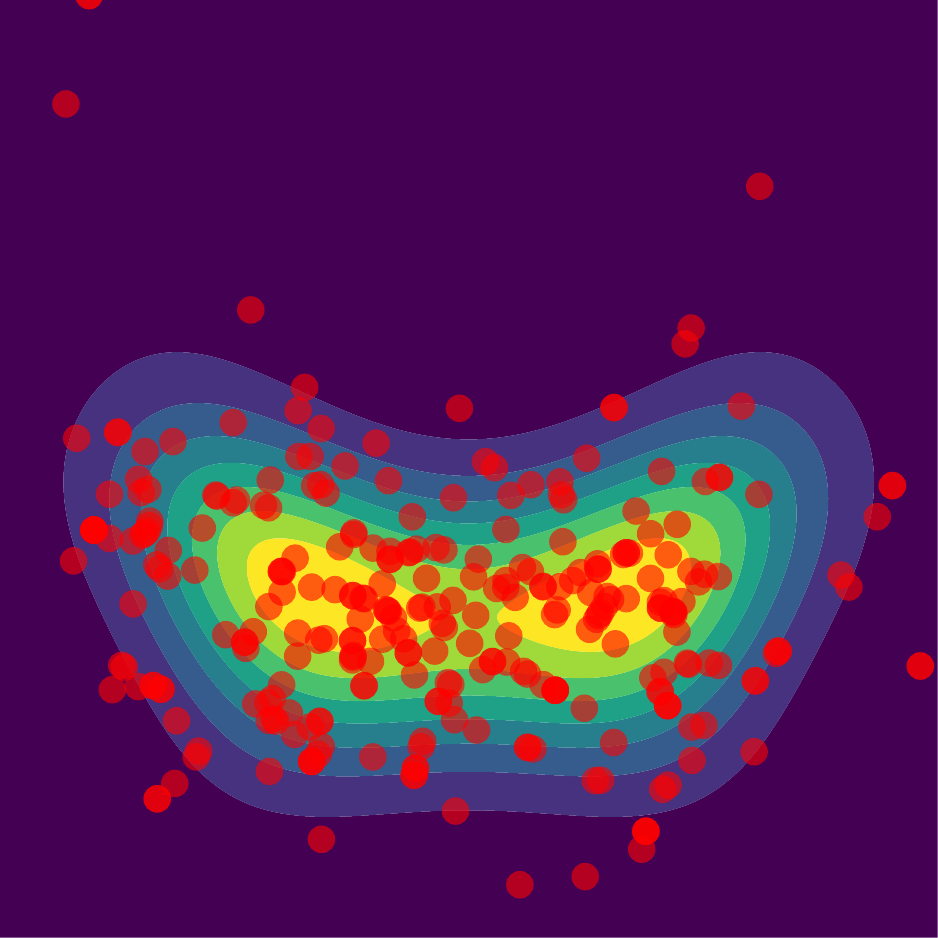}
			\end{subfigure}
			\begin{subfigure}{\samplewidth}
				\includegraphics[width=\linewidth]{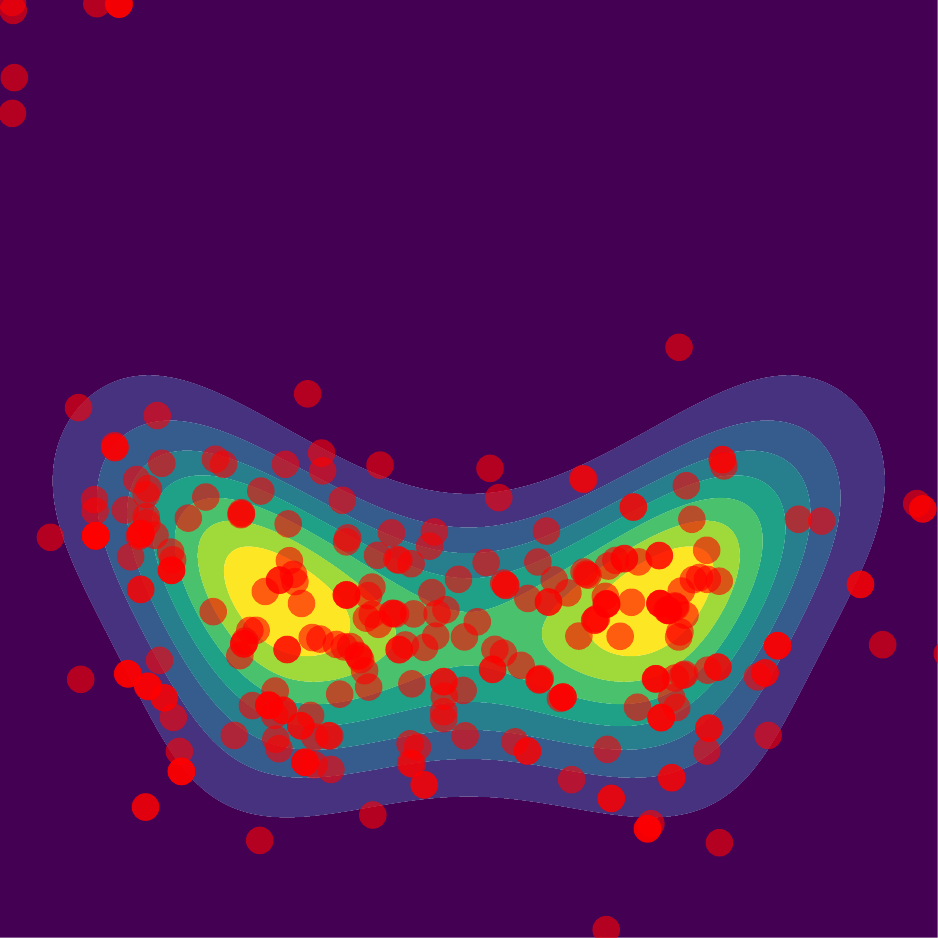}
			\end{subfigure}
			\begin{subfigure}{\samplewidth}
				\includegraphics[width=\linewidth]{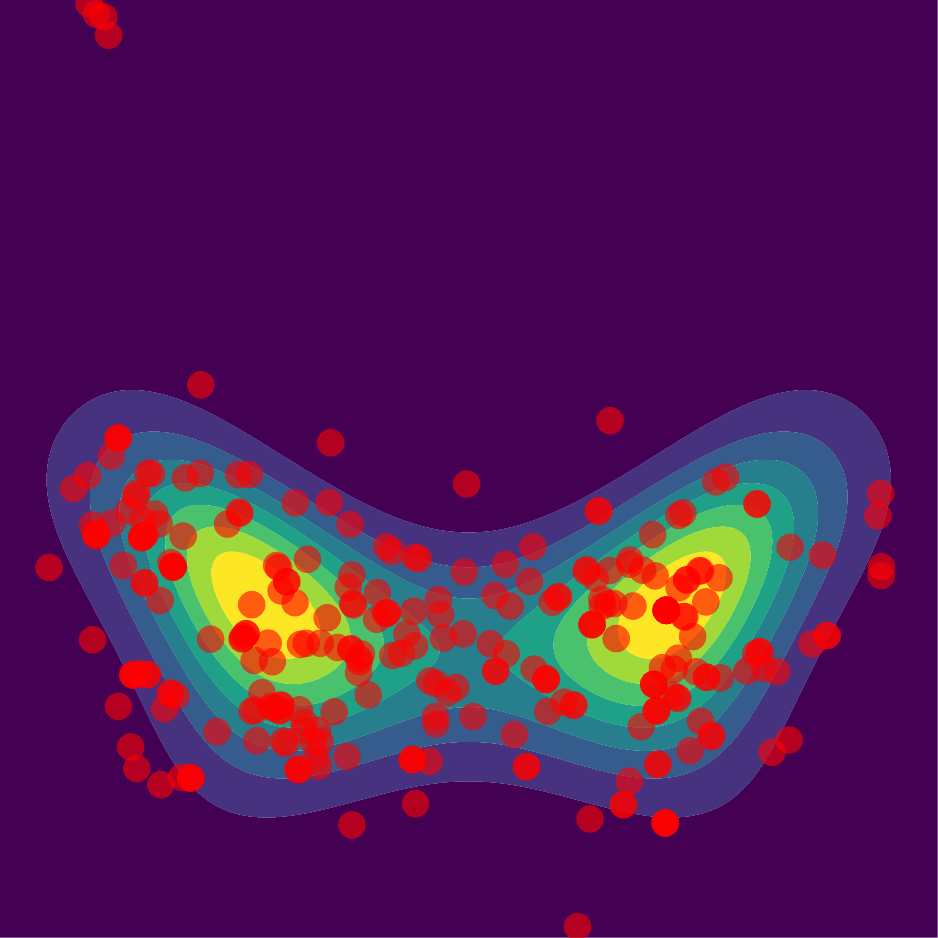}
			\end{subfigure}
			\\
			\begin{subfigure}{\samplewidth}
				\includegraphics[width=\linewidth]{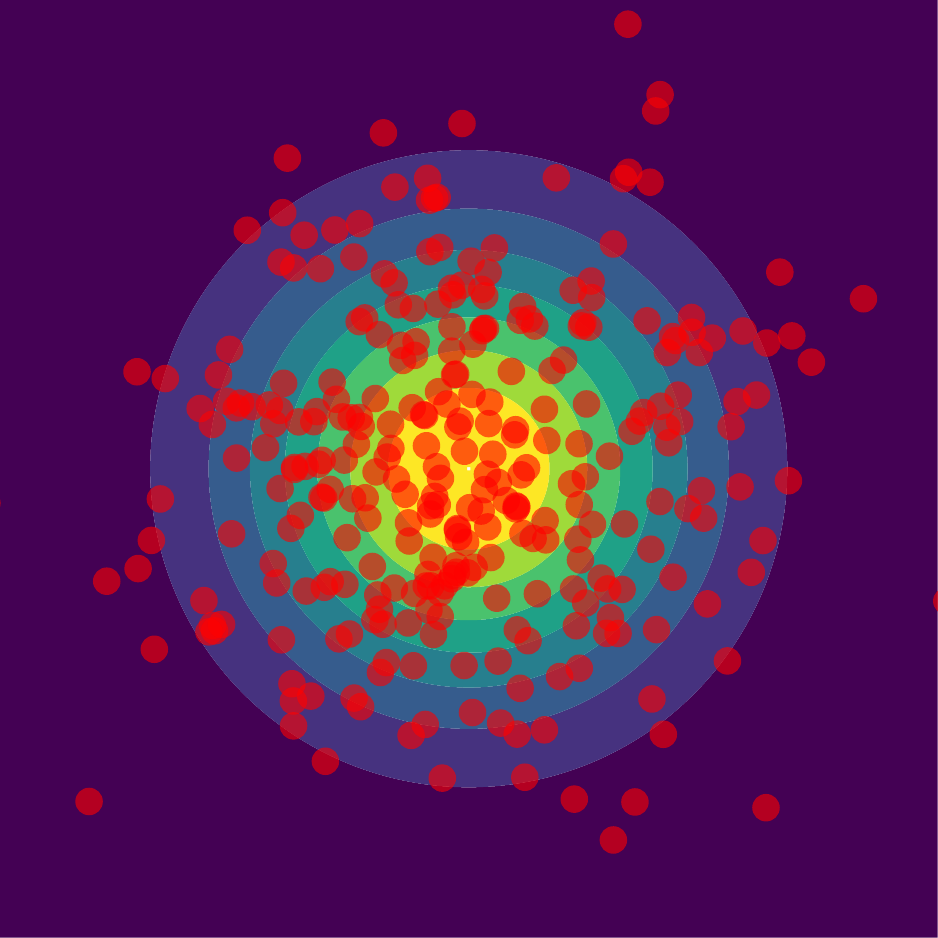}
				\subcaption*{$t=0$}
			\end{subfigure}
			\begin{subfigure}{\samplewidth}
				\includegraphics[width=\linewidth]{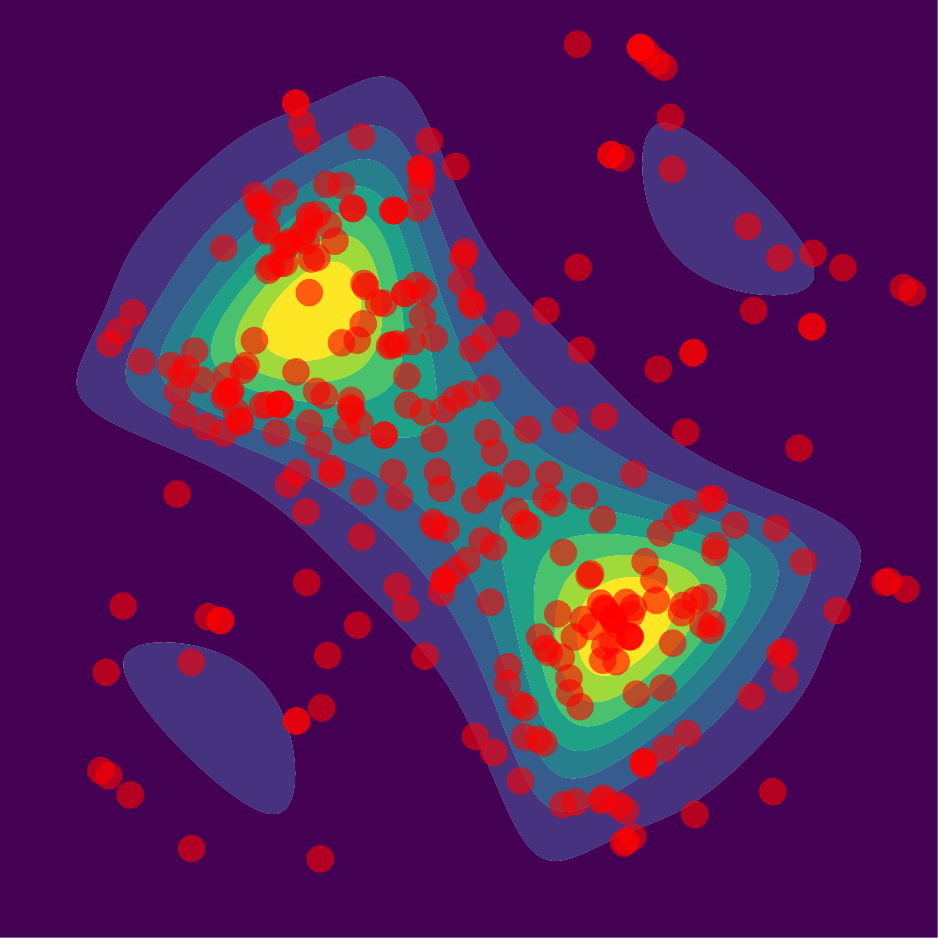}
				\subcaption*{$t=0.25$}
			\end{subfigure}
			\begin{subfigure}{\samplewidth}
				\includegraphics[width=\linewidth]{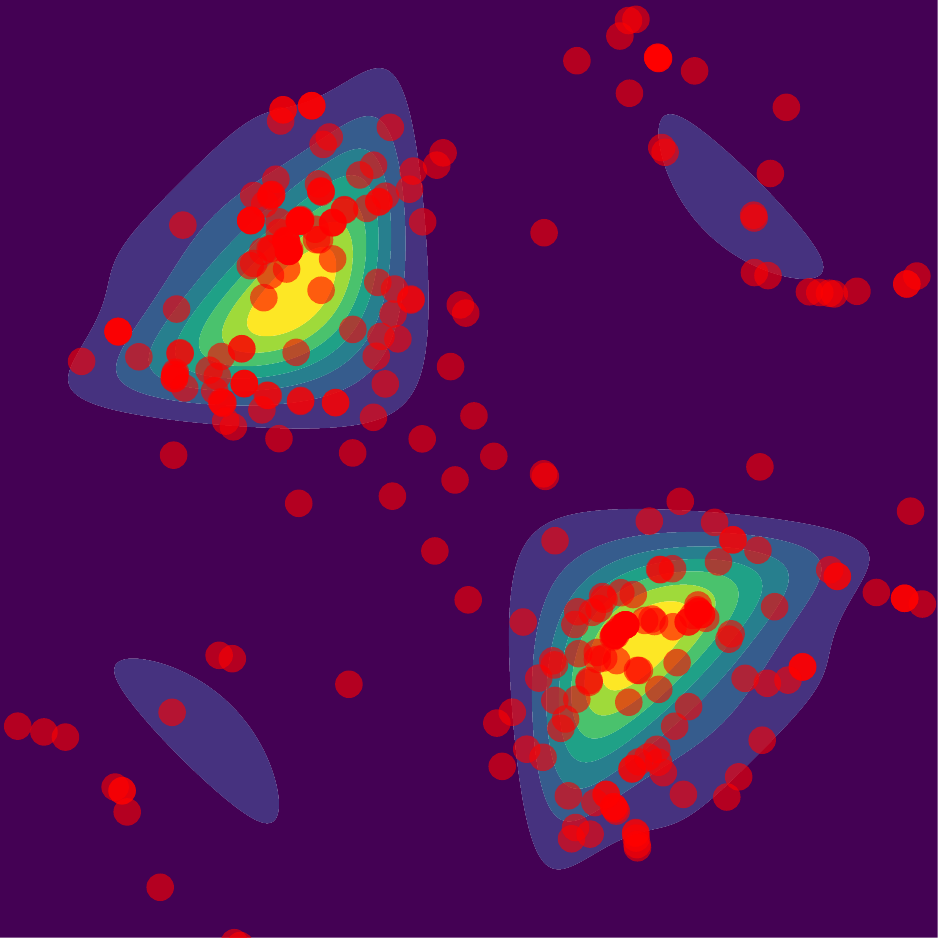}
				\subcaption*{$t=0.5$}
			\end{subfigure}
			\begin{subfigure}{\samplewidth}
				\includegraphics[width=\linewidth]{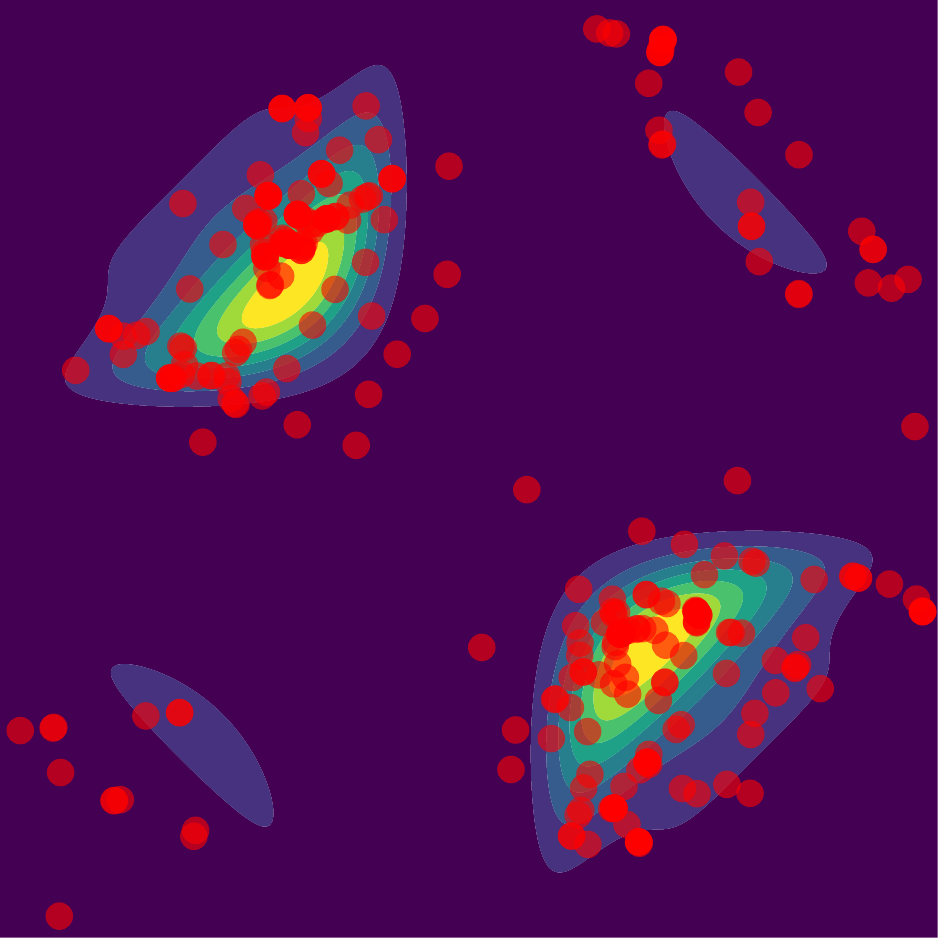}
				\subcaption*{$t=0.75$}
			\end{subfigure}
			\begin{subfigure}{\samplewidth}
				\includegraphics[width=\linewidth]{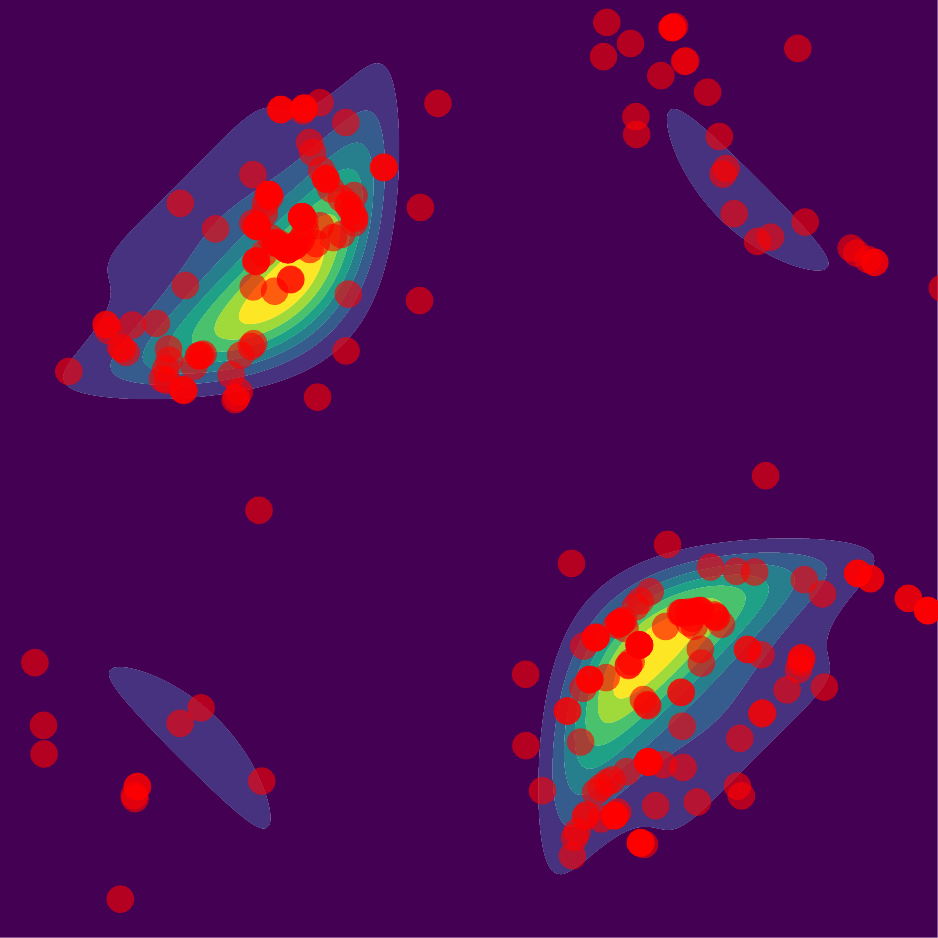}
				\subcaption*{ $t=1$}
			\end{subfigure}
			\caption{\textbf{Two-dimensional posteriors:} samples at $t \in \{0, 0.25, 0.5, 0.75, 1\}$ generated by KFRFlow \eqref{eq:IPS_ode} for the donut (top), butterfly (middle), and spaceships (bottom) examples.}
			\label{fig:threeExamples}
	\end{figure}
	We apply KFRFlow \eqref{eq:IPS_ode} %
 and KFRFlow-I \eqref{eq:IPS_importance} to sample three two-dimensional densities. %
 In all three cases %
    $\pi_1$ is a Bayesian posterior proportional to $\pi_0 \ell$ for a likelihood of the form
	 $
	 \ell(x) \propto \exp\left(-\frac{1}{\sigma_\varepsilon^{2}}\|y^* - G(x) \|_2^2 \right),
	 $
	 i.e., $y^* \in \R$ is Gaussian with mean $G(x)$ and variance $\sigma_\varepsilon^2$. 
	Definitions of the three likelihoods may be found in \cref{app:2d}. 	

\Cref{fig:threeExamples} displays $J = 300$ samples obtained from a forward Euler discretization of KFRFlow %
 with uniform timestep $\Delta t = 0.01$. The samples at $t = 1$ are qualitatively consistent with the target densities for each example. %
 \begin{figure}[h]
 \centering 
\small $J = 400$ \\[0.1cm] 
    \begin{subfigure}{0.32\linewidth}
    \includegraphics[width=\linewidth]{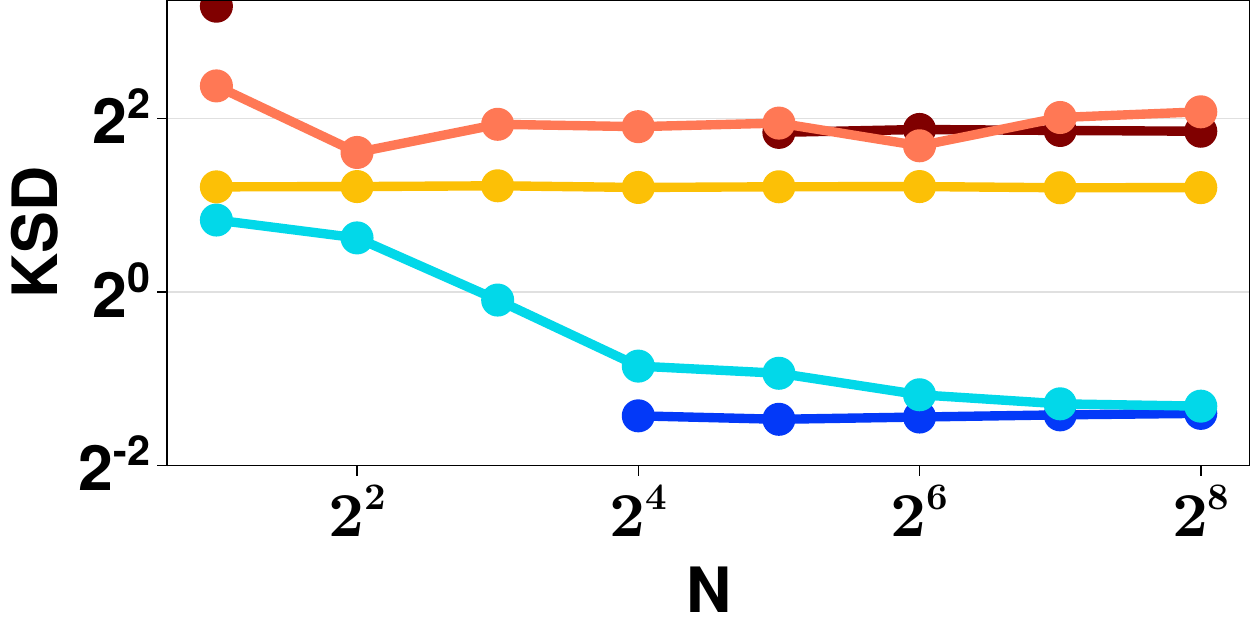}
    \end{subfigure}
    \begin{subfigure}{0.32\linewidth}
    \includegraphics[width=\linewidth]{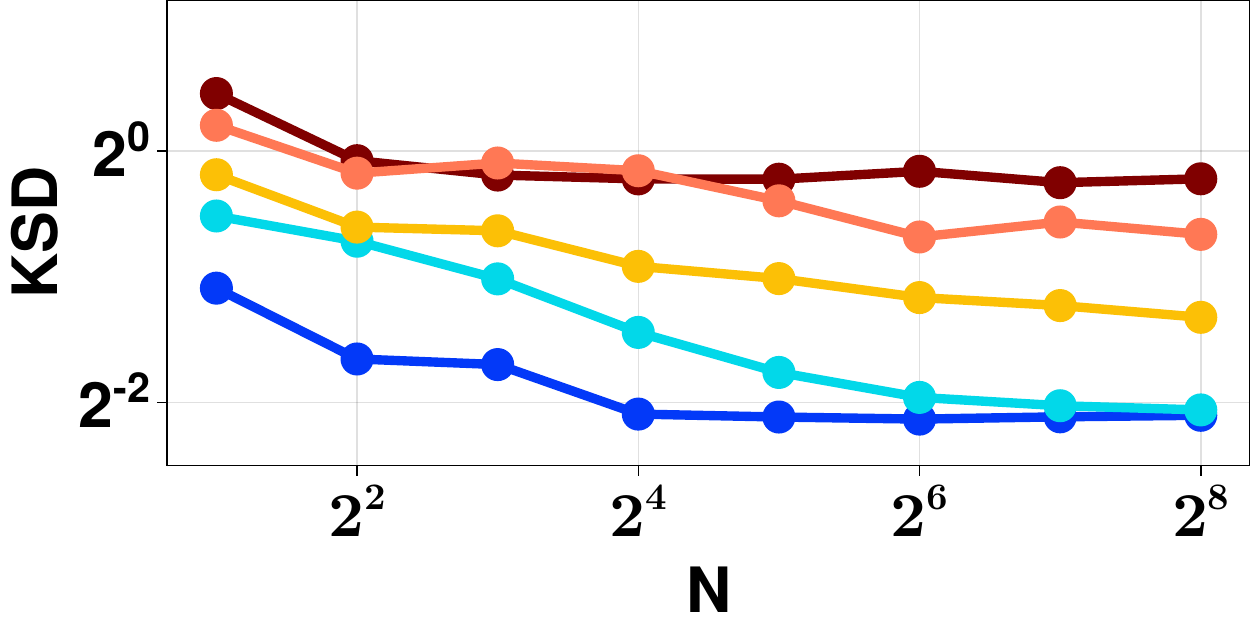}  
    \end{subfigure}
    \begin{subfigure}{0.32\linewidth}
    \includegraphics[width=\linewidth]{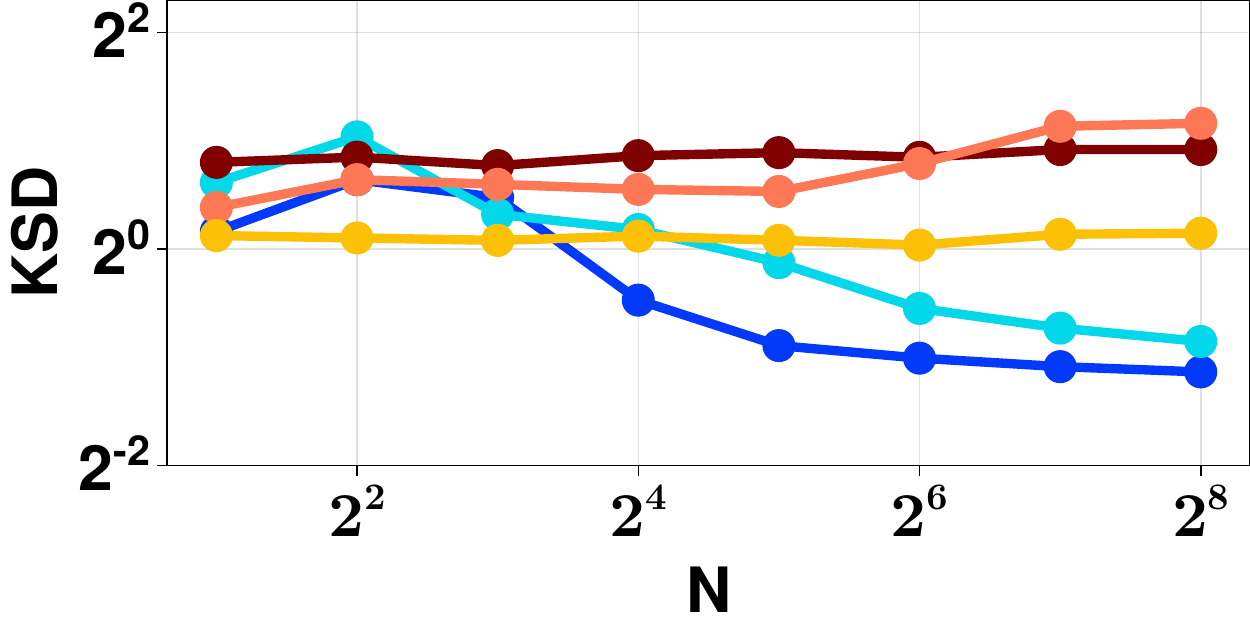}
    \end{subfigure}
    \\ 
\small    $J = 100$ \\[0.1cm] 
 \begin{subfigure}{0.32\linewidth}
    \includegraphics[width=\linewidth]{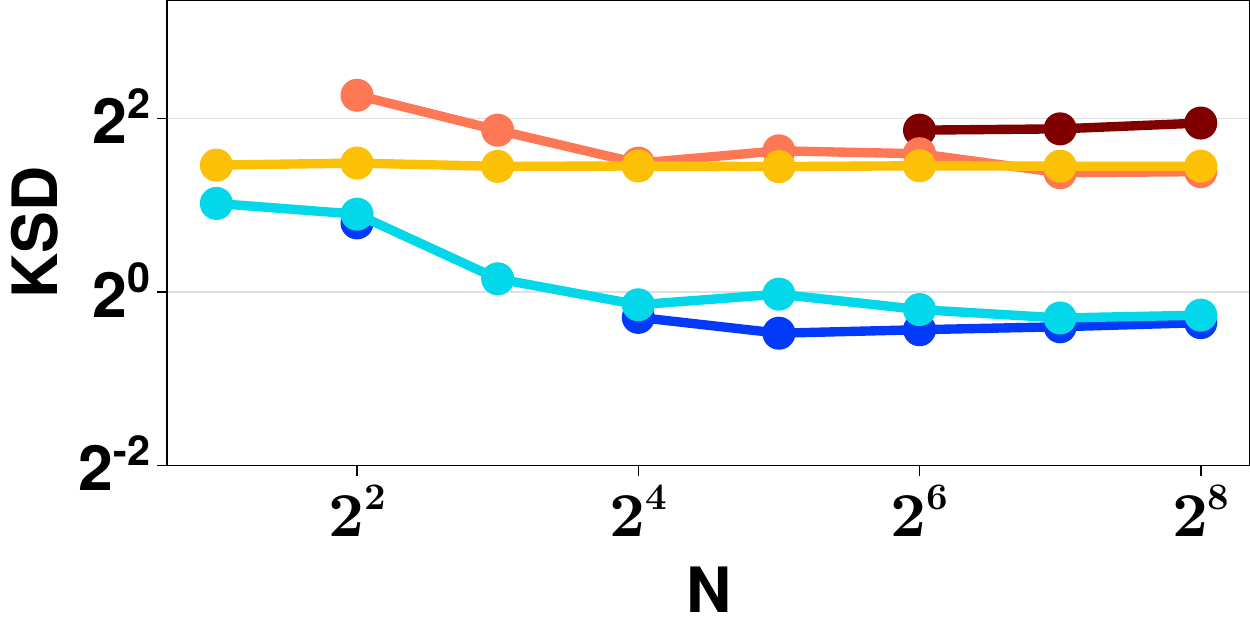}  
    \subcaption*{Donut} 
    \end{subfigure}
    \begin{subfigure}{0.32\linewidth}
    \includegraphics[width=\linewidth]{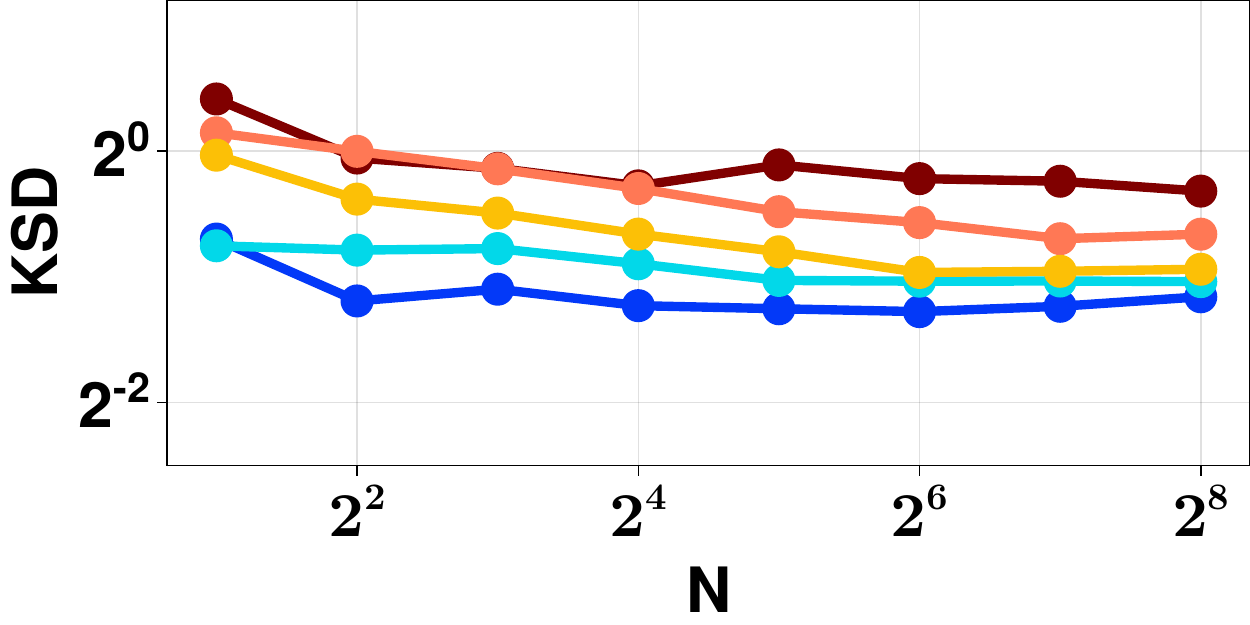}  
    \subcaption*{Butterfly}
    \end{subfigure}
    \begin{subfigure}{0.32\linewidth}
    \includegraphics[width=\linewidth]{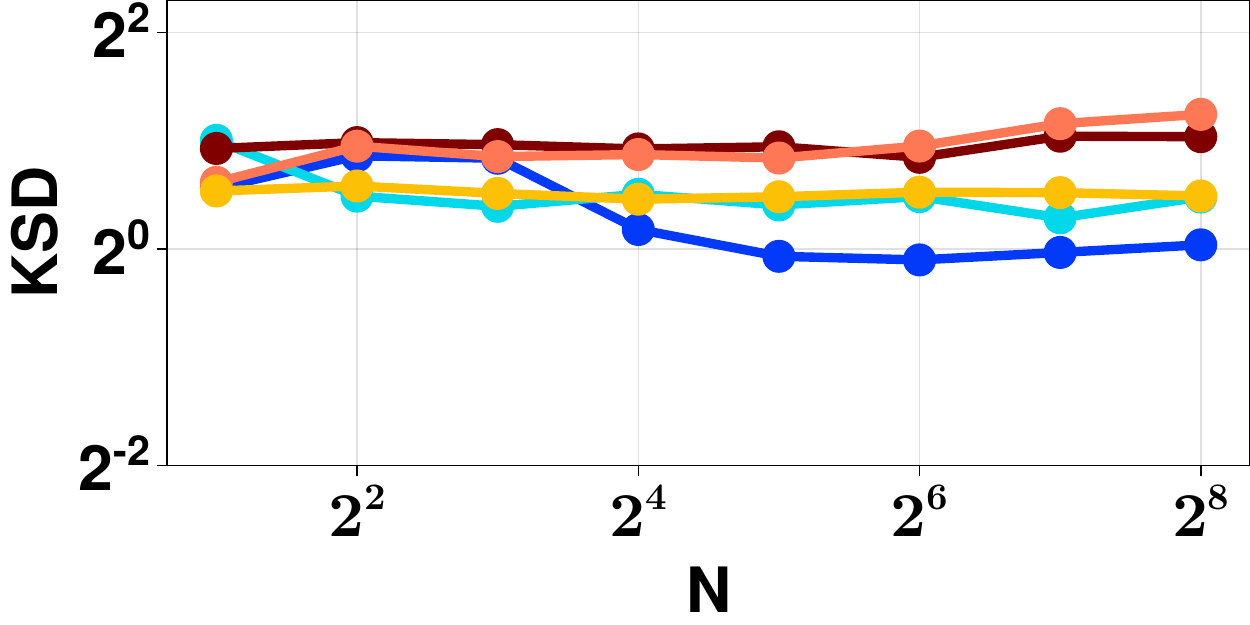}
    \subcaption*{Spaceships}
    \end{subfigure}
    \begin{subfigure}{\linewidth}
    \centering 
    \includegraphics[width=\linewidth]{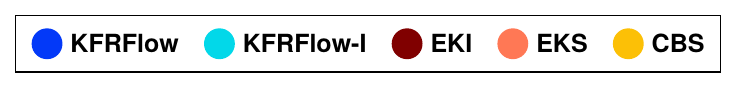}
    \end{subfigure} 
\caption{\textbf{Two-dimensional posteriors:} average KSD at stopping time between $\pi_1$ and ensembles of size $J \in \{100, 400\}$ generated by gradient-free samplers. A missing point indicates that a method was unstable at that setting of $N$.}
\label{fig:KSDvsN_gf_par}
 \end{figure} 

\begin{figure}[h] %
\centering 
\small $J = 400$ \\[0.1cm] 
    \begin{subfigure}{0.32\linewidth}
    \includegraphics[width=\linewidth]{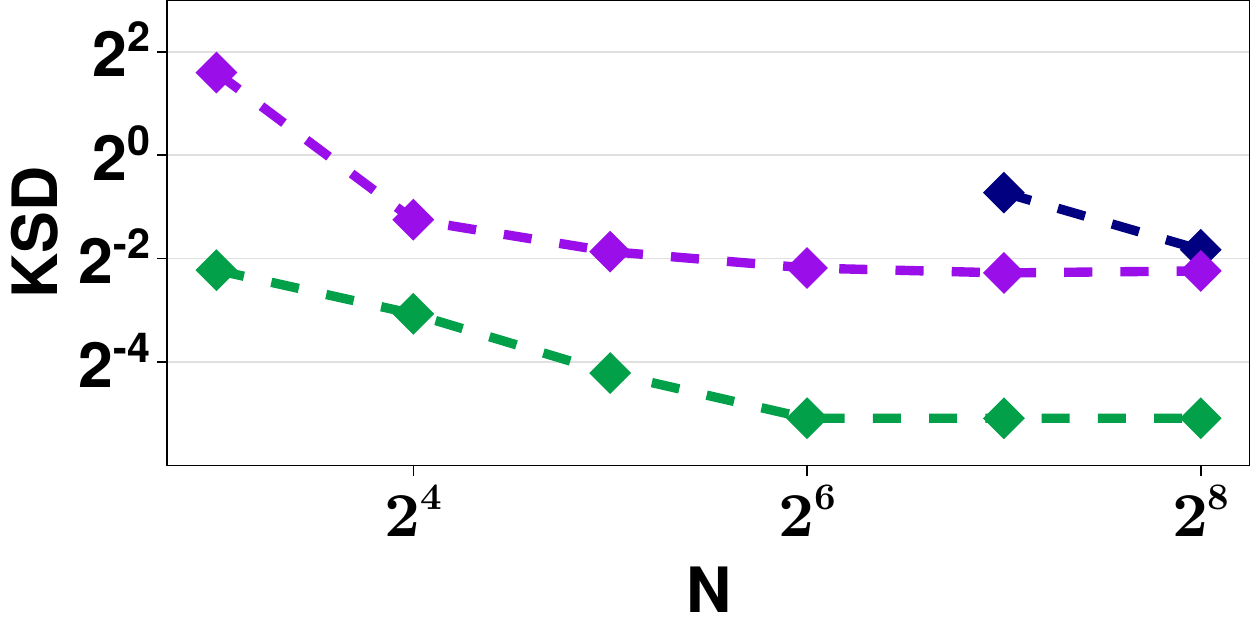}  
    \end{subfigure}
    \begin{subfigure}{0.32\linewidth}
    \includegraphics[width=\linewidth]{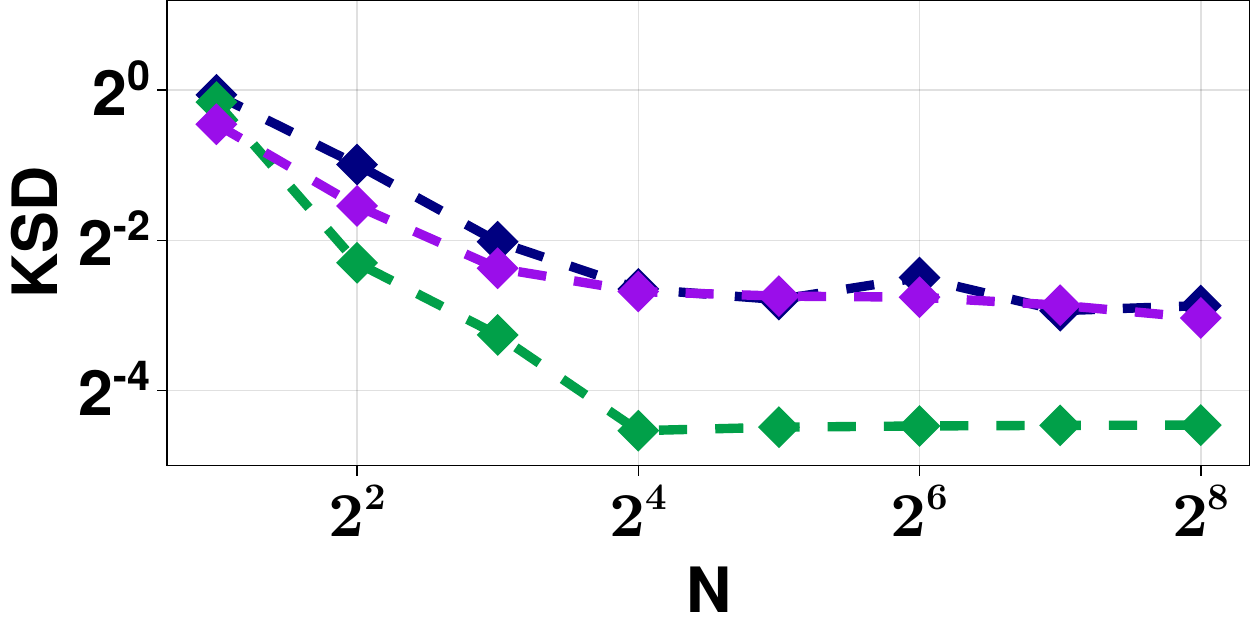}  
    \end{subfigure}
    \begin{subfigure}{0.32\linewidth}
    \includegraphics[width=\linewidth]{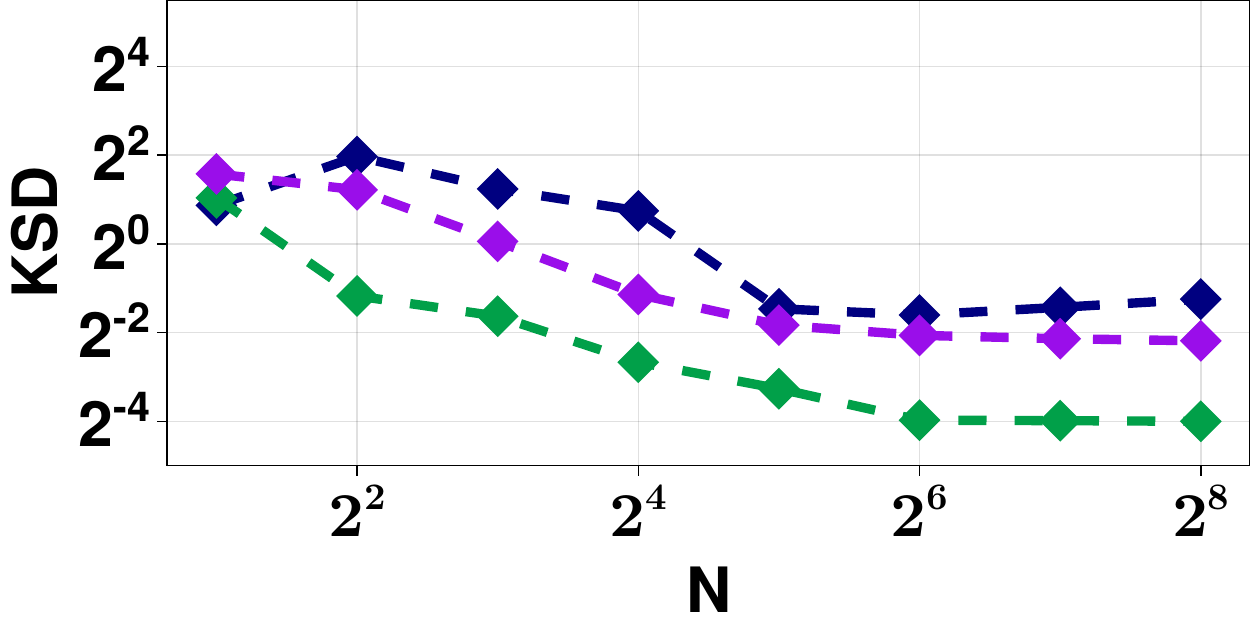}  
    \end{subfigure}
    \\ 
\small    $J = 100$ \\[0.1cm] 
 \begin{subfigure}{0.32\linewidth}
    \includegraphics[width=\linewidth]{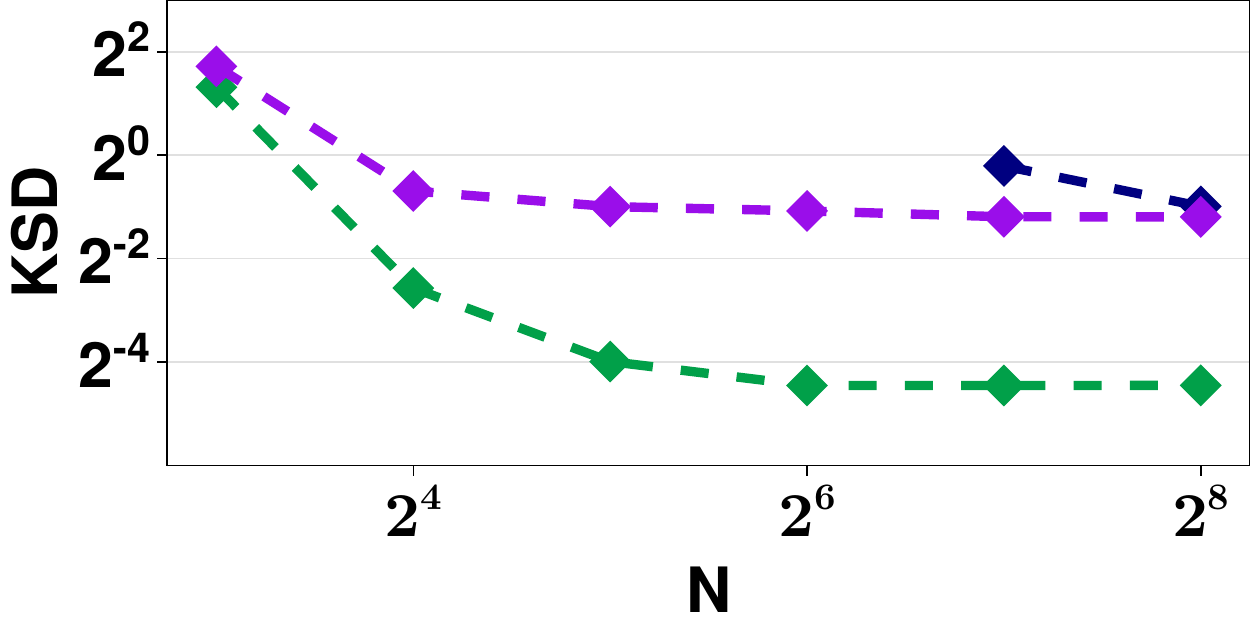}  
    \subcaption*{Donut} 
    \end{subfigure}
    \begin{subfigure}{0.32\linewidth}
    \includegraphics[width=\linewidth]{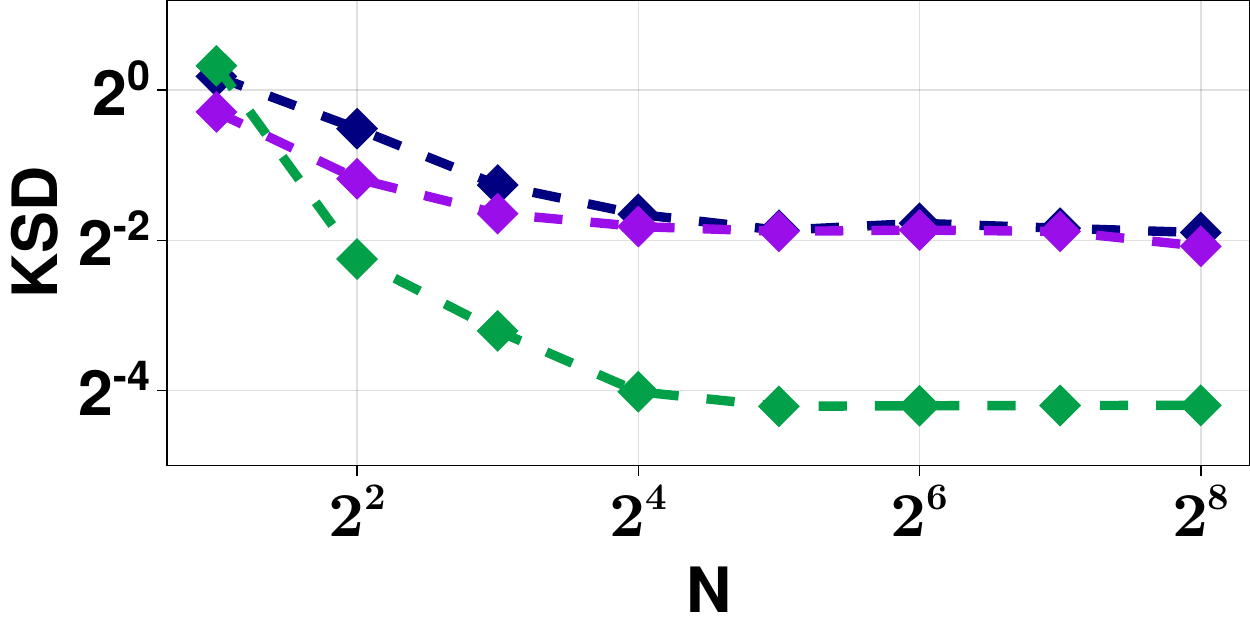}  
    \subcaption*{Butterfly}
    \end{subfigure}
    \begin{subfigure}{0.32\linewidth}
    \includegraphics[width=\linewidth]{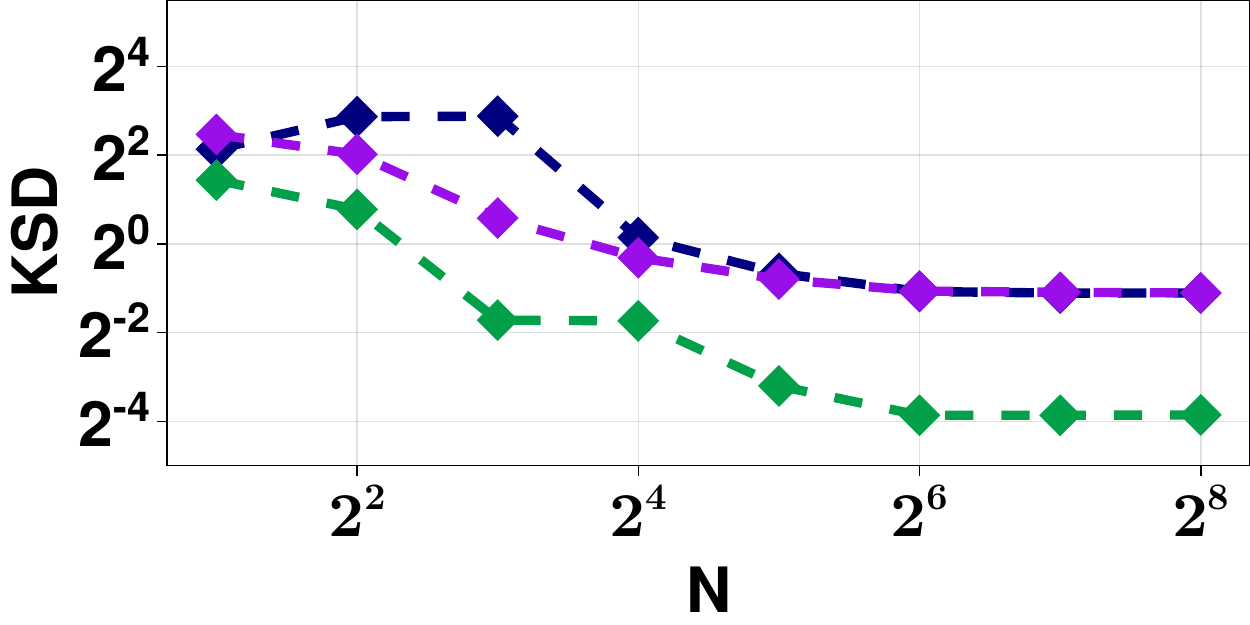}  
    \subcaption*{Spaceships}
    \end{subfigure}
    \begin{subfigure}{\linewidth}
    \centering 
    \includegraphics[width=0.5\linewidth]{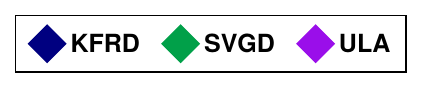}
    \end{subfigure} 
    \caption{\textbf{Two-dimensional posteriors:} average KSD at stopping time between $\pi_1$ and ensembles of size $J \in \{100, 400\}$ generated by gradient-based samplers. A missing point indicates that a method was unstable at that setting of $N$. %
    }
    \label{fig:KSDvsN_gb_parallel}
\end{figure}

\begin{figure}[h]
\centering 
\small $J = 400$ \\[0.1cm] 
    \begin{subfigure}{0.32\linewidth}
    \centering  
    \includegraphics[width=\linewidth]{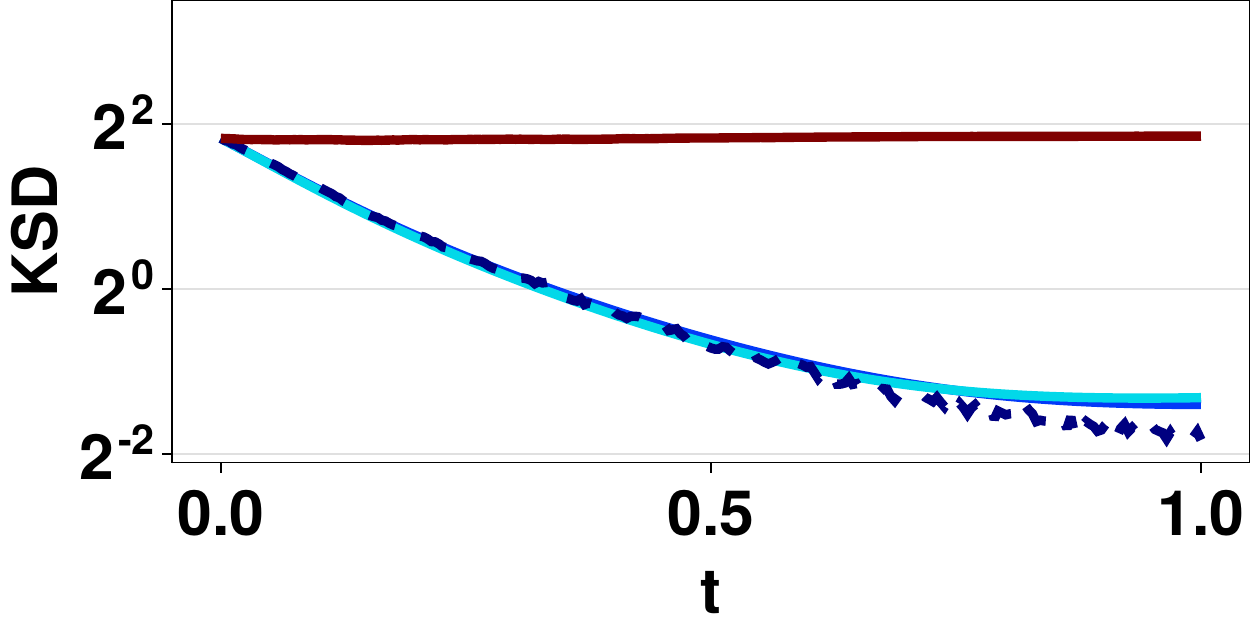}        
    \end{subfigure}
    \begin{subfigure}{0.32\linewidth}
    \centering 
    \includegraphics[width=\linewidth]{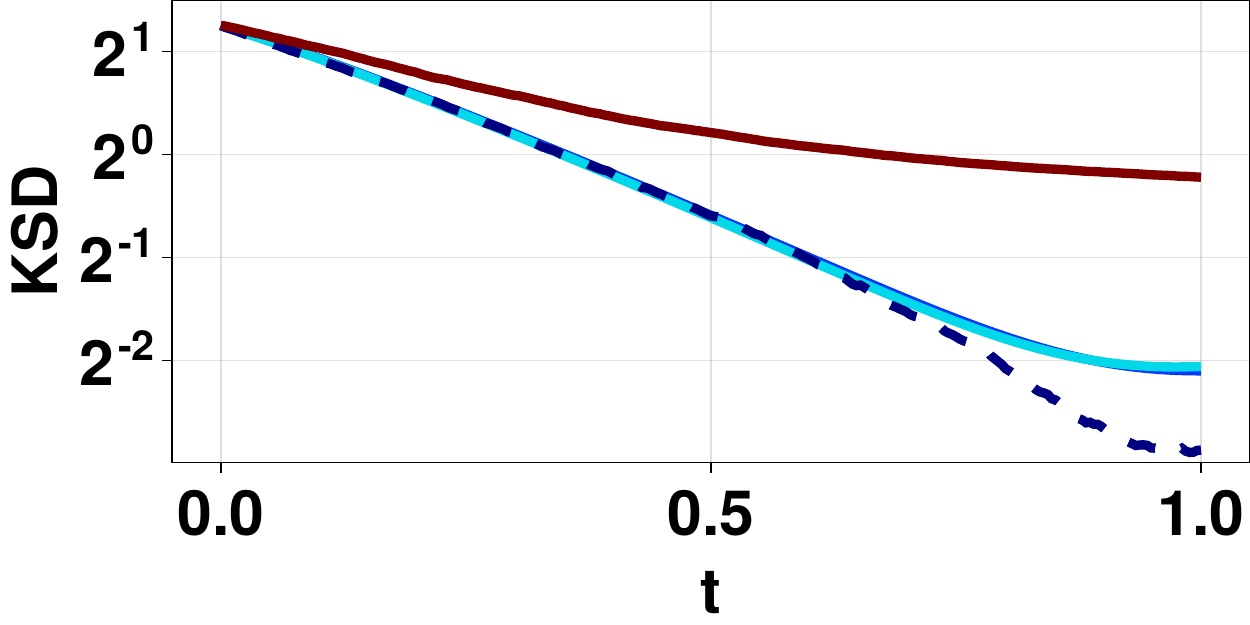}        
    \end{subfigure}
    \begin{subfigure}{0.32\linewidth}
    \centering 
    \includegraphics[width=\linewidth]{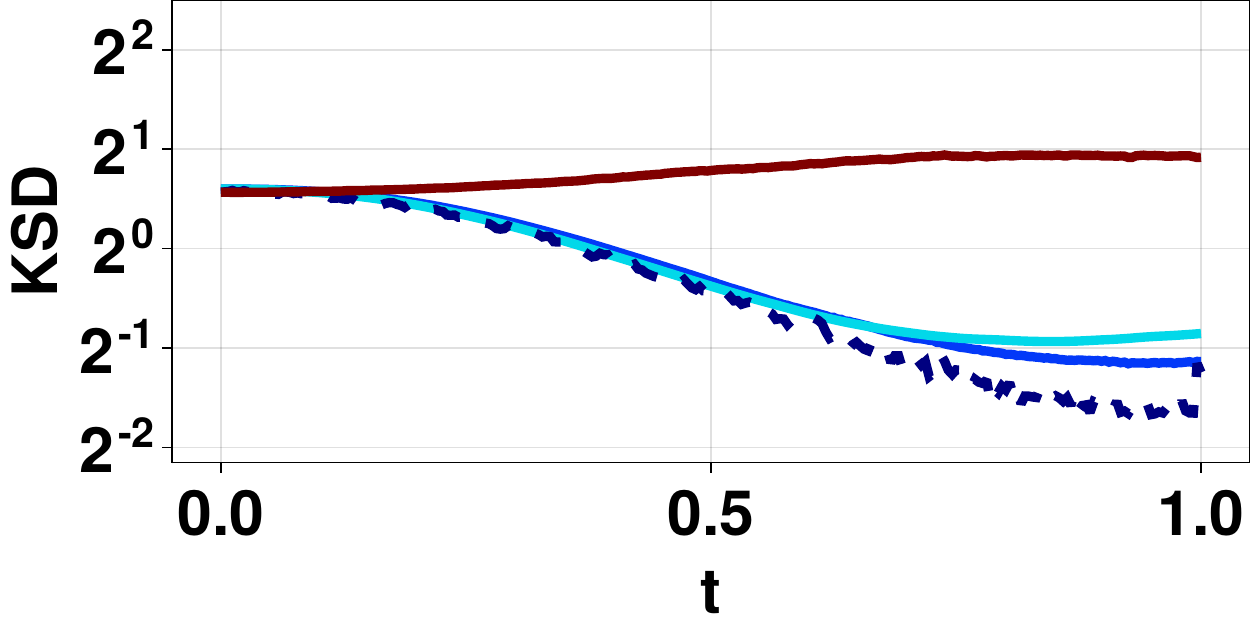}        
    \end{subfigure}
    \\
   \small  $J = 100$ \\[0.1cm] 
    \begin{subfigure}{0.32\linewidth}
    \includegraphics[width=\linewidth]{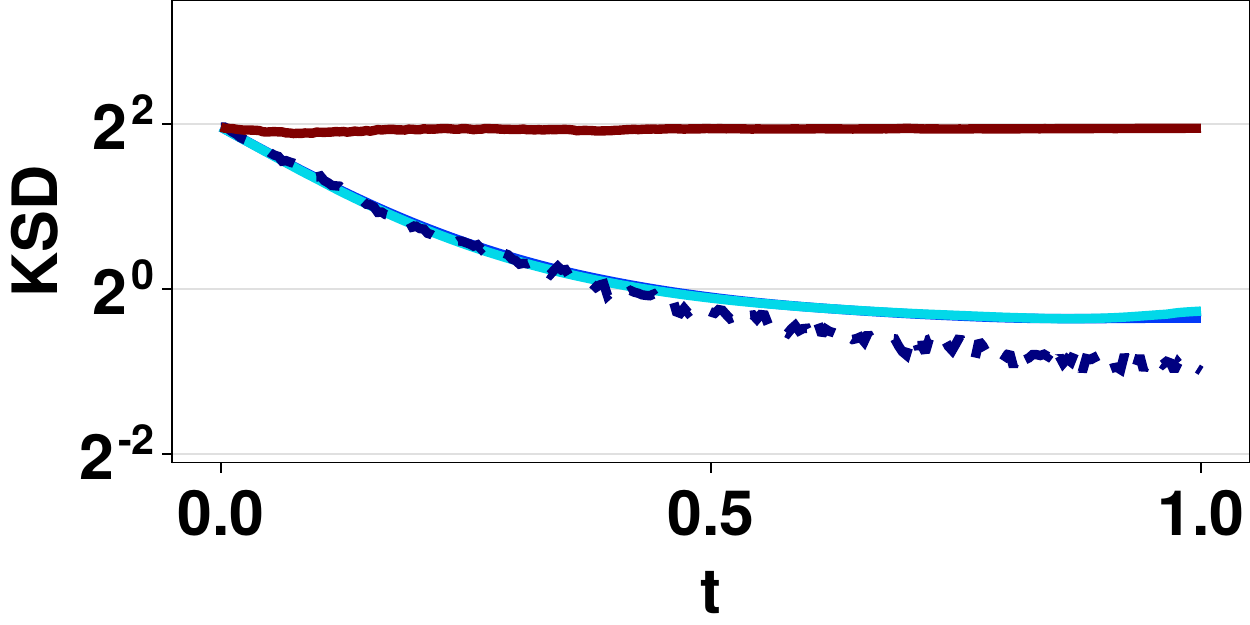}        \subcaption*{Donut}
    \end{subfigure}
    \begin{subfigure}{0.32\linewidth}
    \includegraphics[width=\linewidth]{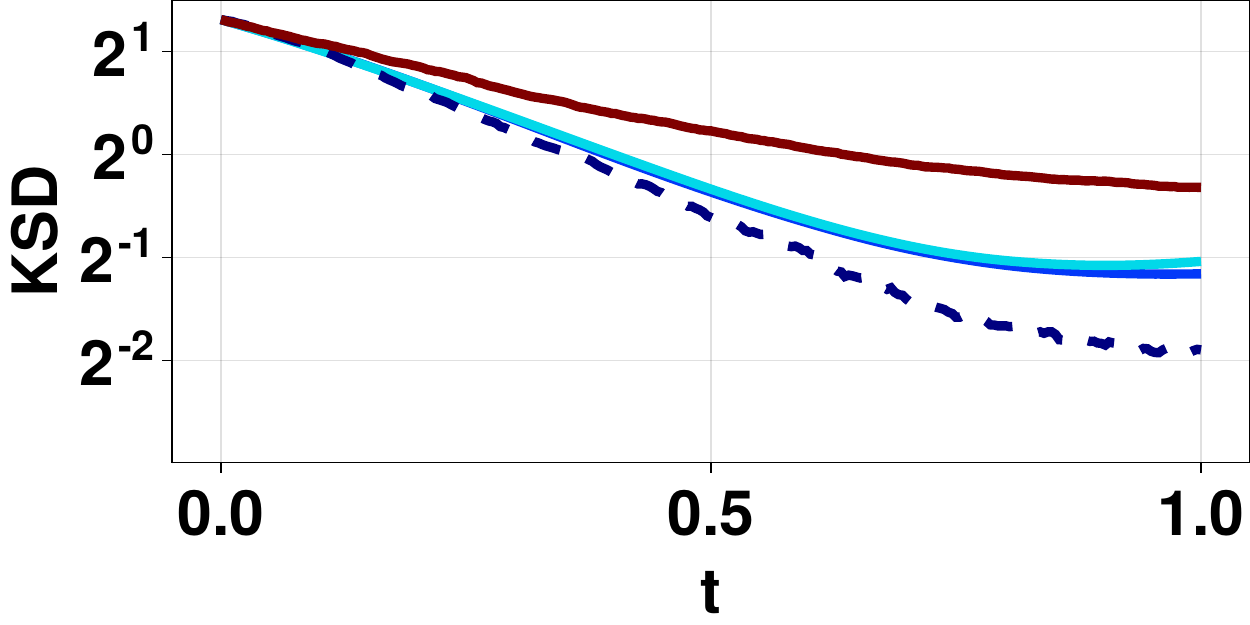}        \subcaption*{Butterfly}
    \end{subfigure}
    \begin{subfigure}{0.32\linewidth}
    \includegraphics[width=\linewidth]{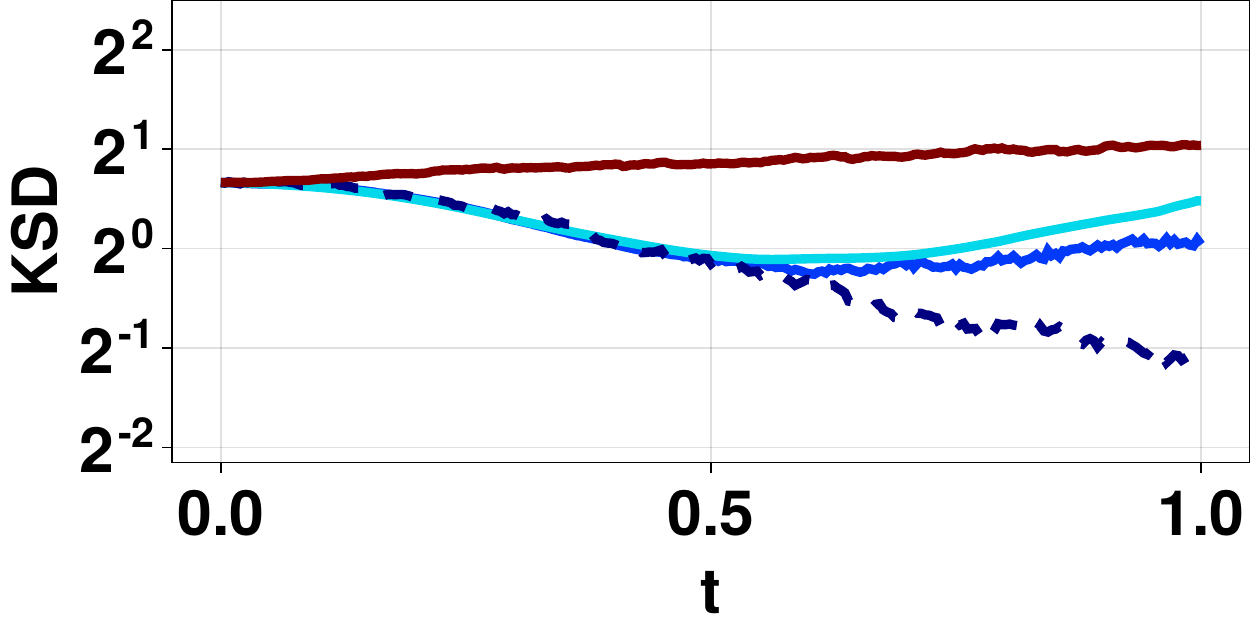}        \subcaption*{Spaceships}
    \end{subfigure}
    \begin{subfigure}{\linewidth}
    \centering 
    \includegraphics[width=0.95\linewidth]{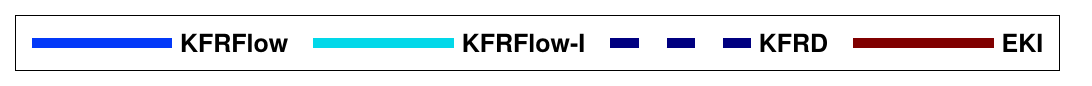}
\end{subfigure}
    \caption{\textbf{Two-dimensional posteriors:} evolution of KSD with $t$ for the unit-time methods KFRFlow, KFRFlow-I, KFRD, and EKI for ensembles of $J=400$ and $J = 100$ with $\Delta t = 2^{-8}$. KFRD is plotted with dashed lines because it requires gradients, whereas KFRFlow(-I) and EKI are gradient-free.} %
    \label{fig:ksdEvol2D}
\end{figure}

 In \cref{fig:KSDvsN_gf_par} we compare the performances of KFRFlow and KFRFlow-I to those of fellow gradient-free sampling algorithms EKI, EKS, and CBS, and in \cref{fig:KSDvsN_gb_parallel} we compare the performances of gradient-based KFRD, SVGD, and ULA. %
 We use the algorithms to generate target ensembles of size $J \in \{25, 50, 100, 200, 400\}$ with number of steps $N \in \{2^1, 2^2, \dots, 2^8\}$. For unit-time KFRFlow(-I), KFRD, and EKI, the resulting step-size is $1/N$, but for infinite-time EKS, CBS, SVGD, and ULA we must choose a stopping time $T > 0$, resulting in a step-size of $T/N$. We test a range of stopping times $T$ for EKS, CBS, SVGD, and ULA, regularization levels $\lambda$ for KFRFlow(-I), noise levels $\epsilon$ for KFRD, and temperature $\beta$ for CBS and report  
KSD corresponding to the best parameter settings for each algorithm and each $(J, N)$. %
 We use a fourth-order Adams--Bashforth discretization of KFRFlow because we find that it generates better samples than forward Euler at little additional cost, while by contrast we use forward Euler for SVGD because we find that SVGD does not benefit from multistep discretizations in these examples. 
 We use Euler--Maruyama discretizations for KFRD, EKI, EKS, CBS, and ULA.  The values of KSD we report are averages over 30 trials. %

 In \cref{fig:KSDvsN_gf_par} we see that KFRFlow and KFRFlow-I generally produce better-quality samples, as measured with KSD, than EKI, EKS, and CBS. %
 There are some settings in the donut example at which KFRFlow or EKI is unstable, but, interestingly, KFRFlow-I is stable across all settings of $(J, N)$, even in some cases when the gradient-based algorithms KFRD and SVGD (\cref{fig:KSDvsN_gb_parallel}) are not. In \cref{fig:KSDvsN_gb_parallel} we see that the performance of KFRD is comparable to that of ULA and generally exceeded by SVGD in these examples, though for small $N$  it is often the case that KFRFlow or KFRFlow-I yields comparable or better performance than SVGD. For additional details and results see \cref{app:2d}. %

 \subsection{Higher-Dimensional Funnel Distributions}
\label{sec:funnel}
Here for dimension $d \in \{5, 10, 15, \rev{20}\}$ we compare the performance of KFRFlow-I, KFRD, \rev{CBS, SVGD, and ULA} in sampling from ``funnel'' distributions of the form 
$
\pi_1(\bfx) = \calN(x_1; 0, 9)\calN(\bfx_{2:d}; \mathbf{0}, \exp(x_1) \mathbf{I}).
$
This family of distributions appears in \citet{neal2003slice} and is a common benchmark for sampling algorithms, e.g., \citet{arbel2021annealed, zhangDiffusionGenerativeFlow2023, xu2023mixflows}. 

\begin{figure}[h]
    \centering
    \includegraphics[width=\linewidth]{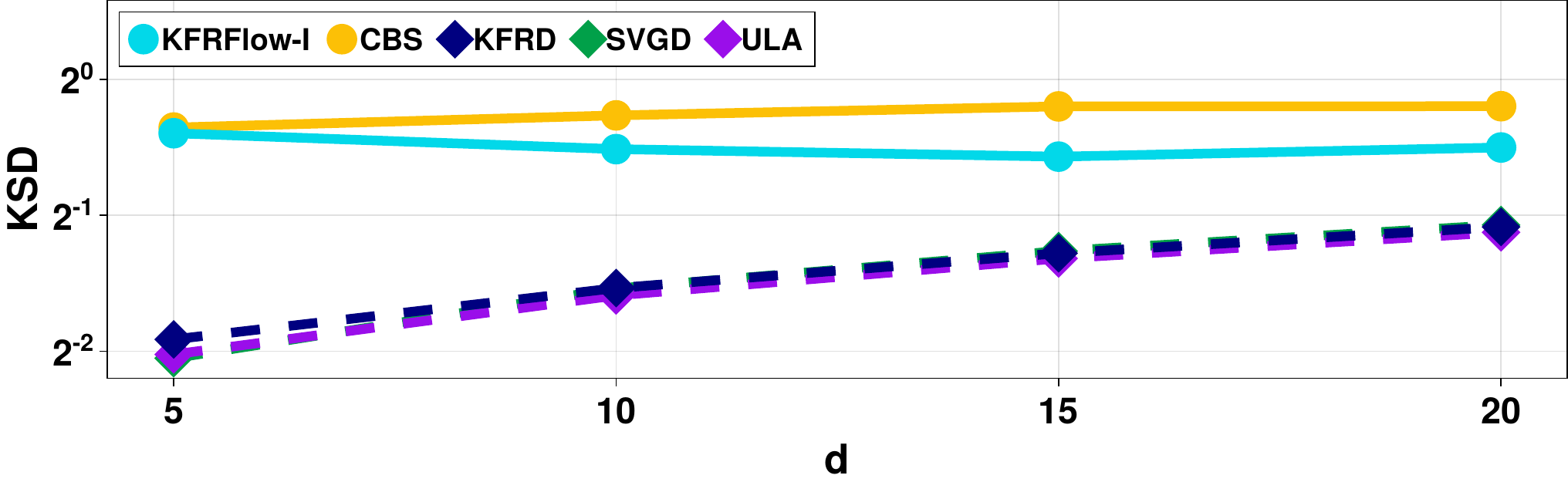}
    \caption{\textbf{Funnels:} average KSD at stopping time between $\pi_1$ and samples generated by KFRFlow-I, KFRD, CBS, SVGD, and ULA for $d \in \{5, 10, 15, 20\}$. Gradient-free algorithms are plotted with solid lines, while gradient-based algorithms are plotted with dashed lines.} %
    \label{fig:ksdVdim_funnel_serial}
\end{figure}

\begin{figure}[h] 
\centering 
    \begin{subfigure}{0.49\linewidth}
    \centering 
\scriptsize    KFRFlow-I \\[0.1cm]
        \includegraphics[width=\linewidth]{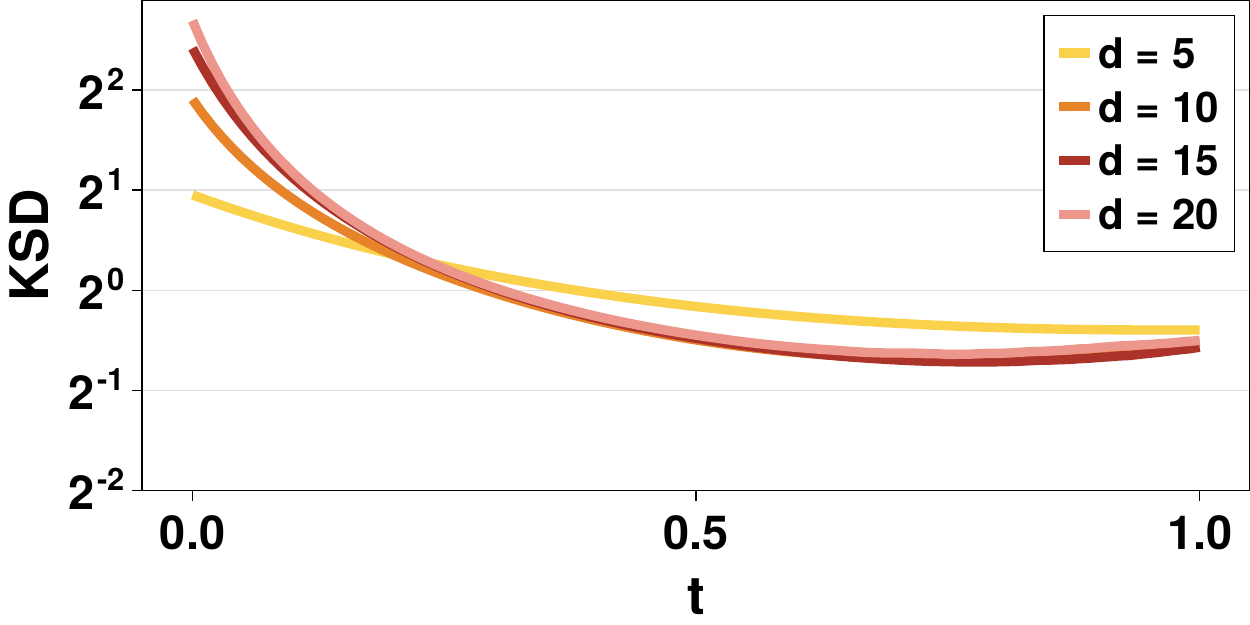}
    \end{subfigure}
    \begin{subfigure}{0.49\linewidth}
    \centering 
  \scriptsize  KFRD \\[0.1cm] 
        \includegraphics[width=\linewidth]{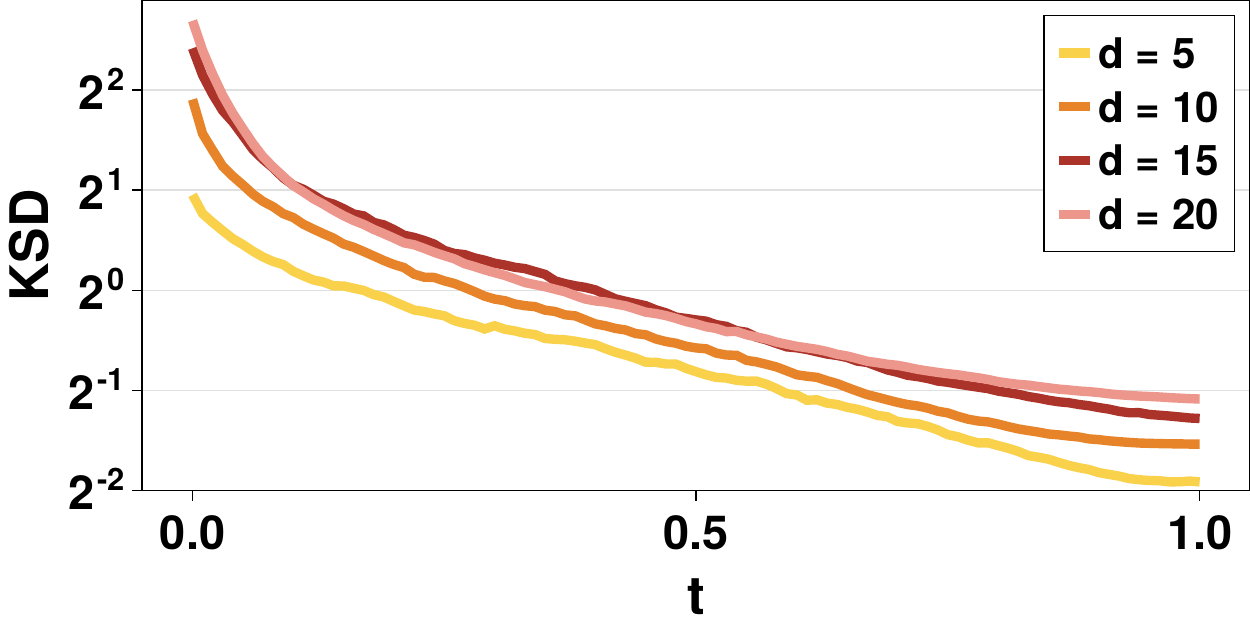}
    \end{subfigure}
    \caption{\textbf{Funnels:} evolution of KSD between $\pi_1$ and samples at time $t$ for KFRFlow-I (left) and KFRD (right). %
    }
    \label{fig:ksdEvolution_funnel}
\end{figure}

For each setting of $d$ we apply these algorithms 
to generate $J = 100$ samples from $\pi_1$.  For KFRFlow-I and KFRD we set $\Delta t = 0.01$, corresponding to $N = 100$ steps for the infinite-time algorithms CBS, SVGD, and ULA. As in \cref{sec:2Ddists} we optimize the hyperparameters for KFRFlow-I, KFRD, CBS, SVGD, and ULA via coarse direct search to minimize KSD between the final samples and $\pi_1$. %
EKI \rev{and EKS are} not applicable to the funnel because it is not a Bayesian posterior with a Gaussian likelihood, and we focus on KFRFlow-I rather than KFRFlow due to its demonstrated stability. The data in \cref{fig:ksdVdim_funnel_serial} and \cref{fig:ksdEvolution_funnel} are reflective of averaging the results of 30 independent trials.

In \cref{fig:ksdVdim_funnel_serial} we plot KSD between the ensembles and the funnel targets %
as a function of dimension $d$. We see that KSD increases with dimension for the gradient-based algorithms and that KFRD is competitive with SVGD and ULA at all values of $d$. We also see that KFRFlow-I generates better quality samples than CBS and, interestingly, that the quality of the samples generated by both of these gradient-free IPS algorithms does not seem to be meaningfully impacted by dimension. A more thorough investigation of this phenomenon, also visible in \cref{fig:ksdEvolution_funnel}, is a topic for future work. For additional details and results see \cref{app:funnel}.

	\section{Discussion and Future Work}
	\label{sec:discussion}
    We have introduced a mean-field ODE and corresponding interacting particle systems which approximately transport samples from $\pi_0$ to $\pi_1$ in unit time. We obtain the mean-field ODE by solving an elliptic PDE arising from the Fisher--Rao gradient flow of the negative log likelihood under the ansatz that the solution lies in a reproducing kernel Hilbert space. The RKHS form of the mean-field ODE gives rise to tractable, gradient-free interacting particle systems for sampling.
    
    \rev{Several lines of inquiry would enhance our understanding of the mean-field model and KFRFlow interacting particle systems. The tempered likelihood path $\pi_t \propto \pi_0^{1-t} \pi_1^t$ employed here is itself an interesting object of study: in addition to being a segment of the Fisher-Rao gradient flow of $\calF(\mu) = -\E_{\mu}[\log \frac{\pi_1}{\pi_0}]$, $(\pi_t)_{t \in [0,1]}$ can also be characterized as a unit-time rescaling of a Fisher-Rao gradient flow of $\mu \mapsto D_{\rm KL}(\mu \| \pi_1)$ \citep{domingo-enrichExplicitExpansionKullbackLeibler2023} and as a path of Kullback--Leibler divergence barycenters (e.g., \citet[Theorem 4.9]{amari2016information}); that is, 
    \begin{equation*}
    \pi_t = \argmin_{\mu} (1-t) D_{\rm KL} (\mu \| \pi_0) + t D_{\rm KL}(\mu \| \pi_1), \ \  t \in [0,1].
    \end{equation*}
    Thus, KFRFlow %
    can be equally well viewed as (i) early-stopping of maximum likelihood, (ii) gradient descent of $D_{\rm KL}(\cdot \| \pi_1)$, or (iii) a continuation method for minimizing $D_{\rm KL}(\cdot \| \pi_1)$. Perhaps equally intriguing is that to produce the KFRFlow mean-field model \eqref{eq:meanfield}, we take the path $(\pi_t)_{t \in [0,1]}$ natively corresponding to Fisher-Rao gradient flow and ``Wasserstein-ize'' it: we seek potentials $u_t: \R^d \to \R$ such that $- \nabla \cdot(\pi_t \nabla u_t) = \partial_t \pi_t$, leading to particle algorithms that transport samples collectively along $(\pi_t)_{t \in [0,1]}$ via the action of a gradient velocity field.
    Understanding how to interpret the path $(\pi_t)_{t \in [0,1]}$ %
    through a (kernelized) Wasserstein lens is thus of great interest.
    } 

    \rev{
    On a less abstract level, KFRFlow has appealing properties for practical scientific sampling applications---it is gradient-free, closed-form, inherently finite-time, and only requires an unnormalized likelihood---but focused effort is needed to move from ``vanilla'' algorithms as in \eqref{eq:IPS_importance} and \eqref{eq:ode_rehash} to more sophisticated implementations suitable for challenging problems in high dimensions. We are currently investigating computational strategies for this purpose, including reducing computational complexity through kernel appproximations and dimension reduction, exploiting or imposing structure and sparsity through well-designed choices of kernel, and adaptive time-stepping for the resulting systems of ODEs. %
    Finally,  %
    strengthening the numerical and statistical analysis of KFRFlow
    by understanding questions of approximation error and sample complexity, 
    particularly as these quantities relate to choice of kernel and time-stepping schedule, is a closely related and important area for future work, which will 
    inform the development of KFRFlow generalizations.}

\section*{Acknowledgements}
The authors thank the anonymous reviewers for their helpful suggestions and discussion. 
	AM and YM were supported by the Office of Naval Research, SIMDA (Sea Ice Modeling and Data Assimilation) MURI, award number N00014-20-1-2595 (Dr.~Reza Malek-Madani and Dr.~Scott Harper). AM was additionally supported by the NSF Graduate Research Fellowship under Grant No.\ 1745302.

\section*{Impact Statement}
This paper presents work whose goal is to advance the field of machine learning. There are many potential societal consequences of our work, none of which we feel must be specifically highlighted here.

\bibliography{references}
\bibliographystyle{icml2024}

\newpage
\appendix
\onecolumn
\section{Derivations and Proofs}
	\label{app:proofs}
	\subsection{Derivation of Kernel Fisher-Rao Flow Interacting Particle System}
 \label{app:proofMeanField}
	
 \rev{
We would like to approximate the mean-field ODE \eqref{eq:meanfield}
\begin{equation}
\dot X_t  =  v_t(X_t) = 
\E_{\rho_t}\! \left [ \nabla_1 K(X_{t}, X') M_{\rho_t}\inv K_{\rho_t}\left( \log\tfrac{\pi_1}{\pi_0} - \E_{\rho_t}\!\left[\log \tfrac{\pi_1}{\pi_0}\right]\right)(X') \right ], 
\label{eq:meanfield_fix}
\end{equation} 
with an interacting particle system $\{X_t^{(j)}\}_{j=1}^J$. To obtain such an IPS, we approximate the expectations in \eqref{eq:meanfield_fix} via Monte Carlo. Beginning with the outer expectation, we have  
\begin{equation}
	\dot X_t^{(j)} \approx \fdfrac{1}{J}\sum_{n=1}^J \nabla_1 K(X_t^{(j)}, X_t^{(n)}) M_{\rho_t} \inv K_{\rho_t} \left( \log\tfrac{\pi_1}{\pi_0} - \E_{\rho_t}\!\left[\log \tfrac{\pi_1}{\pi_0}\right]\right)(X_t^{(n)}).
	\label{eq:firstDisc}
\end{equation}
Next, we obtain a Monte Carlo approximation of $f_t \equiv M_{\rho_t}\inv K_{\rho_t} \left( \log\tfrac{\pi_1}{\pi_0} - \E_{\rho_t}\!\left[\log \tfrac{\pi_1}{\pi_0}\right]\right)$ by examining the system $M_{\rho_t}f_t = K_{\rho_t} \left( \log\tfrac{\pi_1}{\pi_0} - \E_{\rho_t}\!\left[\log \tfrac{\pi_1}{\pi_0}\right]\right)$,
\begin{equation*} 
	\iint_{\R^d\times \R^d}  f_t(z) \left\langle \nabla_1 K(y, \cdot), \;  \nabla_1 K(y, z) \right\rangle \, \rmd \pi_t(y)\, \rmd \pi_t(z)  %
	= \int_{\R^d} K(\cdot, y)\left (\log \ltfrac{\pi_1}{\pi_0}(y) - \E_{\pi_t}\left[\log \ltfrac{\pi_1}{\pi_0} \right] \right)\, \rmd\pi_t(y). %
\end{equation*} 
and approximating expectations on both sides via Monte Carlo,
\begin{equation}
	\fdfrac{1}{J^2}\sum_{m=1}^J \sum_{i=1}^J f_t(X_t^{(m)}) \left\langle \nabla_1 K(X_t^{(i)}, \cdot), \;  \nabla_1 K(X_t^{(i)}, X_t^{(m)}) \right\rangle  = \fdfrac{1}{J} \sum_{k=1}^J K(\cdot, X_t^{(k)})\left (\log \ltfrac{\pi_1}{\pi_0}(X_t^{(k)}) - \fdfrac{1}{J}\sum_{i=1}^J \log \ltfrac{\pi_1}{\pi_0}(X_t^{(i)}) \right).
	\label{eq:equationForFt}
\end{equation} 
We enforce \eqref{eq:equationForFt} at $\{X_t^{(\ell)}\}_{\ell=1}^J$ in order to obtain a system of $J$ equations for $f_t(X_t^{(1)}), \dots, f_t(X_t^{(J)})$,
\begin{multline}
	\fdfrac{1}{J^2}\sum_{m=1}^J \sum_{i=1}^J f_t(X_t^{(m)}) \left\langle \nabla_1 K(X_t^{(i)}, X_t^{(\ell)}), \;  \nabla_1 K(X_t^{(i)}, X_t^{(m)}) \right\rangle \\ = \fdfrac{1}{J} \sum_{k=1}^J K(X_t^{(\ell)}, X_t^{(k)})\left (\log \ltfrac{\pi_1}{\pi_0}(X_t^{(k)}) - \fdfrac{1}{J}\sum_{i=1}^J \log \ltfrac{\pi_1}{\pi_0}(X_t^{(i)}) \right), \quad \ell=1,\dots,J, 
	\label{eq:sumSys}
\end{multline} 
which we write succinctly as 
\begin{equation} 
M_t \bff_t = \sum_{k=1}^J \bfK_t(X_t^{(k)})\left (\log \ltfrac{\pi_1}{\pi_0}(X_t^{(k)}) - \fdfrac{1}{J}\sum_{i=1}^J \log \ltfrac{\pi_1}{\pi_0}(X_t^{(i)}) \right), 
\label{eq:discreteSys}
\end{equation} 
where $\bff_t = (f_t(X_t^{(1)}), \dots, f_t(X_t^{(J)}))$, $\bfK_t(\cdot) = (K(\cdot, X_t^{(1)}), \dots, K(\cdot, X_t^{(J)}))$, and $M_t \in \R^{J \times J}$ is given by 
\[
(M_t)_{\ell, m} = \fdfrac{1}{J}\sum_{i=1}^J  \langle\nabla_1 K(X_{t}^{(i)}, X_{t}^{(\ell)}), \, \nabla_1 K(X_{t}^{(i)}, X_{t}^{(m)}) \rangle, \quad \ell, m=1,\dots,J.
\] 
Notice that to arrive at \eqref{eq:discreteSys} we have canceled a common factor of $\frac{1}{J}$ on either side of \eqref{eq:sumSys}. Hence, we have 
\begin{equation*} 
	\bff_t = M_t\inv \sum_{k=1}^J \bfK_t(X_t^{(k)})\left (\log \ltfrac{\pi_1}{\pi_0}(X_t^{(k)}) - \fdfrac{1}{J}\sum_{i=1}^J \log \ltfrac{\pi_1}{\pi_0}(X_t^{(i)}) \right), 
\end{equation*} 
and we approximate $M_{\rho_t} \inv K_{\rho_t} \left( \log\tfrac{\pi_1}{\pi_0} - \E_{\rho_t}\!\left[\log \tfrac{\pi_1}{\pi_0}\right]\right)(X_t^{(\ell)}) $ in \eqref{eq:firstDisc} with $f_t(X_t^{(\ell)})$, writing the result in vector form, 
\begin{equation*}
\dot X_t^{(j)} = \ltfrac{1}{J} \begin{pmatrix}\nabla_1 K(X_t^{(j)}, X_t^{(1)}), \dots, \nabla_1 K(X_t^{(j)}, X_t^{(J)}) \end{pmatrix}  M_t\inv  \sum_{k=1}^J \bfK_t(X_t^{(k)})\left (\log \ltfrac{\pi_1}{\pi_0}(X_t^{(k)}) - \fdfrac{1}{J}\sum_{i=1}^J \log \ltfrac{\pi_1}{\pi_0}(X_t^{(i)}) \right),
\end{equation*}
and obtaining the IPS \eqref{eq:IPS_ode}.
}

	\subsection{Proof of \cref{thm:ctsTime}}
 \label{app:proofCtsTime}
	Notice that time only enters the update equation \eqref{eq:IPS_importance} through the importance weights $w_t^{(k)}$. To obtain the continuous time limiting ODE we rearrange, divide by $\Delta t$ on both sides, and take $\Delta t \to 0$,
	\begin{equation*}
		\lim_{\Delta t \to 0} \frac{X_{t+ \Delta t}^{(j)} - X_{t}^{(j)}}{\Delta t} = %
  \lim_{\Delta t \to 0} - \begin{pmatrix}\nabla_1 K(X_{t}^{(j)}, X_{t}^{(1)}) & \cdots & \nabla_1 K(X_{t}^{(j)}, X_{t}^{(J)}) \end{pmatrix} M_t\inv  \sum_{k=1}^J \frac{\frac{1}{J} - w_t^{(k)}}{\Delta t} \begin{pmatrix} K(X_{t}^{(k)}, X_{t}^{(1)})\\ \vdots \\ K(X_{t}^{(k)}, X_{t}^{(J)}) \end{pmatrix}.
	\end{equation*} 
	Examining the terms above involving $\Delta t$, we see that for $k \in \{1, \dots, J\}$ we have 
		\begin{align*}
			\lim_{\Delta t \to 0} \frac{\frac{1}{J} - w_t^{(k)}}{\Delta t} &= -\lim_{\Delta t \to 0} \frac{ \frac{(\frac{\pi_1}{\pi_0}(X_t^{(k)}))^{\Delta t}}{\sum_{i=1}^J (\frac{\pi_1}{\pi_0}(X_t^{(i)}))^{\Delta t}} - \frac{1}{J}}{\Delta t} = -\lim_{\Delta t \to 0} \frac{\frac{(\frac{\pi_1}{\pi_0}(X_t^{(k)}))^{\Delta t}}{\sum_{i=1}^J (\frac{\pi_1}{\pi_0}(X_t^{(i)}))^{\Delta t}} - \frac{(\frac{\pi_1}{\pi_0}(X_t^{(k)}))^{0}}{\sum_{i=1}^J (\frac{\pi_1}{\pi_0}(X_t^{(i)}))^{0}}}{\Delta t}  \\
			 &= -\frac{\rmd}{\rmd s} \left.  \frac{(\frac{\pi_1}{\pi_0}(X_t^{(k)}))^{s}}{\sum_{i=1}^J (\frac{\pi_1}{\pi_0}(X_t^{(i)}))^{s}} \right |_{s = 0} \\ 
			&= - \left. \frac{ (\frac{\pi_1}{\pi_0}(X_t^{(k)}))^{s}\log\frac{\pi_1}{\pi_0}(X_t^{(k)}){\displaystyle \sum_{i=1}^J} (\frac{\pi_1}{\pi_0}(X_t^{(i)}))^{s} -  (\frac{\pi_1}{\pi_0}(X_t^{(k)}))^{s}{\displaystyle \sum_{i=1}^J} (\frac{\pi_1}{\pi_0}(X_t^{(i)}))^{s} \log \frac{\pi_1}{\pi_0}(X_t^{(i)}) }{\left(\sum_{i=1}^J (\frac{\pi_1}{\pi_0}(X_t^{(i)}))^{s} \right)^2} \right |_{s = 0} \\
			& = -\frac{J \log\frac{\pi_1}{\pi_0} (X_t^{(k)}) - \sum_{i=1}^j \log\frac{\pi_1}{\pi_0} (X_t^{(i)})}{J^2} = -\frac{1}{J} \left(\log\frac{\pi_1}{\pi_0} (X_t^{(k)}) - \frac{1}{J}\sum_{i=1}^J \log\frac{\pi_1}{\pi_0} (X_t^{(i)})\right).
		\end{align*}
	Hence, the ODE arising from the limit of \eqref{eq:IPS_importance} as $\Delta t \to 0$ is the KFRFlow interacting particle system \eqref{eq:IPS_ode}
 \begin{small} 
	\begin{equation*} 
		\dot X_t^{(j)} = \begin{pmatrix}\nabla_1 K(X_{t}^{(j)}, X_{t}^{(1)}) & \cdots & \nabla_1 K(X_{t}^{(j)}, X_{t}^{(J)}) \end{pmatrix} M_t\inv %
		\fdfrac{1}{J} \sum_{k=1}^J \left(\log\ltfrac{\pi_1}{\pi_0}(X_t^{(k)}) - \fdfrac{1}{J}\sum_{i=1}^J \log\ltfrac{\pi_1}{\pi_0} (X_t^{(i)}) \right) \begin{pmatrix} K(X_{t}^{(k)}, X_{t}^{(1)})\\ \vdots \\ K(X_{t}^{(k)}, X_{t}^{(J)}) \end{pmatrix},
	\end{equation*} 
 \end{small}
	with initial condition $\{X_0^{(j)}\}_{j=1}^J \overset{\rm i.i.d.}{\sim} \pi_0$. %

\section{Kernel Fisher--Rao Diffusion}
\label{app:KFRD}
The continuity equation \eqref{eq:vt_must_satisfy} which we solve for the velocity $v_t$ can equivalently be written for any $\epsilon > 0$ as a {Fokker--Planck equation} 
\begin{equation*} 
 \partial_t\pi_t = -\nabla \cdot (\pi_t (v_t + \epsilon \nabla \log \pi_t)) + \epsilon \nabla^2 \pi_t,
\end{equation*} 
corresponding to the SDE
\begin{equation}
	\rmd X_t = (v_t(X_t) + \epsilon \nabla \log \pi_t(X_t)) \,\rmd t + \sqrt{2\epsilon} \,\rmd W_t, \quad t \in [0,1] 
	\label{eq:SDE}. 
\end{equation}
This SDE possesses the same marginal distributions as the ODE \eqref{eq:dynamics}. Using the same interacting particle approximation for the mean-field velocity $v_t$ \eqref{eq:meanfield} that we use to define deterministic KFRFlow \eqref{eq:IPS_ode}, we can also define a \textit{stochastic} interacting particle system for approximately traversing the geometric mixture $\pi_t \propto \pi_0^t \pi_1^{1-t}$,  
\begin{equation} 
    \rmd X_t^{(j)} = \nabla\bfK_t (X_t^{(j)})\t M_t\inv  
    \fdfrac{1}{J} \sum_{k=1}^J \left(\log\ltfrac{\pi_1}{\pi_0}(X_t^{(k)}) - \fdfrac{1}{J}\sum_{i=1}^J \log\ltfrac{\pi_1}{\pi_0} (X_t^{(i)}) \right)  \bfK_t(X_t^{(k)}) \, \rmd t  
    + \epsilon \nabla \log\pi_t(X_t^{(j)}) \, \rmd t + \sqrt{2\epsilon} \, \rmd W_t ,
    \label{eq:KFRD}
\end{equation} 
where we have used the notation $\mathbf{K}_t(\cdot) = (K(\cdot, X_t^{(1)}), \dots, K(\cdot, X_t^{(J)}))\t$ and $\nabla \bfK_t \in \R^{J \times d}$ is the Jacobian of $\bfK_t$. \Cref{eq:KFRD} is obtained by applying Monte Carlo approximations to $v_t$ in the SDE \eqref{eq:SDE}, where $v_t$ is as in \eqref{eq:meanfield}. Because we know $\pi_t \propto \pi_0^{1-t}\pi_1^t$ explicitly, the score of $\pi_t$ can be computed directly as $\nabla \log\pi_t = (1-t) \nabla \log\pi_0 + t \nabla \log\pi_1$. We refer to the interacting particle system \eqref{eq:KFRD} as \textbf{Kernel Fisher--Rao Diffusion} (KFRD) and note that it can be simulated with any off-the-shelf SDE solver, for example Euler--Maruyama. 
\newpage 

\section{Additional Numerical Results}
\label{app:numerics} 
\subsection{Two-Dimensional Bayesian Posteriors}
\label{app:2d} 
\subsubsection{Experimental setup}
	We apply KFRFlow \eqref{eq:IPS_ode}, KFRFlow-I \eqref{eq:IPS_importance}, and KFRD to sample three two-dimensional densities. %
 In all three cases %
    $\pi_1$ is a Bayesian posterior proportional to $\pi_0 \ell$ for a likelihood $\ell = \pi(y^* \mid \cdot)$ of the form
	 \[
	 \ell(x) \propto \exp\left(-\frac{1}{\sigma_\varepsilon^{2}}\|y^* - G(x) \|_2^2 \right),
	 \]
	 i.e., $y^* \in \R$ is Gaussian with mean $G(x)$ and variance $\sigma_\varepsilon^2$. 
	Definitions of the three likelihoods and descriptions of the deformation behavior they entail may be found in \cref{tab:bayesianPosteriors}.
	\begin{table}[H]
		\centering 
  		\caption{Likelihoods for the two-dimensional Bayesian example problems} 
		\begin{tabular}{ccccc}
			$G(x)$ & $y^*$ & $\sigma_\epsilon^2$ & Behavior & Nickname \\ 
			\toprule
			$\sqrt{x_1^2 + x_2^2}$  & 2 & $0.25^2 $ & Concentration & Donut \\ 
			\midrule 
			\makecell{$\sin(x_2)$ \\ $+$ \\ $\cos(x_1)$} & -1 & $0.6^2$ & Bimodality & Butterfly \\ 
			\midrule 
			\makecell{$\sin(x_1x_2)$ \\$+$ \\ $\cos(x_1x_2)$} & -1 & $0.5^2$ & Multimodality & Spaceships \\ %
			\midrule 
		\end{tabular}
		\label{tab:bayesianPosteriors}
	\end{table}
We compare the sampling performances of KFRFlow, KFRFlow-I, and KFRD to those of EKI, EKS, CBS, SVGD, and ULA.   We use the algorithms to generate target ensembles of size $J \in \{25, 50, 100, 200, 400\}$ with number of steps $N \in \{2^1, 2^2, \dots, 2^8\}$. For unit-time KFRFlow(-I), KFRD, and EKI, the resulting step-size is $1/N$, but for infinite-time EKS, CBS, SVGD, and ULA we must choose a stopping time $T > 0$, resulting in a step-size of $T/N$. We test a range of stopping times $T$ for EKS, CBS, SVGD, and ULA, regularization levels $\lambda$ for KFRFlow(-I), noise levels $\epsilon$ for KFRD, and temperature $\beta$ for CBS and report KSD corresponding to the best parameter settings for each $(J, N)$. %
We do not inflate $M_t$ in KFRD in these two-dimensional examples (i.e., $\lambda = 0$ for KFRD). The resulting choices of parameters for each algorithm and setting of $(J, N)$ can be seen in \cref{fig:params2D}. 

\begin{figure}[h]
 \centering 
 Donut \\ 
\begin{subfigure}{0.19\linewidth}
\includegraphics[width=\linewidth]{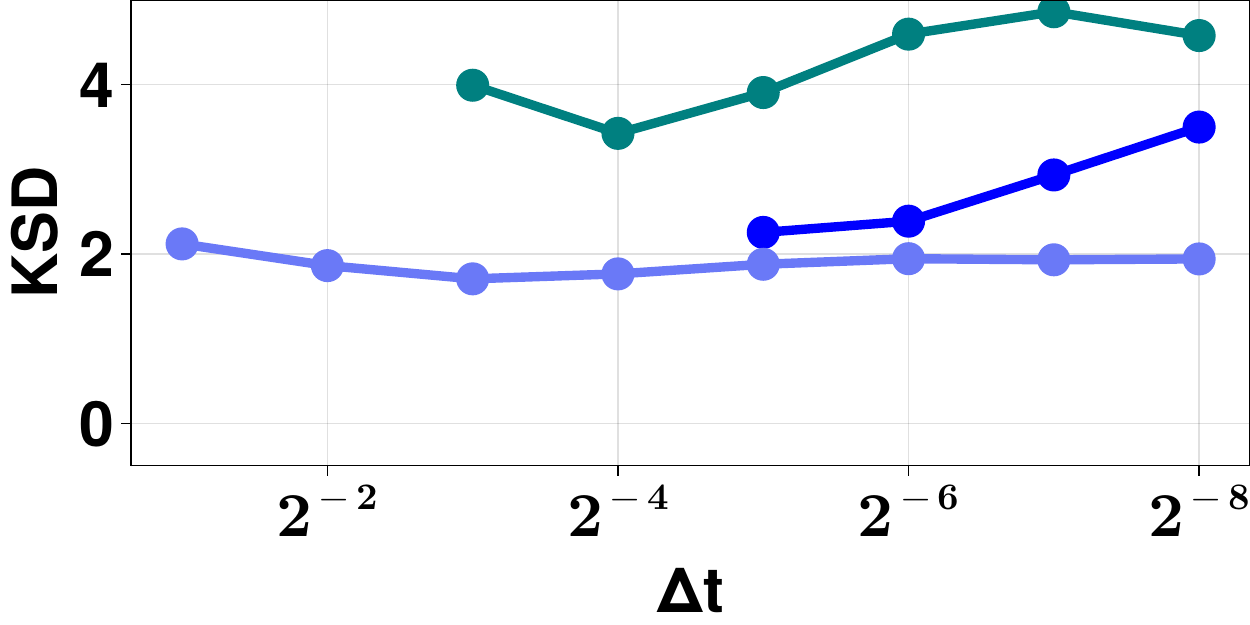}    
\end{subfigure}
\begin{subfigure}{0.19\linewidth}
\includegraphics[width=\linewidth]{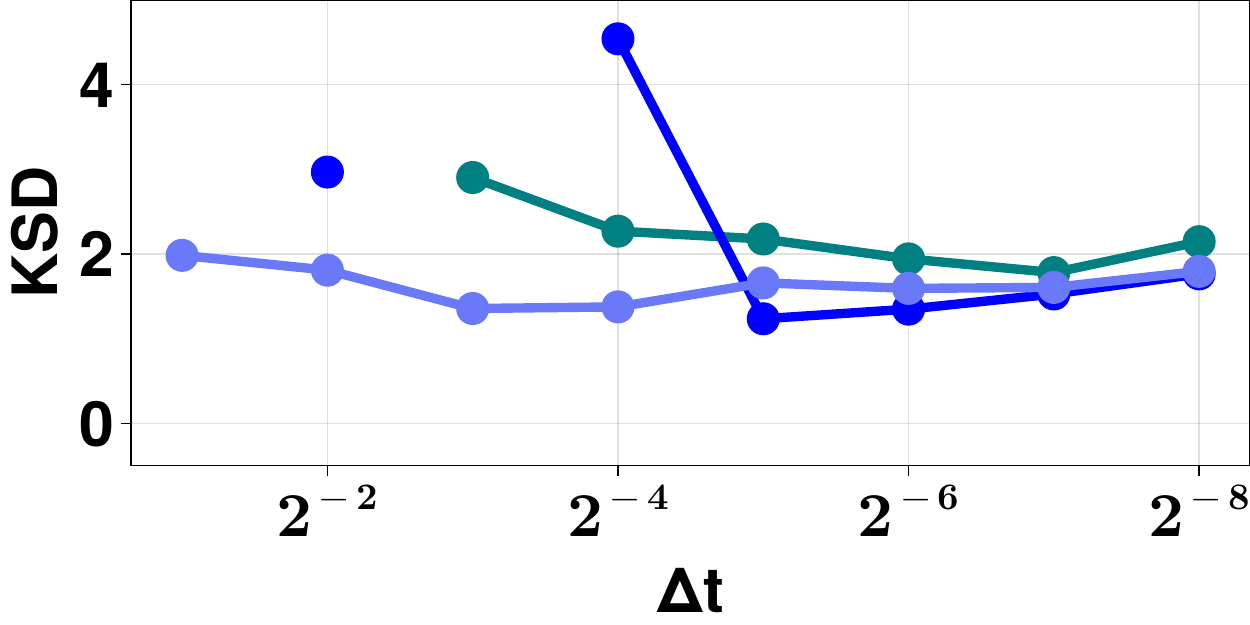}    
\end{subfigure}
\begin{subfigure}{0.19\linewidth}
\includegraphics[width=\linewidth]{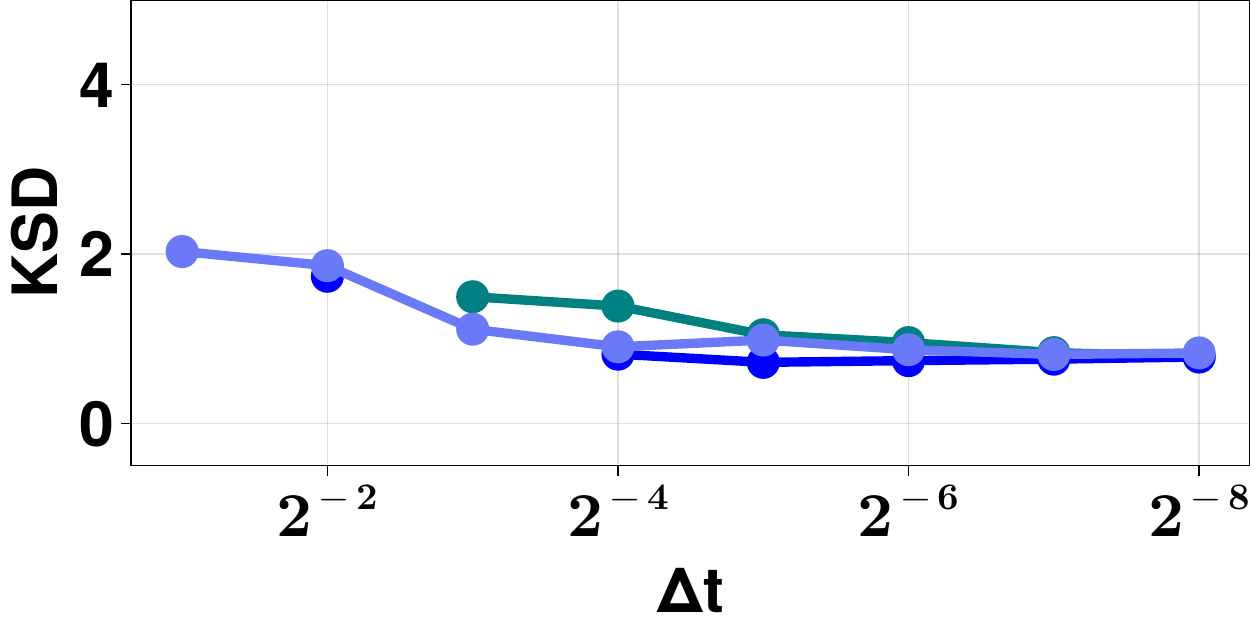}    
\end{subfigure}
\begin{subfigure}{0.19\linewidth}
\includegraphics[width=\linewidth]{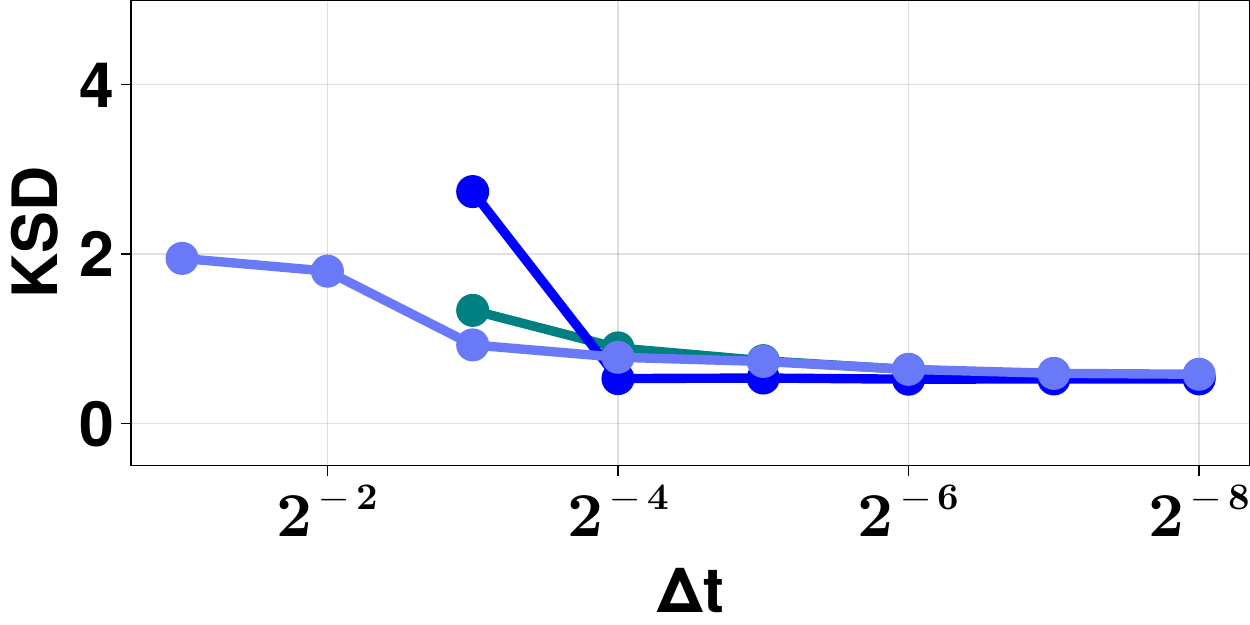}    
\end{subfigure}
\begin{subfigure}{0.19\linewidth}
\includegraphics[width=\linewidth]{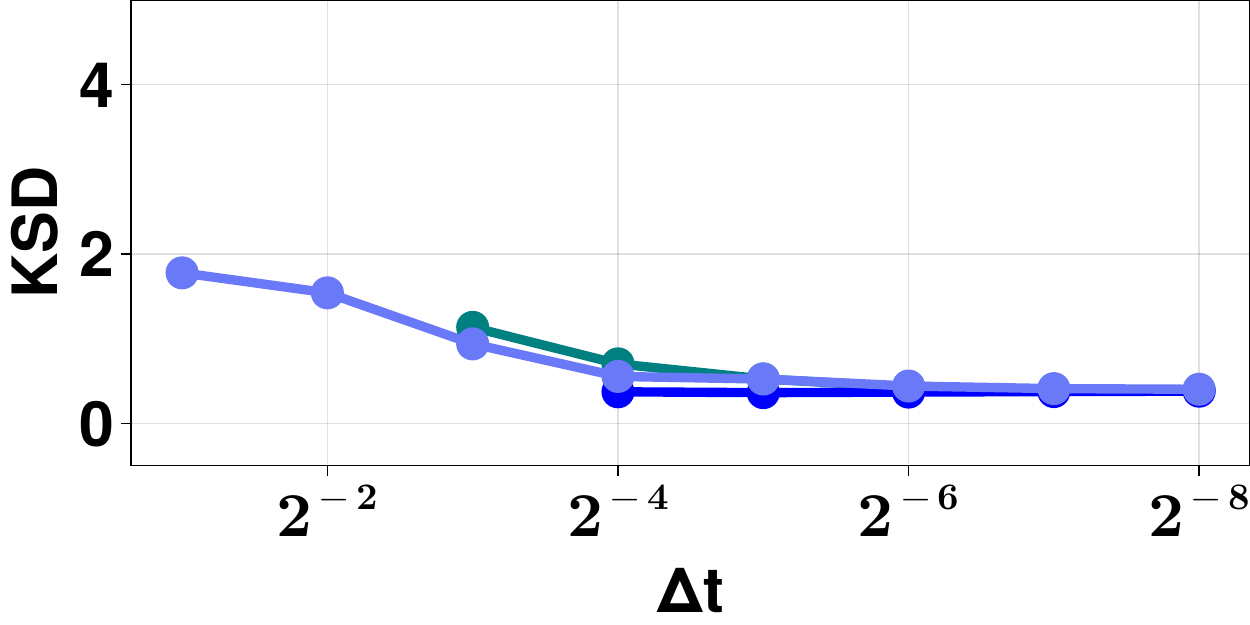}    
\end{subfigure}
\\ 
 Butterfly \\ 
\begin{subfigure}{0.19\linewidth}
\includegraphics[width=\linewidth]{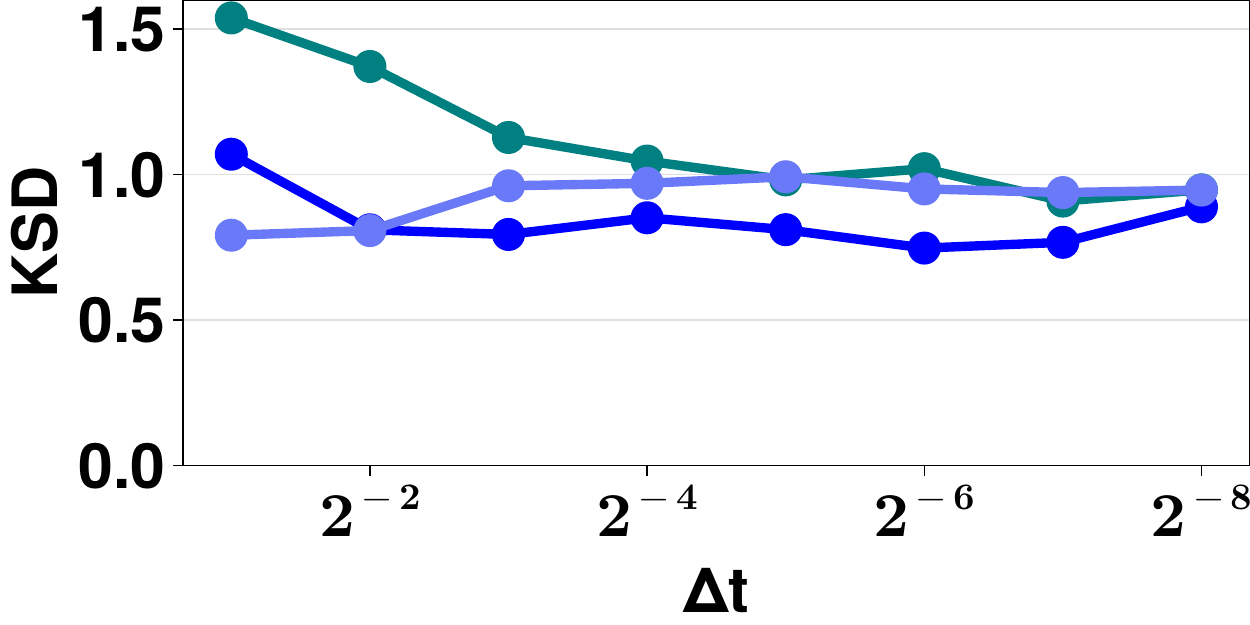}    
\end{subfigure}
\begin{subfigure}{0.19\linewidth}
\includegraphics[width=\linewidth]{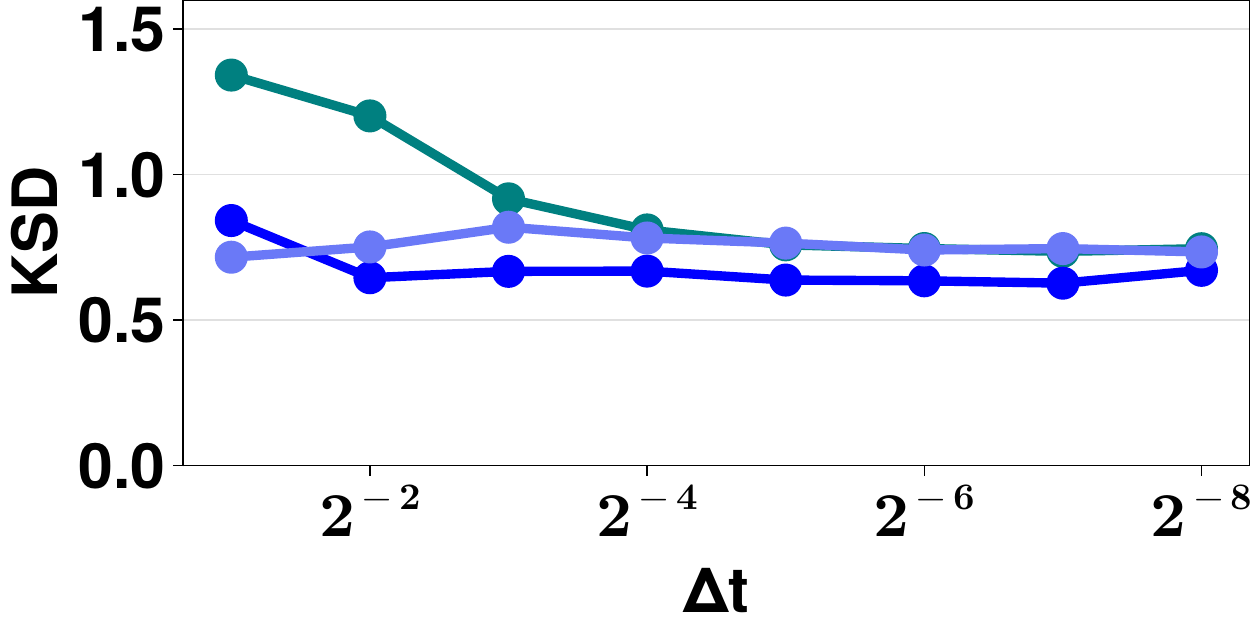}    
\end{subfigure}
\begin{subfigure}{0.19\linewidth}
\includegraphics[width=\linewidth]{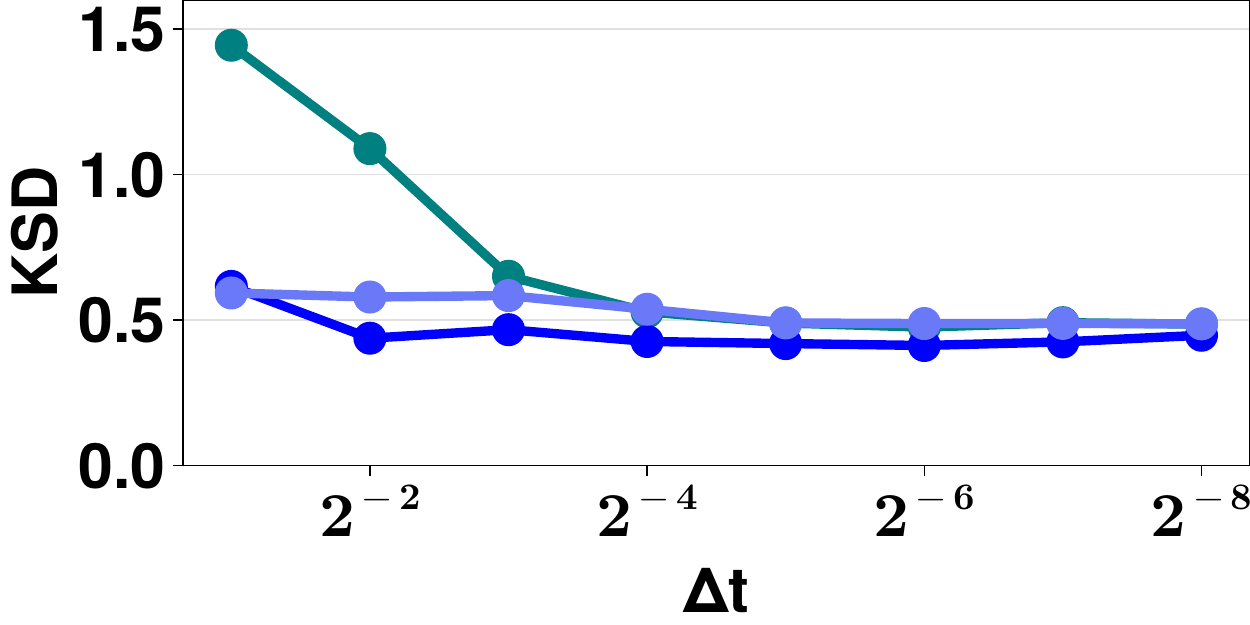}    
\end{subfigure}
\begin{subfigure}{0.19\linewidth}
\includegraphics[width=\linewidth]{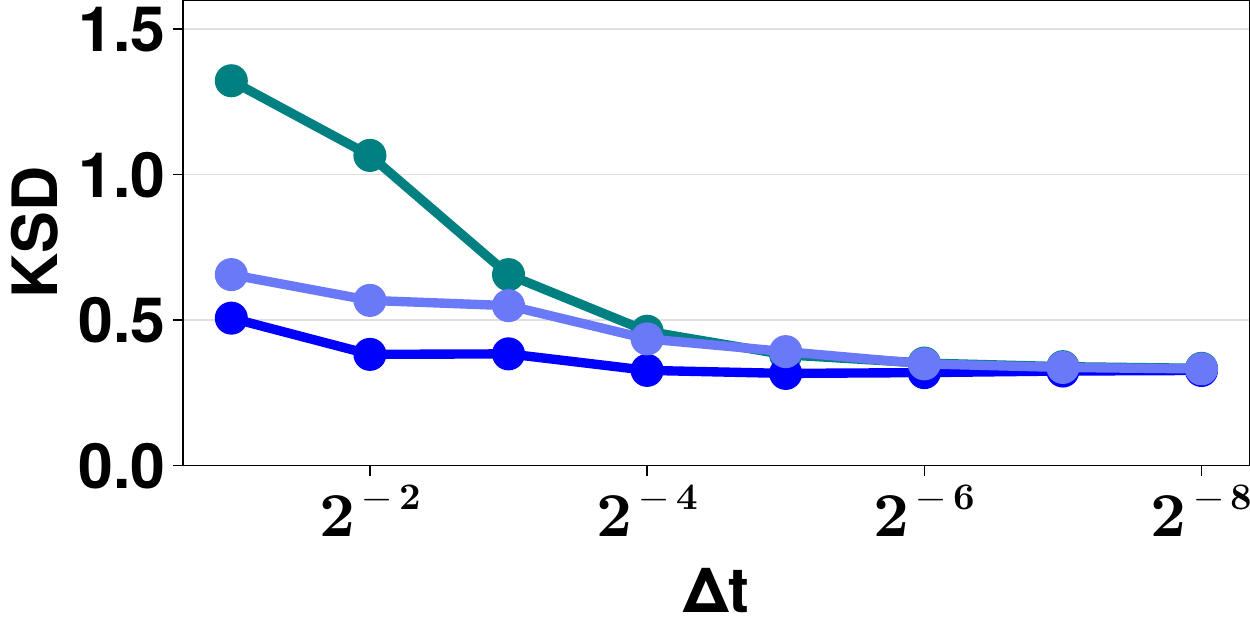}    
\end{subfigure}
\begin{subfigure}{0.19\linewidth}
\includegraphics[width=\linewidth]{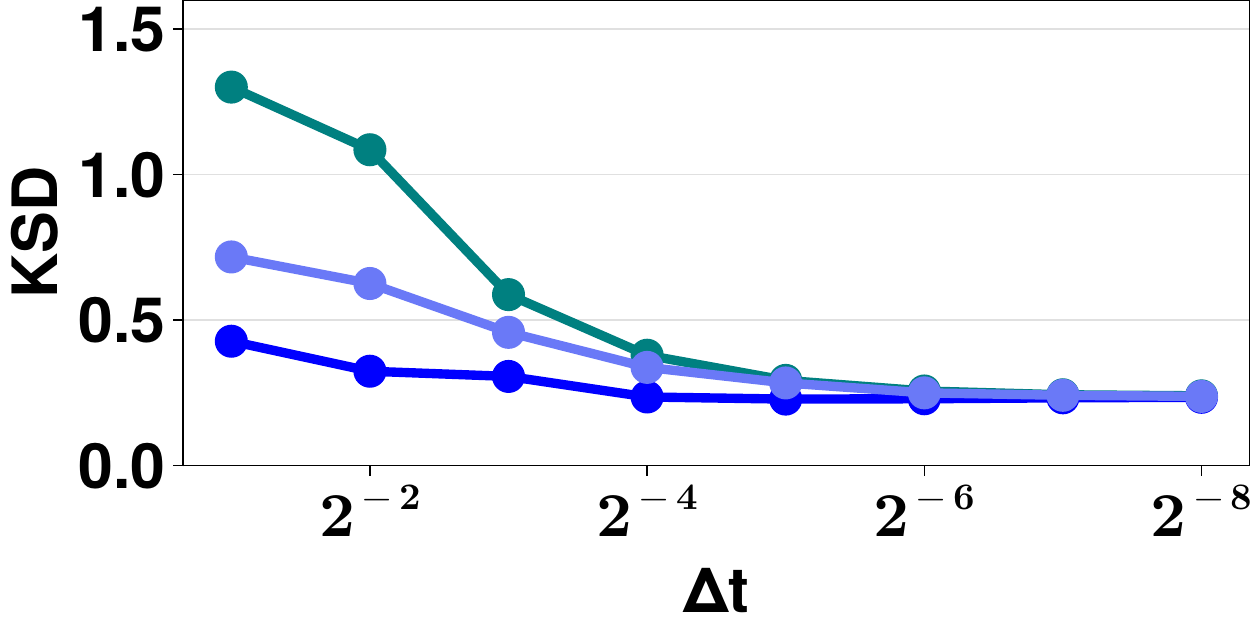}    
\end{subfigure}
\\
 Spaceships \\ 
\begin{subfigure}{0.19\linewidth}
\includegraphics[width=\linewidth]{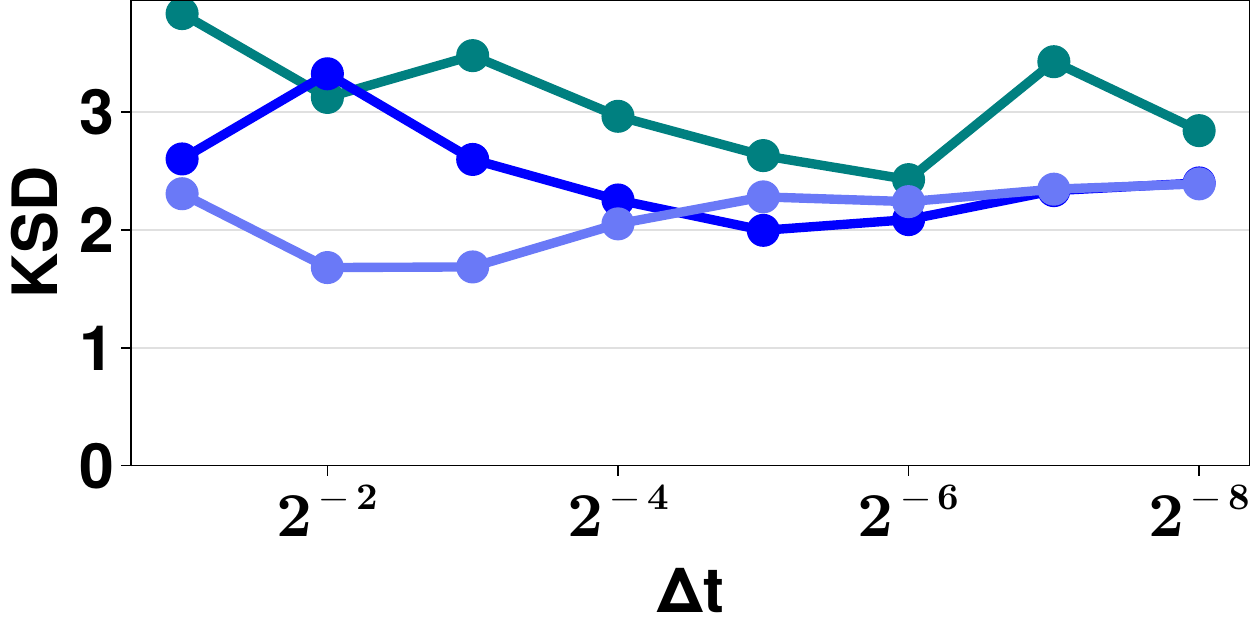}   
\subcaption*{J=25}
\end{subfigure}
\begin{subfigure}{0.19\linewidth}
\includegraphics[width=\linewidth]{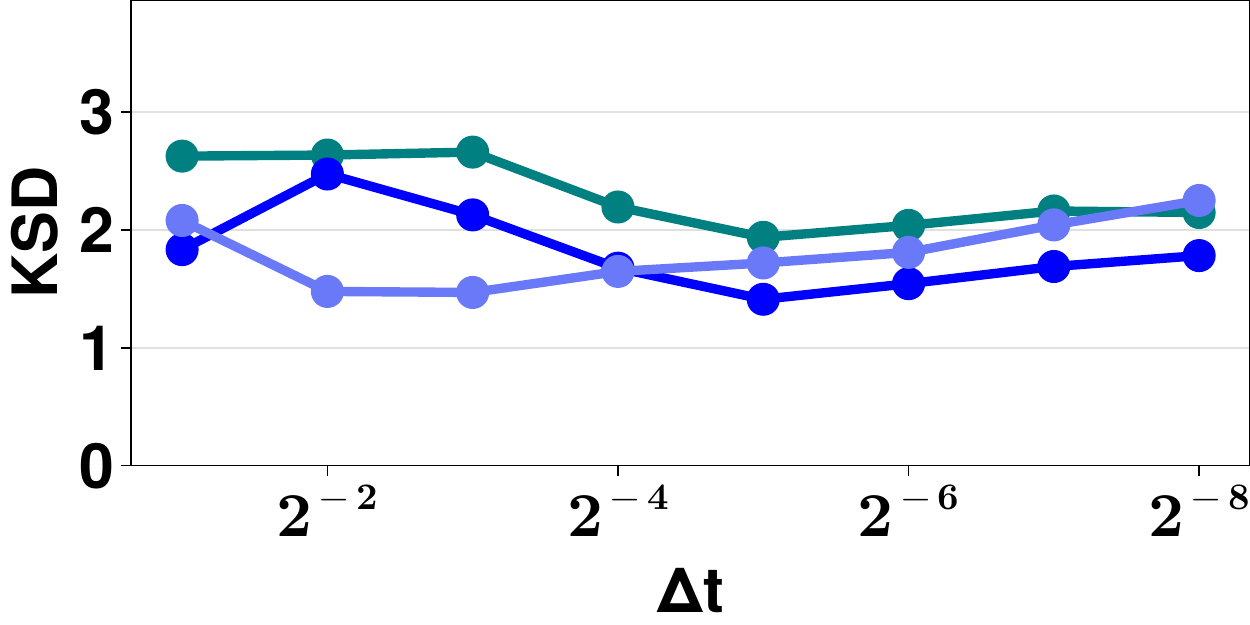}  
\subcaption*{J=50}
\end{subfigure}
\begin{subfigure}{0.19\linewidth}
\includegraphics[width=\linewidth]{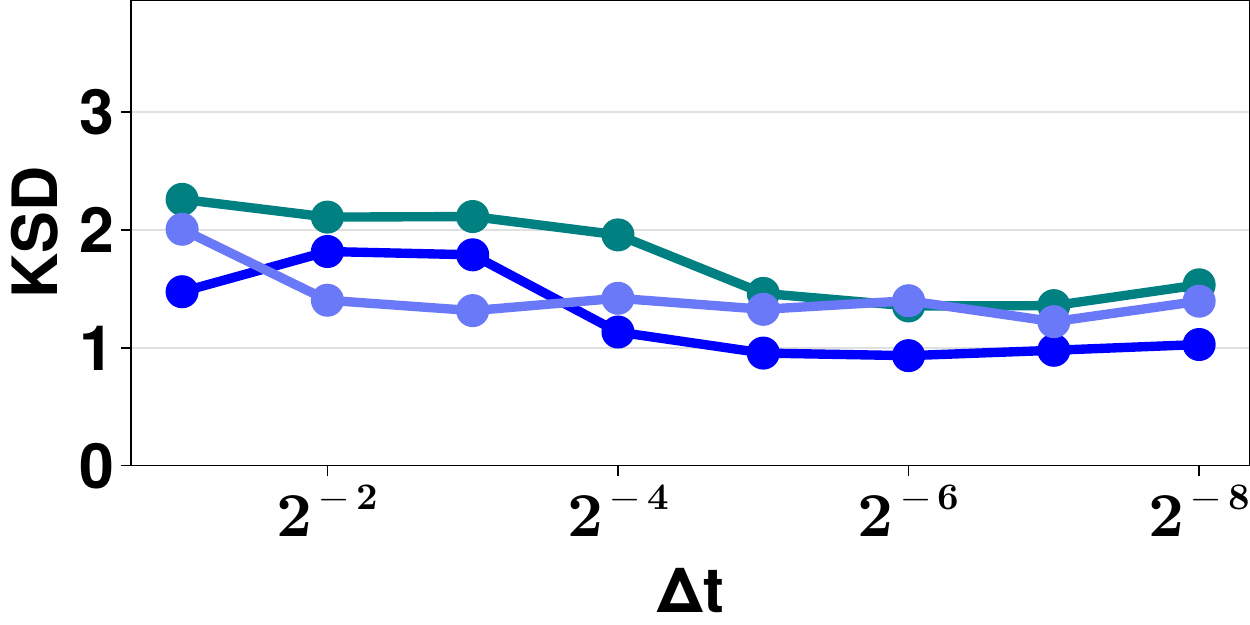}    
\subcaption*{J=100}
\end{subfigure}
\begin{subfigure}{0.19\linewidth}
\includegraphics[width=\linewidth]{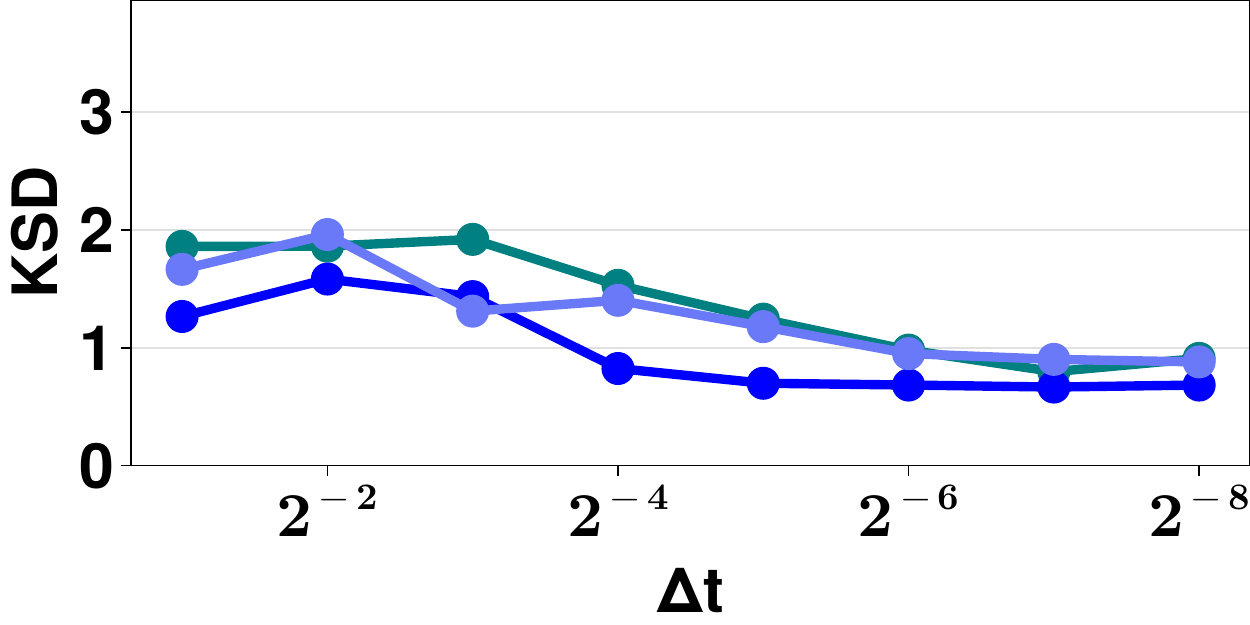}    
\subcaption*{J=200}
\end{subfigure}
\begin{subfigure}{0.19\linewidth}
\includegraphics[width=\linewidth]{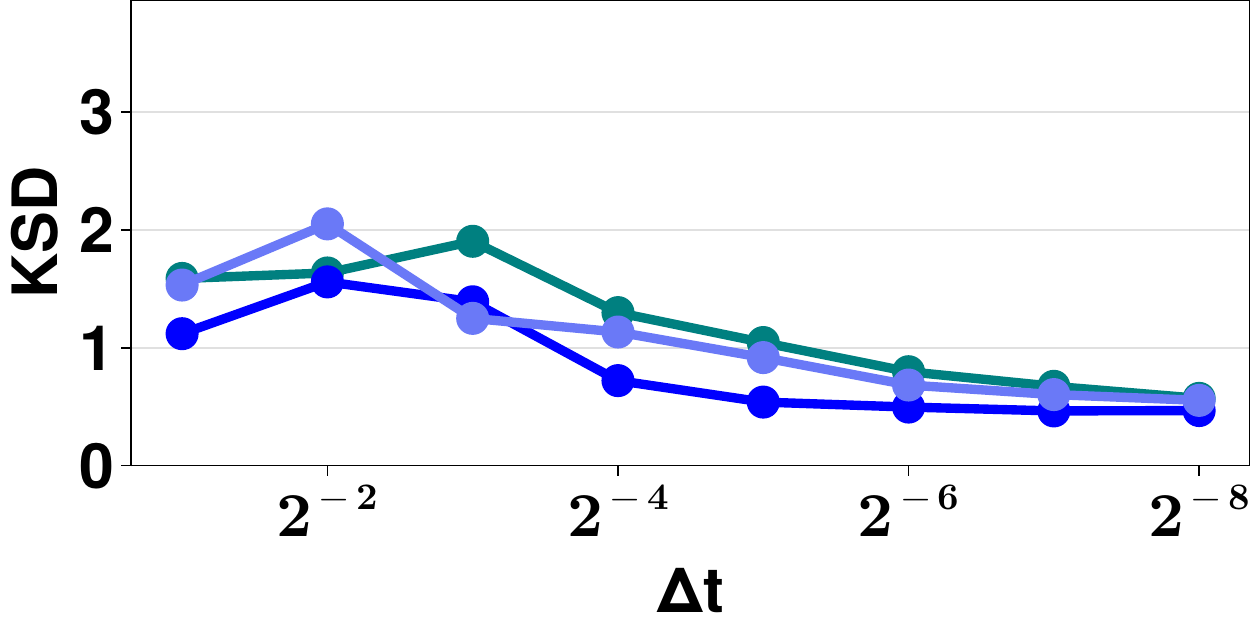}    
\subcaption*{J=400}
\end{subfigure}
\\
\begin{subfigure}{\linewidth}
\centering 
    \includegraphics[width=0.35\linewidth]{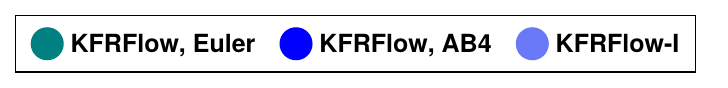}
\end{subfigure}
\caption{\textbf{Two-dimensional posteriors:} average KSD of ensembles generated by forward Euler and AB4 discretizations of KFRFlow and by KFRFlow-I for varying ensemble size and $\Delta t$. KFRFlow-I and the AB4 discretization of KFRFlow generally outperform the forward Euler discretization of KFRFlow.}
\label{fig:ODEdisc}
 \end{figure}

\begin{figure}[h]
 \centering 
\centering 
 Donut \\ 
\begin{subfigure}{0.19\linewidth}
\includegraphics[width=\linewidth]{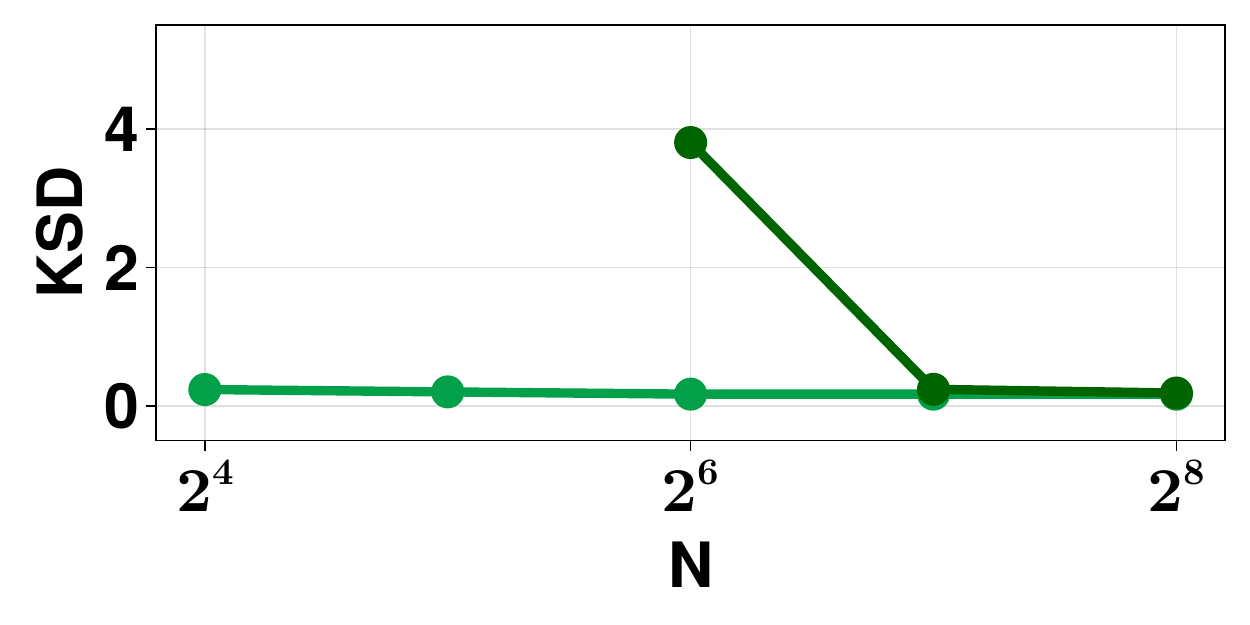}    
\end{subfigure}
\begin{subfigure}{0.19\linewidth}
\includegraphics[width=\linewidth]{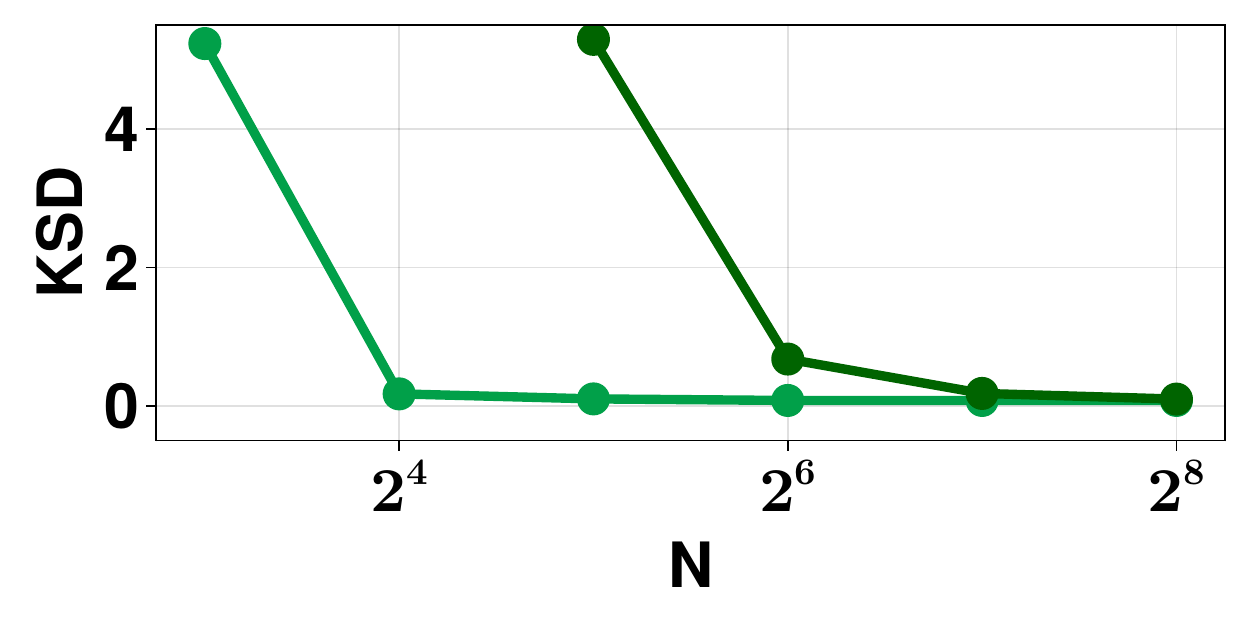}    
\end{subfigure}
\begin{subfigure}{0.19\linewidth}
\includegraphics[width=\linewidth]{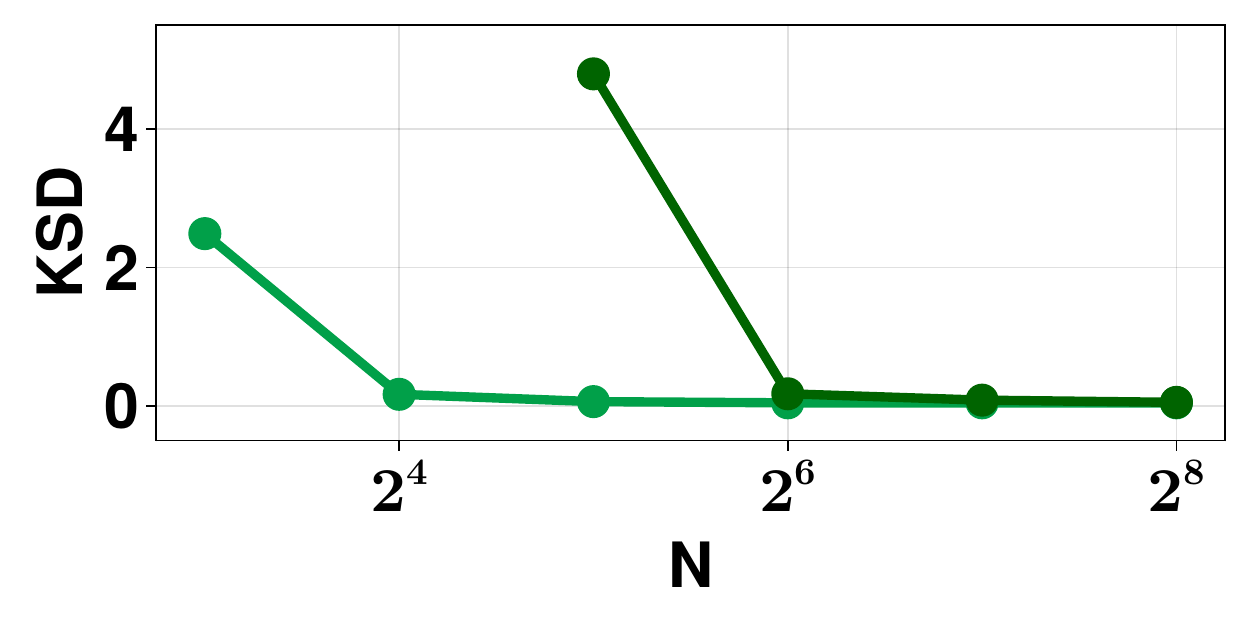}    
\end{subfigure}
\begin{subfigure}{0.19\linewidth}
\includegraphics[width=\linewidth]{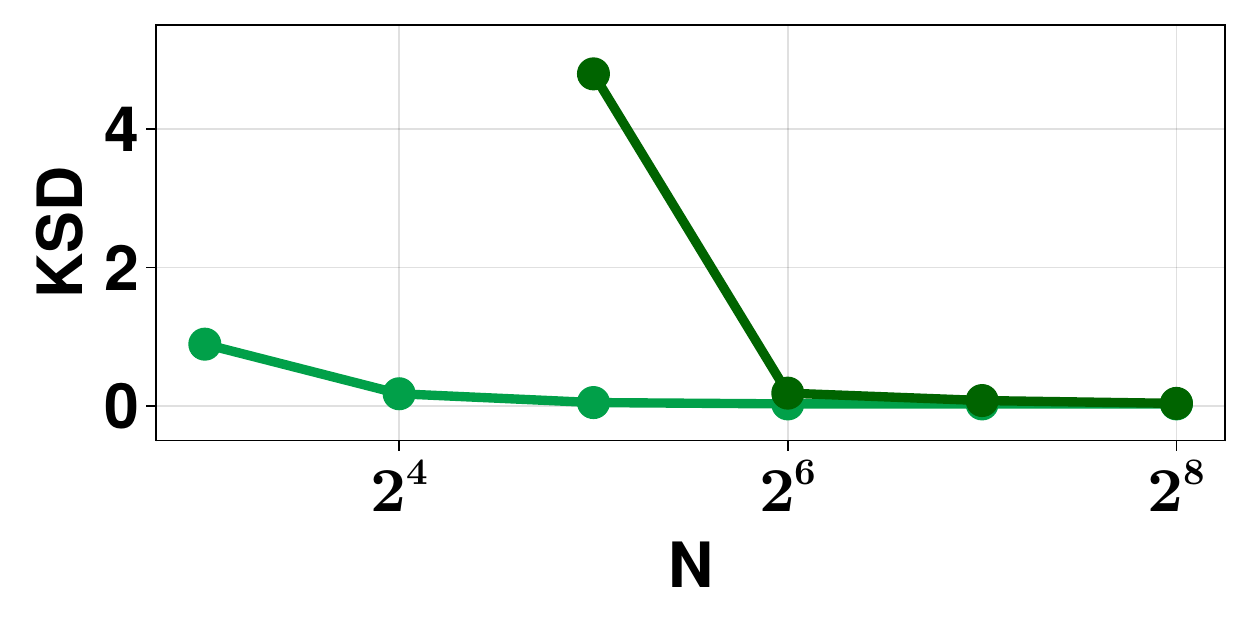}    
\end{subfigure}
\begin{subfigure}{0.19\linewidth}
\includegraphics[width=\linewidth]{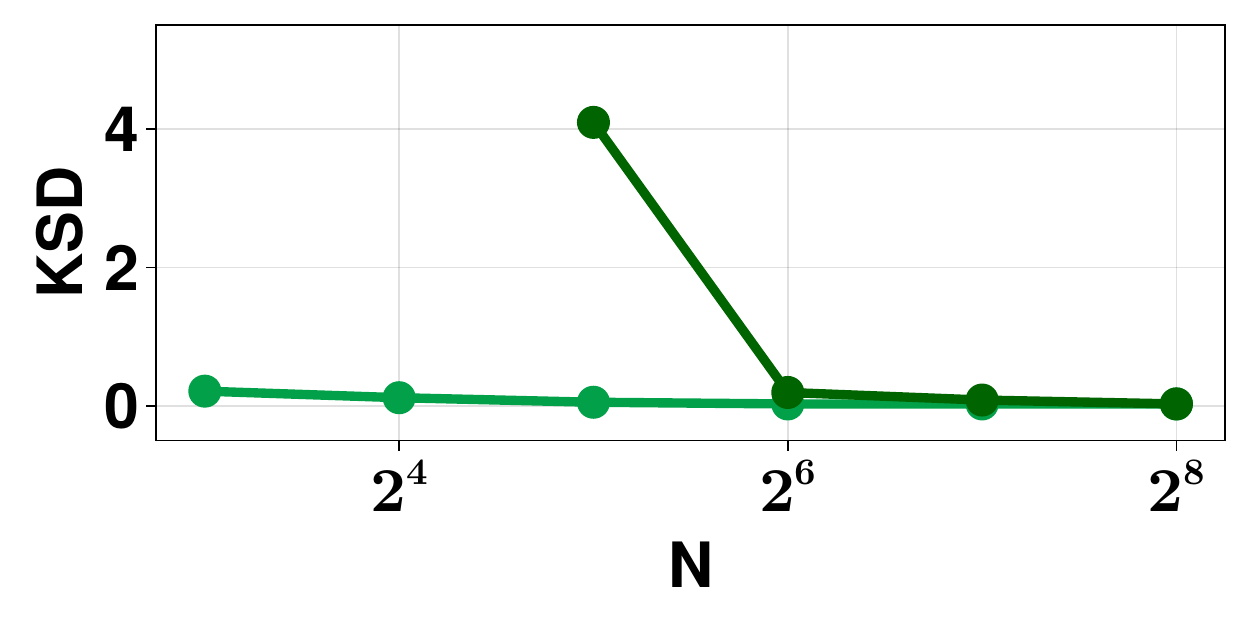}    
\end{subfigure}
\\ 
 Butterfly \\ 
\begin{subfigure}{0.19\linewidth}
\includegraphics[width=\linewidth]{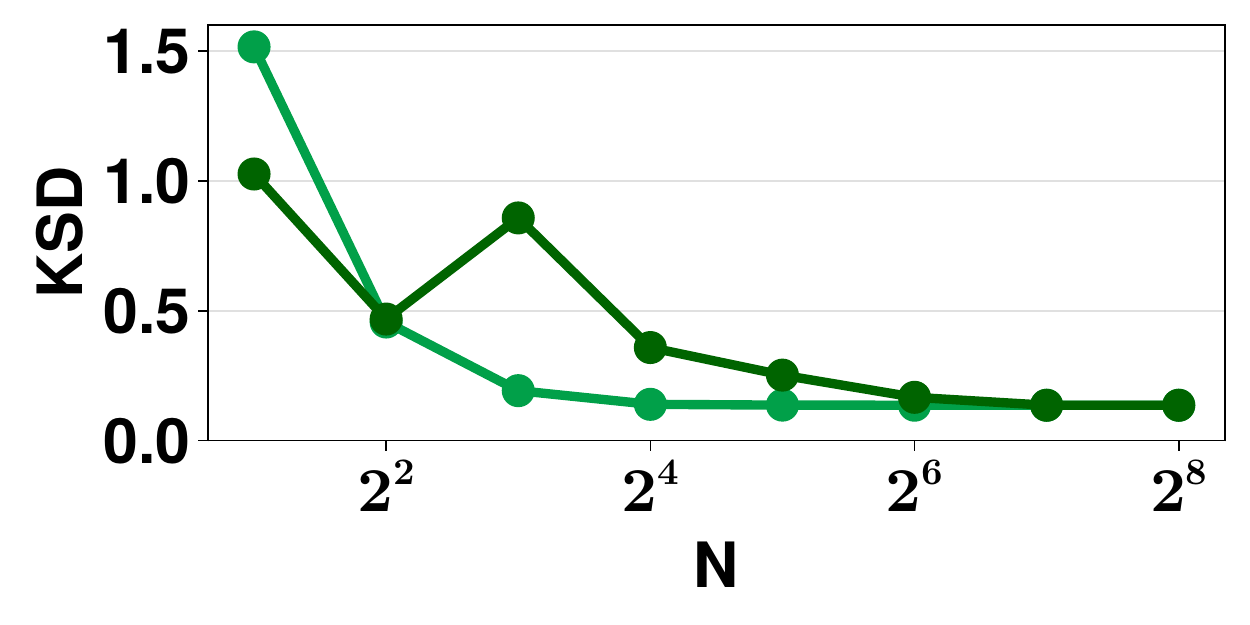}    
\end{subfigure}
\begin{subfigure}{0.19\linewidth}
\includegraphics[width=\linewidth]{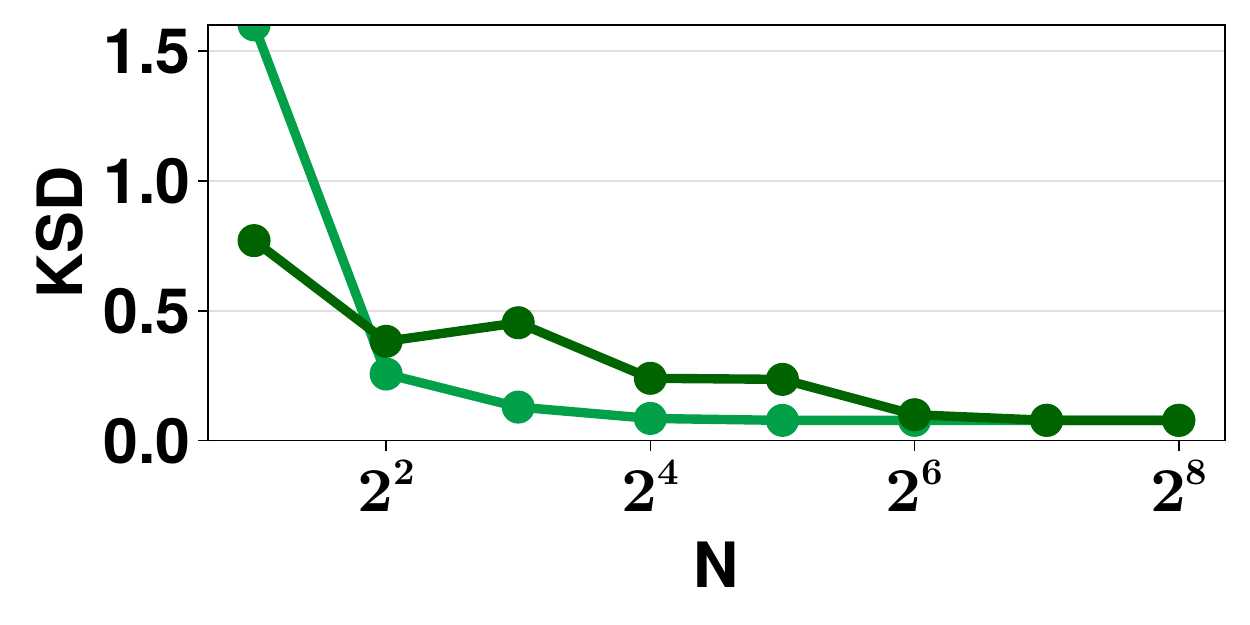}    
\end{subfigure}
\begin{subfigure}{0.19\linewidth}
\includegraphics[width=\linewidth]{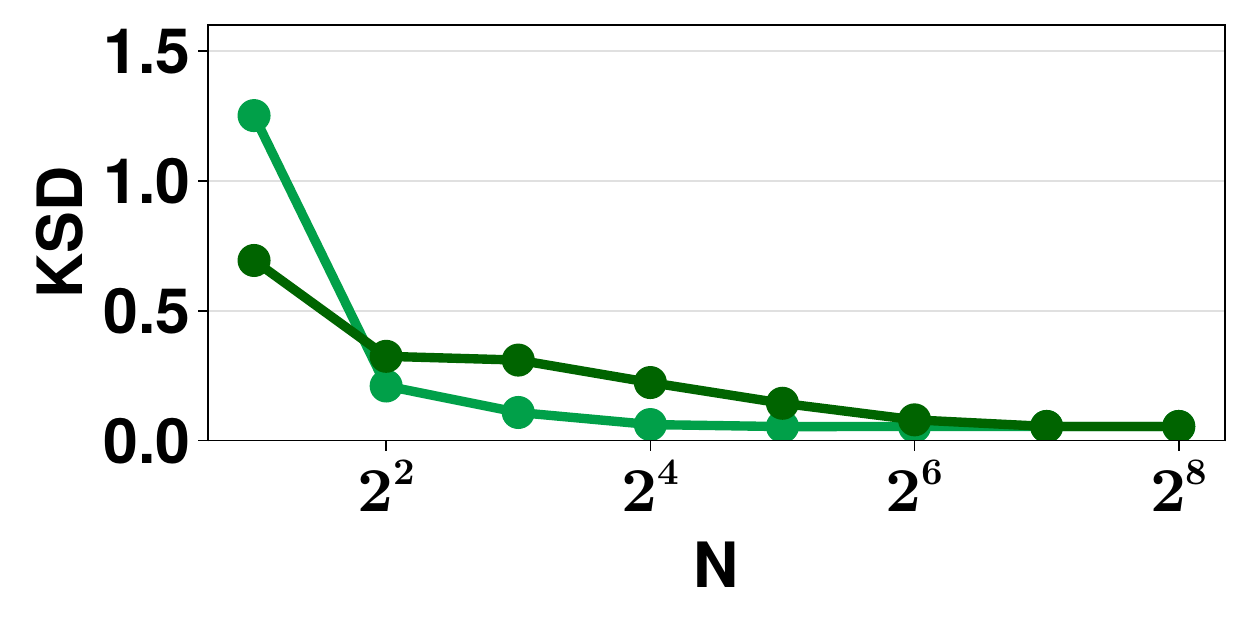}    
\end{subfigure}
\begin{subfigure}{0.19\linewidth}
\includegraphics[width=\linewidth]{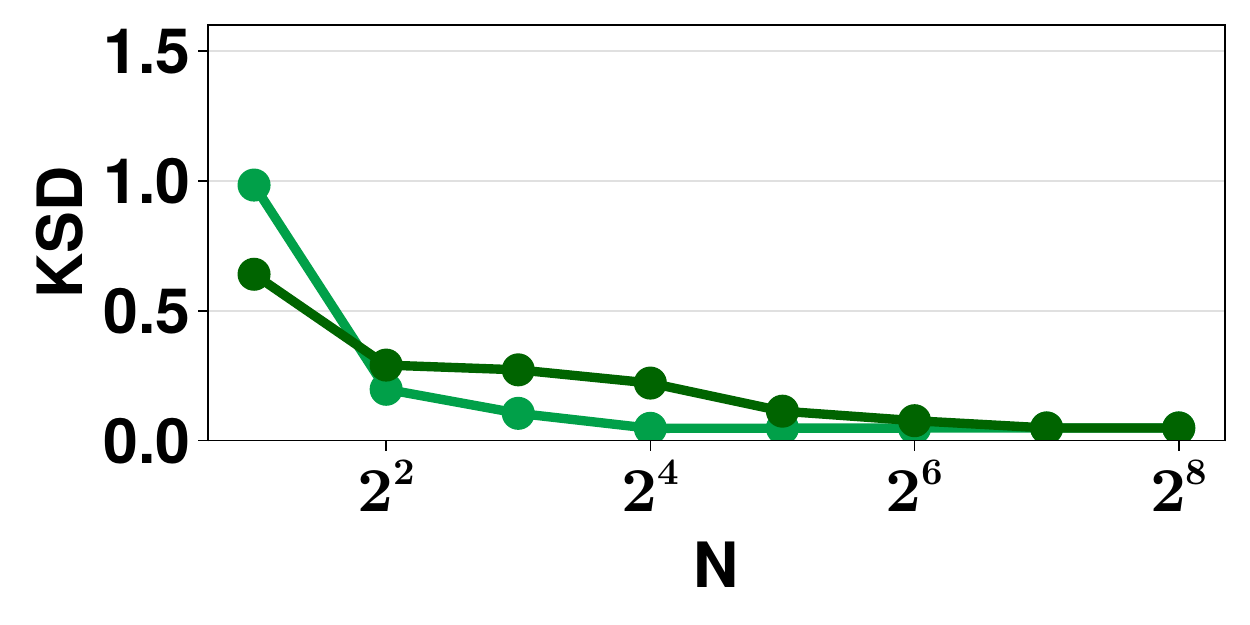}    
\end{subfigure}
\begin{subfigure}{0.19\linewidth}
\includegraphics[width=\linewidth]{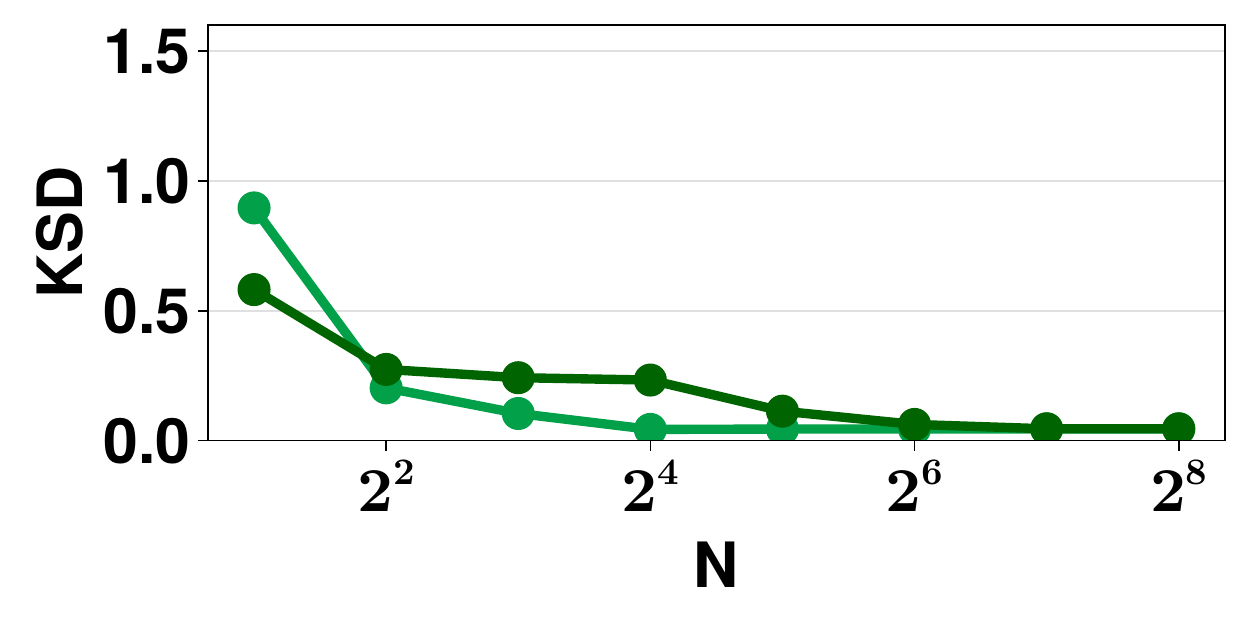}    
\end{subfigure}
\\
 Spaceships \\ 
\begin{subfigure}{0.19\linewidth}
\includegraphics[width=\linewidth]{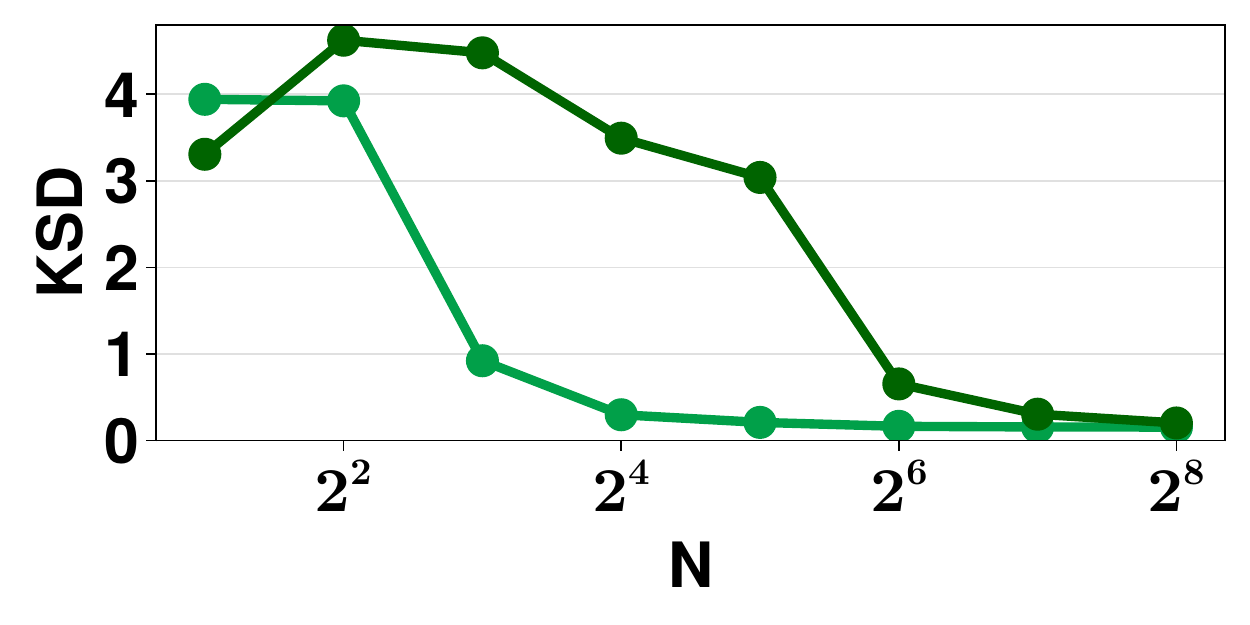}   
\subcaption*{J=25}
\end{subfigure}
\begin{subfigure}{0.19\linewidth}
\includegraphics[width=\linewidth]{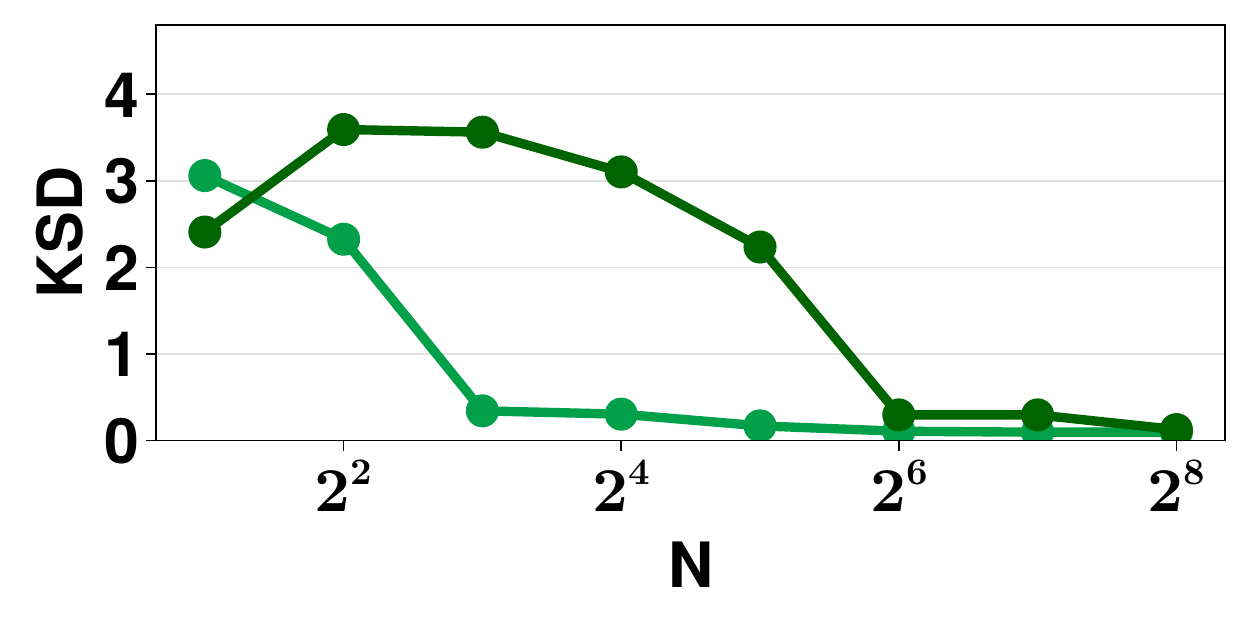}  
\subcaption*{J=50}
\end{subfigure}
\begin{subfigure}{0.19\linewidth}
\includegraphics[width=\linewidth]{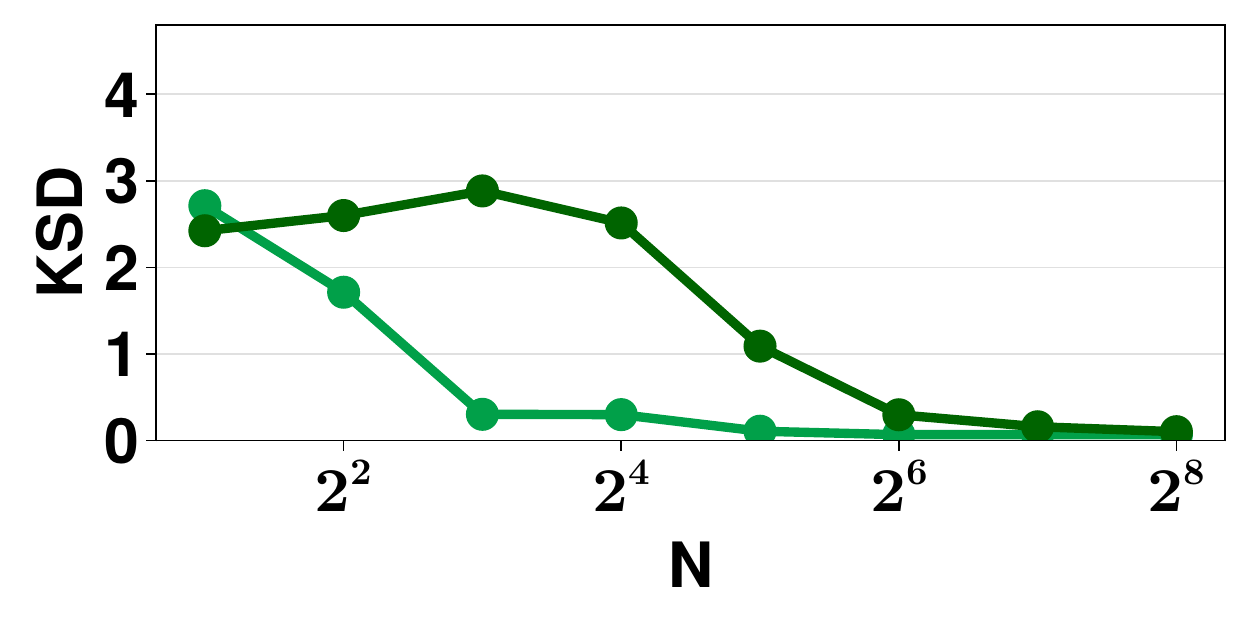}    
\subcaption*{J=100}
\end{subfigure}
\begin{subfigure}{0.19\linewidth}
\includegraphics[width=\linewidth]{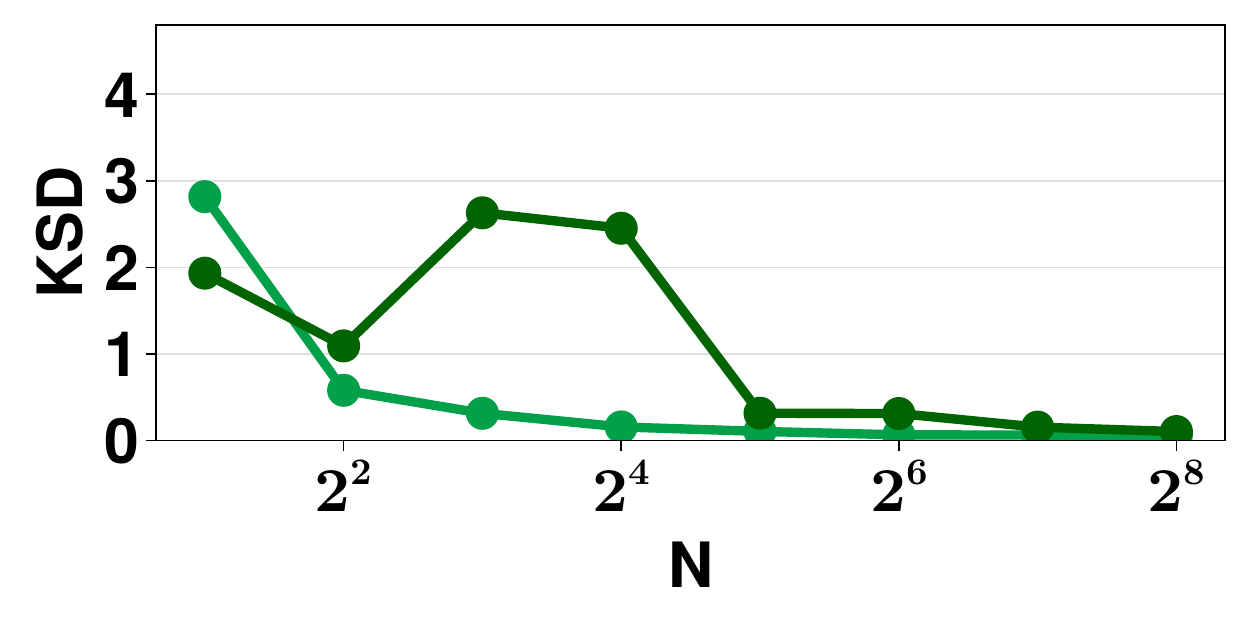}    
\subcaption*{J=200}
\end{subfigure}
\begin{subfigure}{0.19\linewidth}
\includegraphics[width=\linewidth]{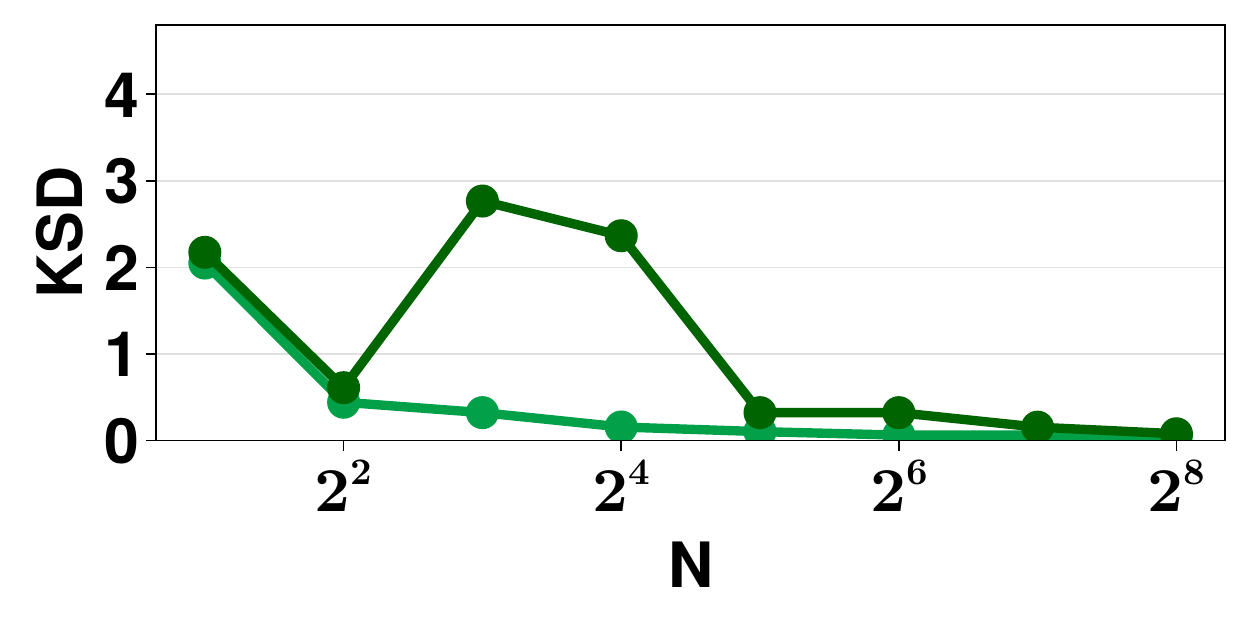}    
\subcaption*{J=400}
\end{subfigure}
\\
\begin{subfigure}{\linewidth}
    \centering 
    \includegraphics[width=0.15\linewidth]{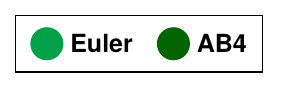}
\end{subfigure}
\caption{\textbf{Two-dimensional posteriors:} average KSD of ensembles generated by forward Euler and AB4 discretizations of SVGD. Using AB4 in place of forward Euler with SVGD tends only to make things worse. }
\label{fig:SVGDdisc}
 \end{figure}

 We use a fourth-order Adams--Bashforth discretization of KFRFlow because we find that it generates better samples than forward Euler at little additional cost, while by contrast we use forward Euler for SVGD because we find that SVGD does not benefit from multistep discretizations in these examples. For comparisons of the sampling performance of these different ODE discretization methods, see \cref{fig:ODEdisc,fig:SVGDdisc}. We use Euler--Maruyama discretizations for KFRD, EKI, EKS, CBS, and ULA.  The values of KSD we report are averages over 30 trials. 

\newcommand{\scalefactor}{0.9}
\begin{figure}[h]
\centering 
\begin{subfigure}{0.32\linewidth}
\centering 
Donut 
\includegraphics[width=\scalefactor\linewidth]{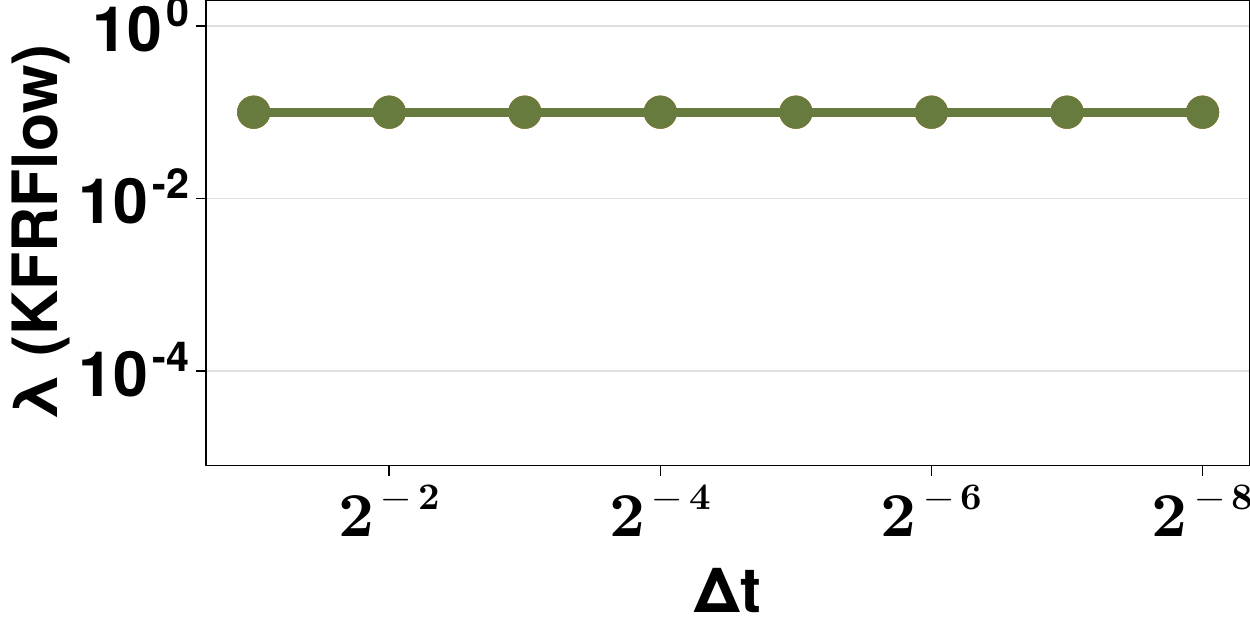}    
\end{subfigure}
\begin{subfigure}{0.32\linewidth}
\centering 
Butterfly 
\includegraphics[width=\scalefactor\linewidth]{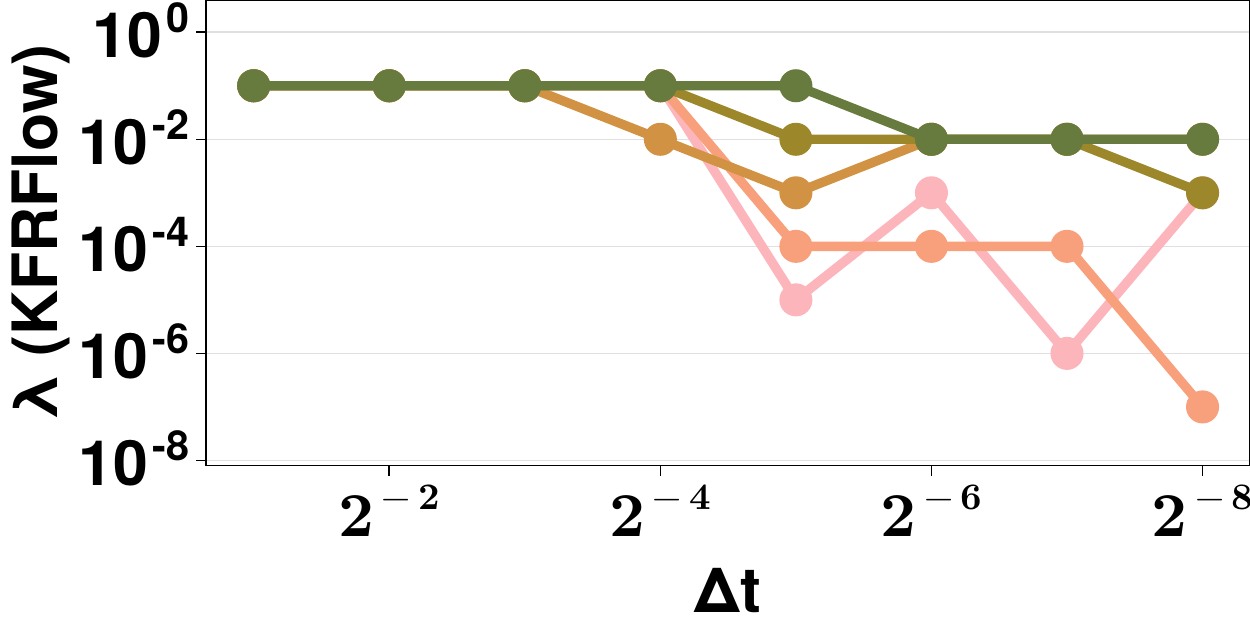}    
\end{subfigure}
\begin{subfigure}{0.32\linewidth}
\centering 
Spaceships 
\includegraphics[width=\scalefactor\linewidth]{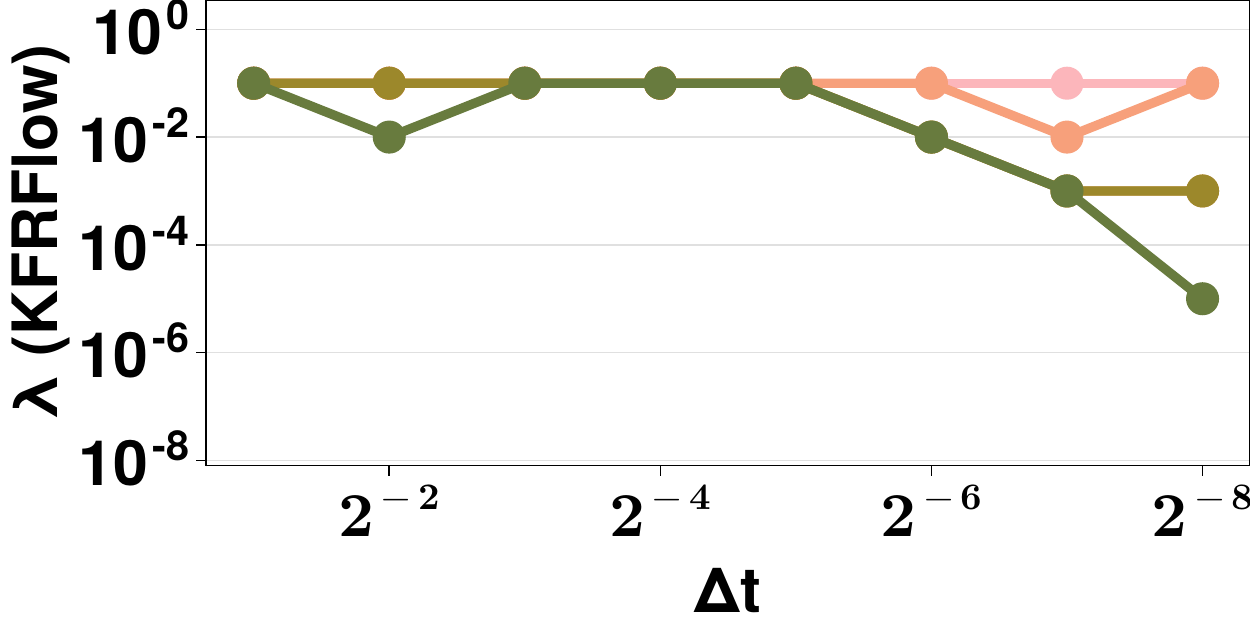}    
\end{subfigure}
\\
\begin{subfigure}{0.32\linewidth}
\centering 
\includegraphics[width=\scalefactor\linewidth]{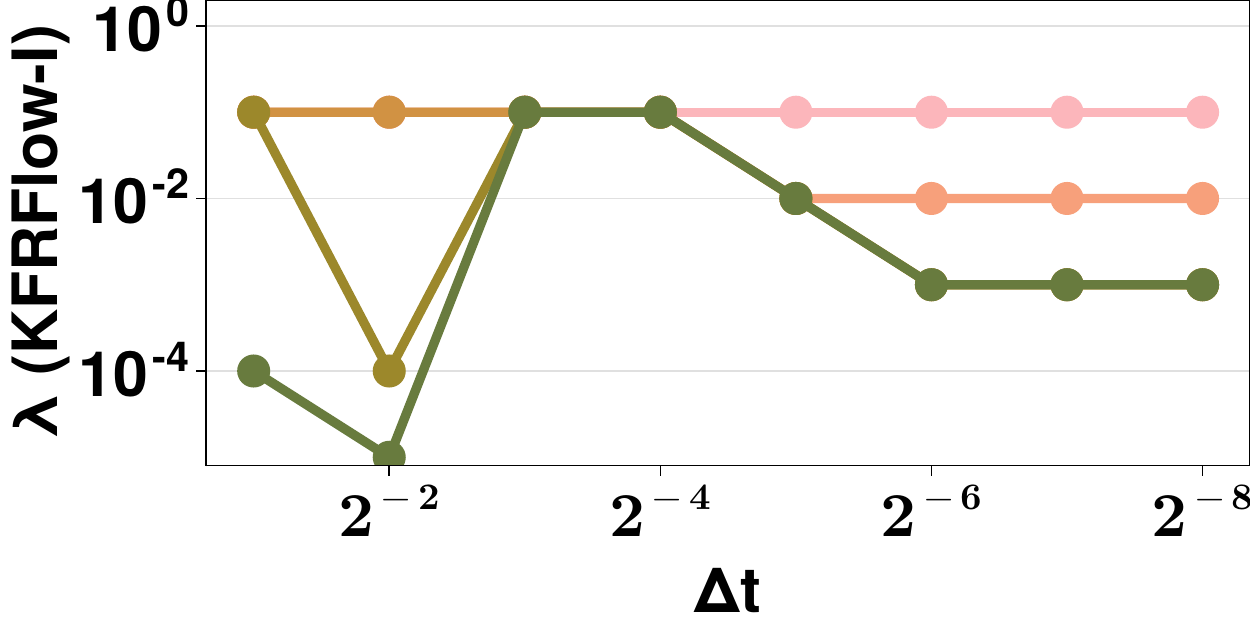}    
\end{subfigure}
\begin{subfigure}{0.32\linewidth}
\centering 
\includegraphics[width=\scalefactor\linewidth]{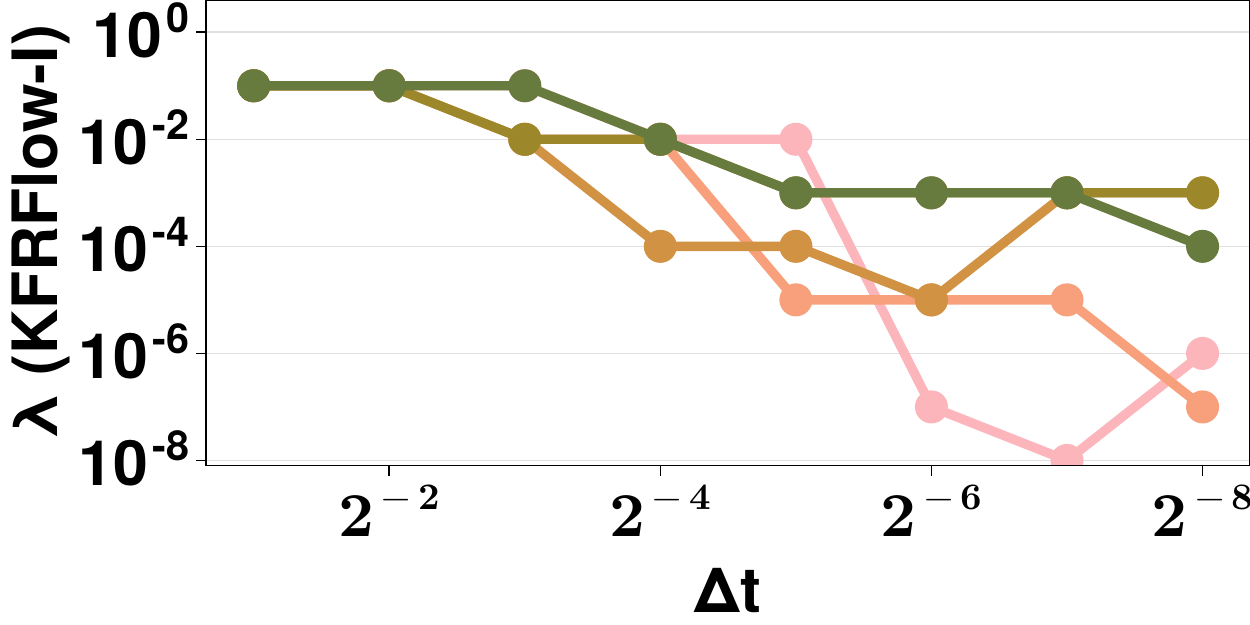}    
\end{subfigure}
\begin{subfigure}{0.32\linewidth}
\centering 
\includegraphics[width=\scalefactor\linewidth]{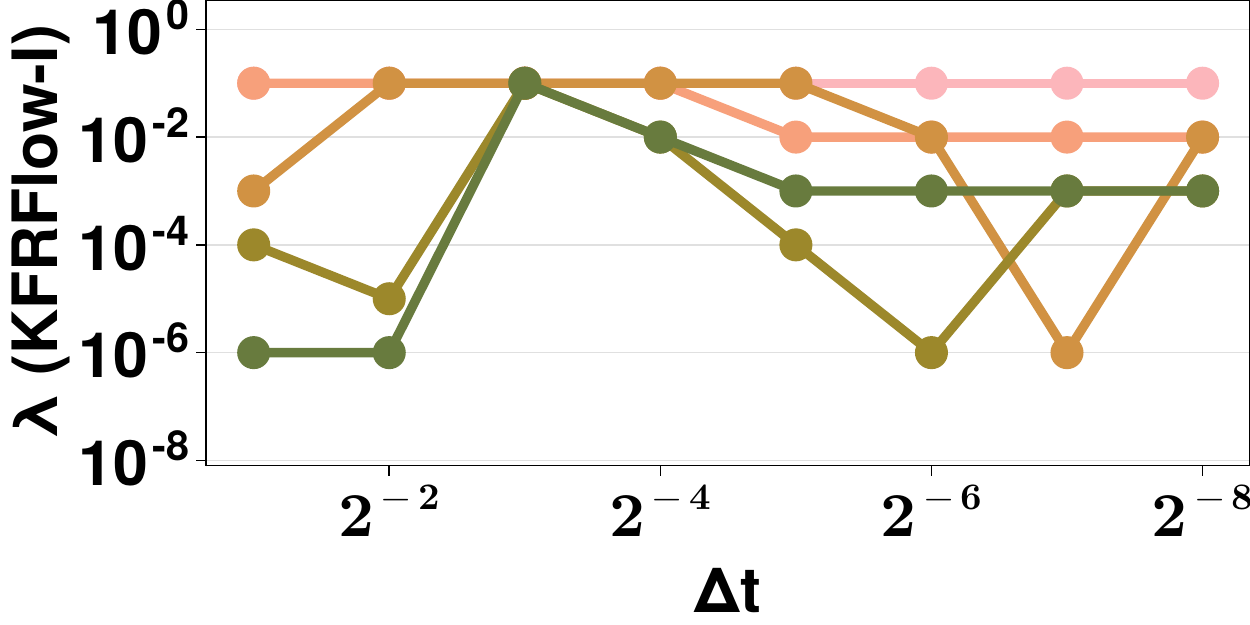}    
\end{subfigure}
 \\
\begin{subfigure}{0.32\linewidth}
\centering 
\includegraphics[width=\scalefactor\linewidth]{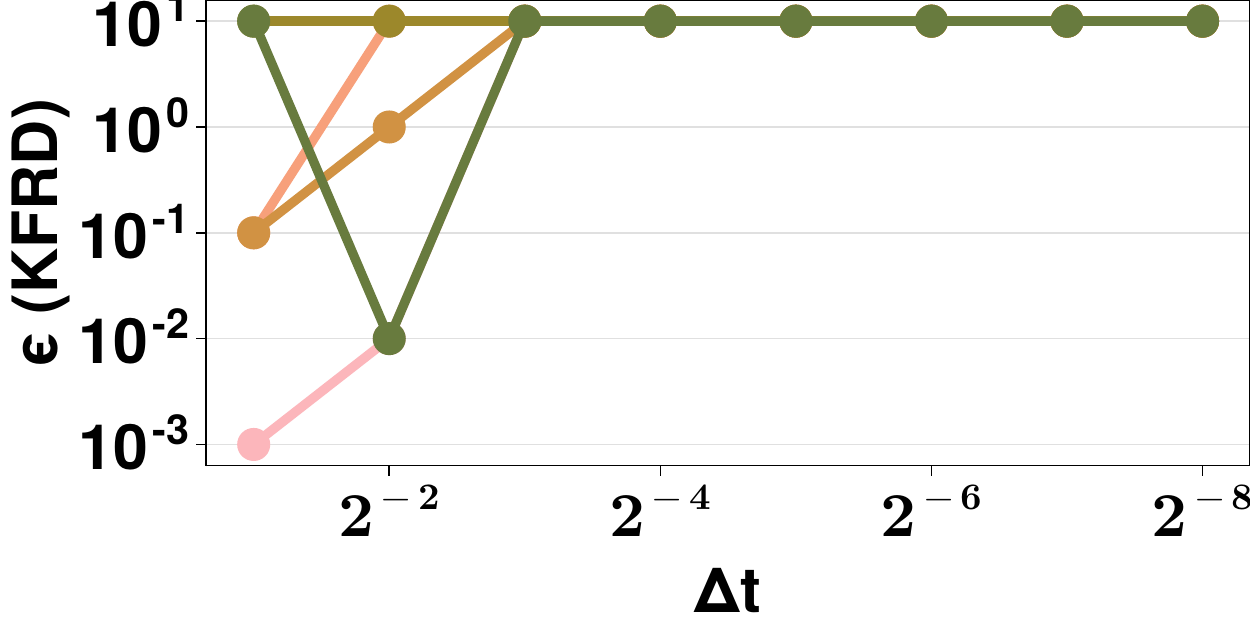}    
\end{subfigure}
\begin{subfigure}{0.32\linewidth}
\centering 
\includegraphics[width=\scalefactor\linewidth]{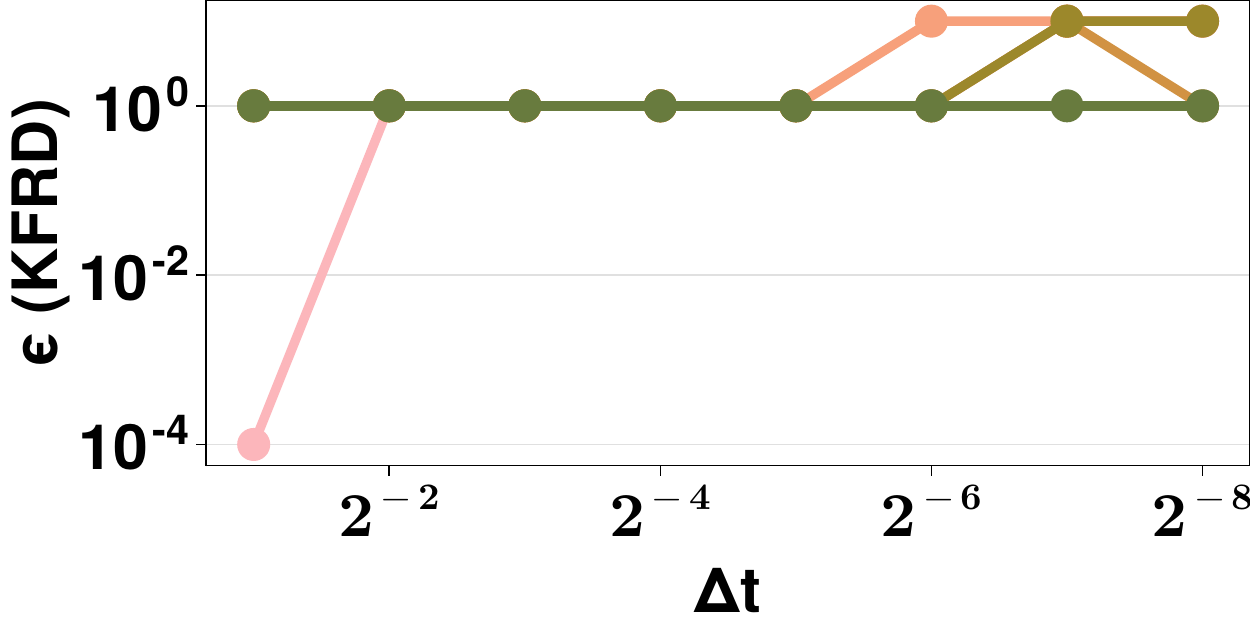}    
\end{subfigure}
\begin{subfigure}{0.32\linewidth}
\centering 
\includegraphics[width=\scalefactor\linewidth]{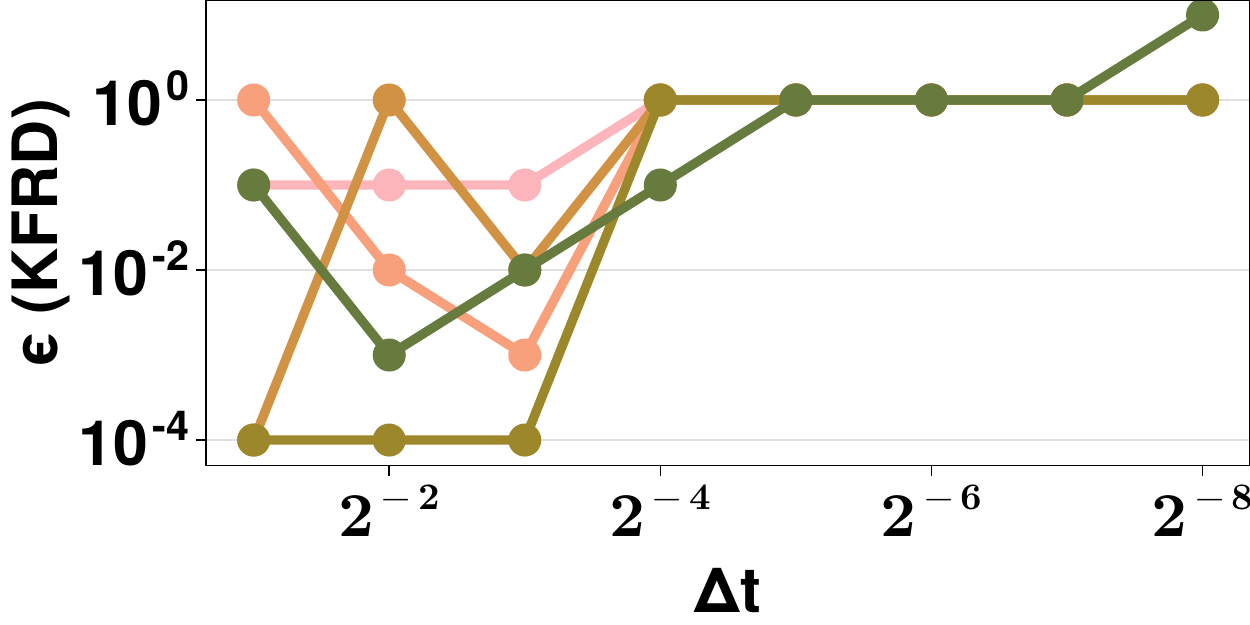}    
\end{subfigure}
\\
\begin{subfigure}{0.32\linewidth}
\centering 
\includegraphics[width=\scalefactor\linewidth]{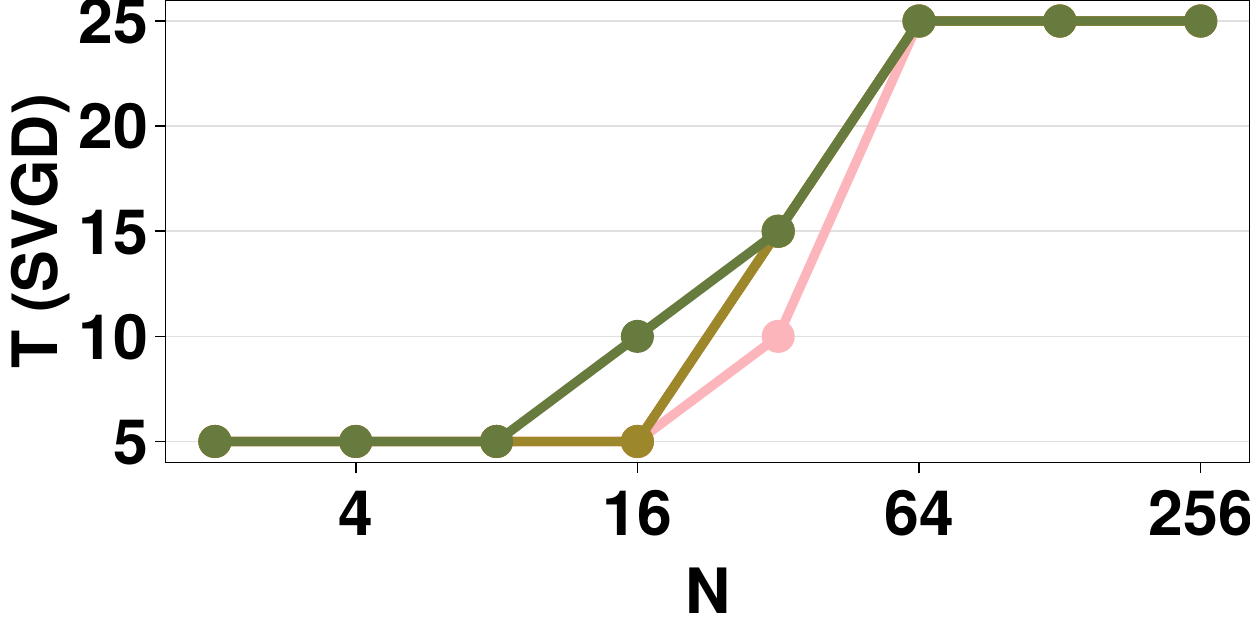}  
\end{subfigure}
\begin{subfigure}{0.32\linewidth}
\centering 
\includegraphics[width=\scalefactor\linewidth]{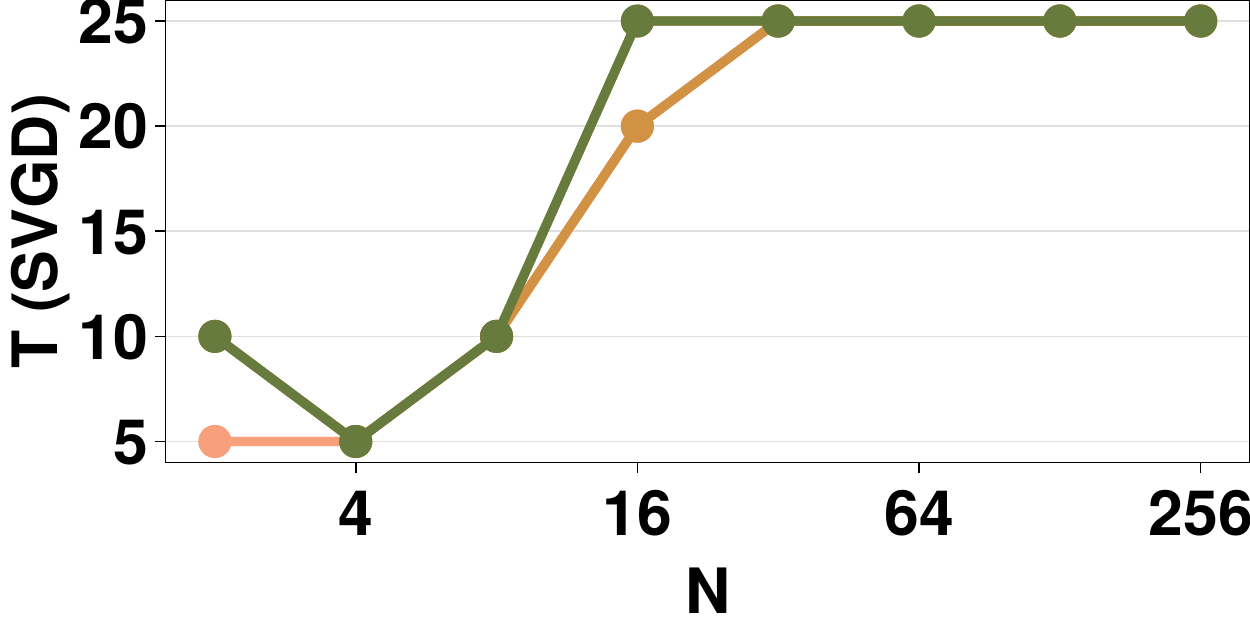}  
\end{subfigure}
\begin{subfigure}{0.32\linewidth}
\centering 
\includegraphics[width=\scalefactor\linewidth]{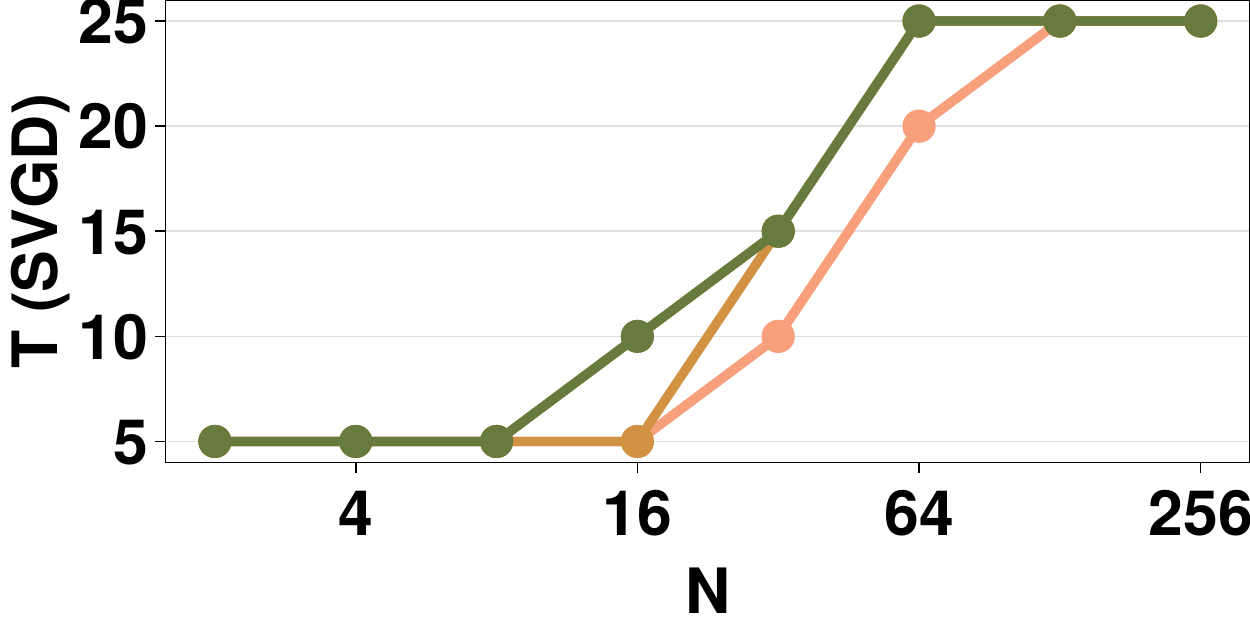}  
\end{subfigure}
\\
\begin{subfigure}{0.32\linewidth}
\centering 
\includegraphics[width=\scalefactor\linewidth]{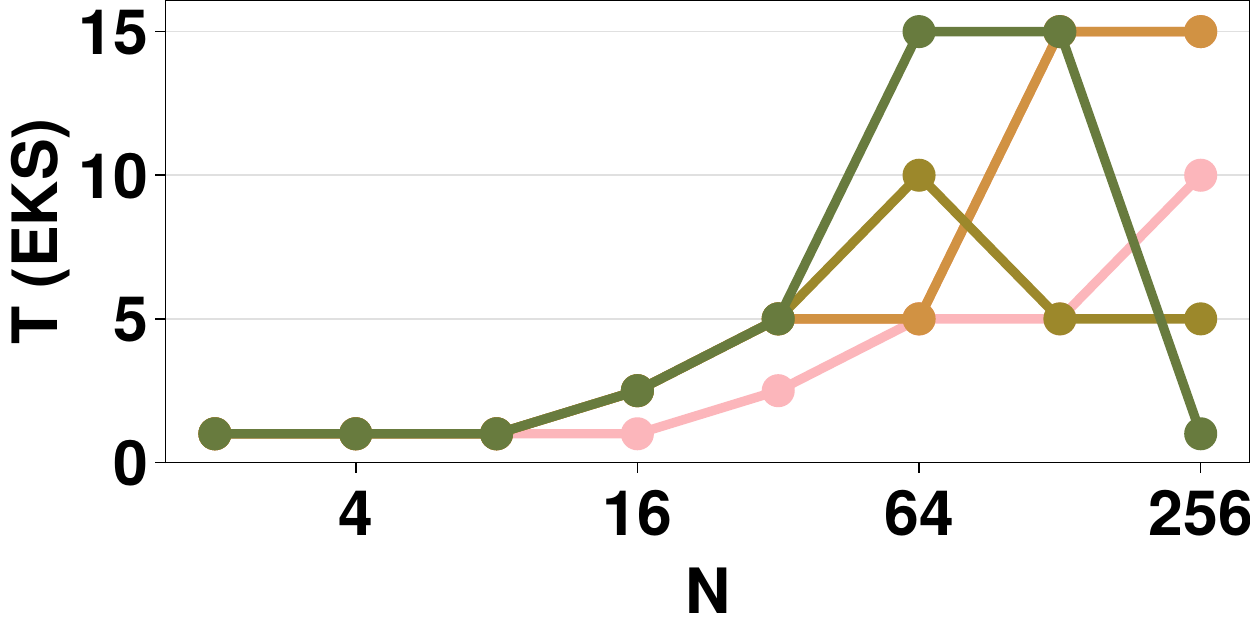} 
\end{subfigure}
\begin{subfigure}{0.32\linewidth}
\centering 
\includegraphics[width=\scalefactor\linewidth]{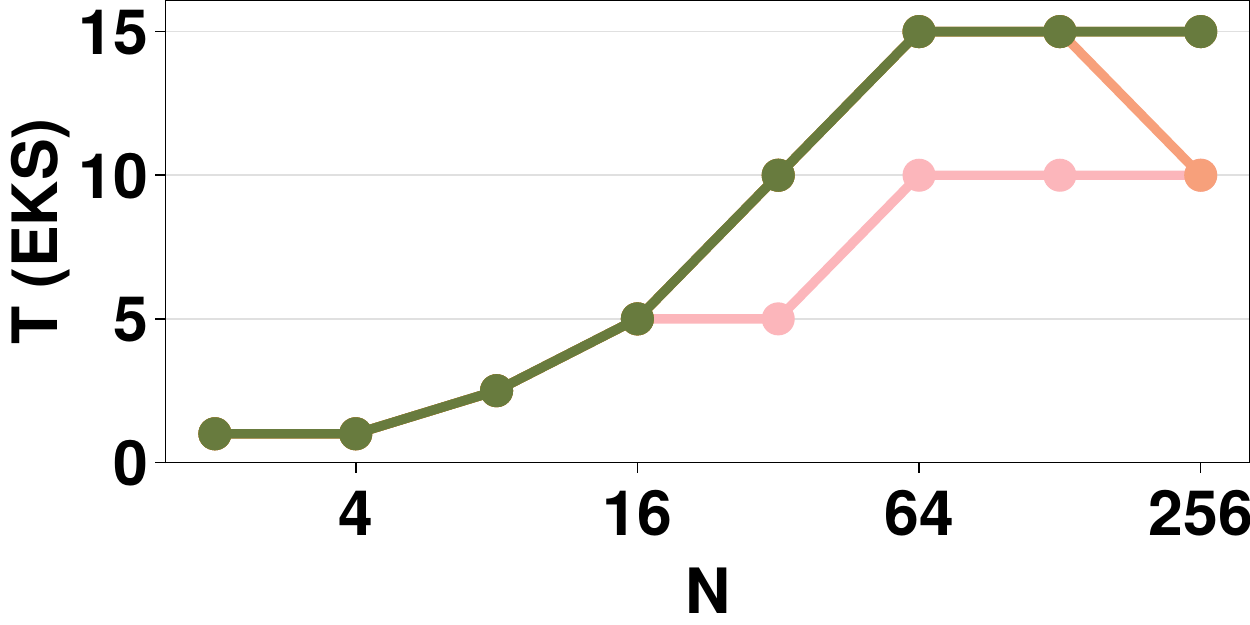}  
\end{subfigure}
\begin{subfigure}{0.32\linewidth}
\centering 
\includegraphics[width=\scalefactor\linewidth]{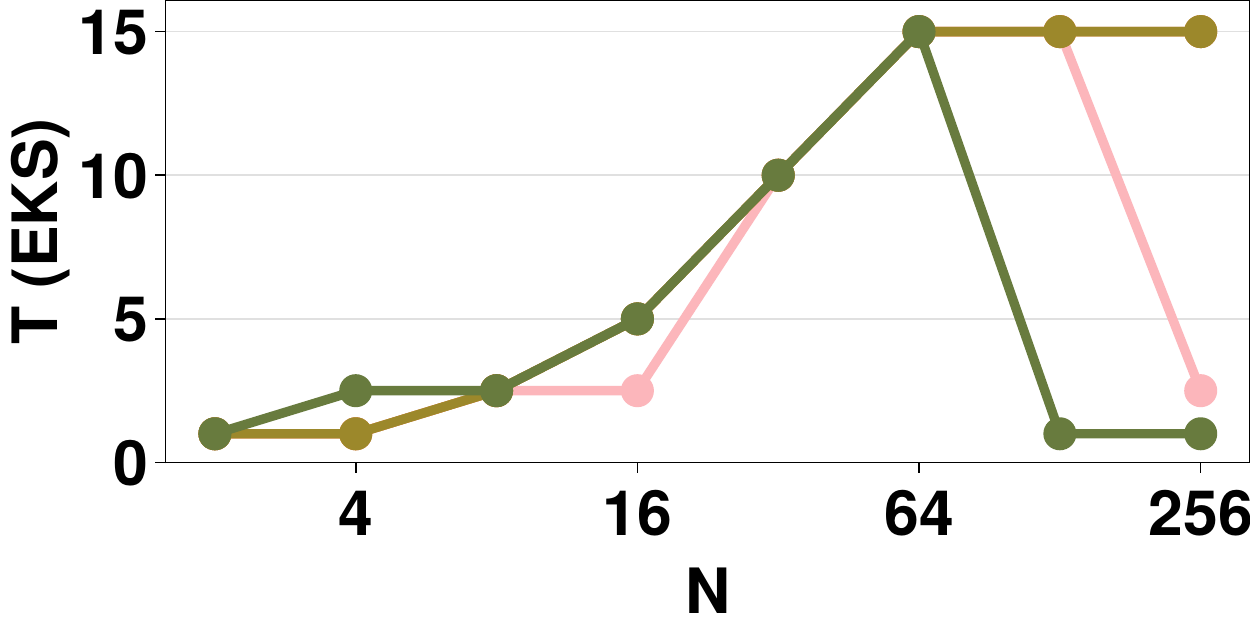}
\end{subfigure} 
\\
\begin{subfigure}{0.32\linewidth}
\centering 
\includegraphics[width=\scalefactor\linewidth]{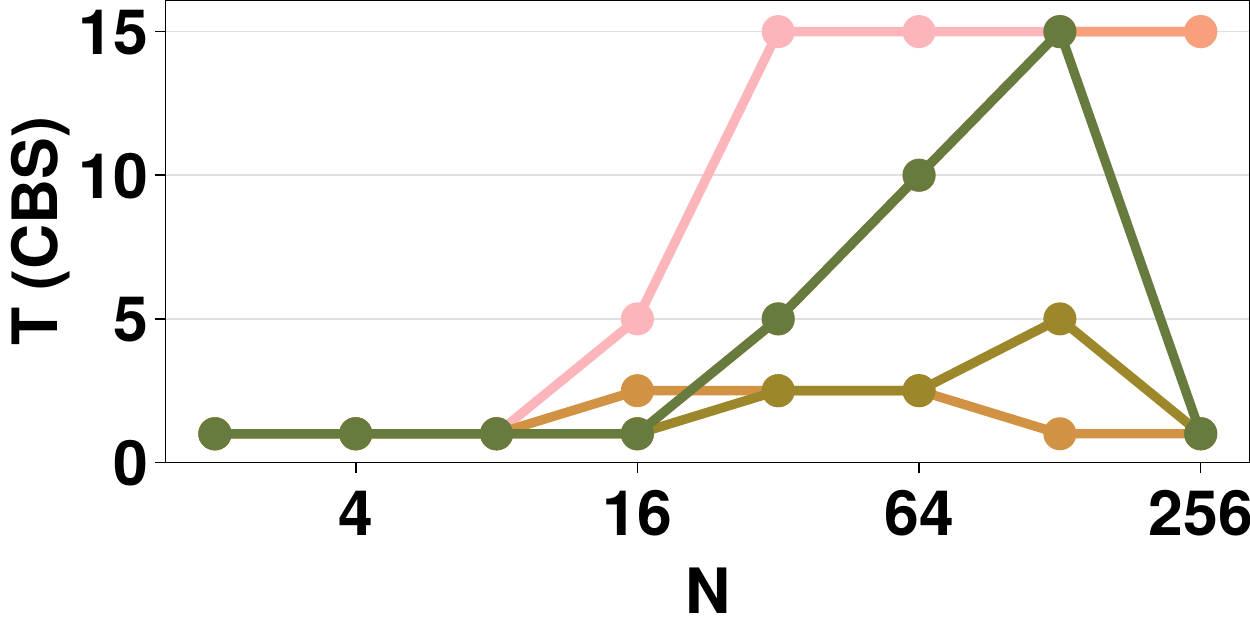} 
\end{subfigure}
\begin{subfigure}{0.32\linewidth}
\centering 
\includegraphics[width=\scalefactor\linewidth]{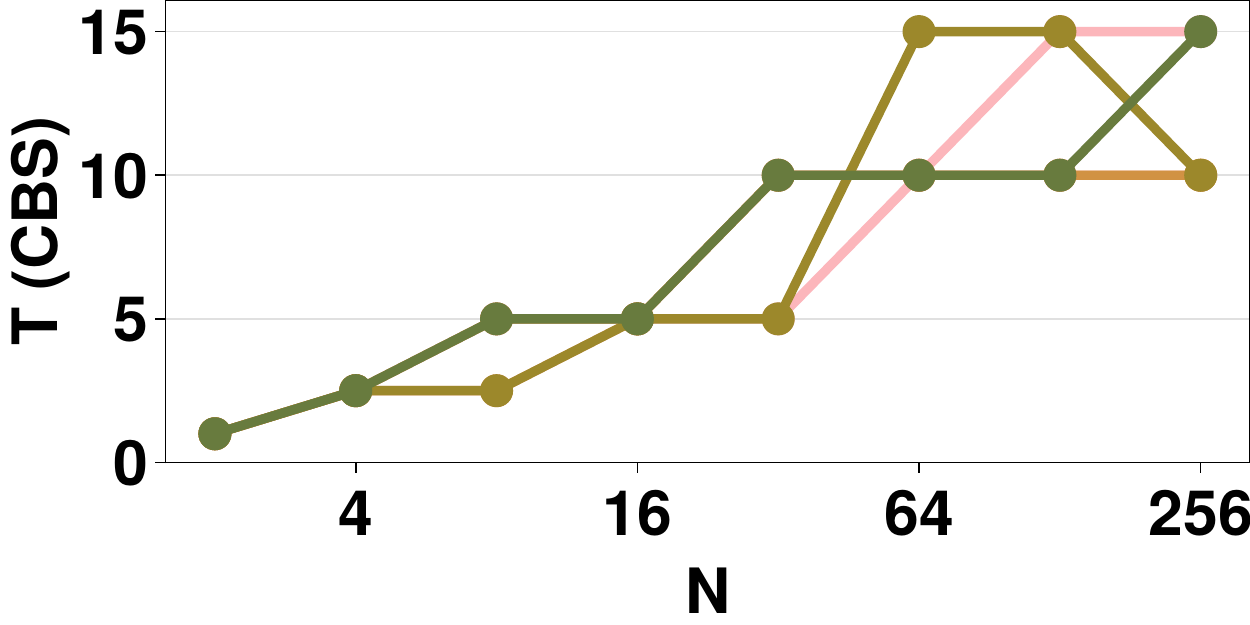}  
\end{subfigure}
\begin{subfigure}{0.32\linewidth}
\centering 
\includegraphics[width=\scalefactor\linewidth]{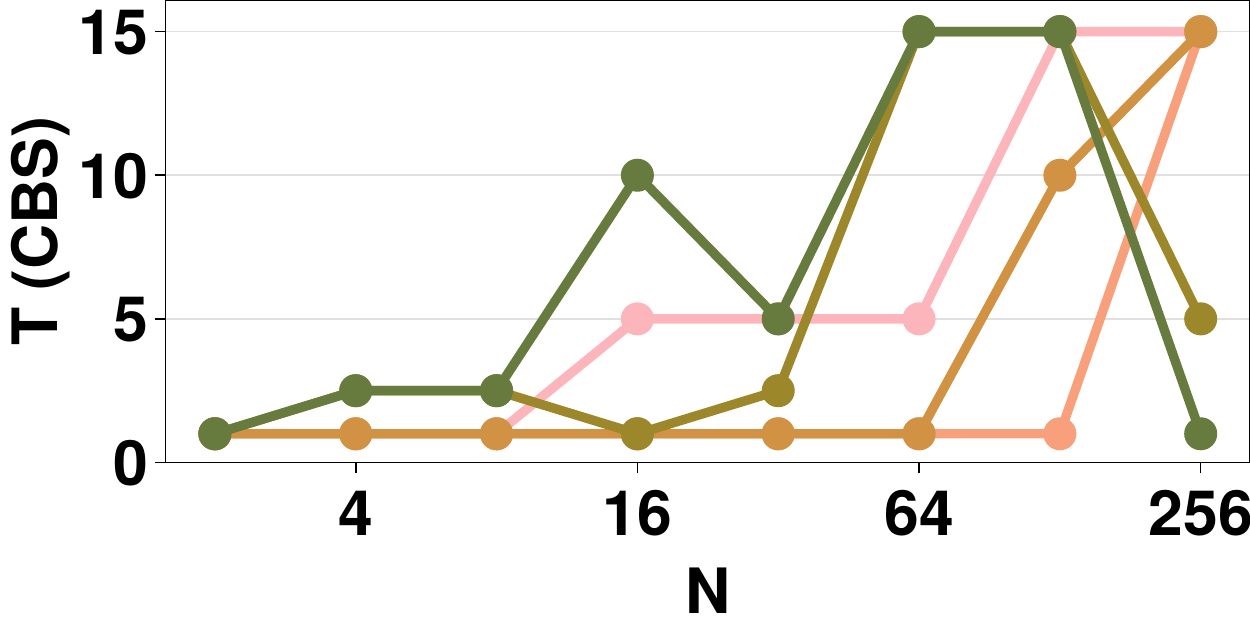}
\end{subfigure} 
\\
\begin{subfigure}{0.32\linewidth}
\centering 
\includegraphics[width=\scalefactor\linewidth]{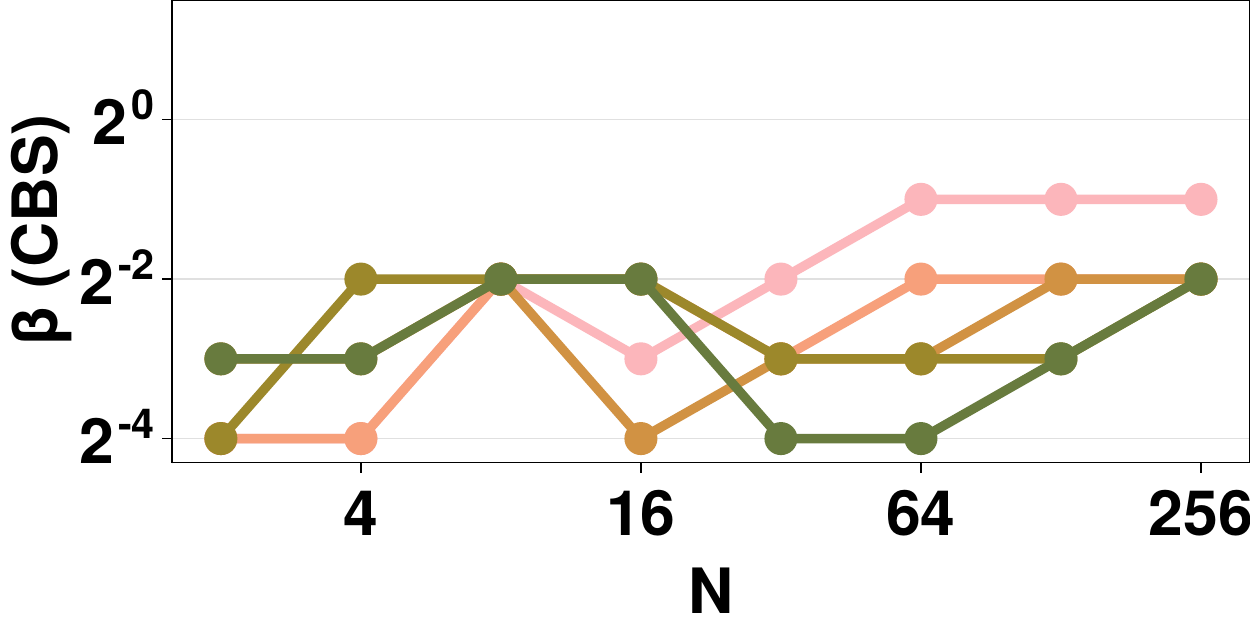} 
\end{subfigure}
\begin{subfigure}{0.32\linewidth}
\centering 
\includegraphics[width=\scalefactor\linewidth]{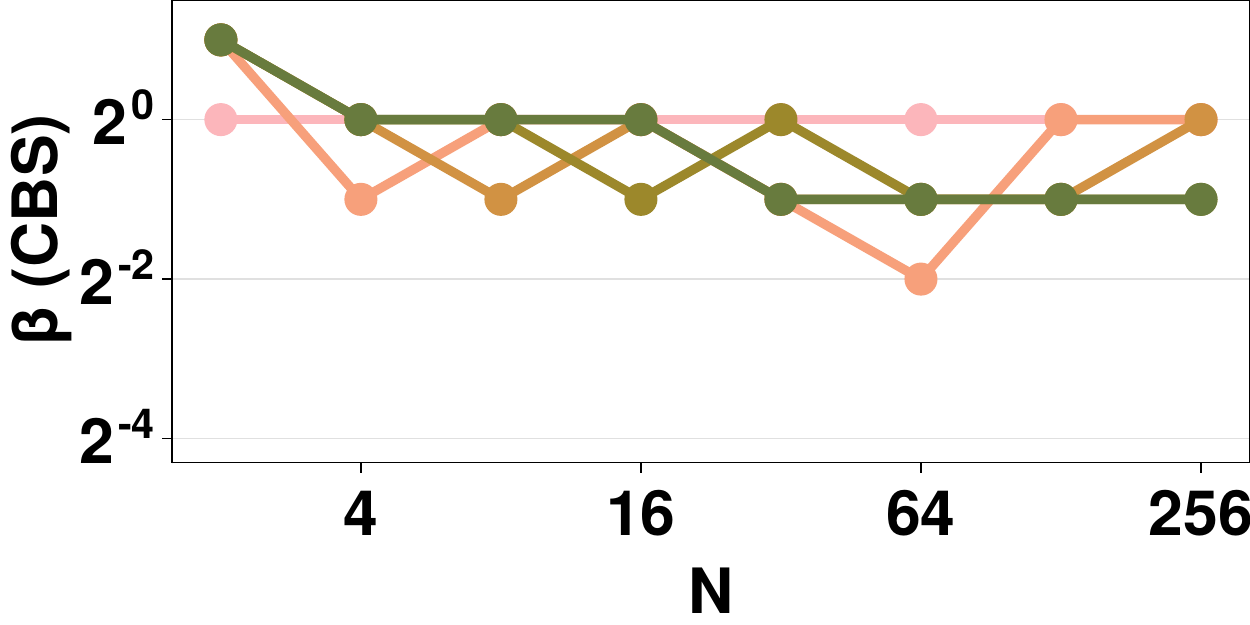}  
\end{subfigure}
\begin{subfigure}{0.32\linewidth}
\centering 
\includegraphics[width=\scalefactor\linewidth]{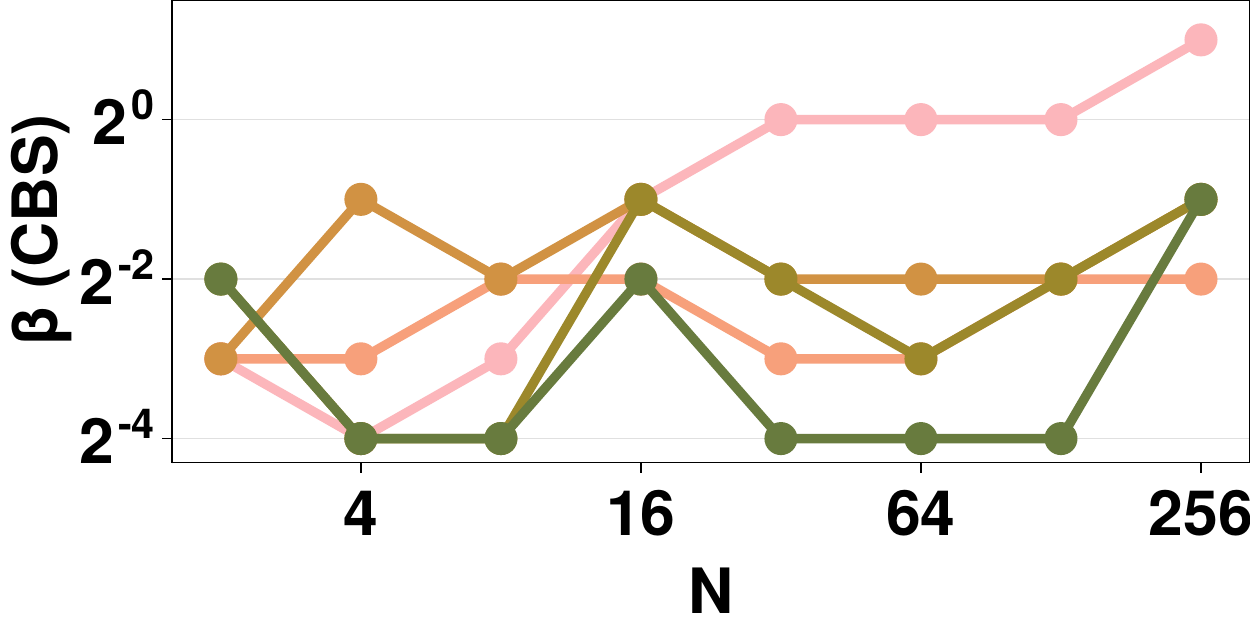}
\end{subfigure} 
\\
\begin{subfigure}{0.32\linewidth}
\centering 
\includegraphics[width=\scalefactor\linewidth]{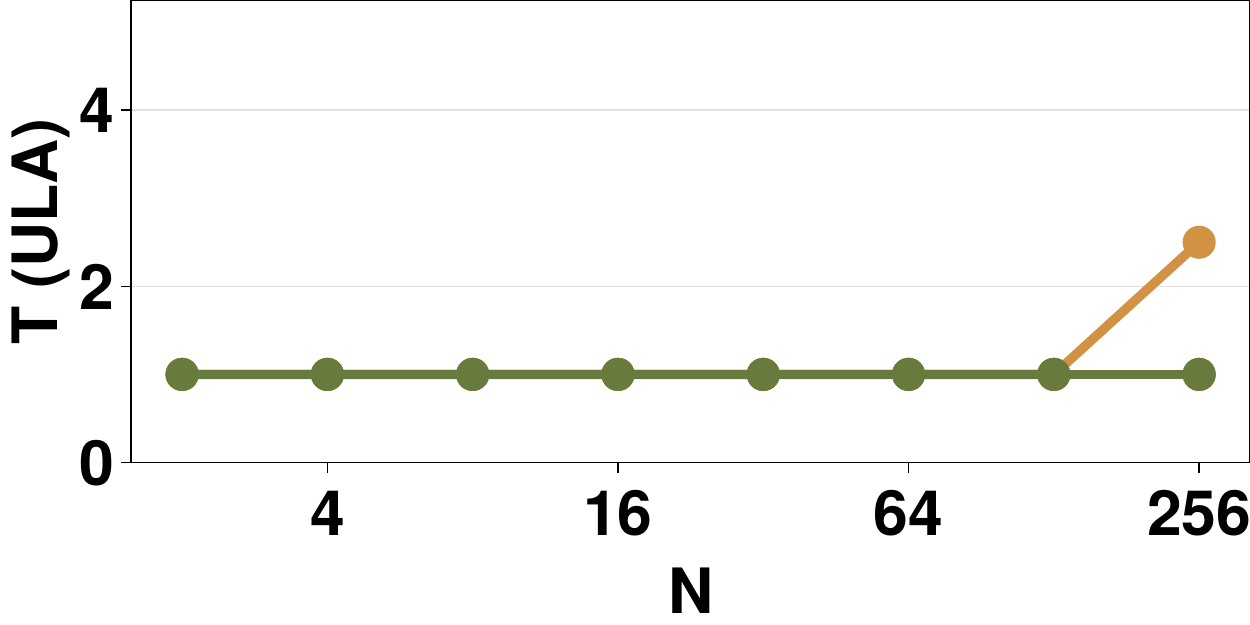} 
\end{subfigure}
\begin{subfigure}{0.32\linewidth}
\centering 
\includegraphics[width=\scalefactor\linewidth]{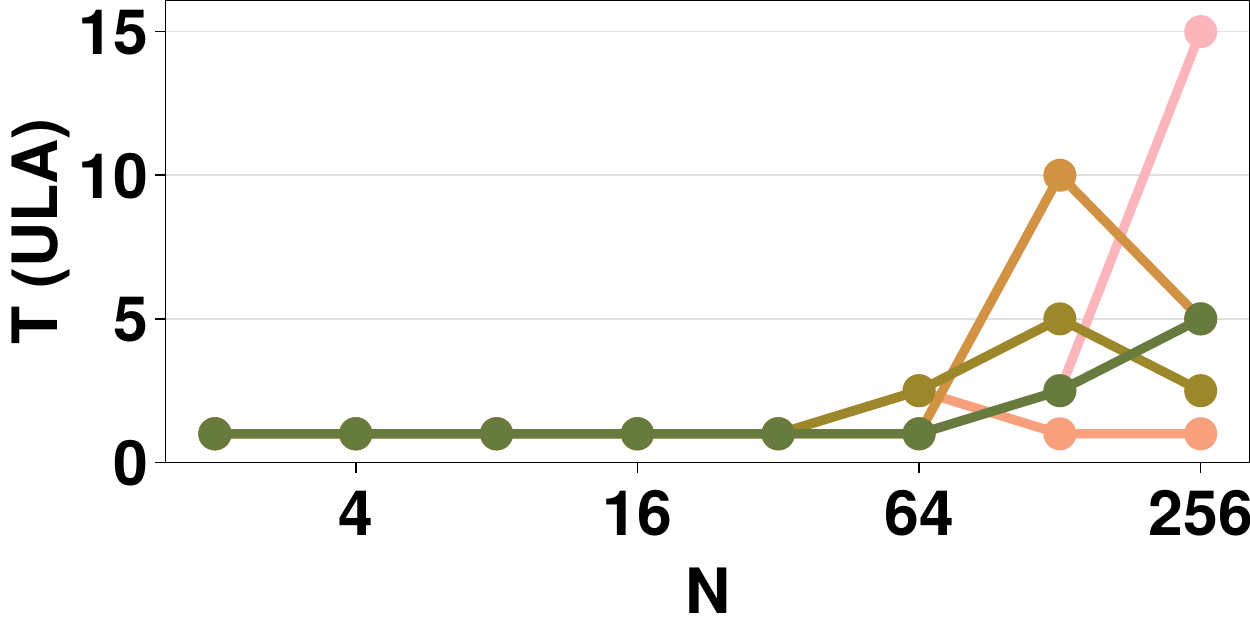}  
\end{subfigure}
\begin{subfigure}{0.32\linewidth}
\centering 
\includegraphics[width=\scalefactor\linewidth]{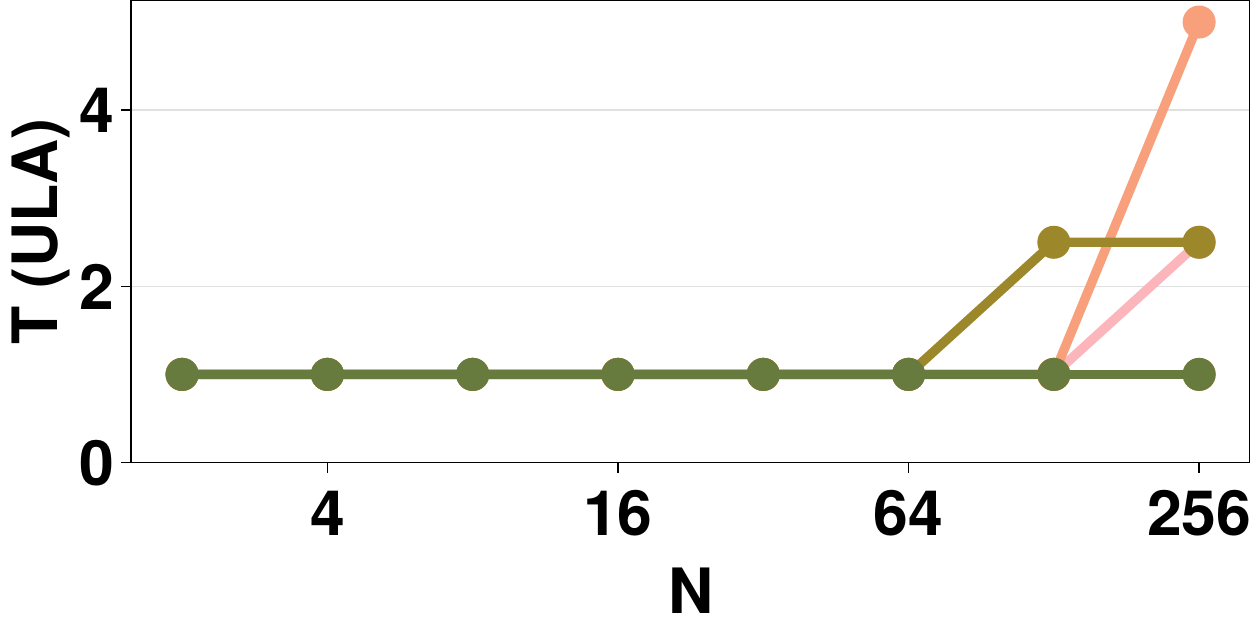}
\end{subfigure} 
\\
\begin{subfigure}{\linewidth}
\centering 
    \includegraphics[width=0.4\linewidth]{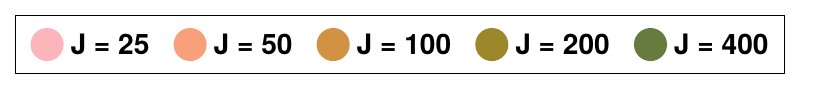}
\end{subfigure}
\caption{Optimal parameter choices for (from top to bottom) KFRFlow, KFRFlow-I, KFRD, SVGD, EKS, CBS, and ULA for the donut (left), butterfly (middle), and spaceships (right) examples.}
\label{fig:params2D}
\end{figure} 

\clearpage 

\subsubsection{Extended Values of $J$} 
\begin{figure}[h]
\centering 
 Donut \\[0.1cm] 
\begin{subfigure}{0.19\linewidth}
\includegraphics[width=\linewidth]{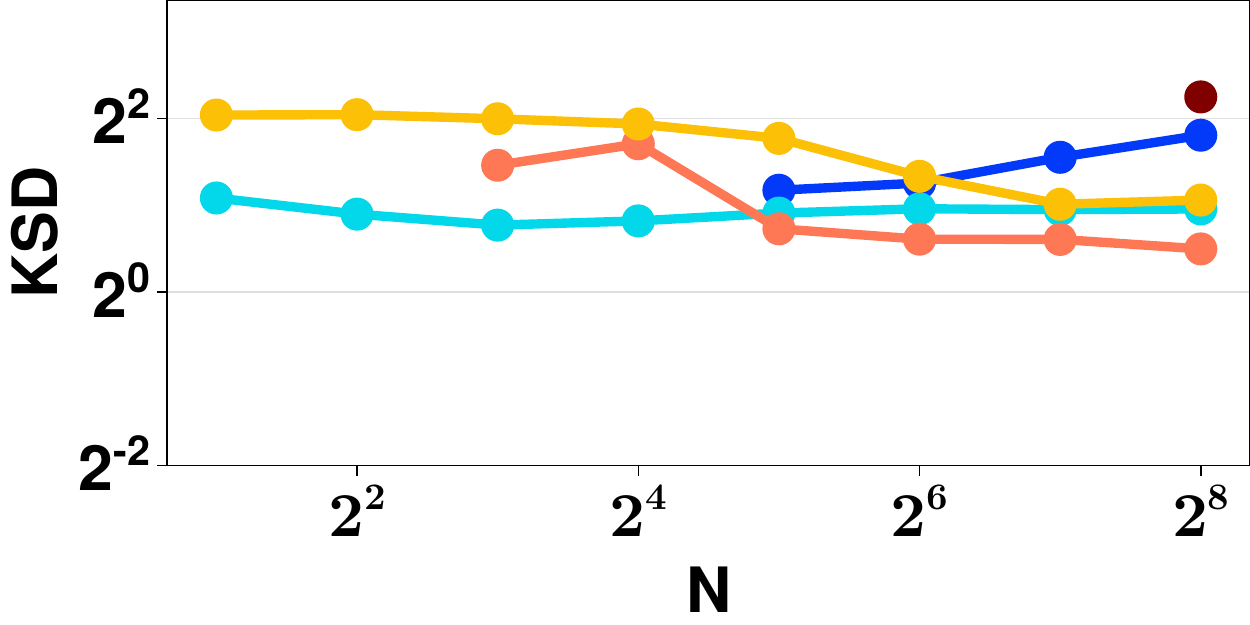}    
\end{subfigure}
\begin{subfigure}{0.19\linewidth}
\includegraphics[width=\linewidth]{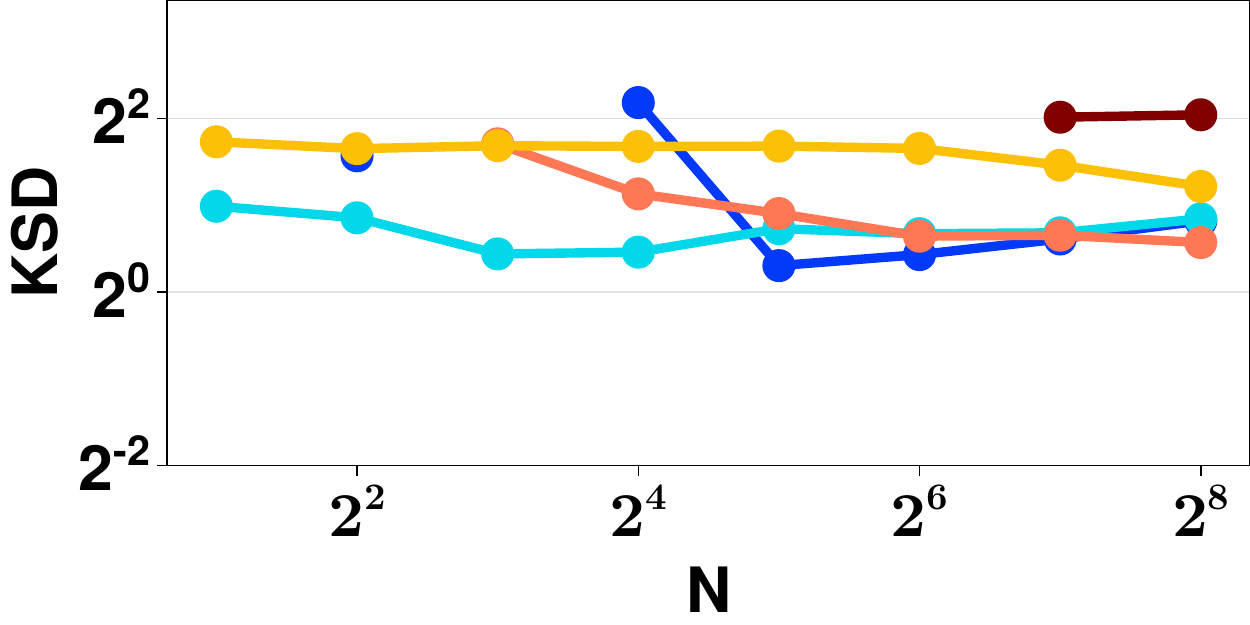}    
\end{subfigure}
\begin{subfigure}{0.19\linewidth}
\includegraphics[width=\linewidth]{figures/donut/gf_KSDvsNlogScale_noTitleJ100.pdf}    
\end{subfigure}
\begin{subfigure}{0.19\linewidth}
\includegraphics[width=\linewidth]{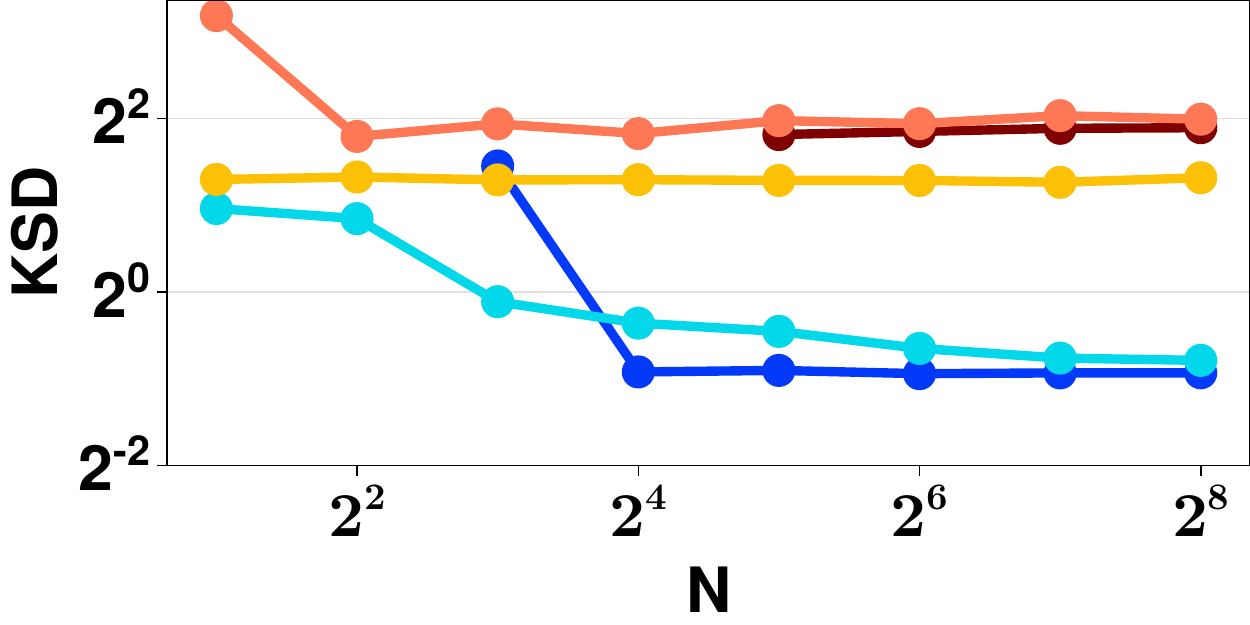}    
\end{subfigure}
\begin{subfigure}{0.19\linewidth}
\includegraphics[width=\linewidth]{figures/donut/gf_KSDvsNlogScale_noTitleJ400.pdf}    
\end{subfigure}
\\ 
 Butterfly \\[0.1cm]
\begin{subfigure}{0.19\linewidth}
\includegraphics[width=\linewidth]{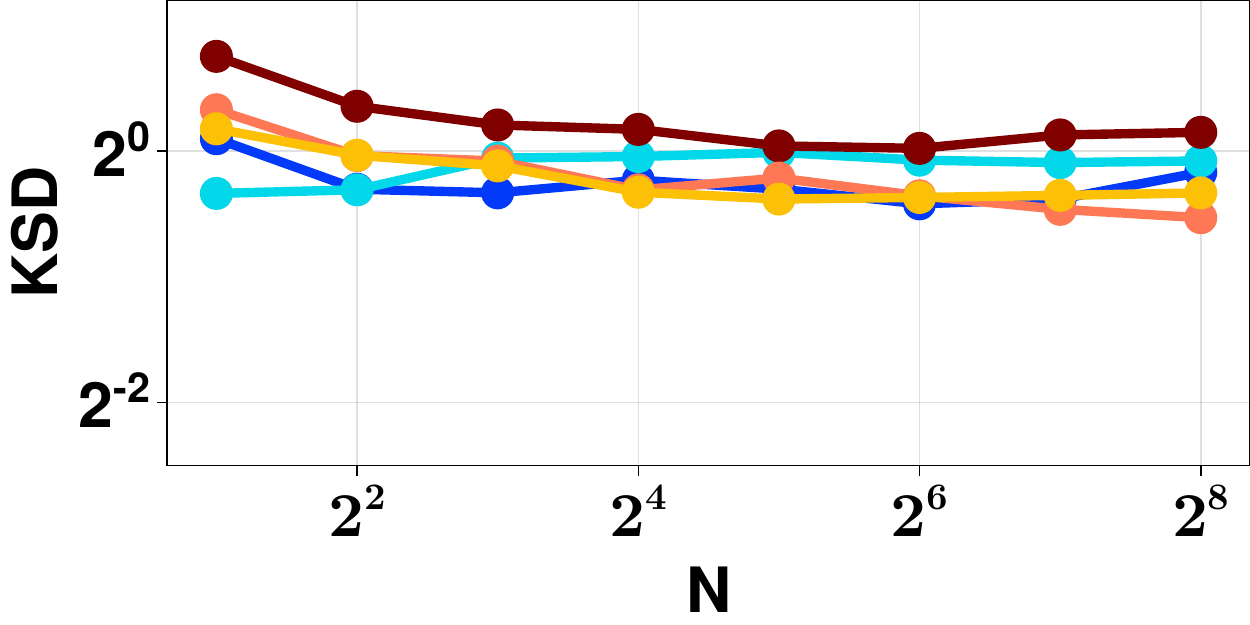}    
\end{subfigure}
\begin{subfigure}{0.19\linewidth}
\includegraphics[width=\linewidth]{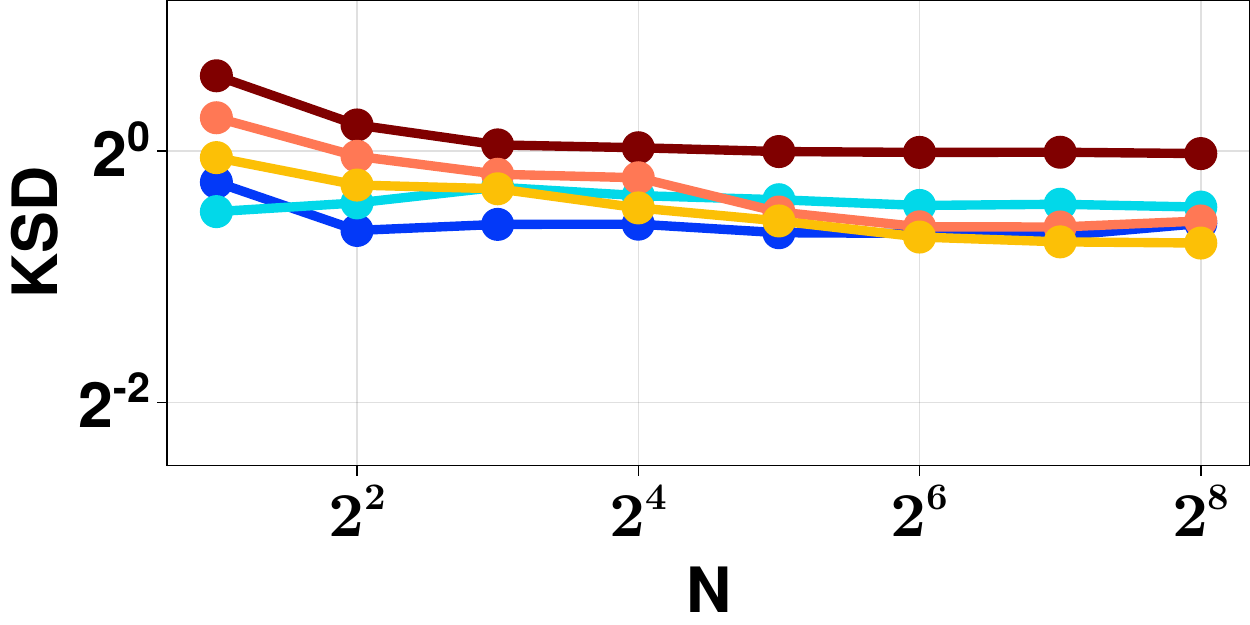}    
\end{subfigure}
\begin{subfigure}{0.19\linewidth}
\includegraphics[width=\linewidth]{figures/butterfly/gf_KSDvsNlogScale_noTitleJ100.pdf}    
\end{subfigure}
\begin{subfigure}{0.19\linewidth}
\includegraphics[width=\linewidth]{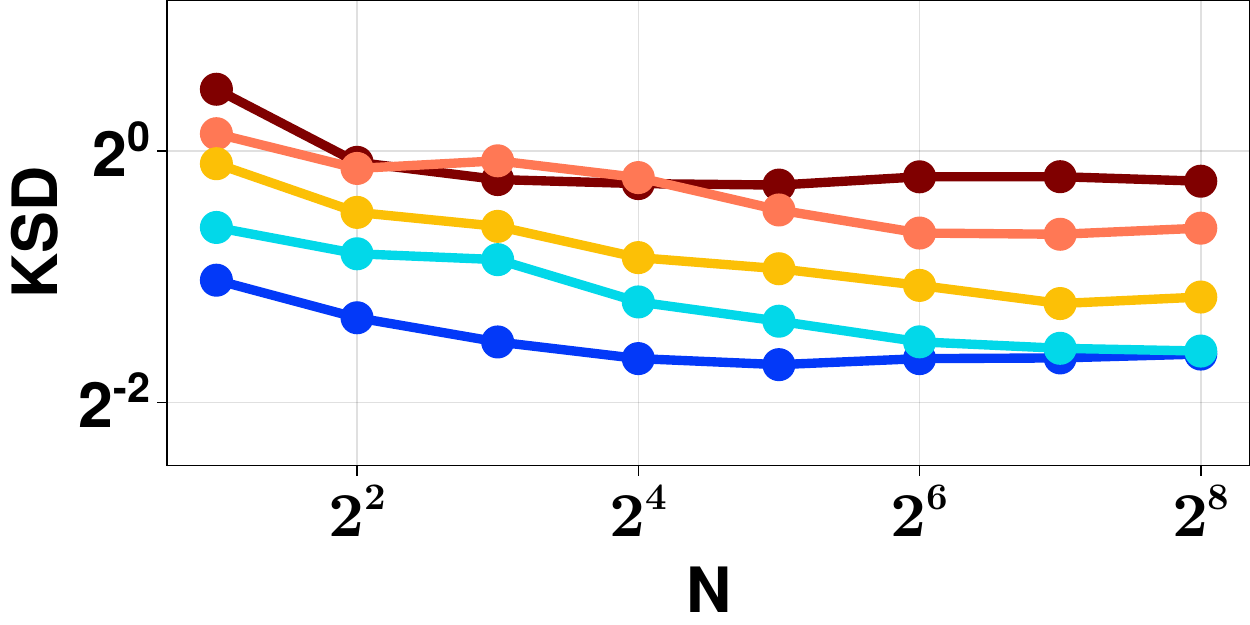}    
\end{subfigure}
\begin{subfigure}{0.19\linewidth}
\includegraphics[width=\linewidth]{figures/butterfly/gf_KSDvsNlogScale_noTitleJ400.pdf}    
\end{subfigure}
\\
 Spaceships \\[0.1cm] 
\begin{subfigure}{0.19\linewidth}
\includegraphics[width=\linewidth]{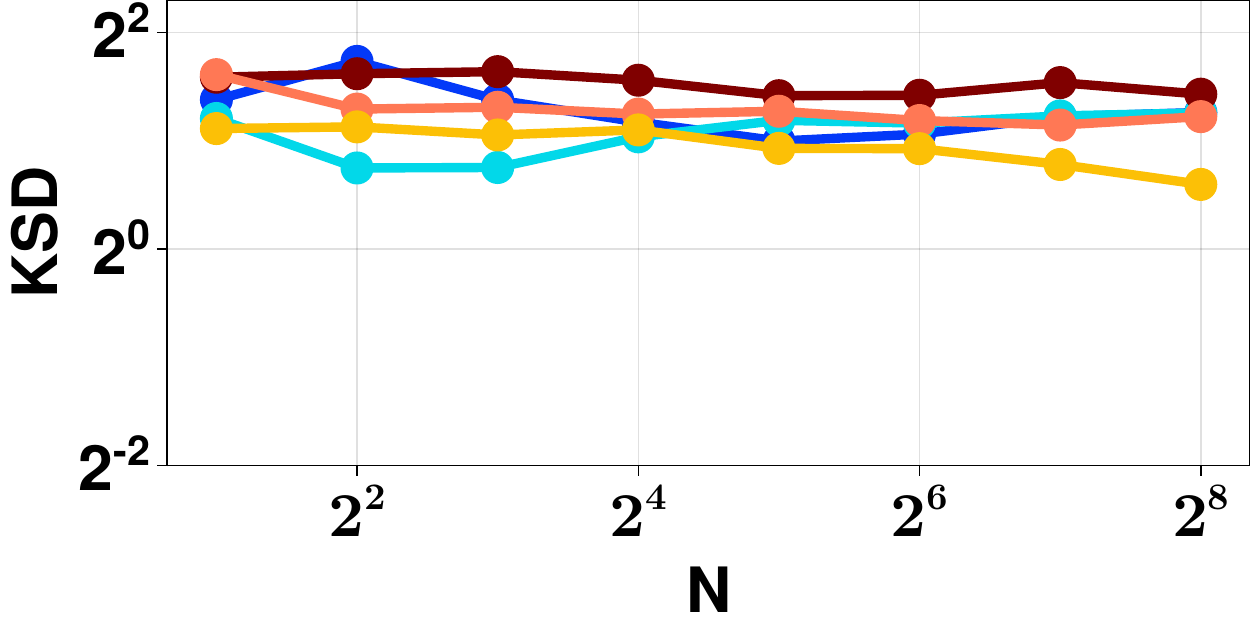}
\subcaption*{J=25}
\end{subfigure}
\begin{subfigure}{0.19\linewidth}
\includegraphics[width=\linewidth]{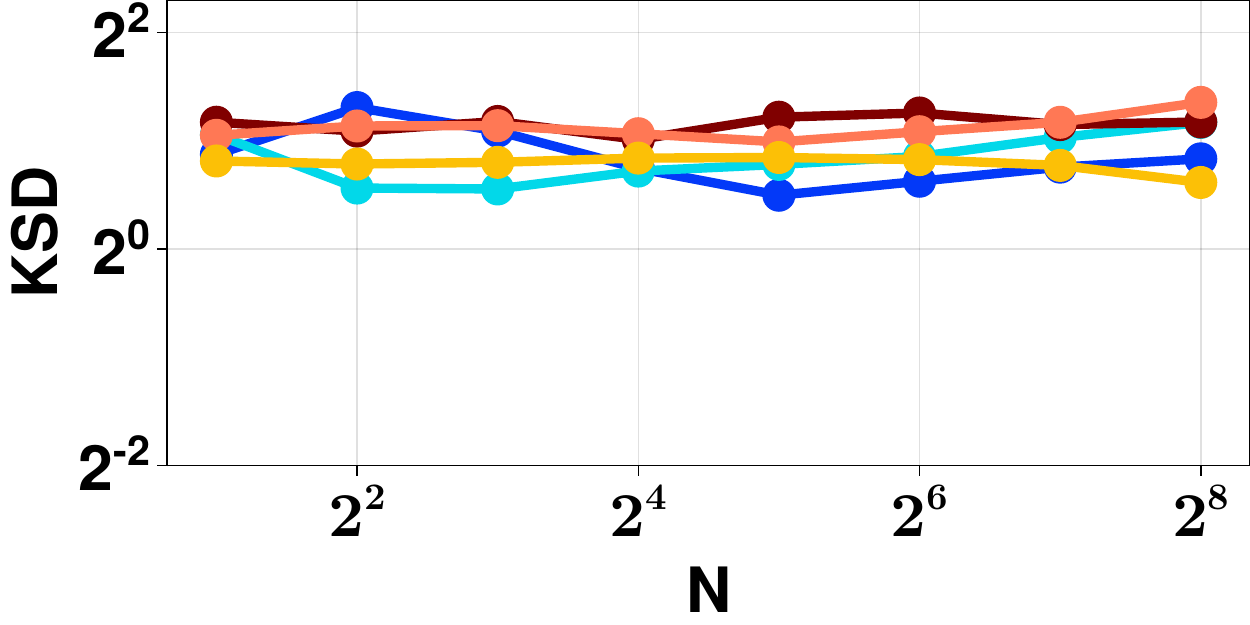}    
\subcaption*{J=50}
\end{subfigure}
\begin{subfigure}{0.19\linewidth}
\includegraphics[width=\linewidth]{figures/spaceships/gf_KSDvsNlogScale_noTitleJ100.pdf}    
\subcaption*{J=100}
\end{subfigure}
\begin{subfigure}{0.19\linewidth}
\includegraphics[width=\linewidth]{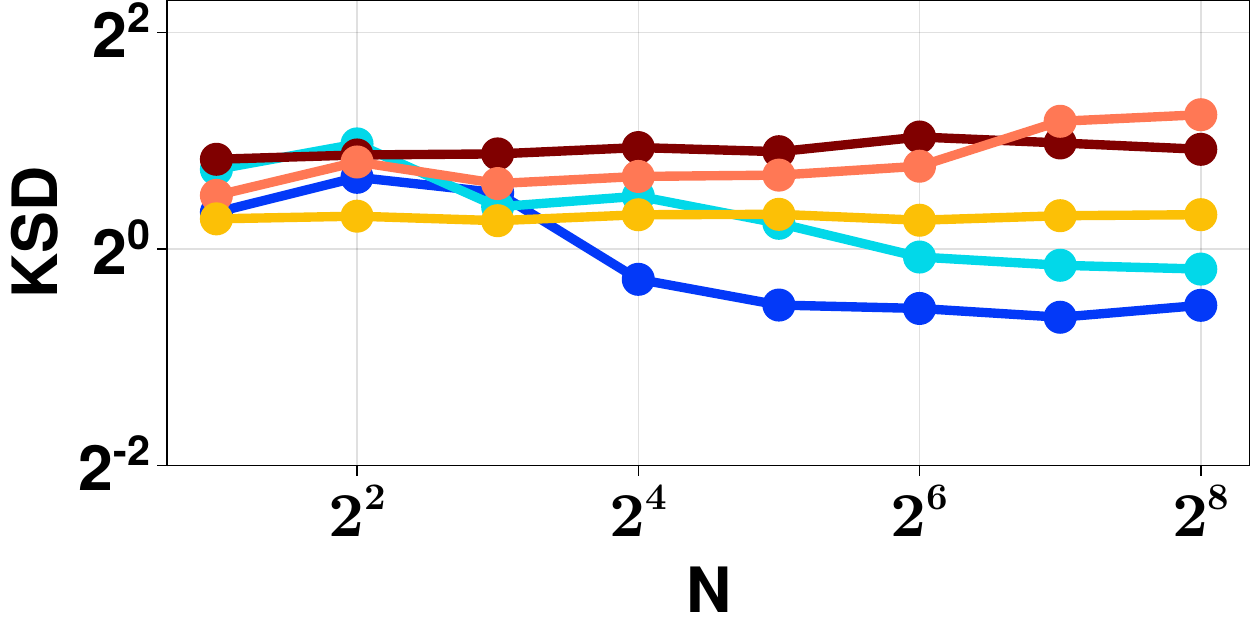}    
\subcaption*{J=200}
\end{subfigure}
\begin{subfigure}{0.19\linewidth}
\includegraphics[width=\linewidth]{figures/spaceships/gf_KSDvsNlogScale_noTitleJ400.pdf}    
\subcaption*{J=400}
\end{subfigure}
\\
\begin{subfigure}{\linewidth}
\centering 
\includegraphics[width=0.5\linewidth]{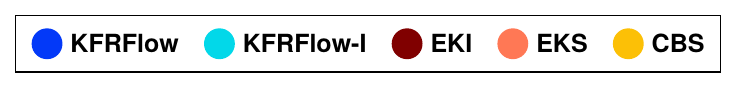}
\end{subfigure} %
\caption{\textbf{Two-dimensional posteriors:} average KSD at stopping time between $\pi_1$ and ensembles of size $J \in \{25, 50, 100, 200, 400\}$ generated by gradient-free samplers. A missing point indicates that a method was unstable at that setting of $N$.}
\label{fig:KSDvsNlogScale_gf_app}
 \end{figure}

\begin{figure}[H]
\centering 
 Donut \\[0.1cm] 
\begin{subfigure}{0.19\linewidth}
\includegraphics[width=\linewidth]{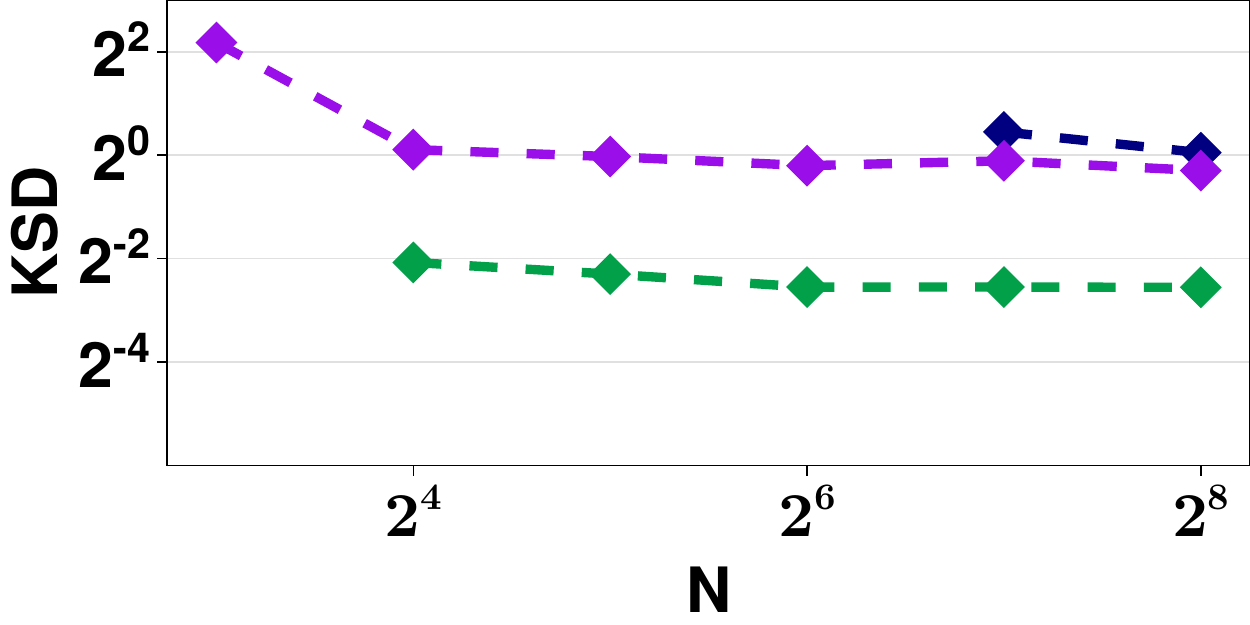}    
\end{subfigure}
\begin{subfigure}{0.19\linewidth}
\includegraphics[width=\linewidth]{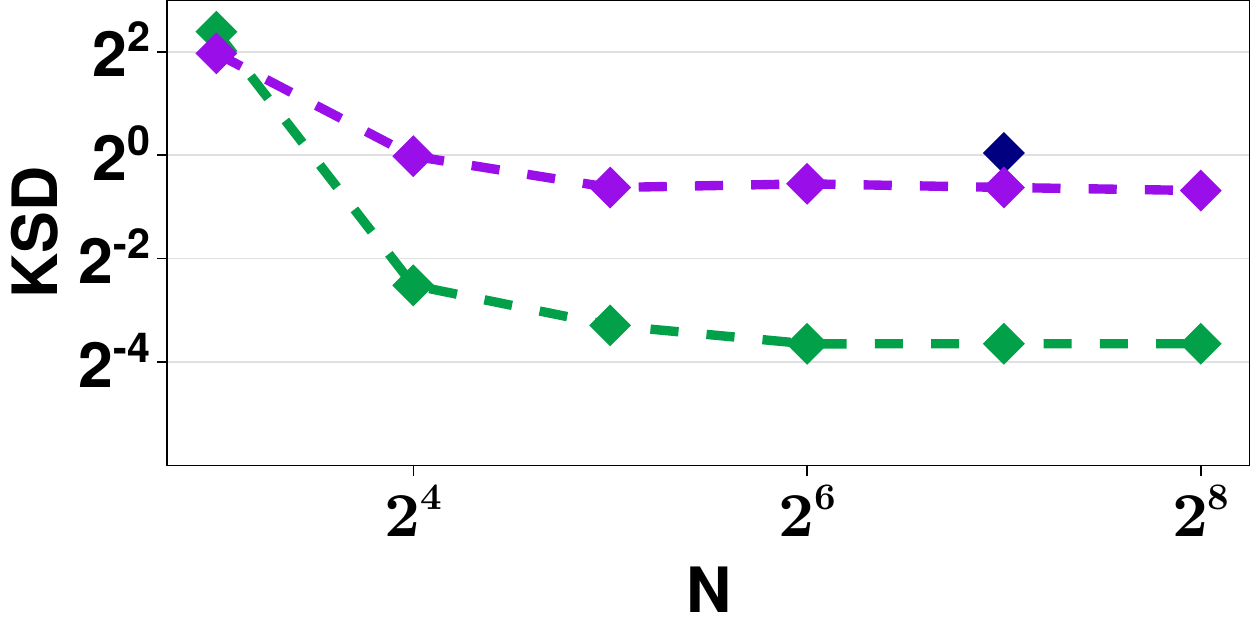}    
\end{subfigure}
\begin{subfigure}{0.19\linewidth}
\includegraphics[width=\linewidth]{figures/donut/gb_KSDvsNlogScale_noTitleJ100.pdf}    
\end{subfigure}
\begin{subfigure}{0.19\linewidth}
\includegraphics[width=\linewidth]{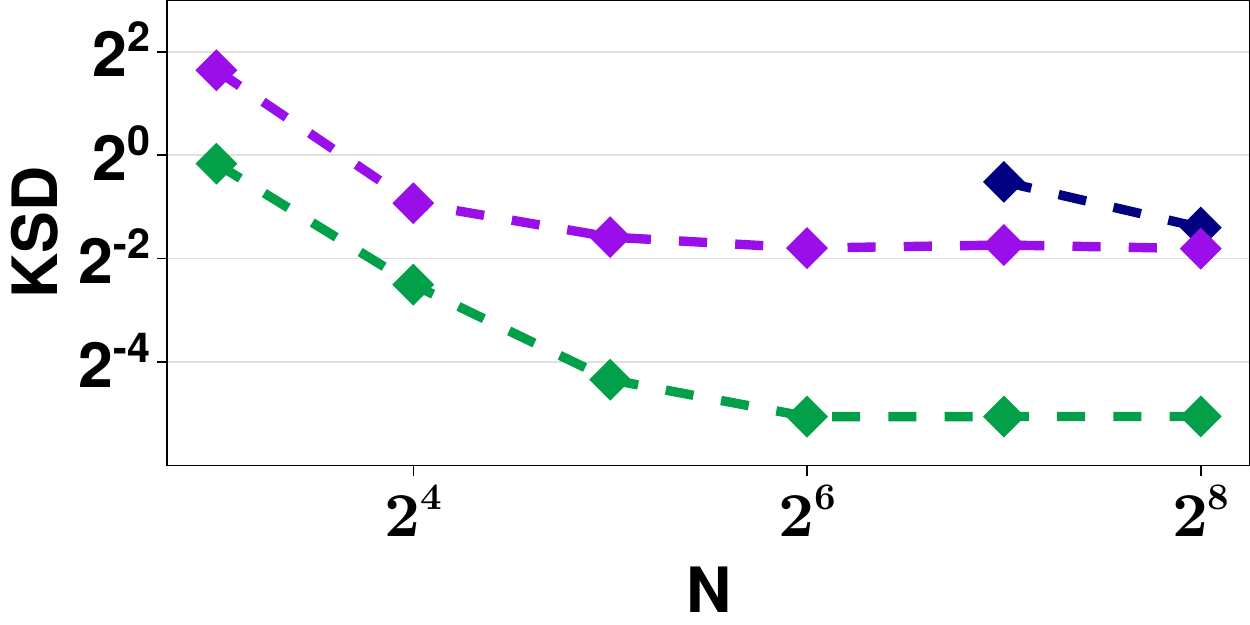}    
\end{subfigure}
\begin{subfigure}{0.19\linewidth}
\includegraphics[width=\linewidth]{figures/donut/gb_KSDvsNlogScale_noTitleJ400.pdf}    
\end{subfigure}
\\ 
 Butterfly \\[0.1cm]
\begin{subfigure}{0.19\linewidth}
\includegraphics[width=\linewidth]{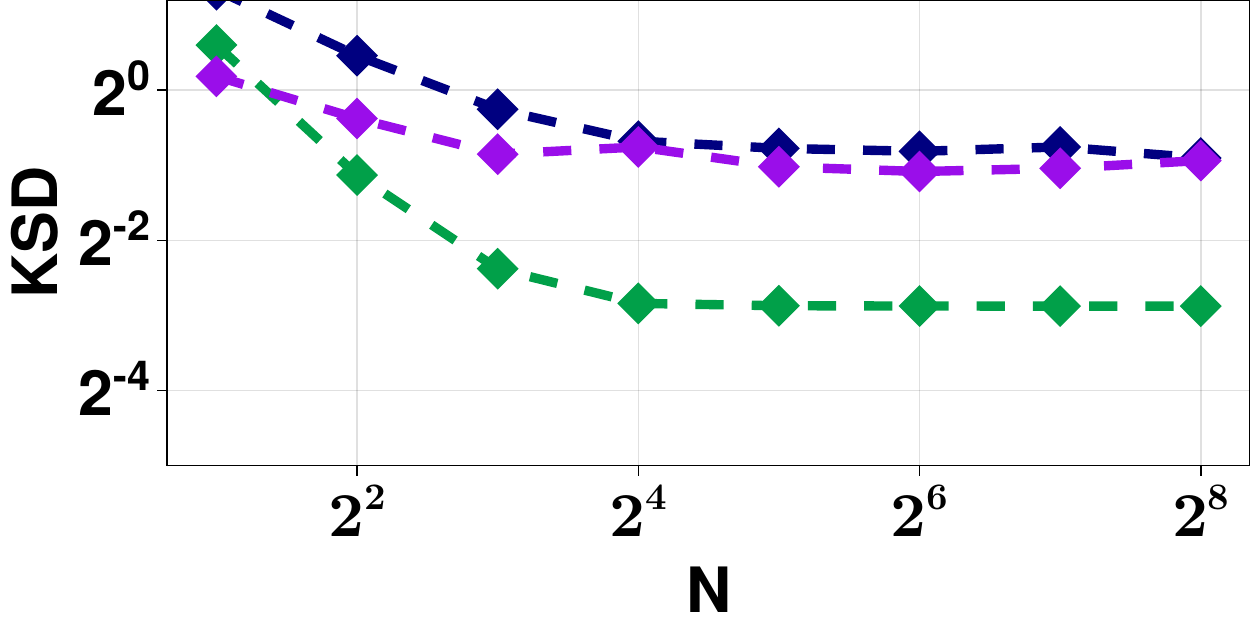}    
\end{subfigure}
\begin{subfigure}{0.19\linewidth}
\includegraphics[width=\linewidth]{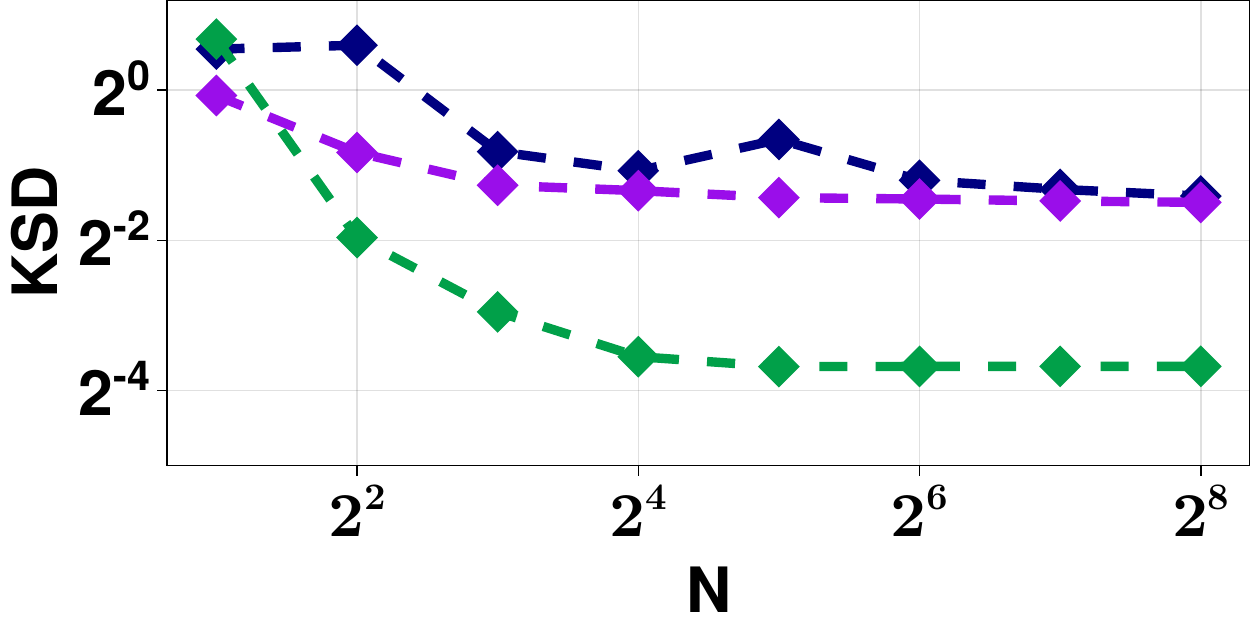}    
\end{subfigure}
\begin{subfigure}{0.19\linewidth}
\includegraphics[width=\linewidth]{figures/butterfly/gb_KSDvsNlogScale_noTitleJ100.pdf}    
\end{subfigure}
\begin{subfigure}{0.19\linewidth}
\includegraphics[width=\linewidth]{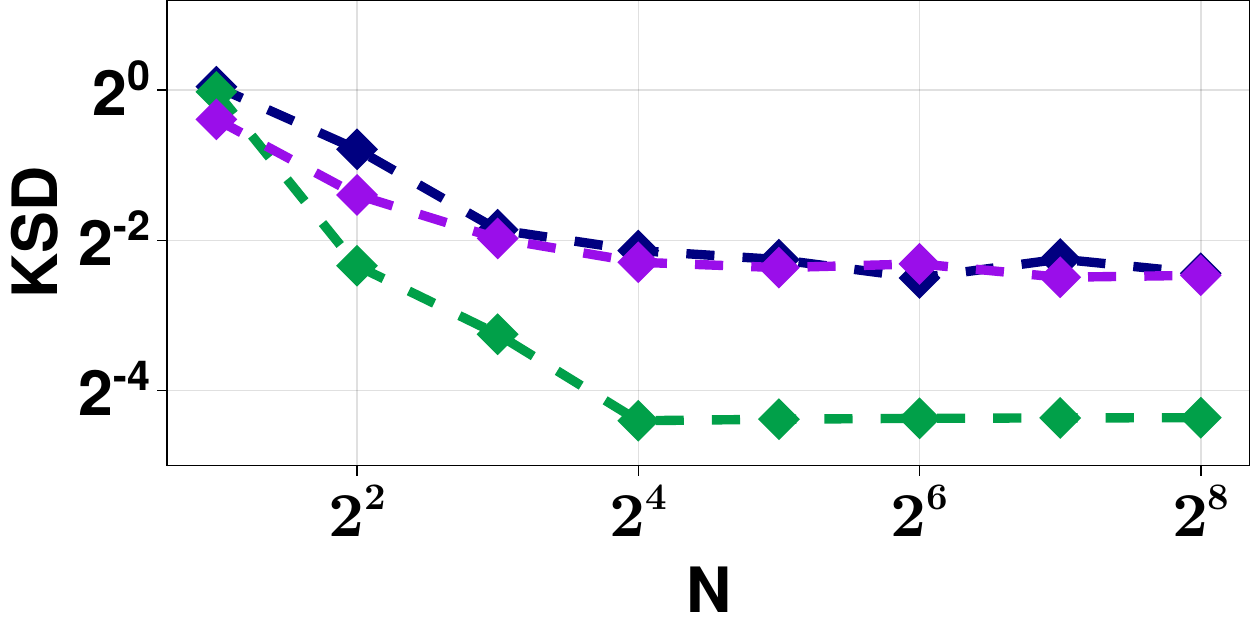}    
\end{subfigure}
\begin{subfigure}{0.19\linewidth}
\includegraphics[width=\linewidth]{figures/butterfly/gb_KSDvsNlogScale_noTitleJ400.pdf}    
\end{subfigure}
\\
 Spaceships \\[0.1cm] 
\begin{subfigure}{0.19\linewidth}
\includegraphics[width=\linewidth]{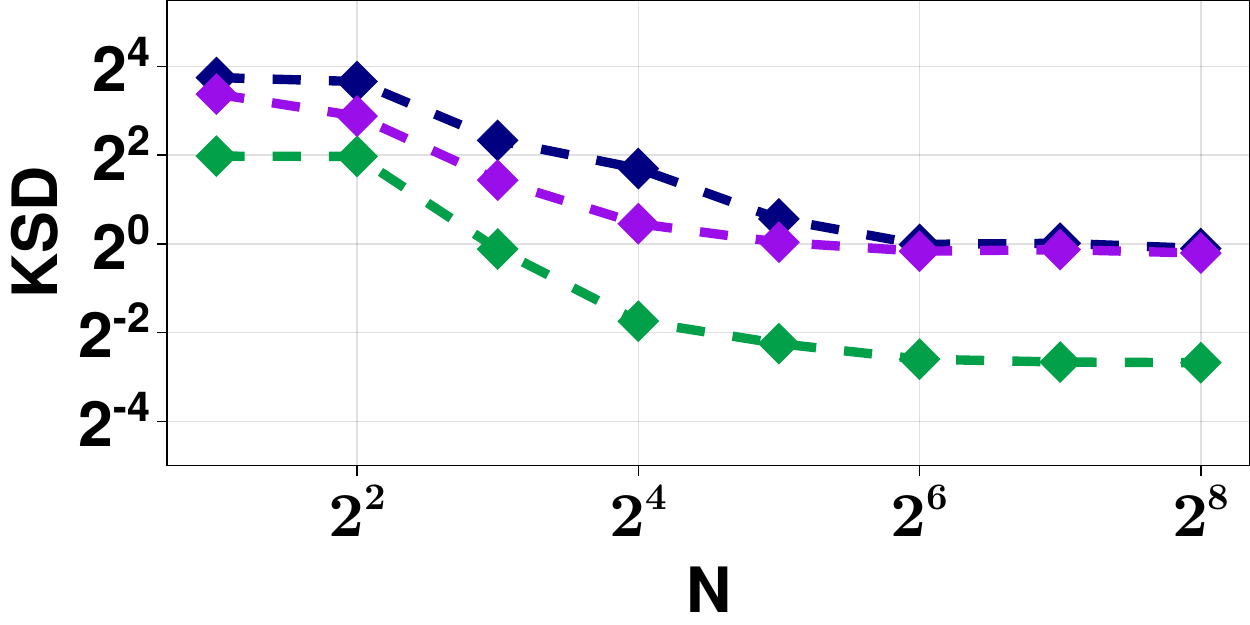}    
\subcaption*{J=25}
\end{subfigure}
\begin{subfigure}{0.19\linewidth}
\includegraphics[width=\linewidth]{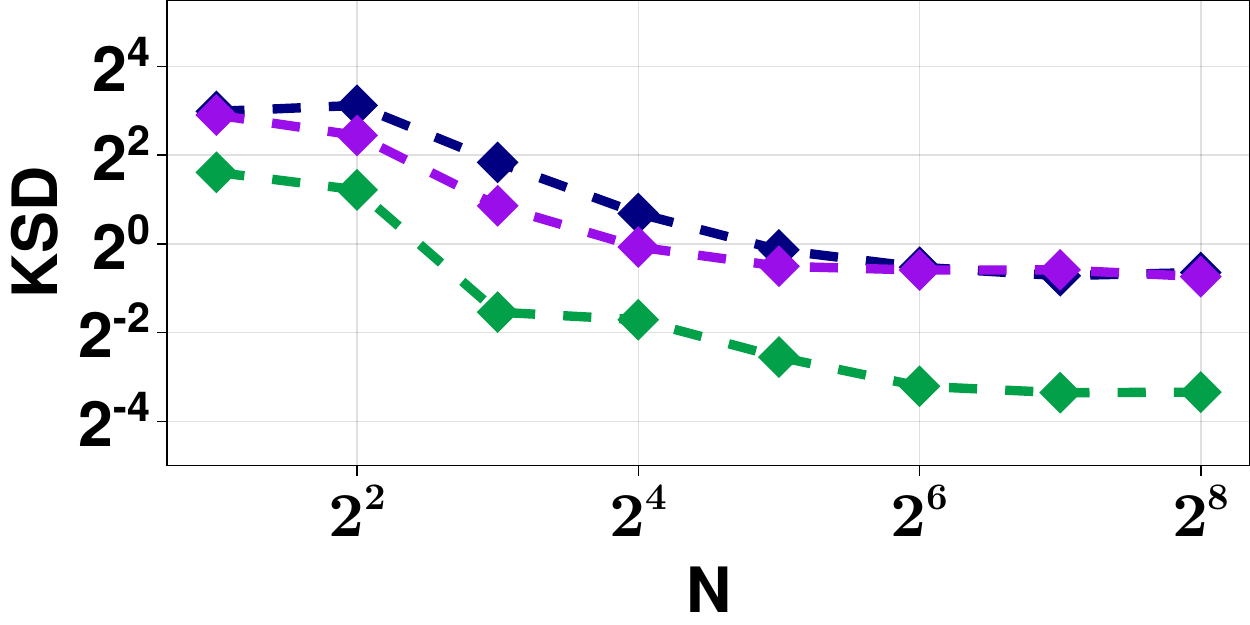}    
\subcaption*{J=50}
\end{subfigure}
\begin{subfigure}{0.19\linewidth}
\includegraphics[width=\linewidth]{figures/spaceships/gb_KSDvsNlogScale_noTitleJ100.pdf}    
\subcaption*{J=100}
\end{subfigure}
\begin{subfigure}{0.19\linewidth}
\includegraphics[width=\linewidth]{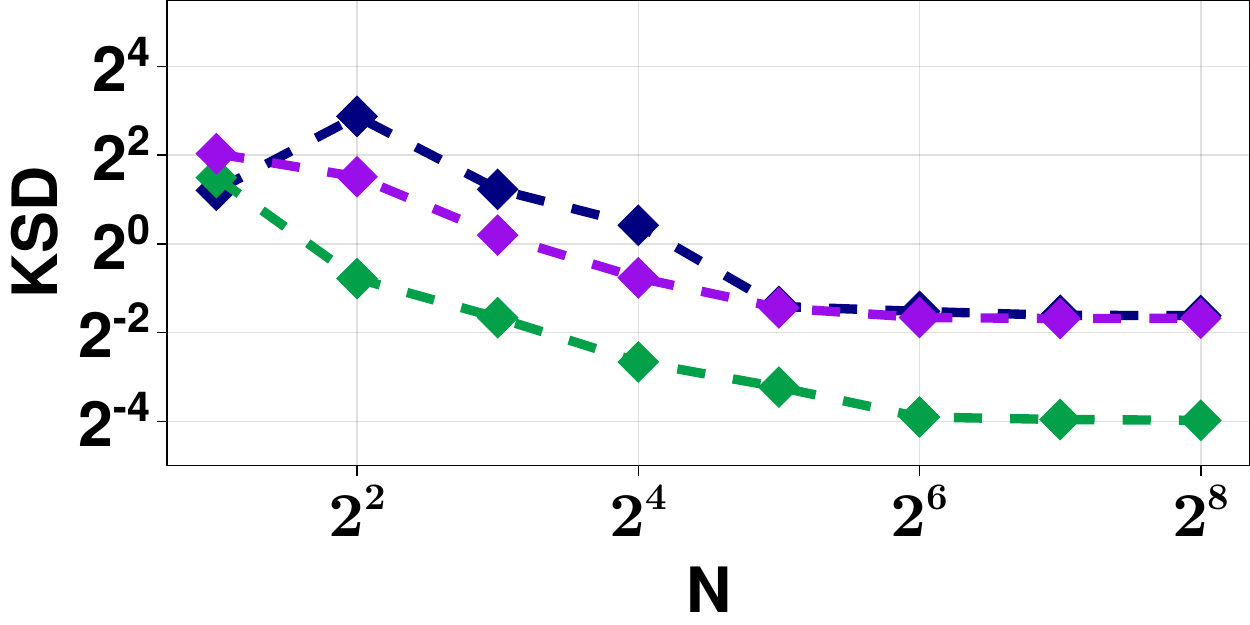}    
\subcaption*{J=200}
\end{subfigure}
\begin{subfigure}{0.19\linewidth}
\includegraphics[width=\linewidth]{figures/spaceships/gb_KSDvsNlogScale_noTitleJ400.pdf}    
\subcaption*{J=400}
\end{subfigure}
\\
\begin{subfigure}{\linewidth}
\centering 
\includegraphics[width=0.3\linewidth]{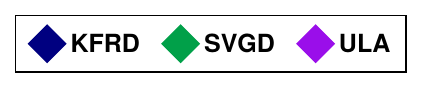}
\end{subfigure} %
\caption{\textbf{Two-dimensional posteriors:} average KSD at stopping time between $\pi_1$ and ensembles of size $J \in \{25, 50, 100, 200, 400\}$ generated by gradient-based samplers. A missing point indicates that a method was unstable at that setting of $N$.}
\label{fig:KSDvsNlogScale_gb_app}
\end{figure}

\begin{figure}[h]
\centering 
 Donut \\[0.1cm] 
\begin{subfigure}{0.19\linewidth}
\includegraphics[width=\linewidth]{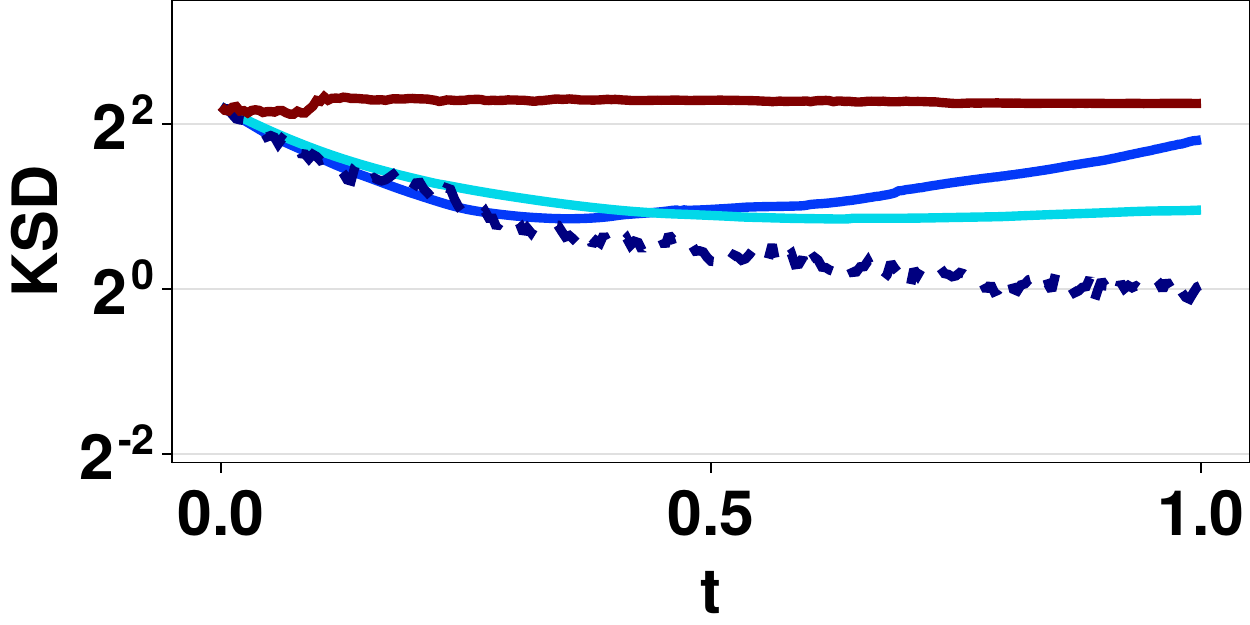}    
\end{subfigure}
\begin{subfigure}{0.19\linewidth}
\includegraphics[width=\linewidth]{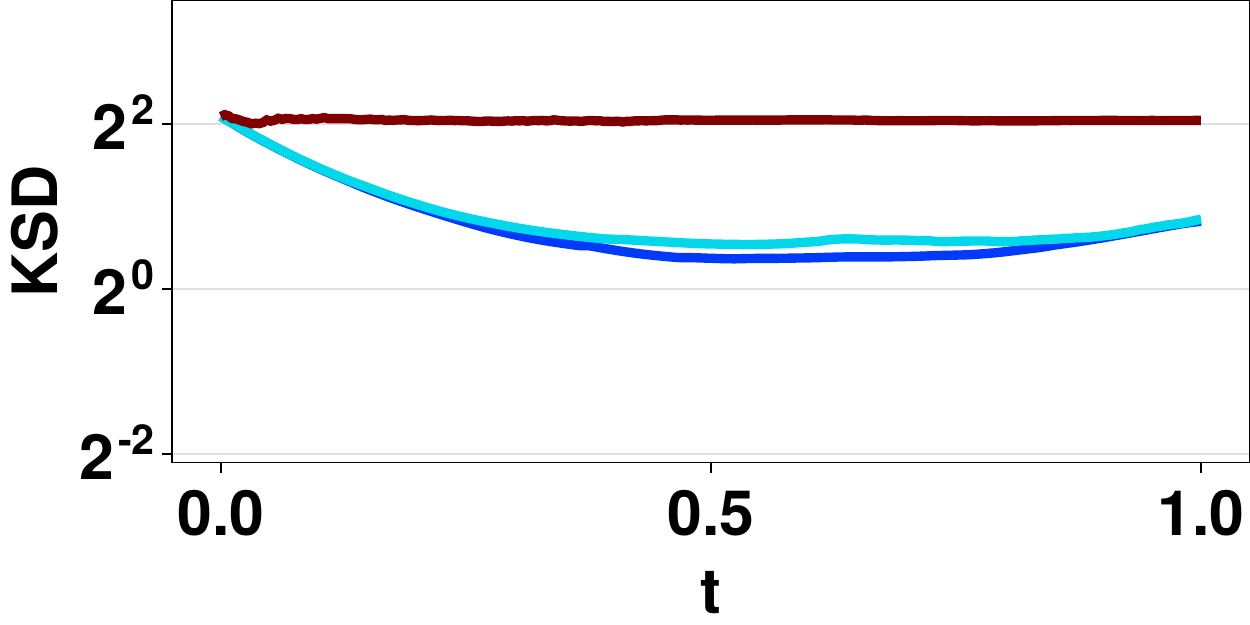}    
\end{subfigure}
\begin{subfigure}{0.19\linewidth}
\includegraphics[width=\linewidth]{figures/donut/KSDevolUTlogScale_noTitleJ100dt2e-8.pdf}    
\end{subfigure}
\begin{subfigure}{0.19\linewidth}
\includegraphics[width=\linewidth]{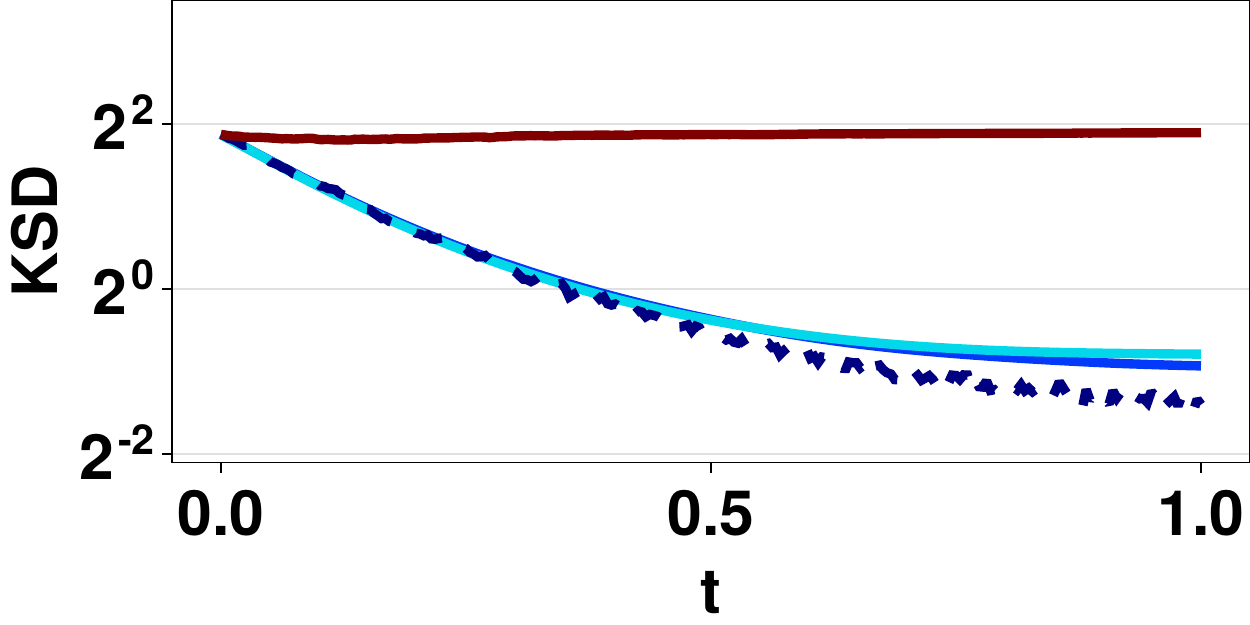}    
\end{subfigure}
\begin{subfigure}{0.19\linewidth}
\includegraphics[width=\linewidth]{figures/donut/KSDevolUTlogScale_noTitleJ400dt2e-8.pdf}    
\end{subfigure}
\\ 
 Butterfly \\[0.1cm]
\begin{subfigure}{0.19\linewidth}
\includegraphics[width=\linewidth]{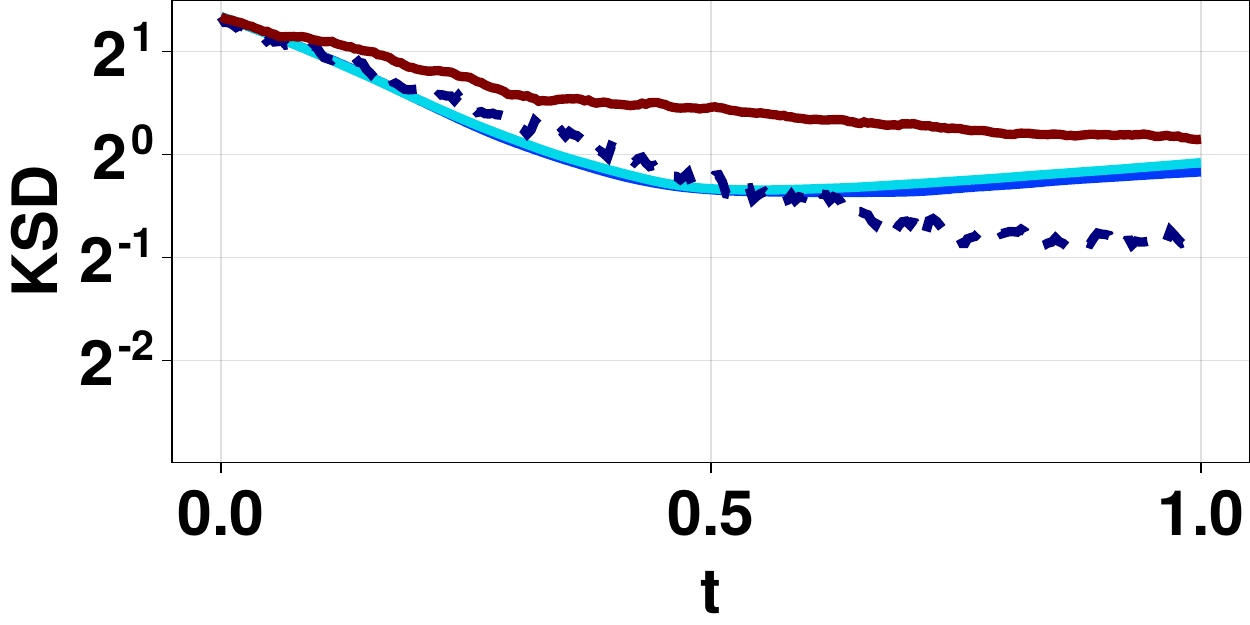}    
\end{subfigure}
\begin{subfigure}{0.19\linewidth}
\includegraphics[width=\linewidth]{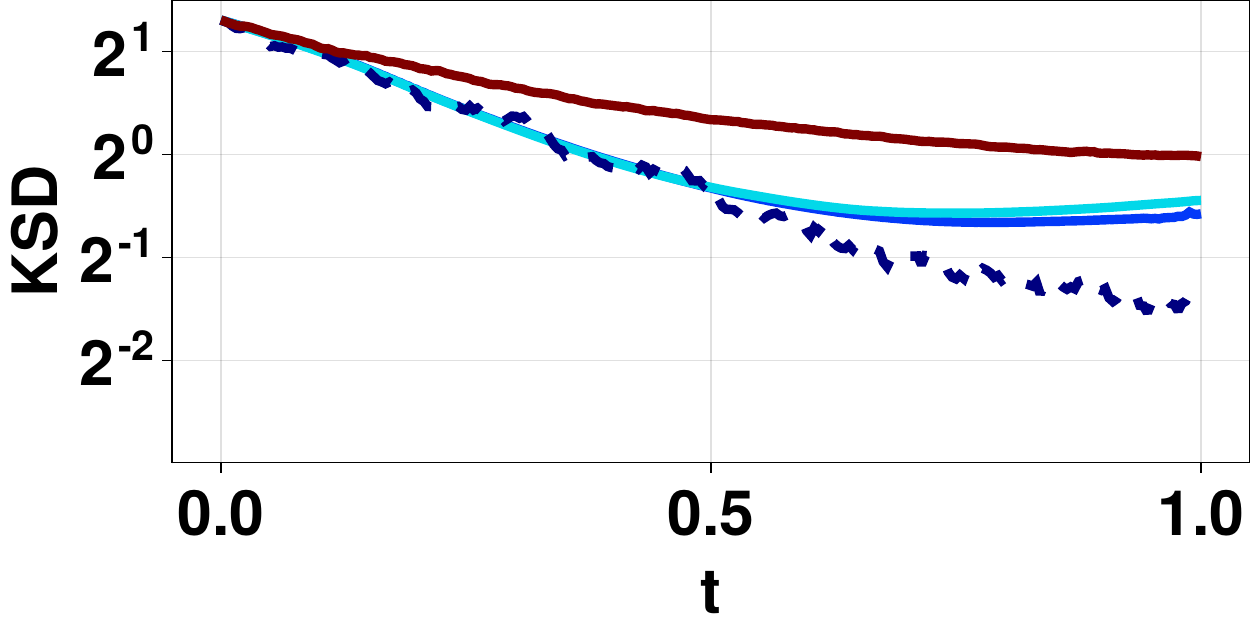}    
\end{subfigure}
\begin{subfigure}{0.19\linewidth}
\includegraphics[width=\linewidth]{figures/butterfly/KSDevolUTlogScale_noTitleJ100dt2e-8.pdf}    
\end{subfigure}
\begin{subfigure}{0.19\linewidth}
\includegraphics[width=\linewidth]{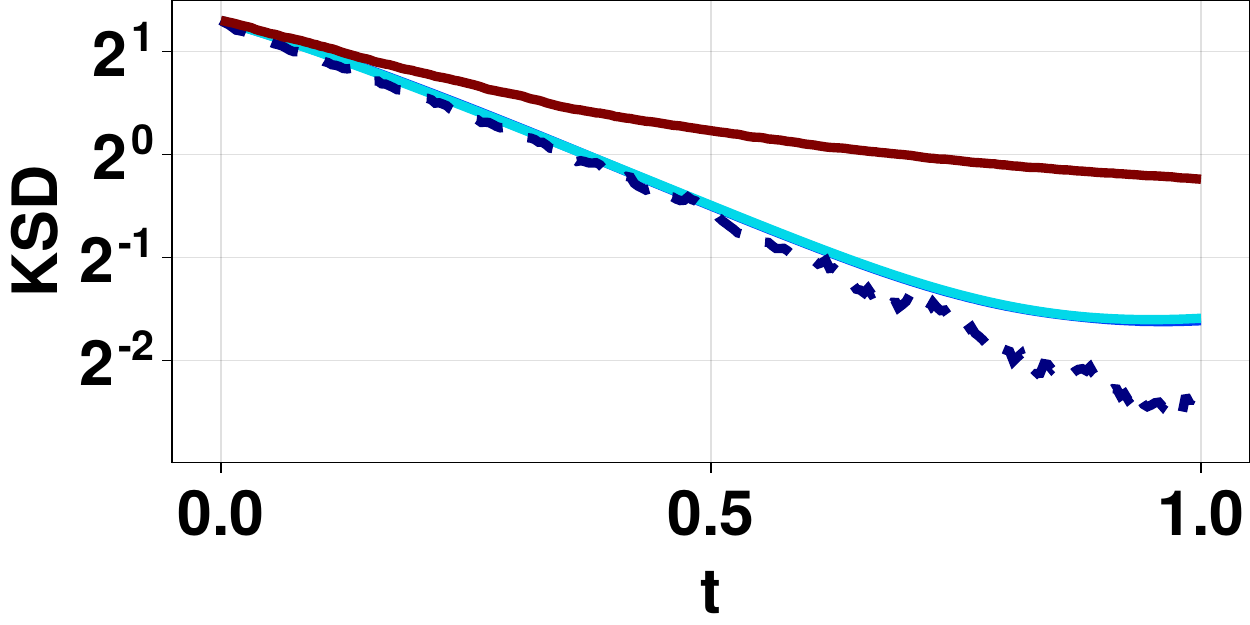}    
\end{subfigure}
\begin{subfigure}{0.19\linewidth}
\includegraphics[width=\linewidth]{figures/butterfly/KSDevolUTlogScale_noTitleJ400dt2e-8.pdf}    
\end{subfigure}
\\
 Spaceships \\[0.1cm] 
\begin{subfigure}{0.19\linewidth}
\includegraphics[width=\linewidth]{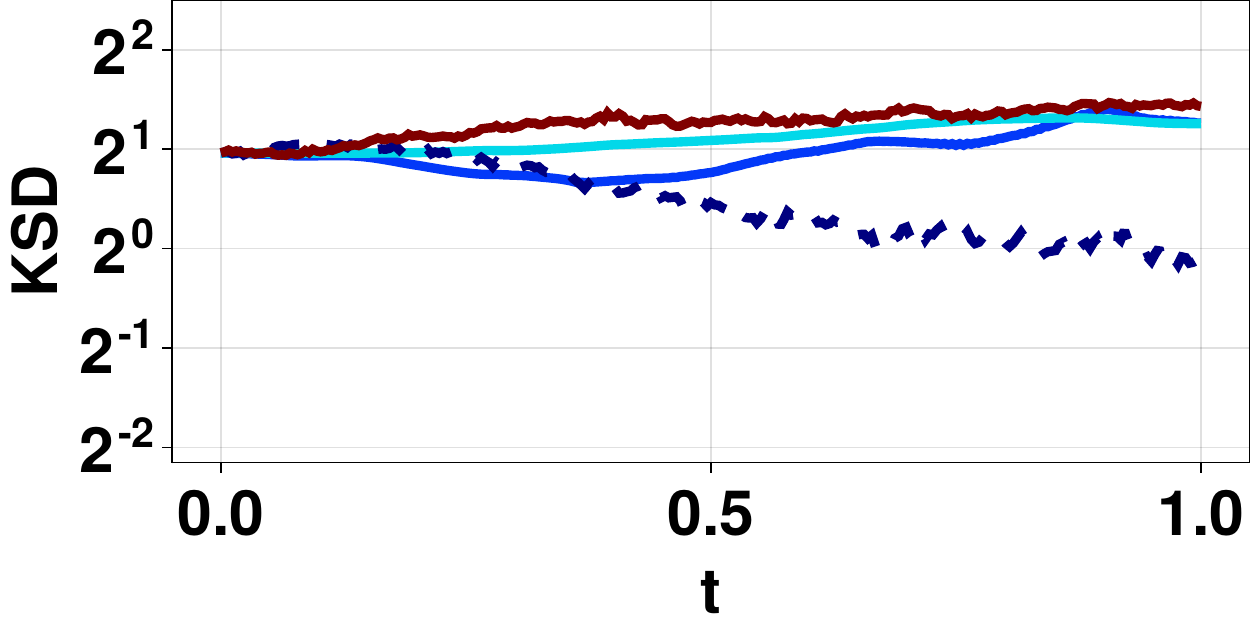}   
\subcaption*{J=25}
\end{subfigure}
\begin{subfigure}{0.19\linewidth}
\includegraphics[width=\linewidth]{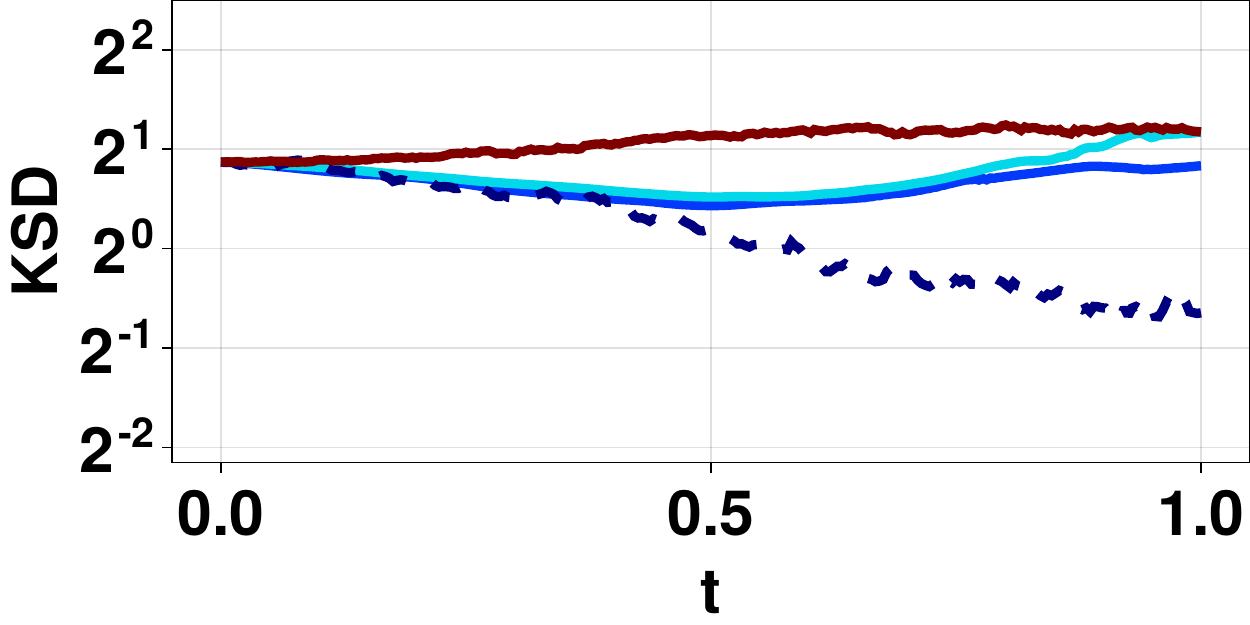}  
\subcaption*{J=50}
\end{subfigure}
\begin{subfigure}{0.19\linewidth}
\includegraphics[width=\linewidth]{figures/spaceships/KSDevolUTlogScale_noTitleJ100dt2e-8.pdf}    
\subcaption*{J=100}
\end{subfigure}
\begin{subfigure}{0.19\linewidth}
\includegraphics[width=\linewidth]{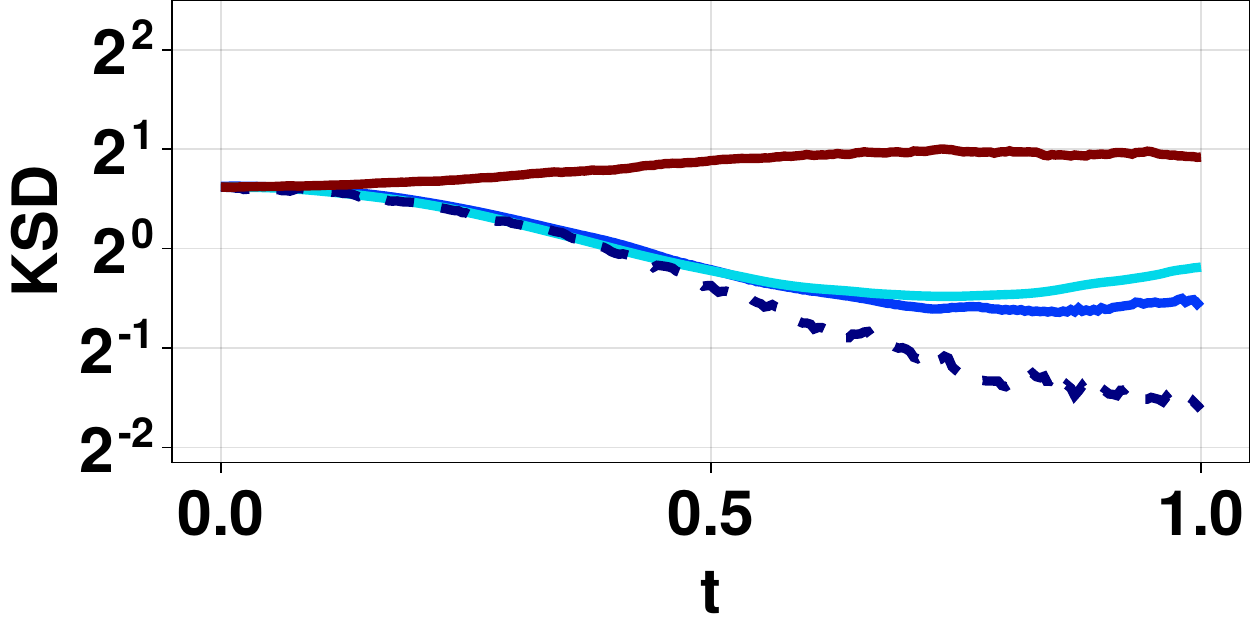}    
\subcaption*{J=200}
\end{subfigure}
\begin{subfigure}{0.19\linewidth}
\includegraphics[width=\linewidth]{figures/spaceships/KSDevolUTlogScale_noTitleJ400dt2e-8.pdf}    
\subcaption*{J=400}
\end{subfigure}
\begin{subfigure}{\linewidth}
    \centering 
    \includegraphics[width=0.7\linewidth]{figures/UnitLegend_horizontal.pdf}
\end{subfigure}
\caption{\textbf{Two-dimensional posteriors:} evolution of KSD between $\pi_1$ and samples generated by the unit-time methods KFRFlow, KFRFlow-I, KFRD, and EKI with $t \in [0,1]$ for ensembles of size $J \in \{25, 50, 100, 200, 400\}$ and $\Delta t = 2^{-8}$. }
\label{fig:KSDevolUTlogScale}
 \end{figure}

In \cref{fig:KSDvsNlogScale_gf_app,fig:KSDvsNlogScale_gb_app} we compare the quality of samples generated by KFRFlow and KFRFlow-I to those of EKI, EKS, and CBS, and the quality of samples generated by KFRD to those of SVGD, and ULA. In \cref{fig:KSDevolUTlogScale} we show the evolution of KSD between $\pi_1$ and samples generated by the unit-time methods KFRFlow, KFRFlow-I, KFRD, and EKI as a function of $t$, with $\Delta t = 2^{-8}$. These figures contain results for a wider range of $J$ than the main-body \cref{fig:KSDvsN_gf_par,fig:KSDvsN_gb_parallel,fig:ksdEvol2D}. \rev{In these and following figures, results corresponding to gradient-free methods are plotted with solid lines and circles, while results corresponding to gradient-based methods are plotted with dashed lines and diamonds.} 

 \subsubsection{KSD between Samples and Intermediate Distributions} 
 \label{sec:discError2D}
\begin{figure}[H]
\centering 
 Donut \\[0.1cm] 
\begin{subfigure}{0.19\linewidth}
\includegraphics[width=\linewidth]{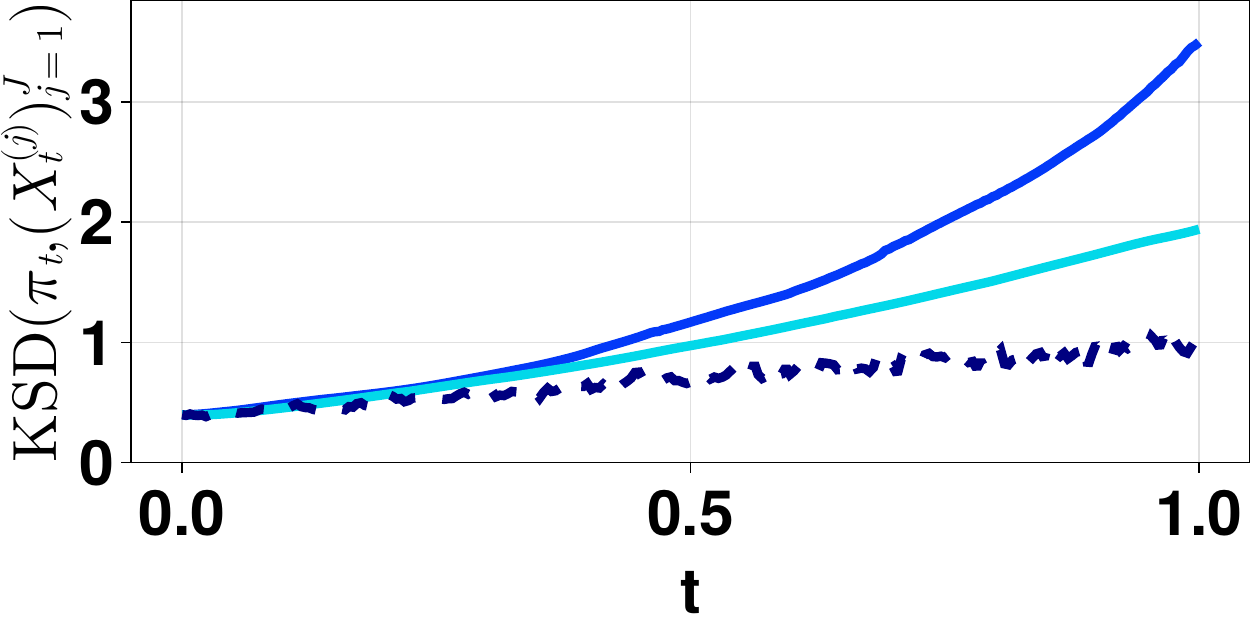}    
\end{subfigure}
\begin{subfigure}{0.19\linewidth}
\includegraphics[width=\linewidth]{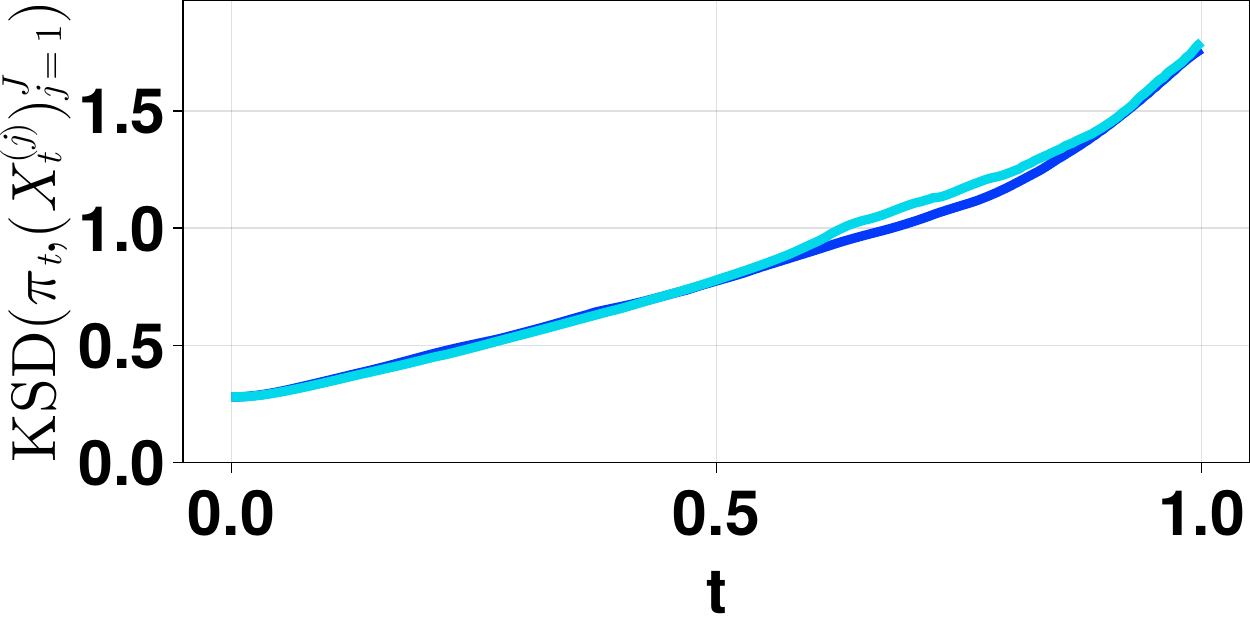}    
\end{subfigure}
\begin{subfigure}{0.19\linewidth}
\includegraphics[width=\linewidth]{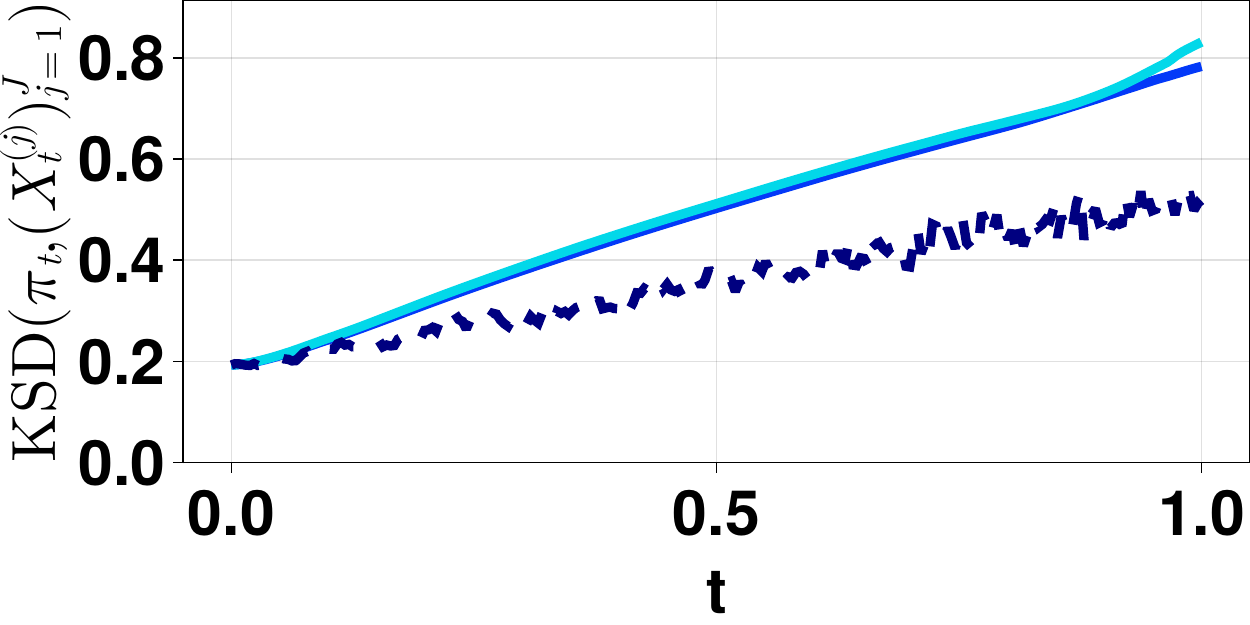}    
\end{subfigure}
\begin{subfigure}{0.19\linewidth}
\includegraphics[width=\linewidth]{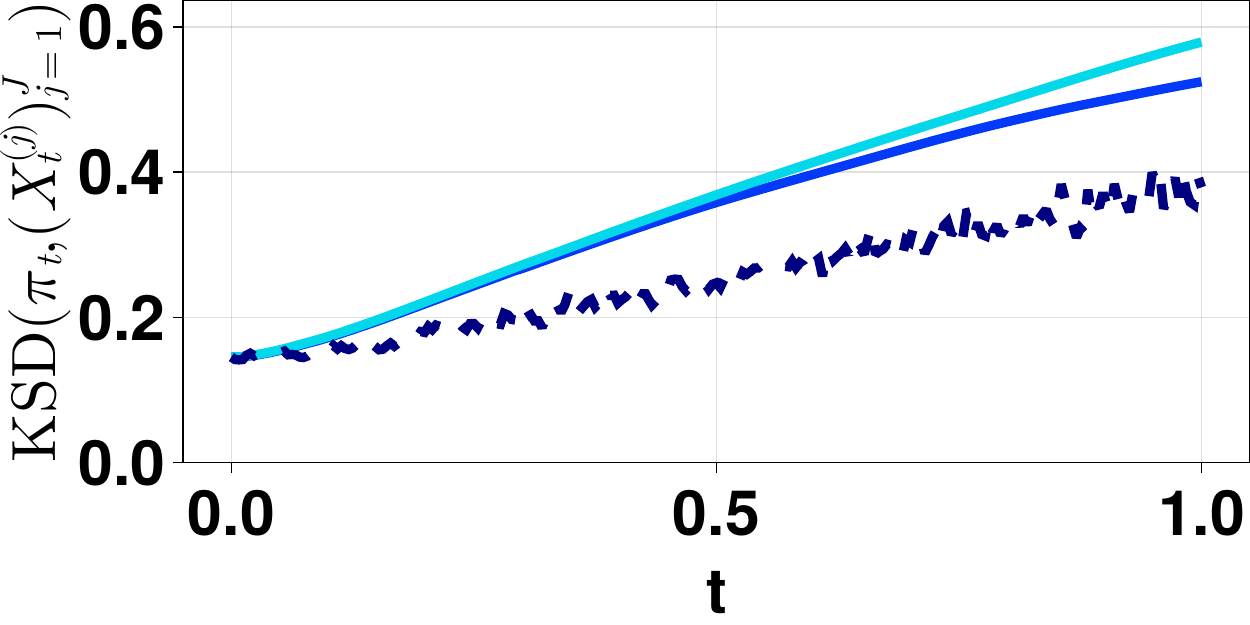}    
\end{subfigure}
\begin{subfigure}{0.19\linewidth}
\includegraphics[width=\linewidth]{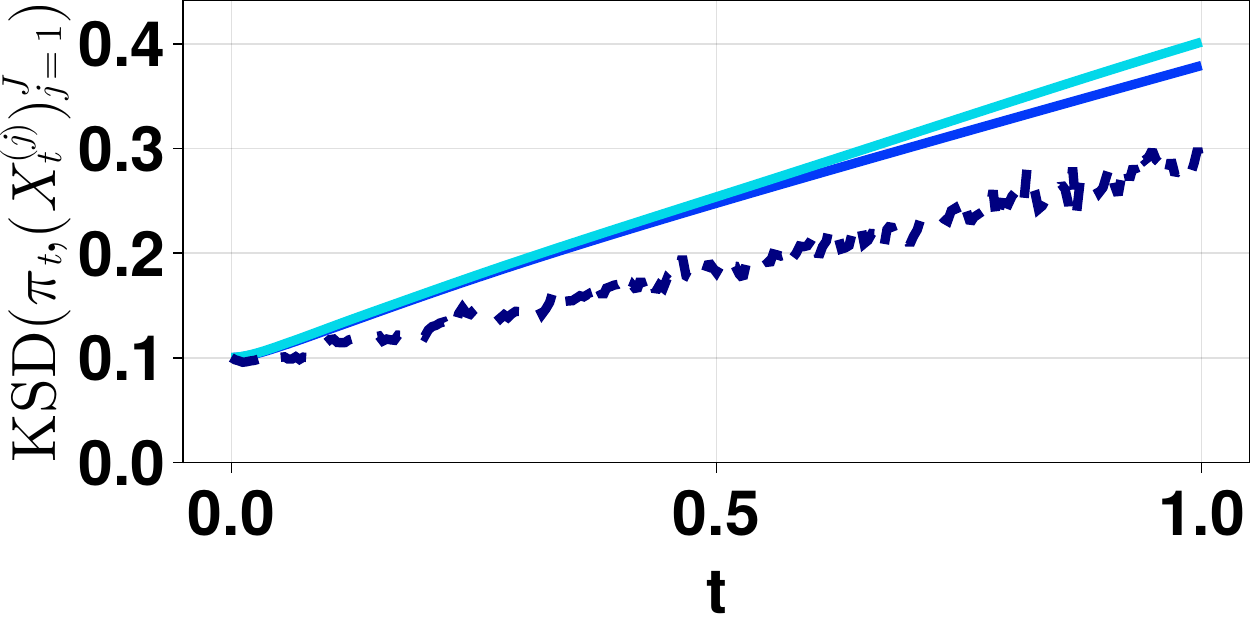}    
\end{subfigure}
\\ 
 Butterfly \\[0.1cm]
\begin{subfigure}{0.19\linewidth}
\includegraphics[width=\linewidth]{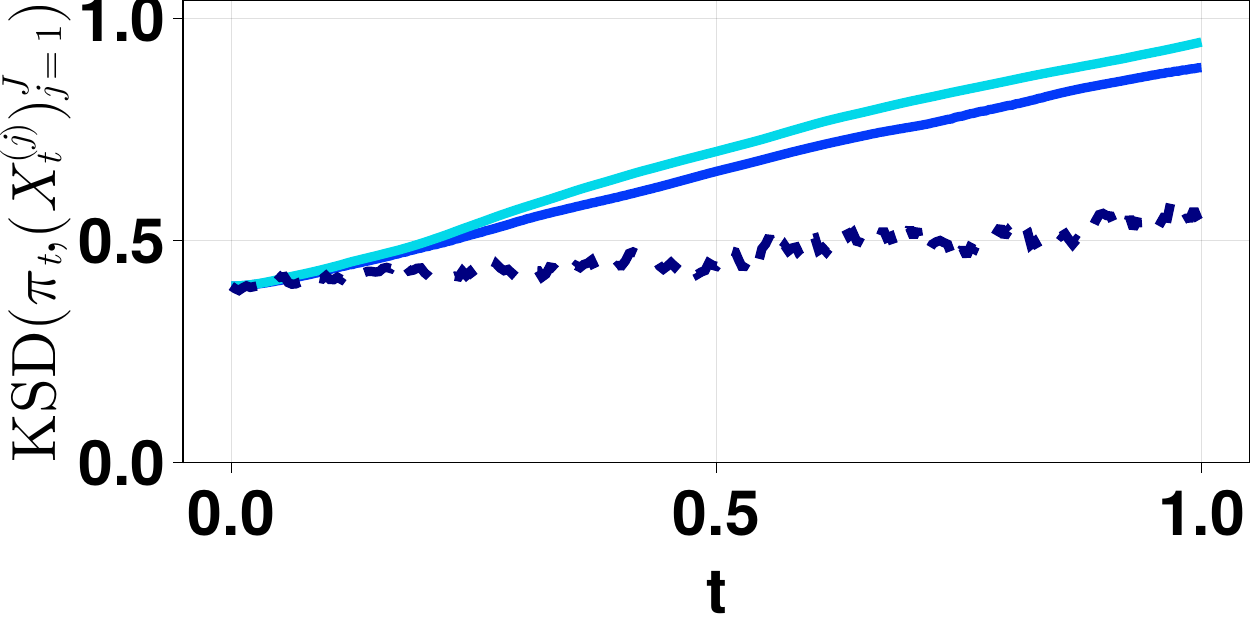}    
\end{subfigure}
\begin{subfigure}{0.19\linewidth}
\includegraphics[width=\linewidth]{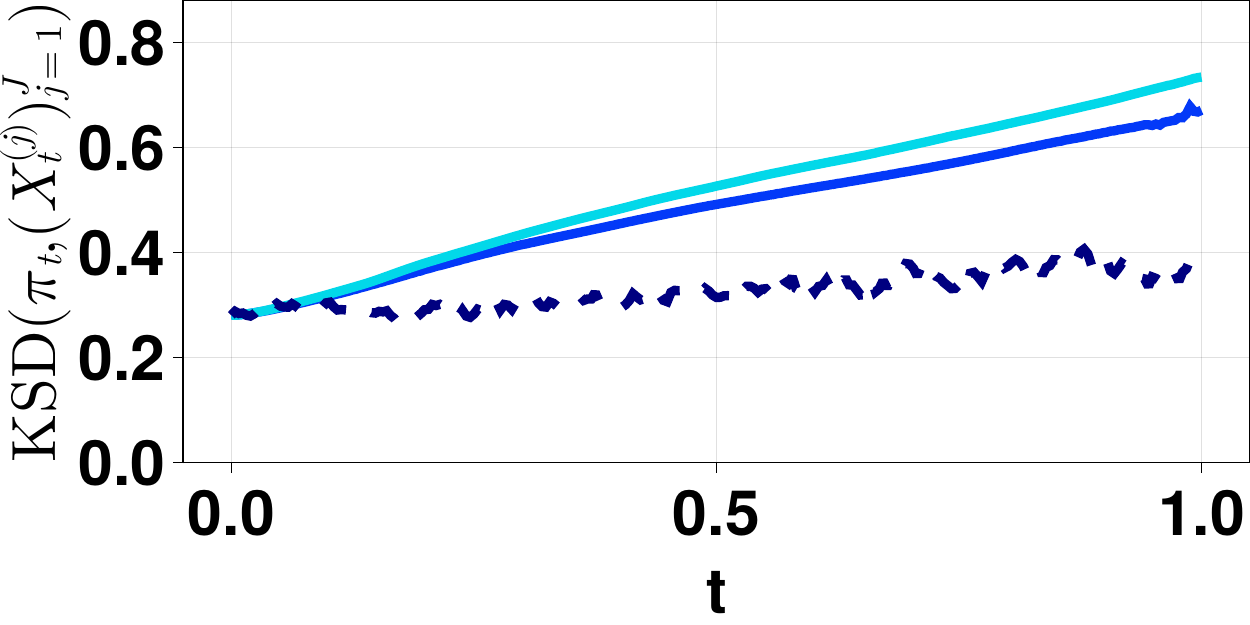}    
\end{subfigure}
\begin{subfigure}{0.19\linewidth}
\includegraphics[width=\linewidth]{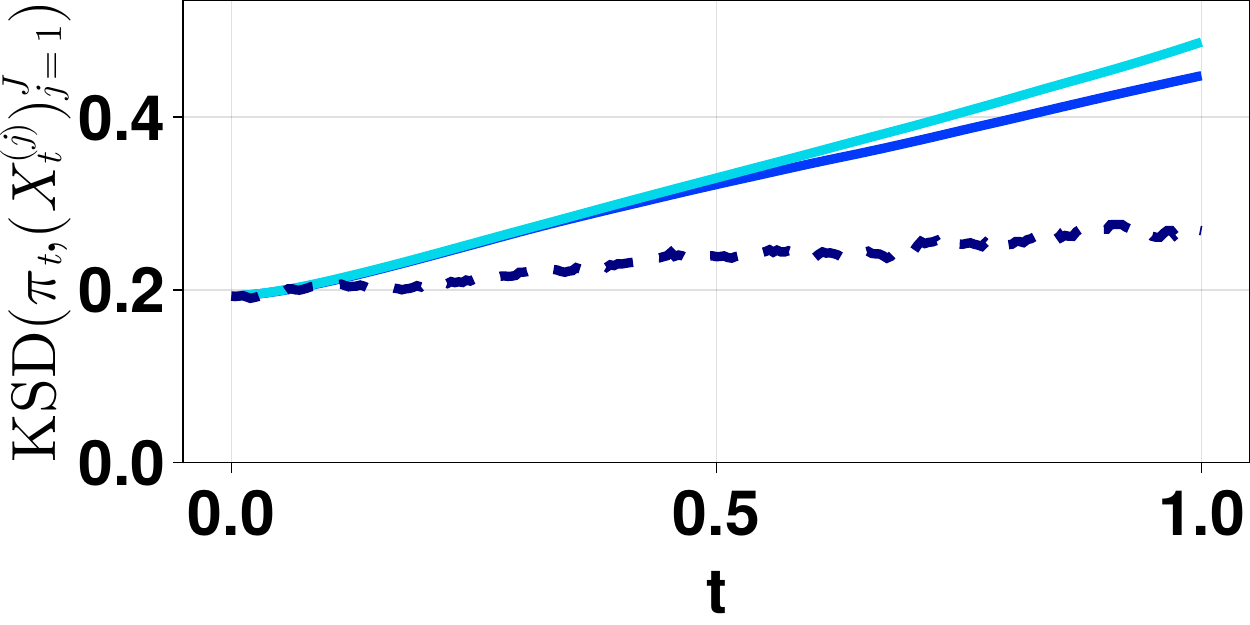}    
\end{subfigure}
\begin{subfigure}{0.19\linewidth}
\includegraphics[width=\linewidth]{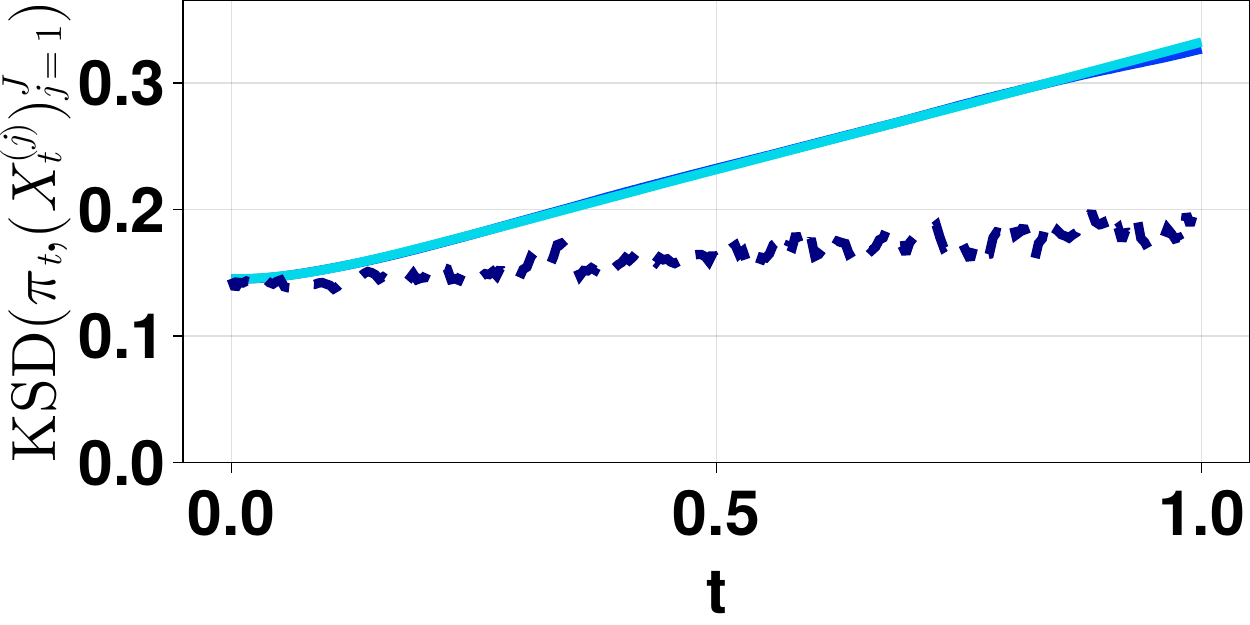}    
\end{subfigure}
\begin{subfigure}{0.19\linewidth}
\includegraphics[width=\linewidth]{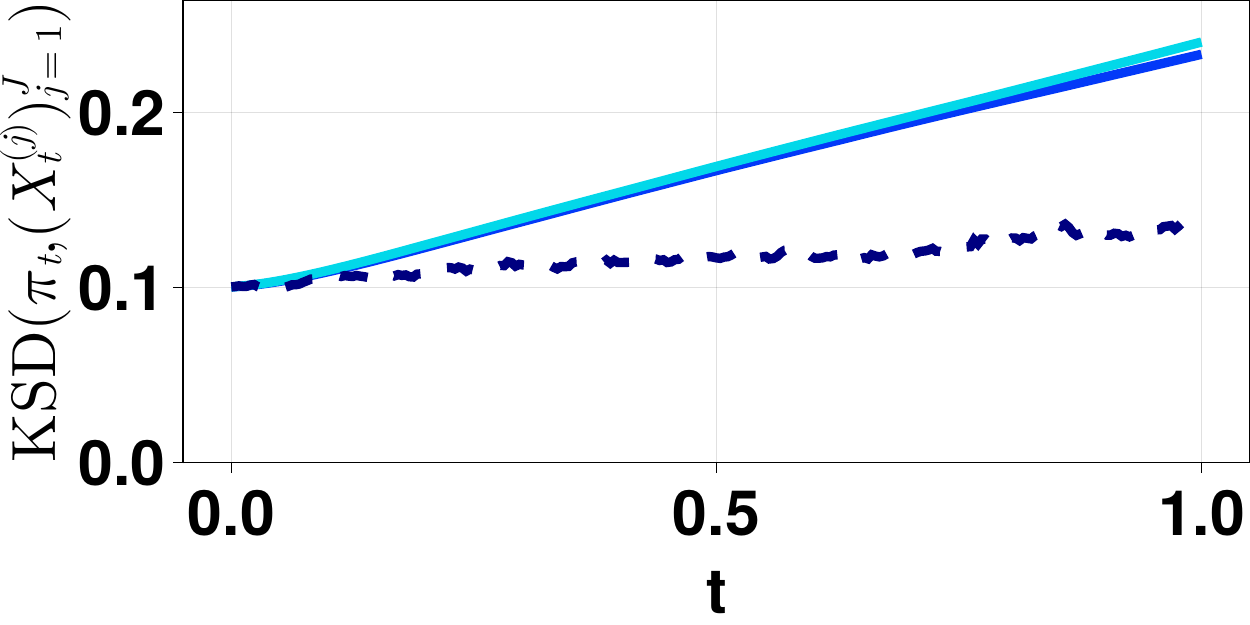}    
\end{subfigure}
\\
 Spaceships \\[0.1cm] 
\begin{subfigure}{0.19\linewidth}
\includegraphics[width=\linewidth]{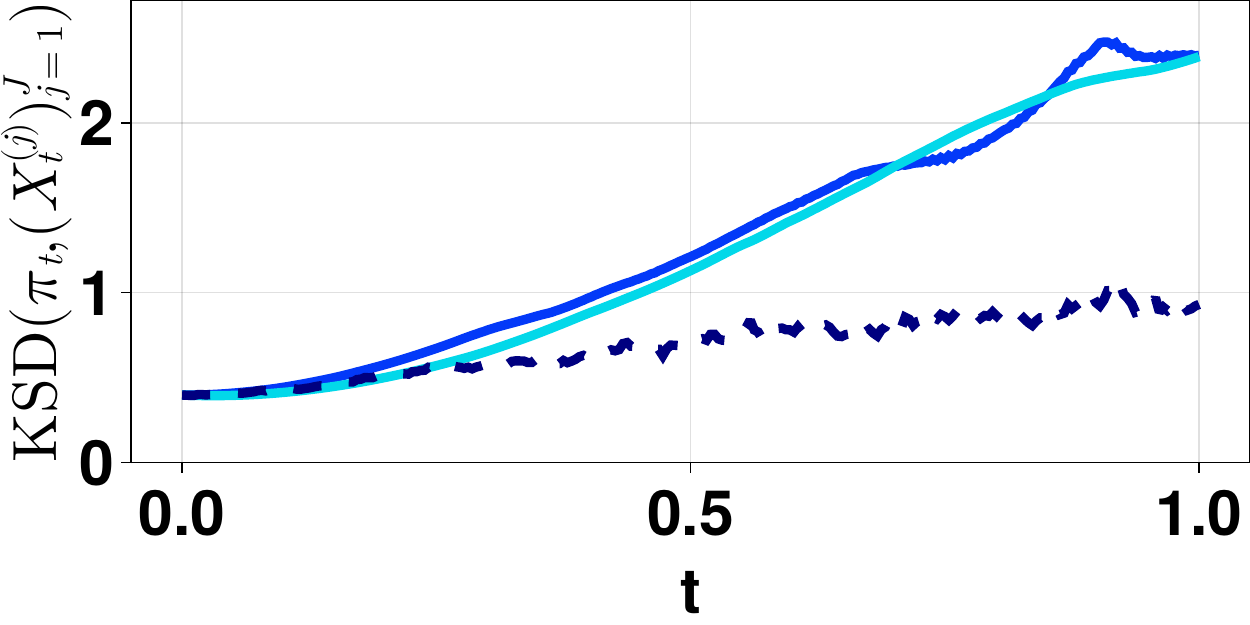}   
\subcaption*{J=25}
\end{subfigure}
\begin{subfigure}{0.19\linewidth}
\includegraphics[width=\linewidth]{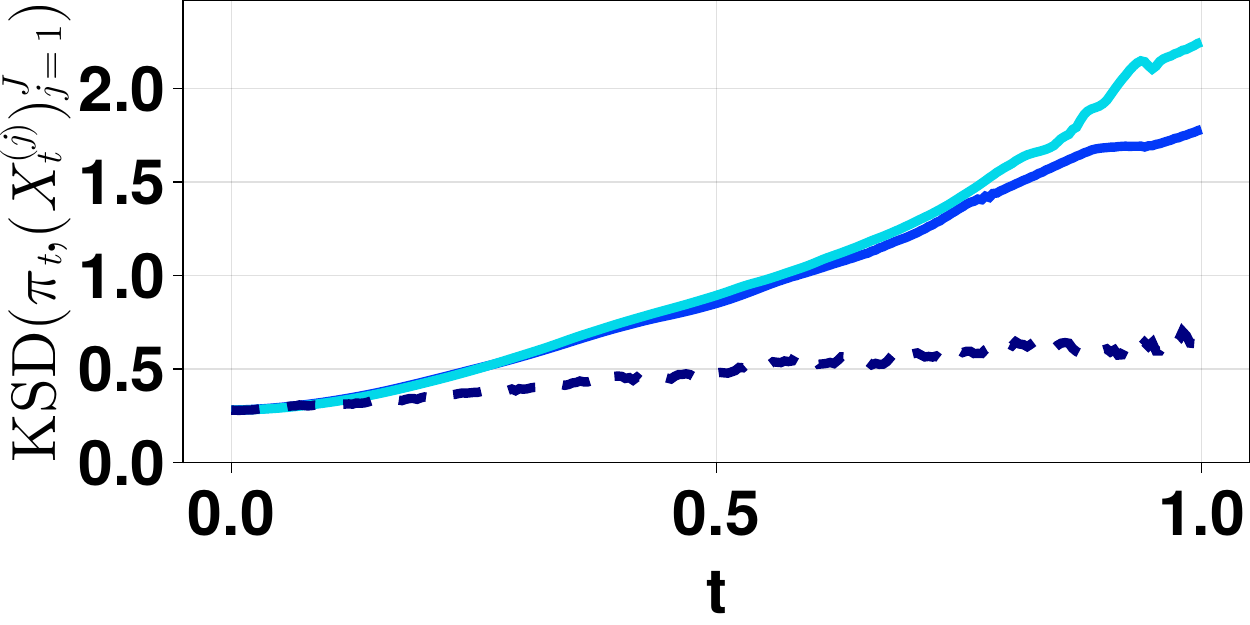}  
\subcaption*{J=50}
\end{subfigure}
\begin{subfigure}{0.19\linewidth}
\includegraphics[width=\linewidth]{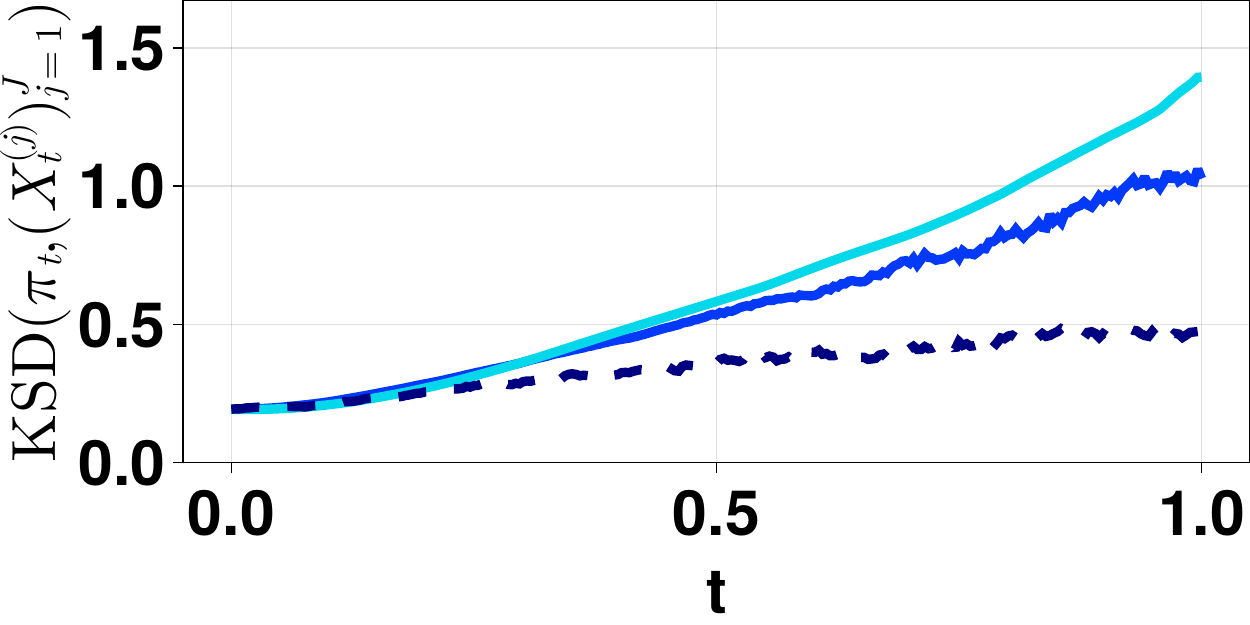}    
\subcaption*{J=100}
\end{subfigure}
\begin{subfigure}{0.19\linewidth}
\includegraphics[width=\linewidth]{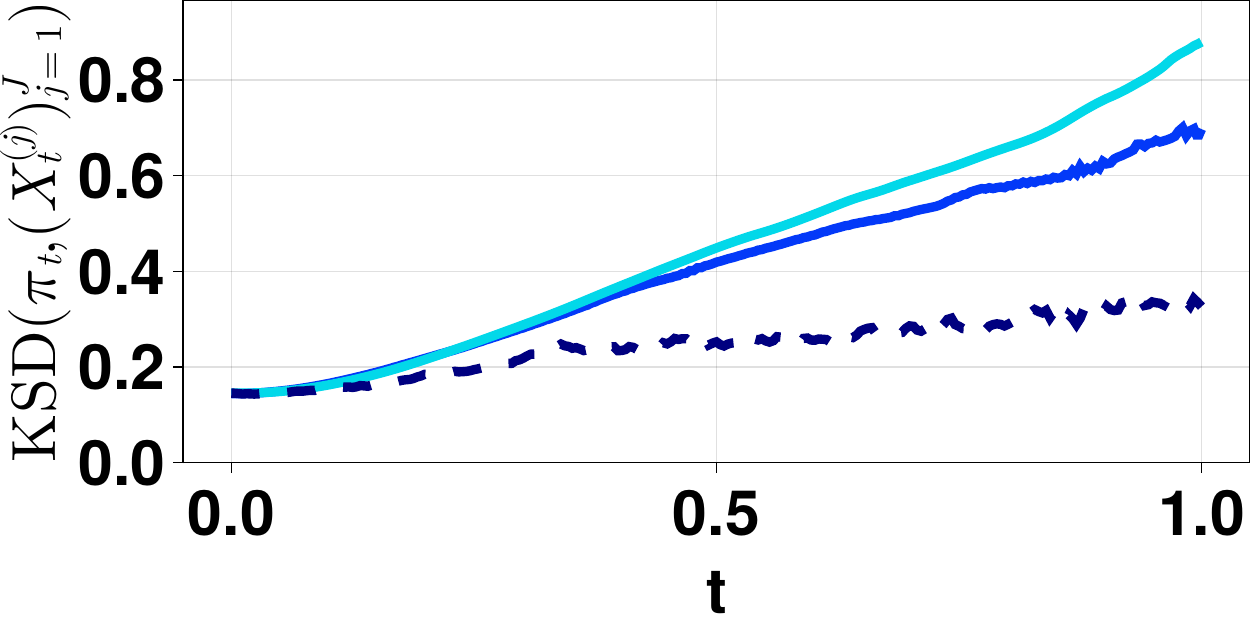}    
\subcaption*{J=200}
\end{subfigure}
\begin{subfigure}{0.19\linewidth}
\includegraphics[width=\linewidth]{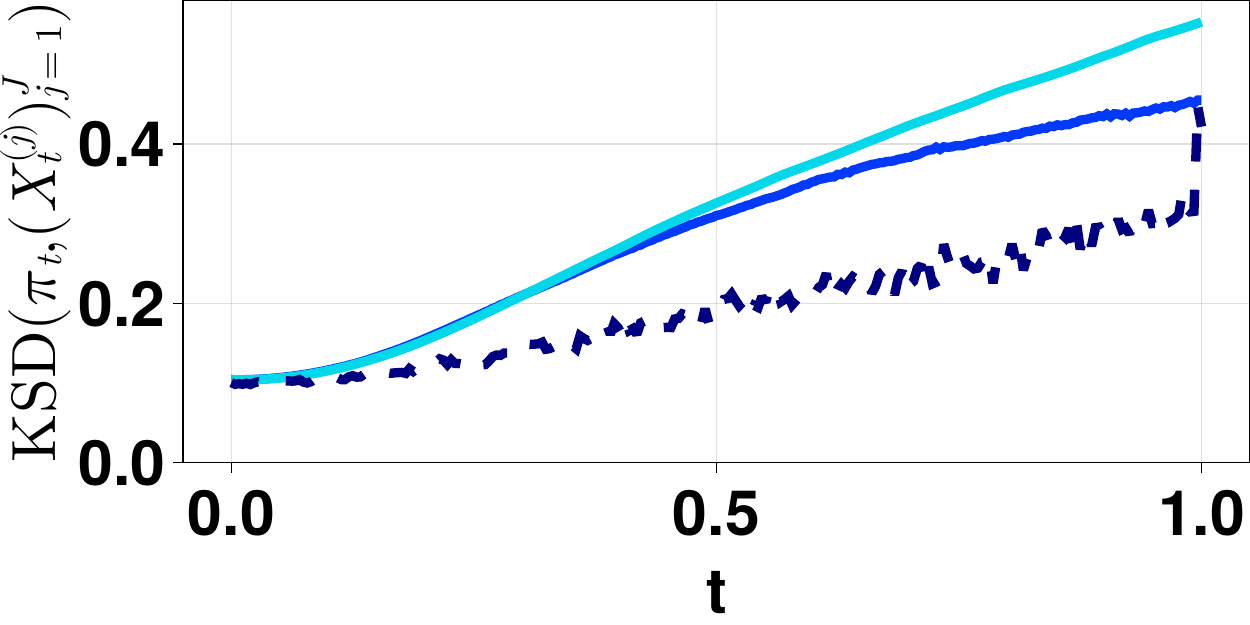}    
\subcaption*{J=400}
\end{subfigure}
\begin{subfigure}{\linewidth}
    \centering 
    \includegraphics[width=0.6\linewidth]{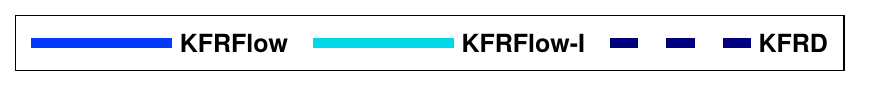}
\end{subfigure}
\caption{\textbf{Two-dimensional posteriors:} KSD between intermediate distributions $\pi_t$ and samples at time $t$ generated by KFRFlow, KFRFlow-I, and KFRD for ensemble sizes $J \in \{25, 50, 100, 200, 400\}$ and $\Delta t = 2^{-8}$. A missing line in a plot indicates that the method was unstable at that setting of $J$. }
\label{fig:KSDdiscUTlogScale_2D}
 \end{figure}

In \cref{fig:KSDdiscUTlogScale_2D} we plot the KSD between samples $\{X_t^{(j)}\}_{j=1}^J$ generated by KFRFlow, KFRFlow-I, and KFRD and the {intermediate distributions} $\pi_t \propto \pi_0^{1-t}\pi_1^t$, for $t \in [0,1]$. $\mathrm{KSD}(\pi_t, \{X_t^{(j)}\}_{j=1}^J)$ can be viewed as a sort of ``discretization error'':  even at $t = 0$ we see that the KSD between $\pi_0$ and the samples $\{X_0^{(j)}\}_{j=1}^J$, which are taken directly from $\pi_0 = \calN(0, I_d)$, is nonzero due to finite $J$.
One naturally expects $\mathrm{KSD}(\pi_t, \{X_t^{(j)}\}_{j=1}^J)$ to increase in time owing to error incurred from kernelization of the solution to \eqref{eq:poisson_u}, Monte Carlo approximation of the mean-field ODE \eqref{eq:meanfield}, and time-discretization of the finite-sample ODE \eqref{eq:IPS_ode}, and we indeed see such increases in \cref{fig:KSDdiscUTlogScale_2D}. The increases are lower for KFRD than KFRFlow and KFRFlow-I, perhaps due to the presence of gradient information in KFRD. Here the increases generally appear to be approximately linear in time. 
 
\subsubsection{Comparison to random walk Metropolis}
\label{sec:rwm_2D}
\begin{figure}[h]
\centering 
 Donut \\[0.1cm] 
\begin{subfigure}{0.19\linewidth}
\includegraphics[width=\linewidth]{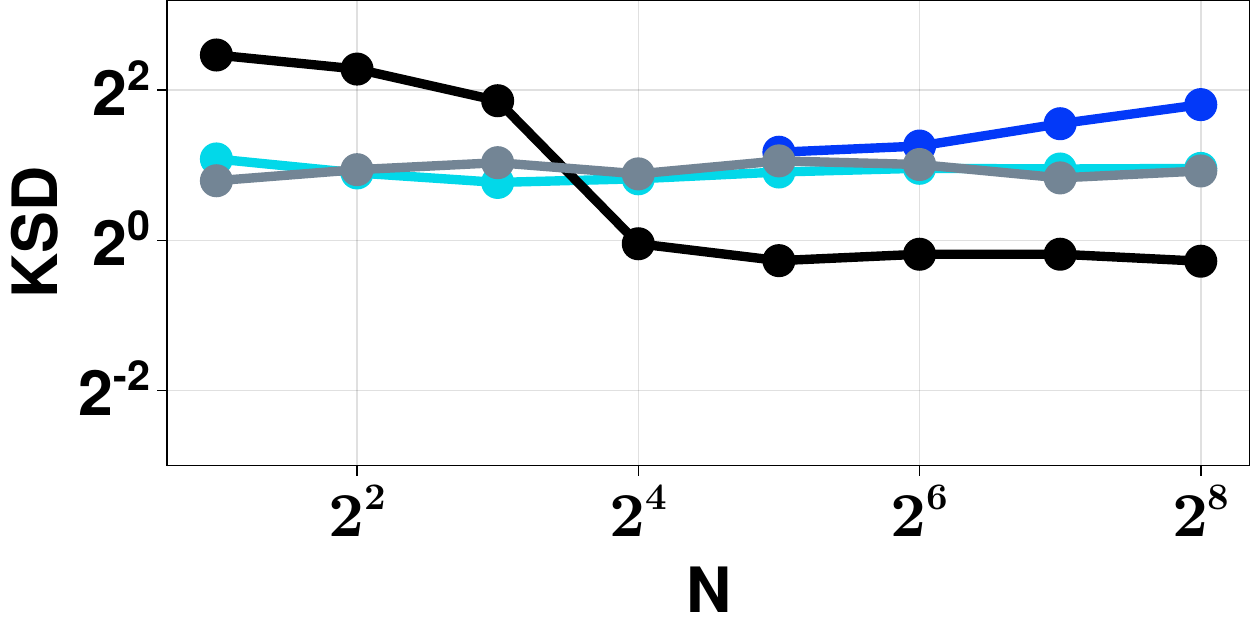}    
\end{subfigure}
\begin{subfigure}{0.19\linewidth}
\includegraphics[width=\linewidth]{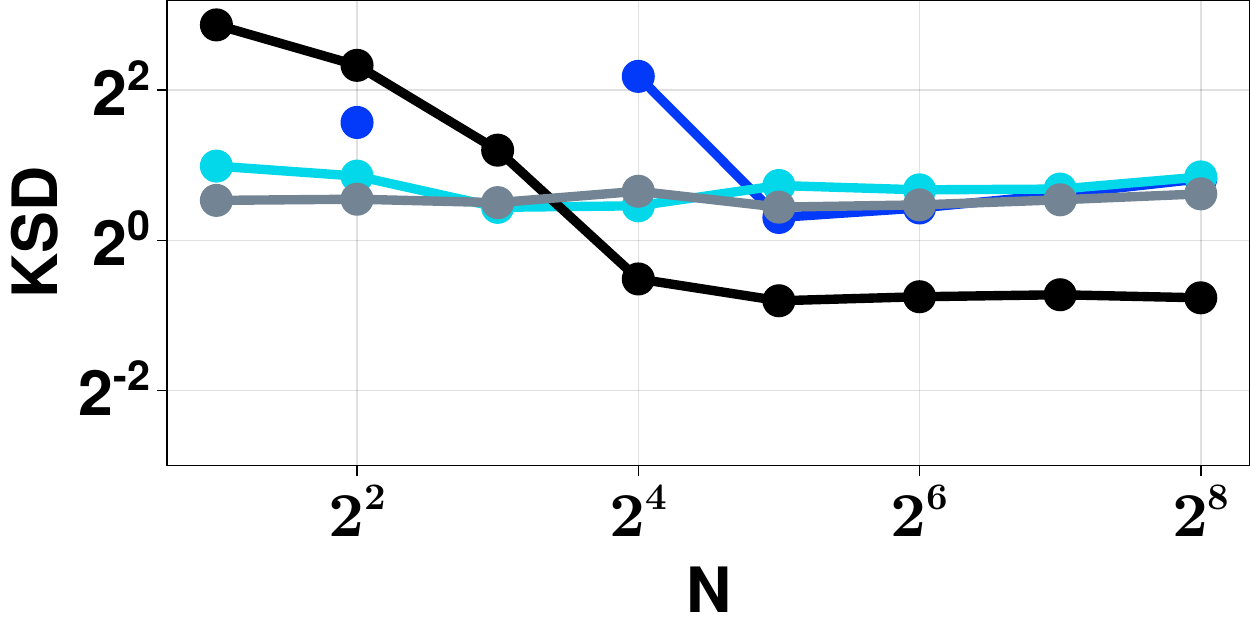}    
\end{subfigure}
\begin{subfigure}{0.19\linewidth}
\includegraphics[width=\linewidth]{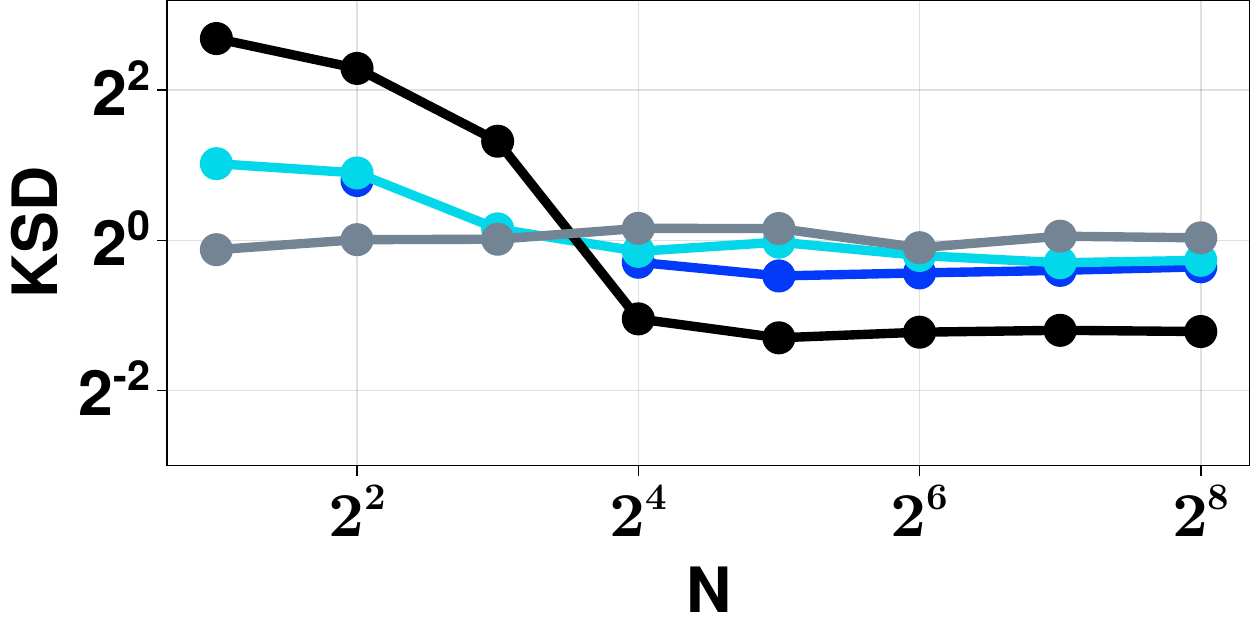}    
\end{subfigure}
\begin{subfigure}{0.19\linewidth}
\includegraphics[width=\linewidth]{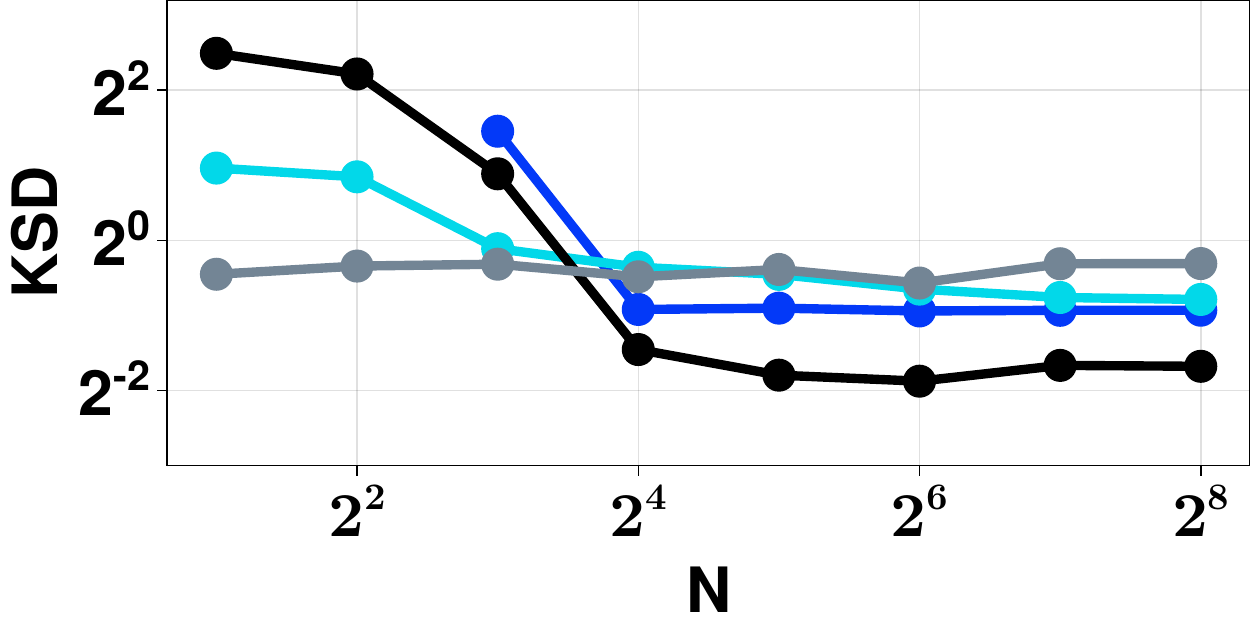}    
\end{subfigure}
\begin{subfigure}{0.19\linewidth}
\includegraphics[width=\linewidth]{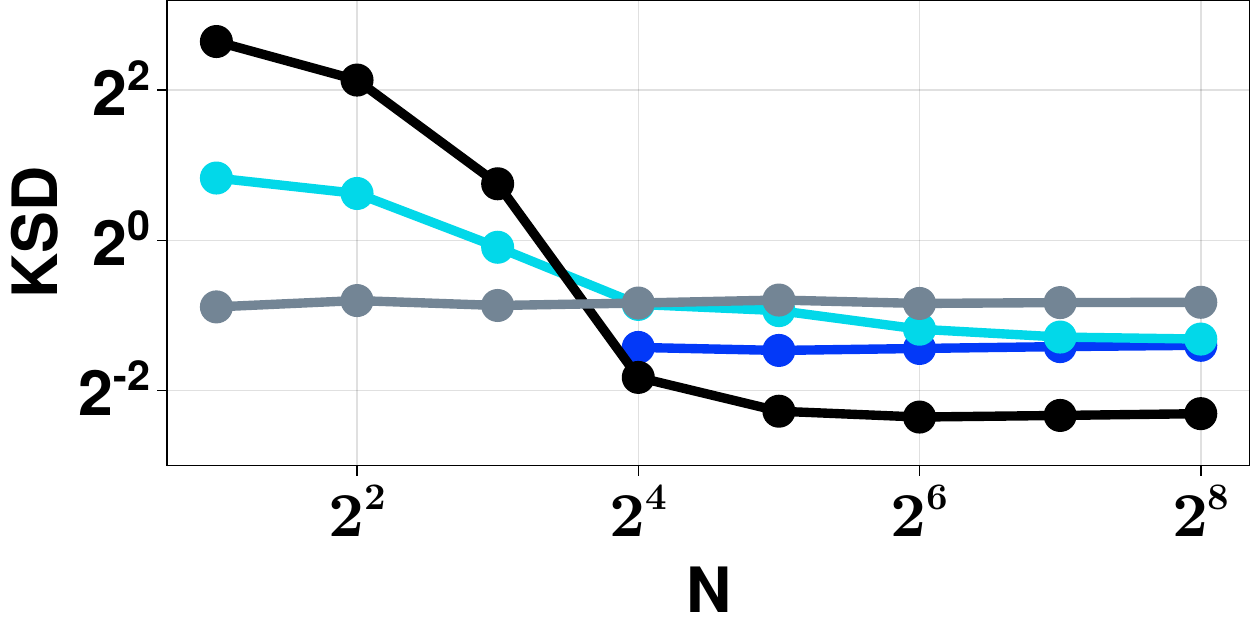}    
\end{subfigure}
\\ 
 Butterfly \\[0.1cm]
\begin{subfigure}{0.19\linewidth}
\includegraphics[width=\linewidth]{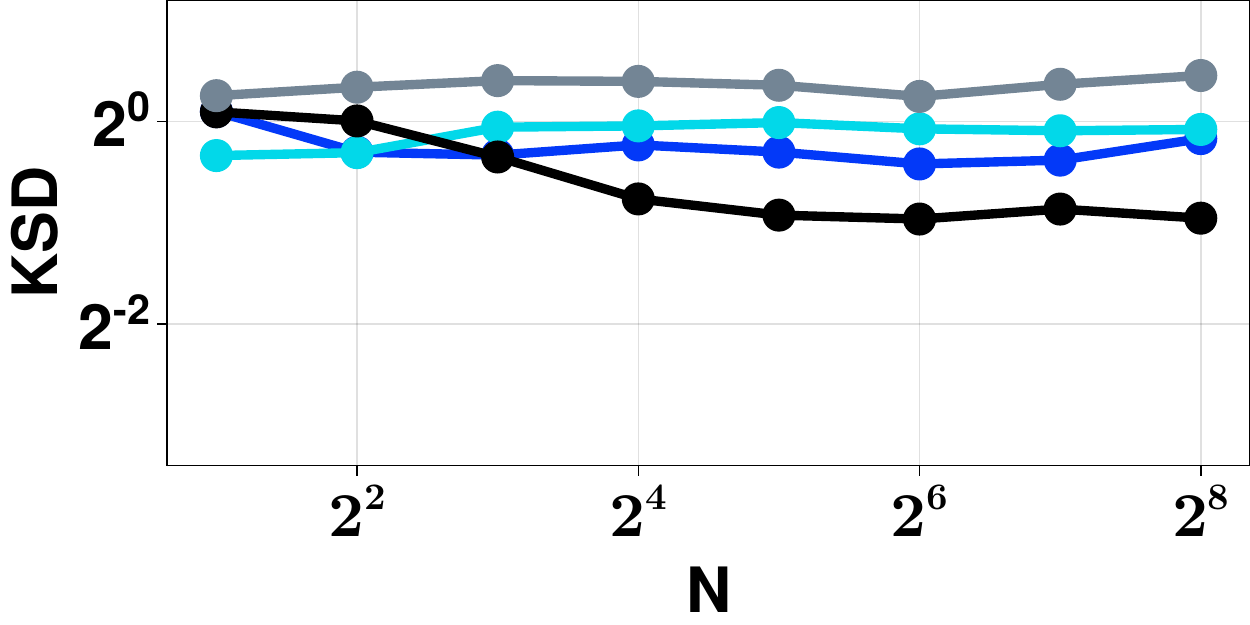}    
\end{subfigure}
\begin{subfigure}{0.19\linewidth}
\includegraphics[width=\linewidth]{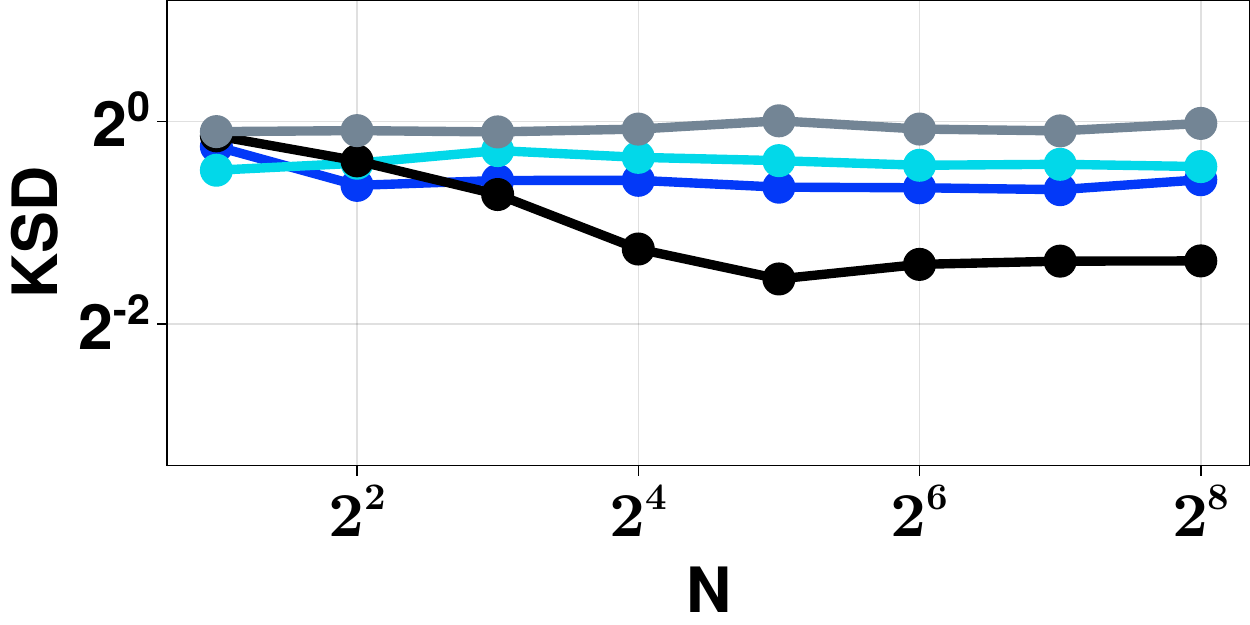}    
\end{subfigure}
\begin{subfigure}{0.19\linewidth}
\includegraphics[width=\linewidth]{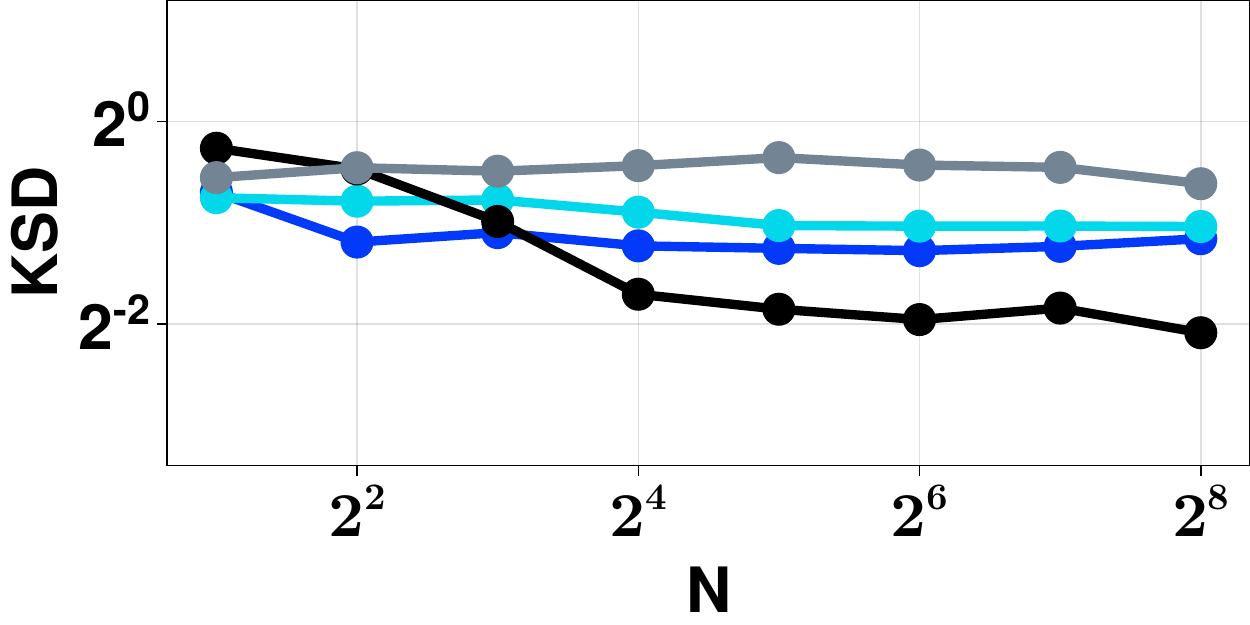}    
\end{subfigure}
\begin{subfigure}{0.19\linewidth}
\includegraphics[width=\linewidth]{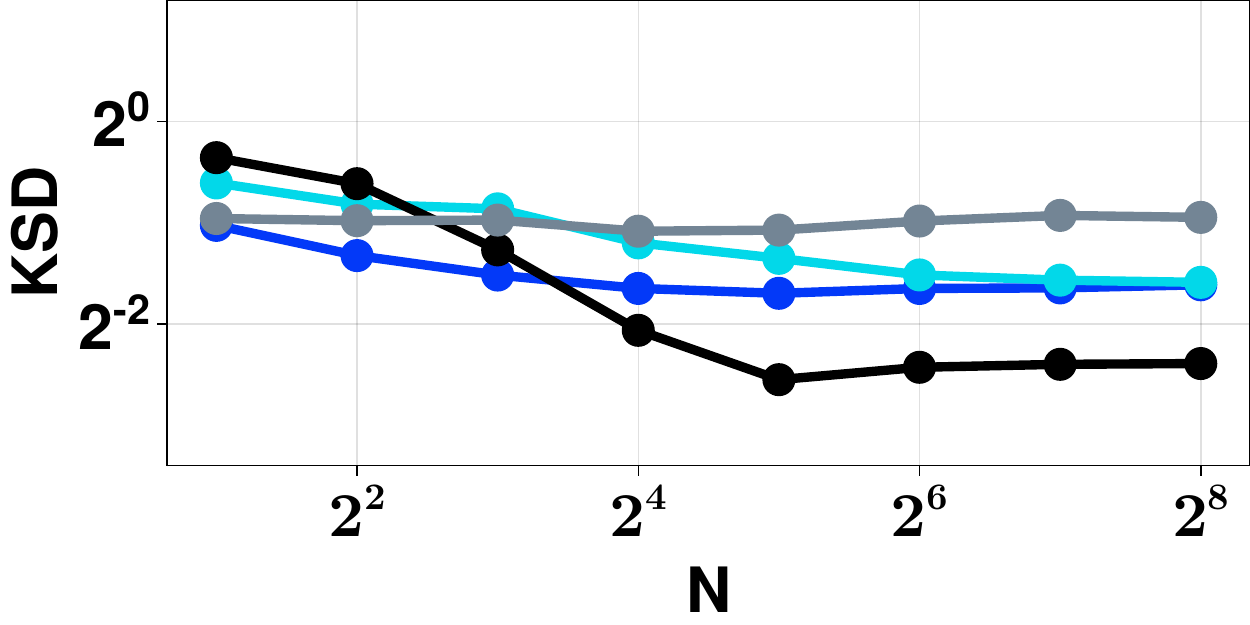}    
\end{subfigure}
\begin{subfigure}{0.19\linewidth}
\includegraphics[width=\linewidth]{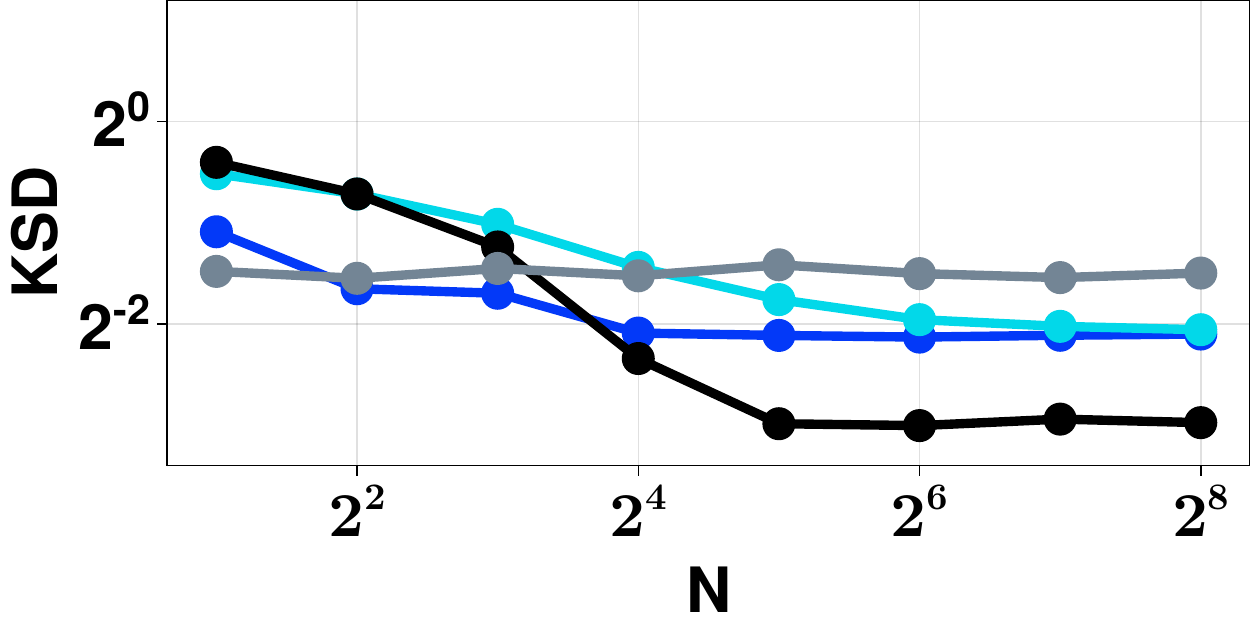}    
\end{subfigure}
\\
 Spaceships \\[0.1cm] 
\begin{subfigure}{0.19\linewidth}
\includegraphics[width=\linewidth]{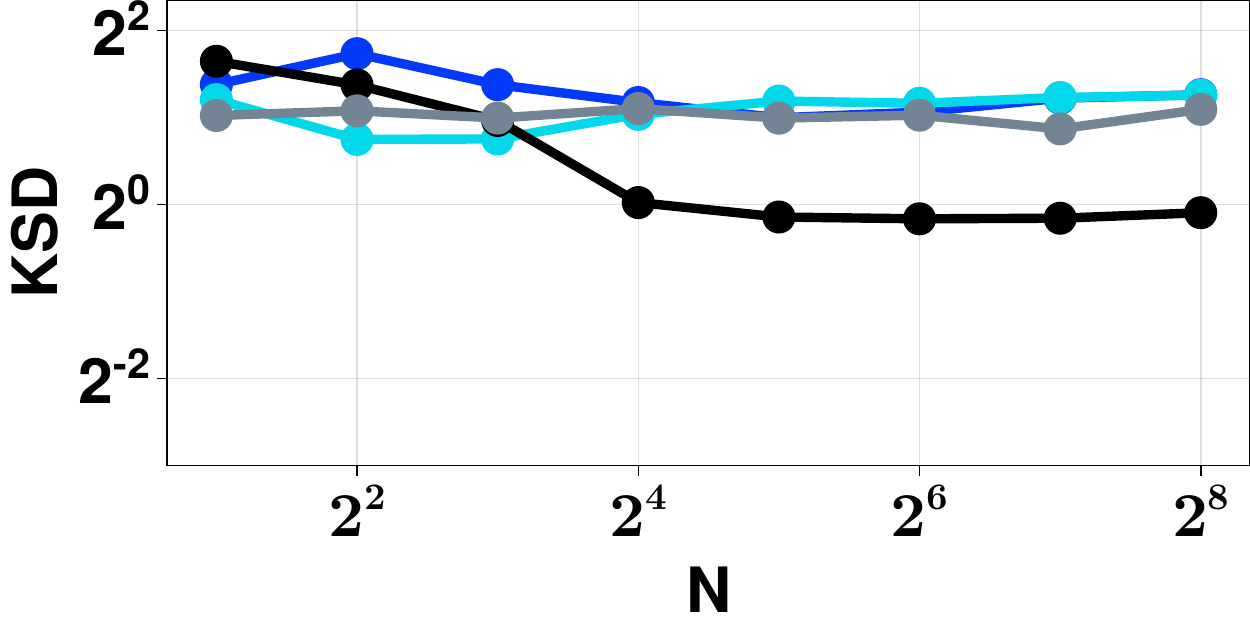}
\subcaption*{J=25}
\end{subfigure}
\begin{subfigure}{0.19\linewidth}
\includegraphics[width=\linewidth]{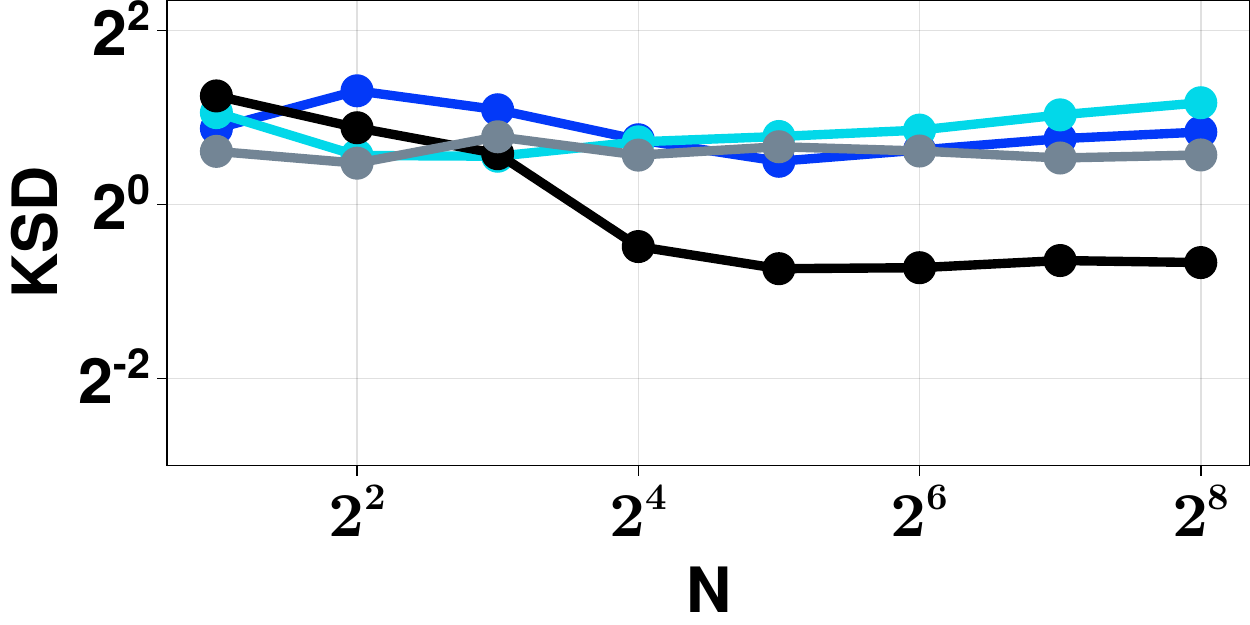}    
\subcaption*{J=50}
\end{subfigure}
\begin{subfigure}{0.19\linewidth}
\includegraphics[width=\linewidth]{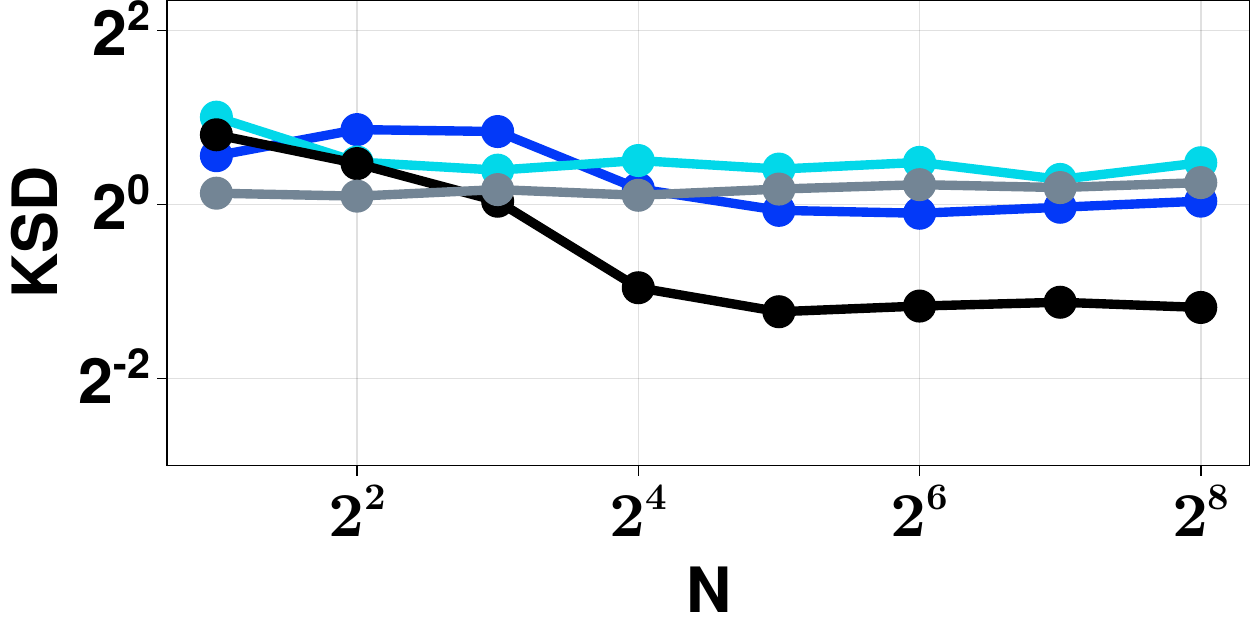}    
\subcaption*{J=100}
\end{subfigure}
\begin{subfigure}{0.19\linewidth}
\includegraphics[width=\linewidth]{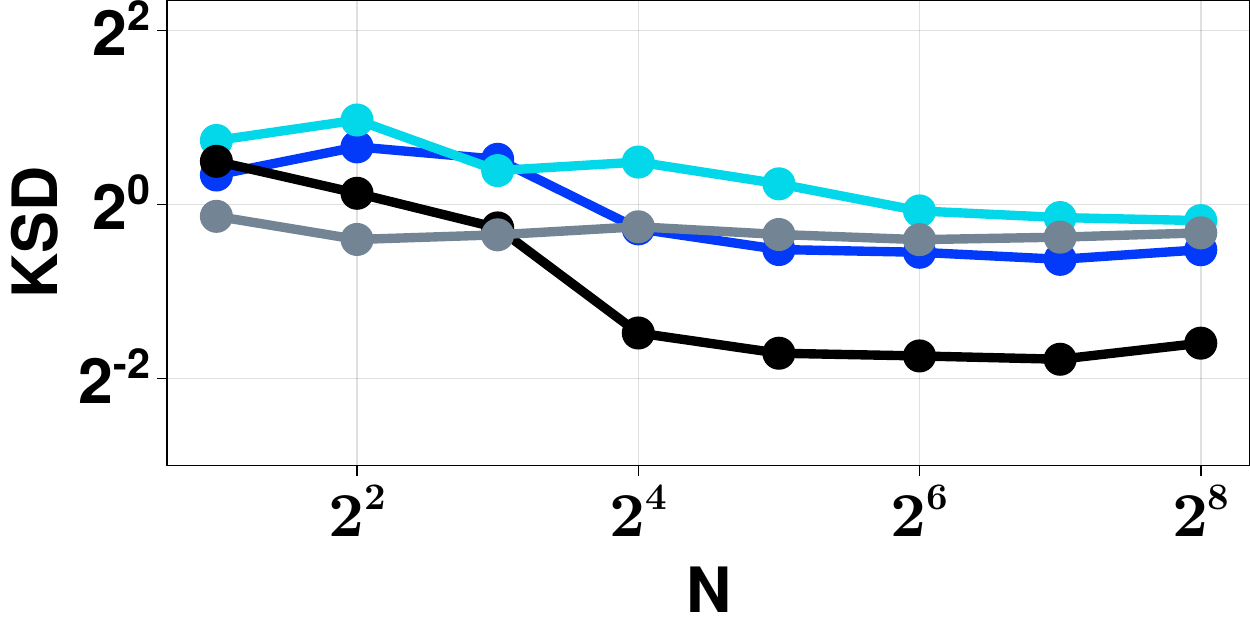}    
\subcaption*{J=200}
\end{subfigure}
\begin{subfigure}{0.19\linewidth}
\includegraphics[width=\linewidth]{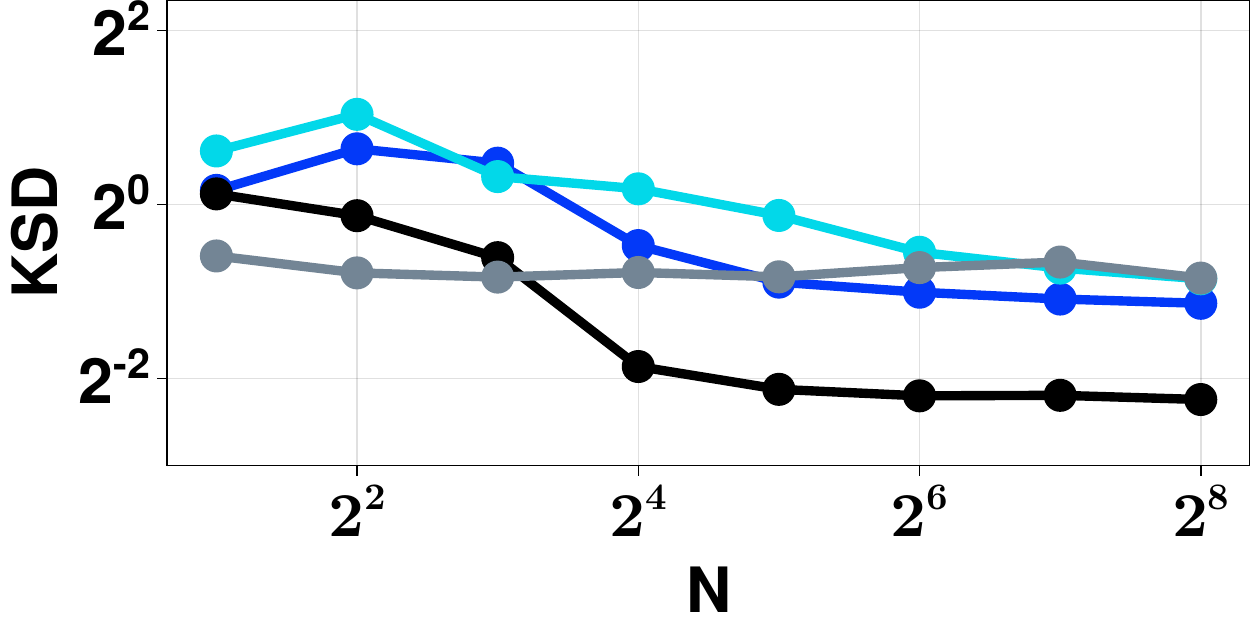}    
\subcaption*{J=400}
\end{subfigure}
\\
\begin{subfigure}{\linewidth}
\centering 
\includegraphics[width=0.5\linewidth]{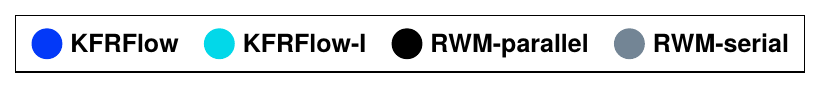}
\end{subfigure} %
\caption{\textbf{Two-dimensional posteriors:} average KSD at stopping time between $\pi_1$ and ensembles of size $J \in \{25, 50, 100, 200, 400\}$ generated by KFRFlow, KFRFlow-I, and RWM. A missing point indicates that a method was unstable at that setting of $N$.}
\label{fig:KSDvsNlogScale_rwmCompare}
 \end{figure}

In \cref{fig:KSDvsNlogScale_rwmCompare} we compare the quality of samples generated by KFRFlow and KFRFlow-I to that of samples generated by random walk Metropolis (RWM, \citet{RobertCasella2004}). We use RWM to sample the target distributions in both ``serial'' mode, in which a single long chain is generated and the last $J$ states of this chain are retained as samples from the target, and ``parallel'' mode, in which $J$ independent chains are run and the last state from each chain is taken to form a set of $J$ samples from $\pi_1$. When we compare the sampling performances of serial and parallel RWM to KFRFlow and KFRFlow-I we compare for equivalent \textit{total} numbers of steps; that is, we run $N$ steps of KFRFlow and KFRFlow-I, $N$ steps on each of the $J$ chains in parallel RWM, and and $NJ$ steps on the single chain in serial RWM. For both RWM settings we tune the variance of the isotropic Gaussian proposal distribution to attain the optimal acceptance rate of 23\% \citep{yang2020optimal}.

Though serial mode is the generally the setting of choice for RWM, owing to the fact that parallel mode inefficiently replicates transient (``burn-in'') behavior across chains, we see in \cref{fig:KSDvsNlogScale_rwmCompare} that for sufficiently large $N$, parallel RWM produces better-quality samples than serial RWM, KFRFlow, and KFRFlow-I. For these large values of $N$, parallel RWM benefits from multiple starting points, which help it collectively sample all regions of the target distributions, and from the fact that each individual chain has moved adequately by step $N$; 
these particular target distributions do not require long burn-in times. By contrast, for small $N$ we deduce from \cref{fig:KSDvsNlogScale_rwmCompare} that the $J$ parallel chains are often not adequately burnt in, as serial RWM (with its $J$-fold longer burn-in time) and KFRFlow-I generally produce better samples in this setting. For most instances of $N > 8$ steps, KFRFlow or KFRFlow-I produces samples of comparable or better quality than serial RWM, as these interacting particle algorithms are able to explore the target distributions more effectively than a single chain of RWM. For $N \leq 8$ the resulting $\Delta t$ is often too large for KFRFlow and KFRFlow-I to be integrated accurately, and serial RWM benefits from the fact that it has $(N-1)J$ burn-in steps. 

\subsubsection{Effect of $\Delta t$} In \cref{fig:KSDvsT} we investigate the impact of step-size $\Delta t$ %
  on the evolution of sample quality, as measured by KSD, with $t \in [0,1]$. 
 For each example and setting of $\Delta t \in \{2^{-1}, 2^{-2}, \dots, 2^{-8} \}$ we generate $J =300$ approximate samples of $\pi_1$ with KFRFlow and KFRFlow-I and compute KSD between the samples and $\pi_1$ at each step of the iterations. We regularize $M_t$ in the Euler discretization of KFRFlow with $\lambda$ set to $10^{-1}$, $10^{-8}$, and $10^{-11}$ for the donut, butterfly, and spaceships examples, respectively, but do not regularize $M_t$ in KFRFlow-I. The data plotted in \cref{fig:KSDvsT} are the result of averaging the values of KSD over 30 repeated trials at each setting of $\Delta t$. 

\newcommand{\dtscale}{0.9}
\begin{figure}[h]
 \centering 
 \small KFRFlow-I \\[0.1cm] 
\begin{subfigure}{0.32\linewidth}
\centering 
    \includegraphics[width=\dtscale\linewidth]{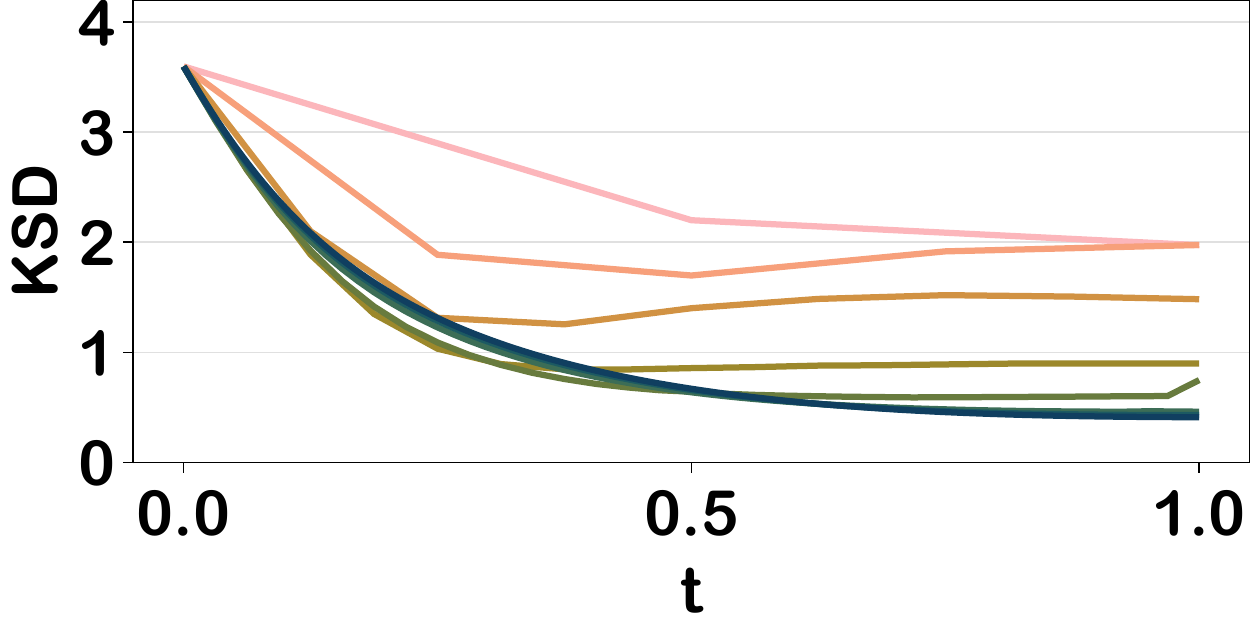}
\end{subfigure}
\begin{subfigure}{0.32\linewidth}
\centering
    \includegraphics[width=\dtscale\linewidth]{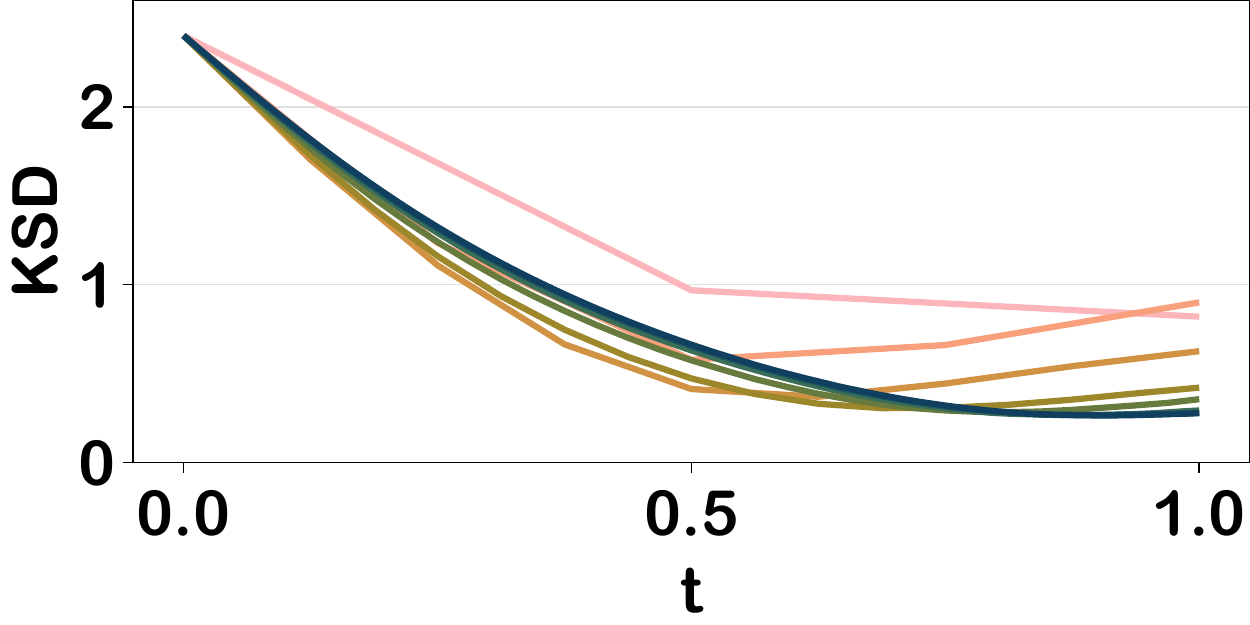}
\end{subfigure}
\begin{subfigure}{0.32\linewidth}
\centering
    \includegraphics[width=\dtscale\linewidth]{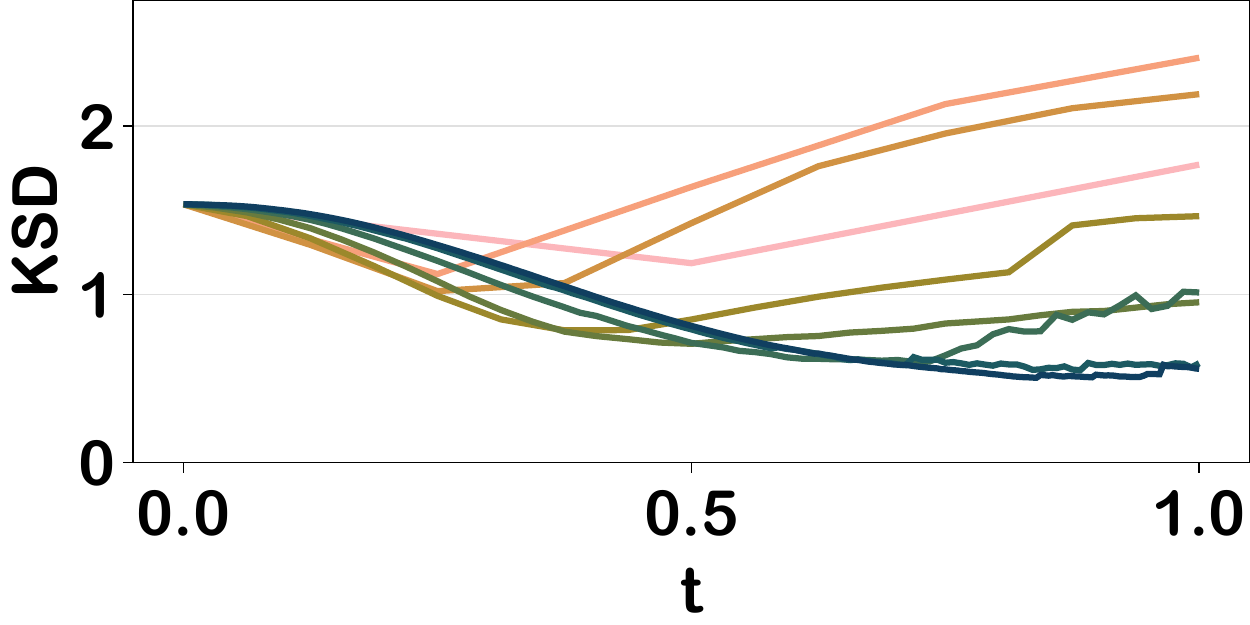}
\end{subfigure}
\\
\small KFRFlow, Euler \\[0.1cm] 
\begin{subfigure}{0.32\linewidth}
\centering
    \includegraphics[width=\dtscale\linewidth]{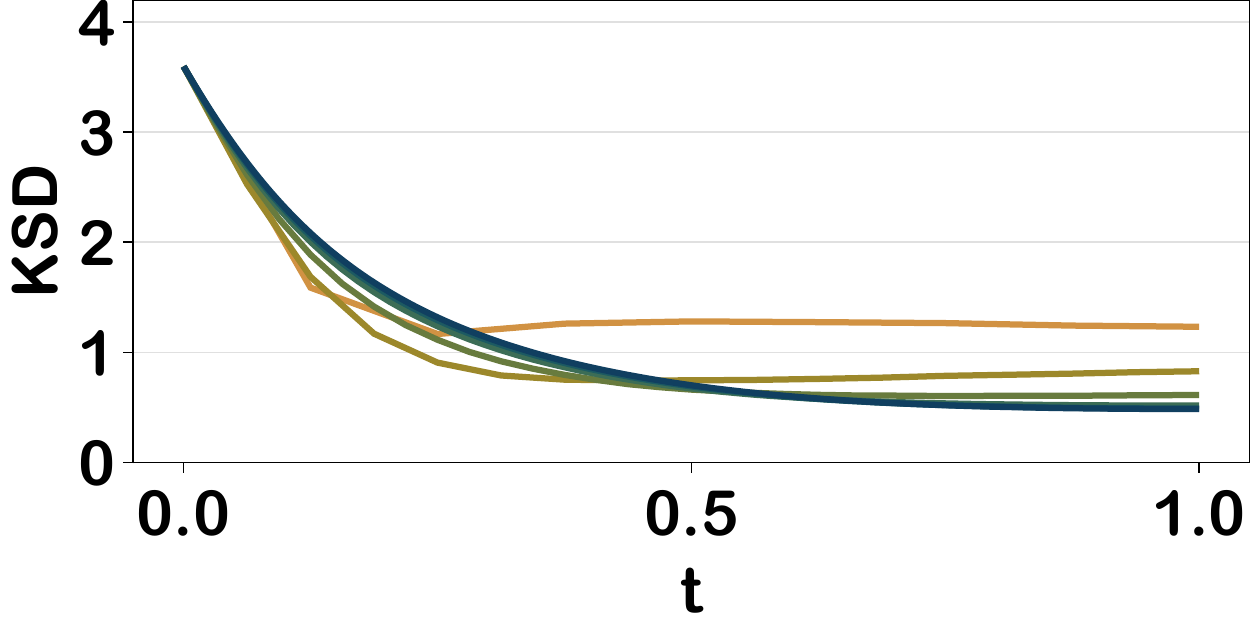}
    \subcaption*{Donut}
\end{subfigure}
\begin{subfigure}{0.32\linewidth}
\centering
    \includegraphics[width=\dtscale\linewidth]{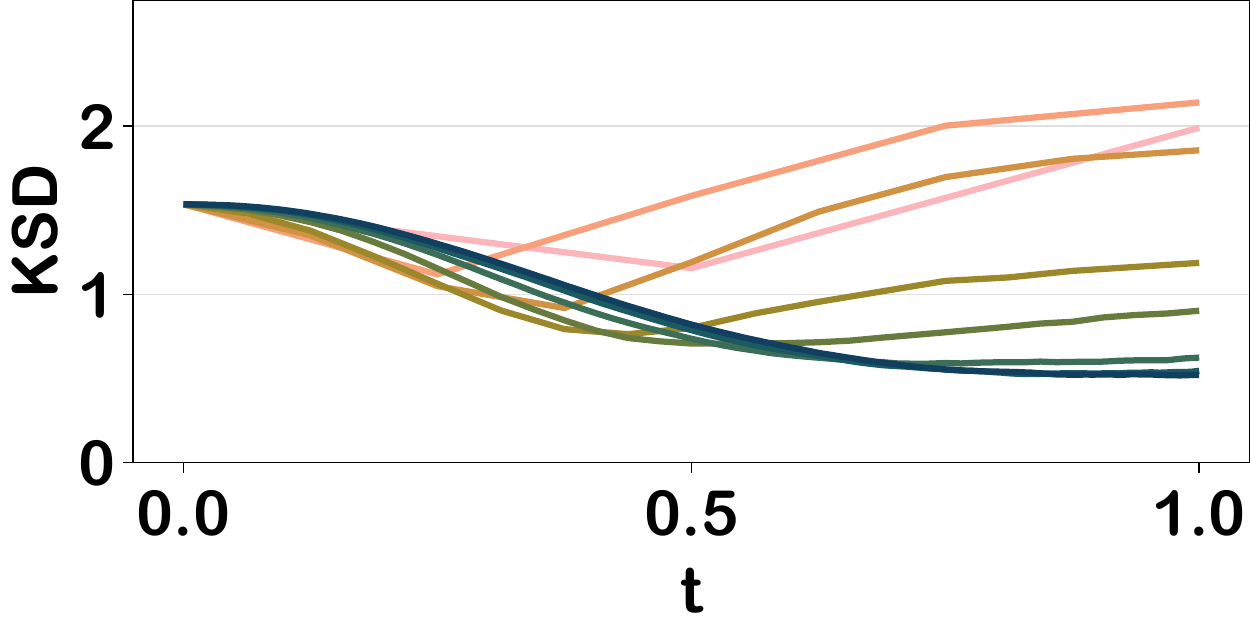}
    \subcaption*{Butterfly}
\end{subfigure}
\begin{subfigure}{0.32\linewidth}
\centering
    \includegraphics[width=\dtscale\linewidth]{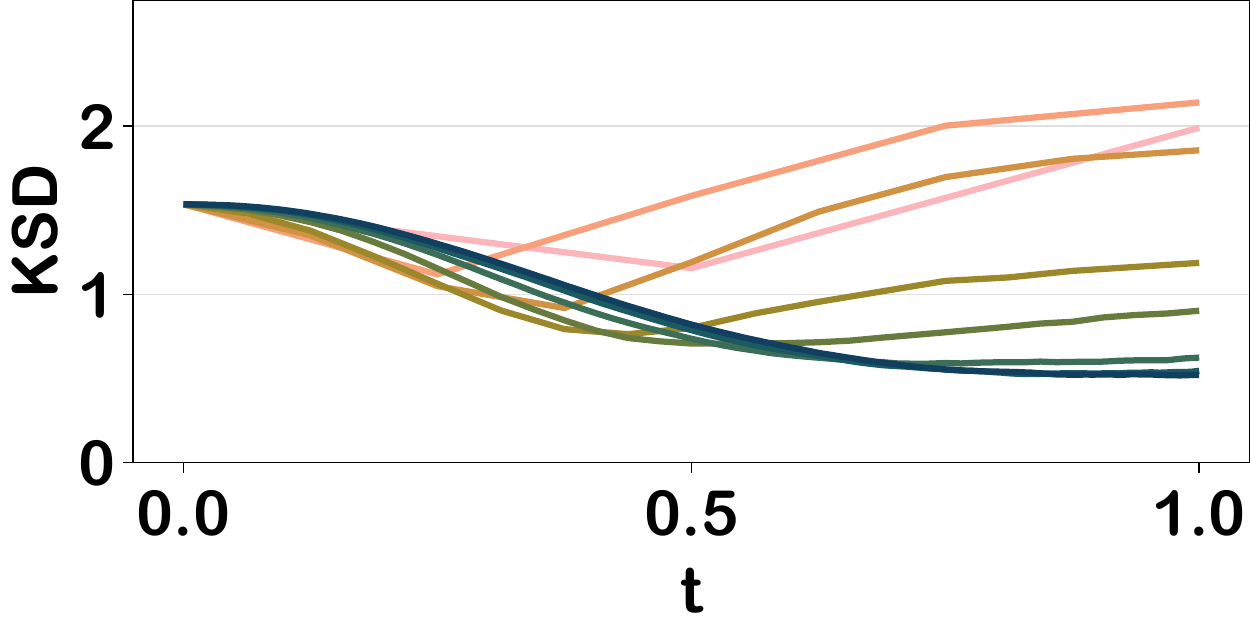}
    \subcaption*{Spaceships}
\end{subfigure}
\\
 \begin{subfigure}{\linewidth}
\centering 
     \includegraphics[width=0.5\linewidth]{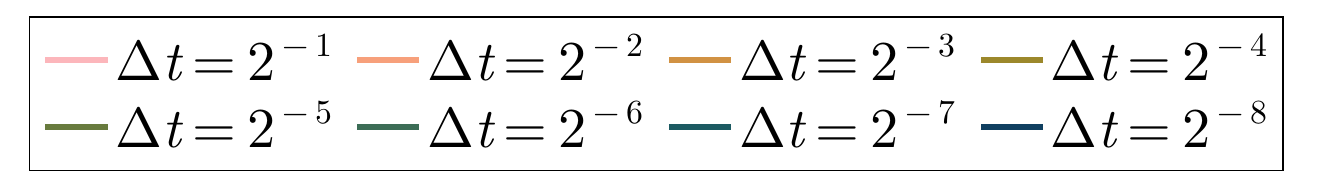}
 \end{subfigure}
\caption{\textbf{Two-dimensional posteriors:} KSD between $\pi_1$ and samples generated by KFRFlow-I (top) and an explicit Euler discretization of KFRFlow (bottom) versus $t$ for various $\Delta t$. In each example $\Delta t$ must be below a certain threshold to ensure that KSD decreases monotonically throughout the iteration. KFRFlow-I is apparently more stable at large $\Delta t$ than the Euler discretization of KFRFlow.}
\label{fig:KSDvsT}
 \end{figure}
\subsubsection{Empirical Runtime}
\label{sec:runtime2D}
\begin{figure}[h]
    \centering
    \includegraphics[width=0.85\linewidth]{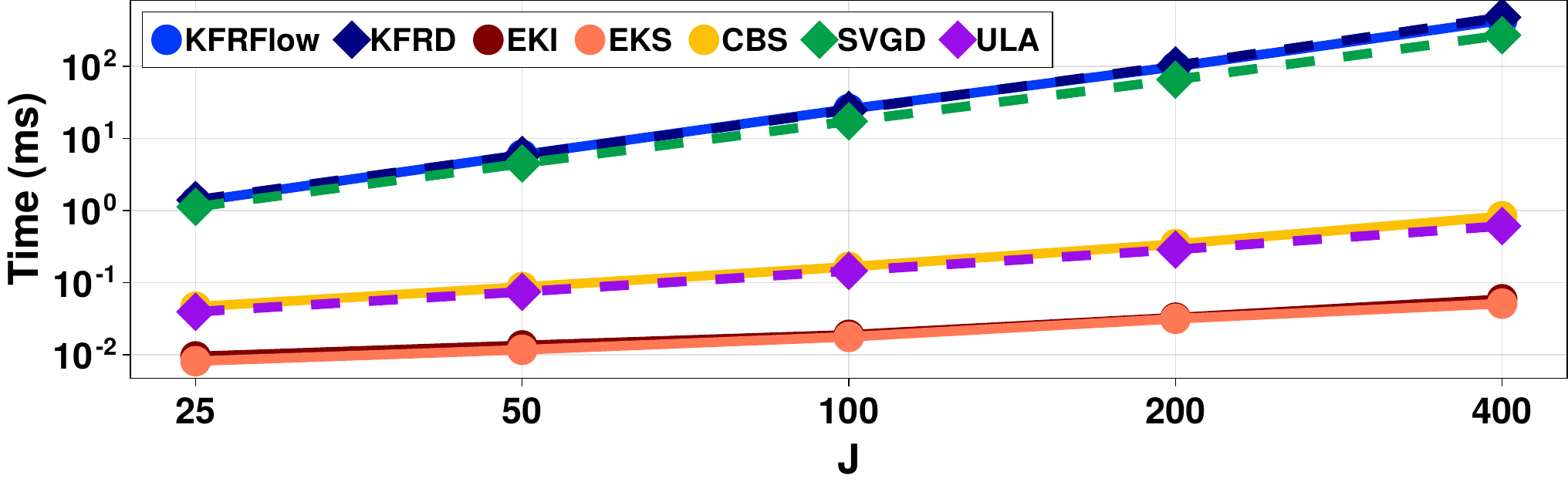}
    \caption{\textbf{Two-dimensional posteriors:} time (ms) taken to compute one ensemble update for each of KFRFlow, KFRD, EKI, EKS, CBS, SVGD, and ULA in two dimensions as a function of ensemble size $J$. }
    \label{fig:2Dtimings}
\end{figure}

In \cref{fig:2Dtimings} we display the median time taken to compute one update (i.e., transform $\{X_t^{(j)}\}_{j=1}^J$ into $\{X_{t + \Delta t}^{(j)}\}_{j=1}^J$) in each of KFRFlow, KFRD, EKI, EKS, CBS, SVGD, and ULA as a function of ensemble size $J$ in the two-dimensional setting. %
Benchmarks were performed in Julia using \texttt{BenchmarkTools.jl} \citep{benchmarkTools} on a 2020 MacBook Air with Apple M1 processor. 

We see in \cref{fig:funnelTimings} that the runtimes increase polynomially with ensemble size $J$, as one would expect based on, e.g., the complexity of KFRFlow (\cref{sec:computationalCost}), and that the methods can be organized into three clusters based on cost: KFRFlow, KFRD, and SVGD are most expensive, EKI and EKS are cheapest, and CBS and ULA fall somewhere in between. As $\nabla \log \pi_1$ is cheap to evaluate in these examples, the data in \cref{fig:2Dtimings} do not capture extra costs that may be incurred by gradient-based methods (SVGD, KFRD, and ULA) in the setting where $\nabla \log \pi_1$ is expensive.

 \subsection{Higher-Dimensional Funnels}
 \label{app:funnel}
 \subsubsection{Experimental Setup}
For dimension $d \in \{5, 10, 15, 20\}$ we compare the performance of KFRFlow-I, KFRD, CBS, SVGD, and ULA in sampling from ``funnel'' distributions of the form 
\[
\pi_1(\bfx) = \calN(x_1; 0, 9)\calN(\bfx_{2:d}; \mathbf{0}, \exp(x_1) \mathbf{I}),
\]
i.e., $X_1$ is distributed normally with mean zero and variance nine, and $(X_2, \dots, X_d)$ are multivariate normal with mean zero and covariance matrix $\exp(X_1)\mathbf{I}$. 
This family of distributions appears in \citet{neal2003slice} and is commonly used as a benchmark for sampling algorithms, e.g., \citet{arbel2021annealed, zhangDiffusionGenerativeFlow2023, xu2023mixflows}. 

\begin{figure}[h] 
\centering 
    \begin{subfigure}{0.49\linewidth}
    \centering 
   KFRFlow-I \\[0.1cm]
        \includegraphics[width=0.9\linewidth]{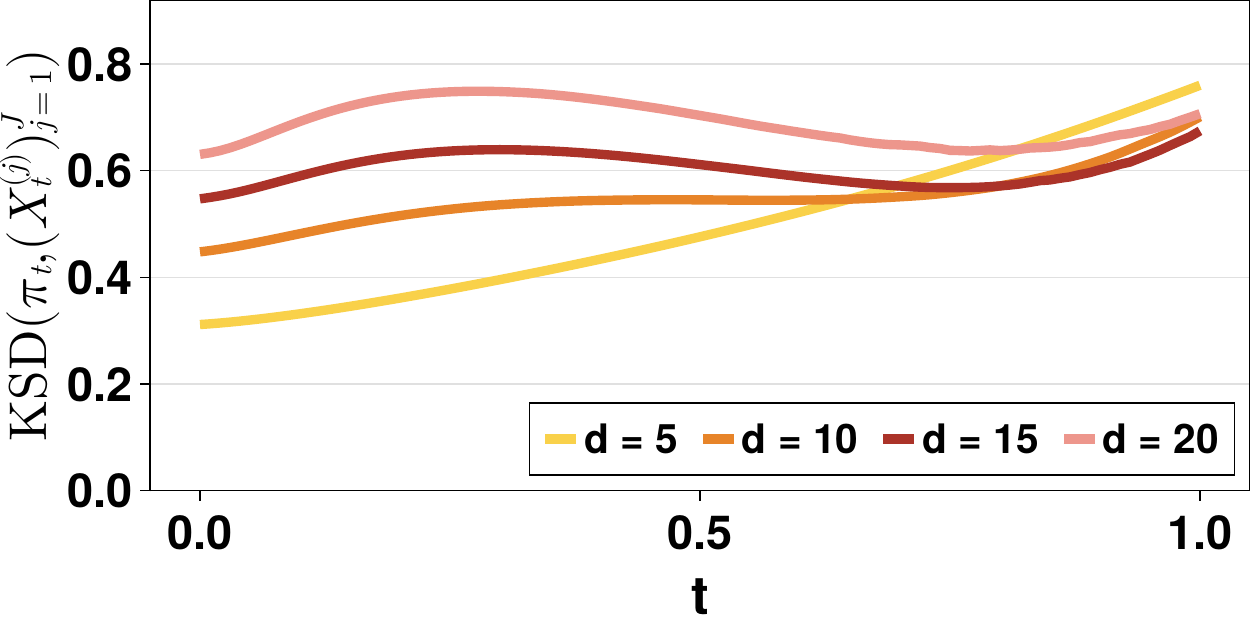}
    \end{subfigure}
    \begin{subfigure}{0.49\linewidth}
    \centering 
  KFRD \\[0.1cm] 
        \includegraphics[width=0.9\linewidth]{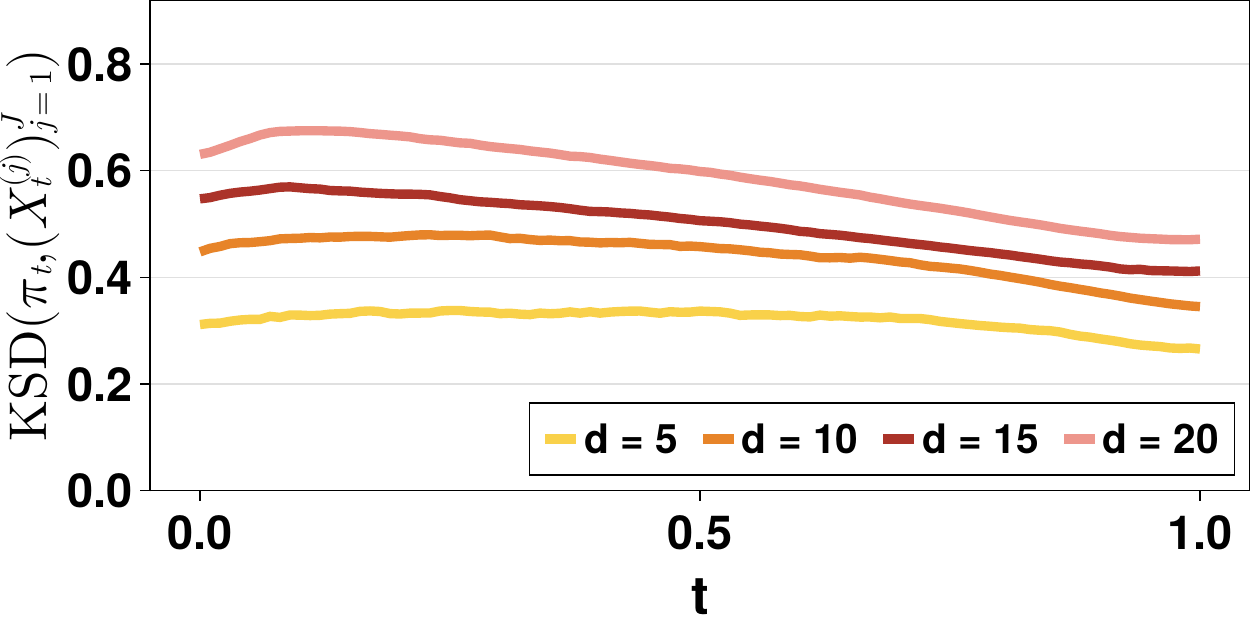}
    \end{subfigure}
    \caption{\textbf{Funnels:} KSD between samples at time $t$ and the intermediate distribution $\pi_t$, for $t \in [0,1]$; KFRFlow-I (left) and KFRD (right). This ``discretization error'' somewhat surprisingly seems to decrease with $t$ for KFRD but is noticeably non-monotonic for KFRFlow-I. 
    }
    \label{fig:ksdDisc_funnel}
\end{figure}

For each setting of $d$ we apply these algorithms 
to generate $J = 100$ samples from $\pi_1$.  For KFRFlow-I and KFRD we set $\Delta t = 0.01$, corresponding to $N = 100$ steps for the infinite-time algorithms CBS, SVGD, and ULA. As in \cref{sec:2Ddists} we optimize the hyperparameters for KFRFlow-I, KFRD, CBS, SVGD, and ULA via coarse direct search to minimize KSD between the final samples and $\pi_1$. The resulting hyperparameter values are shown in \cref{tab:funnelParams}. The data in \cref{fig:ksdVdim_funnel_serial,fig:ksdEvolution_funnel,fig:ksdDisc_funnel,fig:ksdEvolution_funnel,fig:RWMcompare_funnel} are produced by averaging the results of 30 independent trials.
    \begin{table}[h]
    \centering 
\begin{tabular}{ccccc}
			 & $d = 5$ & $d = 10$ & $d = 15$ & $d = 20$ \\ 
			\toprule
   $\lambda$ (KFRFlow-I) & 0.01 & 0.001 & 0.001 & 0.001 \\ 
   \midrule 
   $\epsilon$ (KFRD) & 5 & 5 & 5 & 2.5 \\ 
   $\lambda$ (KFRD) & 0.001 & 0.01 & 0.1 & 0.1 \\
   \midrule %
		$T$ (CBS) & 25 & 12.5 & 25 & 25\\ 
			$\beta$ (CBS) & 0.125 & 0.5 & 0.25 & 0.25\\ 
			\midrule %
   $T$ (SVGD) & 100 & 100 & 100 & 100 \\ 
   \midrule %
   $T$ (ULA) & 12.5 & 12.5 & 25 & 12.5
		\end{tabular}
      \caption{\textbf{Funnels:} Selected hyperparameters for each algorithm and target dimension $d$.}
  \label{tab:funnelParams}
    \end{table}

\subsubsection{KSD between Samples and Intermediate Distributions}
In \cref{fig:ksdDisc_funnel} we plot the KSD between samples $\{X_t^{(j)}\}_{j=1}^J$ generated by KFRFlow and KFRD and the {intermediate distributions} $\pi_t \propto \pi_0^{1-t}\pi_1^t$, $t \in [0,1]$. This quantity can be viewed as a sort of ``discretization error,'' for even at $t = 0$ we see that the KSD between $\pi_0$ and the samples $\{X_0^{(j)}\}_{j=1}^J$, which are sampled directly from $\pi_0 = \calN(0, I_d)$, is nonzero due to the finiteness of $J$.

As mentioned in \cref{sec:discError2D}, one naturally expects $\mathrm{KSD}(\pi_t, \{X_t^{(j)}\}_{j=1}^J)$ to increase in time due to accumulation of error, but interestingly we see in \cref{fig:ksdDisc_funnel} that $\mathrm{KSD}(\pi_t, \{X_t^{(j)}\}_{j=1}^J)$ is not generally monotone increasing in time. This quantity, in fact, is mostly decreasing in time for KFRD and tends to follow an undulating pattern, which becomes more evident with increasing $d$, for KFRFlow-I. Understanding these phenomena and their relationship to choice of IPS (KFRFlow-I vs KFRD) and dimension $d$ is an interesting area for future work.

\subsubsection{Comparison to Random Walk Metropolis}
\begin{figure}[h]
    \centering
    \includegraphics[width=0.85\linewidth]{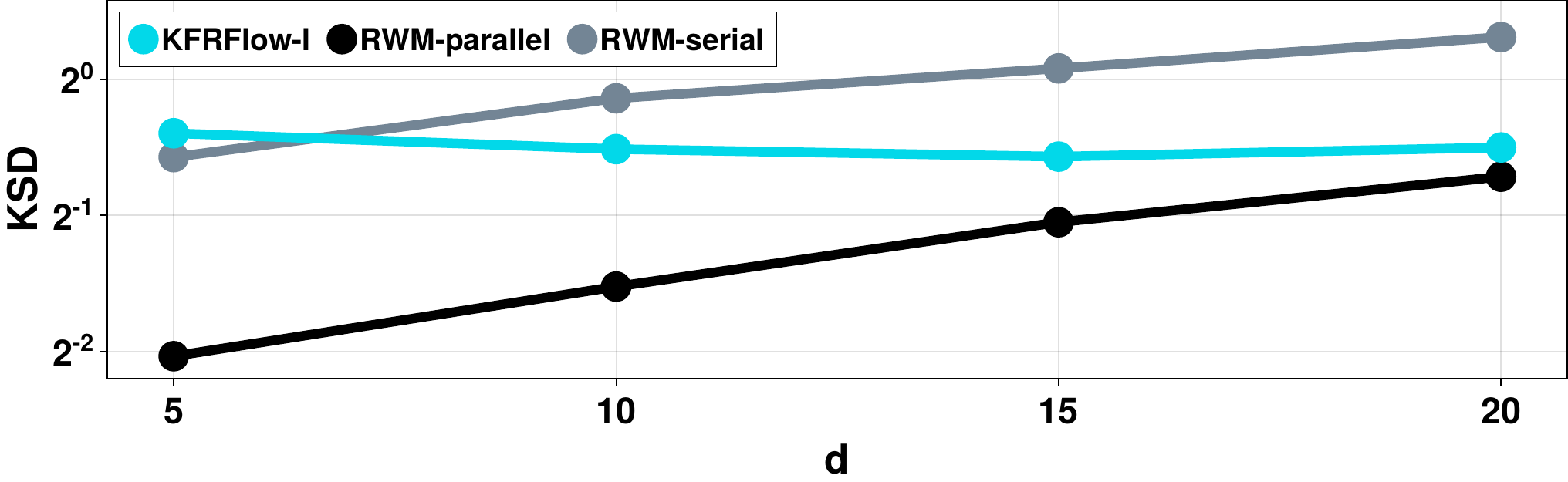}
    \caption{\textbf{Funnels:} average KSD at stopping time between $\pi_1$ and samples generated by KFRFlow-I and RWM for $d \in \{5, 10, 15, 20\}$}
    \label{fig:RWMcompare_funnel}
\end{figure}

In \cref{fig:RWMcompare_funnel} we compare the quality of samples generated by KFRFlow-I to that of samples generated by random walk Metropolis (RWM, \citet{RobertCasella2004}). As in \cref{sec:rwm_2D}, we apply RWM in both serial mode and parallel mode and make comparisons among algorithms for equivalent total numbers of steps. For both RWM settings we tune the variance of the isotropic Gaussian proposal distribution to attain the optimal acceptance rate of 23\% \citep{yang2020optimal}.

Similarly to the behavior in \cref{fig:KSDvsNlogScale_rwmCompare}, we see in \cref{fig:RWMcompare_funnel} that for all settings of $d$ parallel RWM produces better samples, as measured with KSD, than serial RWM. We posit that the burn-in time for RWM is not very long for these funnel distributions, and thus parallel RWM is benefiting from multiple initializations and from the fact that each individual chain is adequately burnt in after $N = 100$ steps. Interestingly, we see in \cref{fig:RWMcompare_funnel} that the quality of samples produced by RWM in both modes degrades with dimension but that the same is not true of KFRFlow-I: for $d = 5$ KFRFlow-I produces samples that are slightly worse than serial RWM, but for $d > 5$ the samples from KFRFlow-I are better than those from serial RWM, with the quality of samples produced by KFRFlow-I becoming comparable to that of samples from parallel RWM by $d = 20$.

\subsubsection{Empirical Runtime}
\begin{figure}[h]
    \centering
    \includegraphics[width=0.85\linewidth]{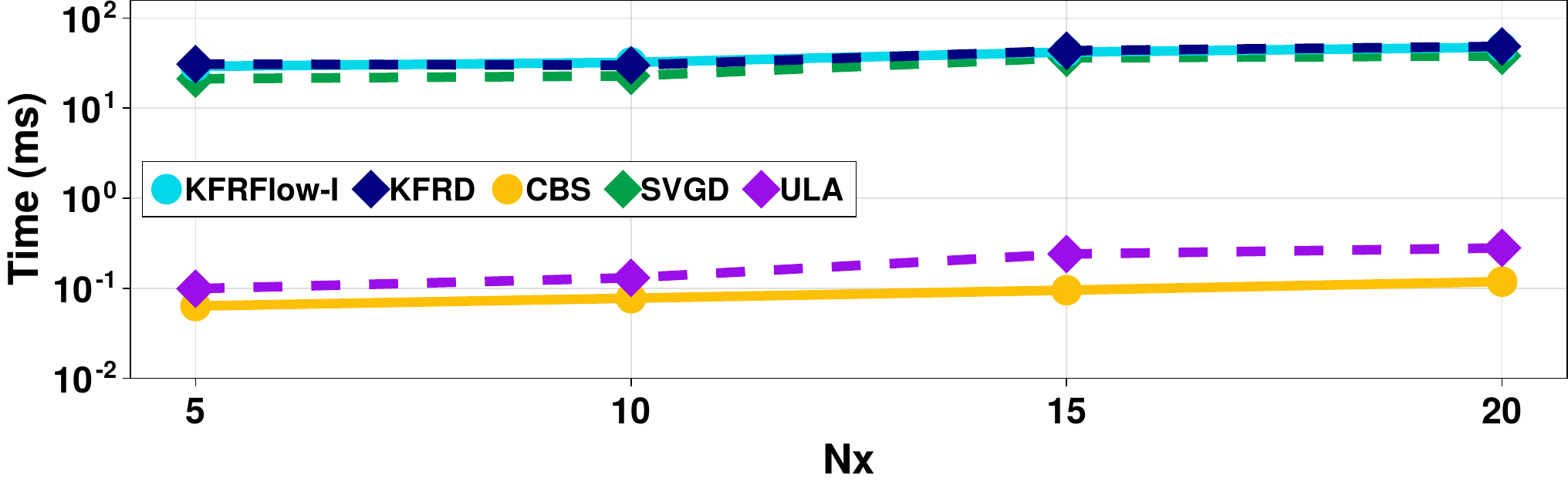}
    \caption{\textbf{Funnels:} median time (ms) taken to compute one ensemble update in each of KFRFlow-I, KFRD, CBS, SVGD, and ULA for the in the funnel example as a function of dimension $d$. Gradient-free methods are plotted with solid lines and circles, while gradient-based methods are plotted with dashed lines and diamonds.}
    \label{fig:funnelTimings}
\end{figure}

In \cref{fig:funnelTimings} we display the median time taken to compute one update (i.e., transform $\{X_t^{(j)}\}_{j=1}^J$ into $\{X_{t + \Delta t}^{(j)}\}_{j=1}^J$) in each of KFRFlow-I, KFRD, CBS, SVGD, and ULA for the settings in the funnel example. %
Benchmarks were performed in Julia using \texttt{BenchmarkTools.jl} \citep{benchmarkTools} on a 2020 MacBook Air with Apple M1 processor. 

We see in \cref{fig:funnelTimings} that the runtimes do not demonstrate distinct dependence on dimension $d$ and that the runtimes of KFRFlow-I, KFRD, and SVGD are comparable, with those of CBS and ULA being significantly lower. Given that the sampling performance of ULA, SVGD, and KFRD were essentially the same in this example, cost considerations suggest that ULA is the best choice of gradient-based method here, but for gradient-free samplers the situation is more nuanced: CBS is cheaper than KFRFlow-I, but KFRFlow-I produces better samples. 

\end{document}